\documentclass[a4paper, english, master]{diploma}

\usepackage[a-1b]{pdfx}
\usepackage[autostyle=true]{csquotes} 
\usepackage[backend=biber, style=iso-numeric, alldates=iso]{biblatex} 
\usepackage{dcolumn} 
\usepackage{caption}    
\usepackage{subcaption} 
\usepackage[cpp]{diplomalst}
\usepackage{amsmath}
\usepackage{amssymb}
\usepackage{tikz}
\usepackage{placeins}
\usetikzlibrary{
    shapes.geometric, 
    arrows.meta,      
    positioning,      
    fit,              
    calc,              
    shadings,      
    fadings,       
    angles,        
    quotes         
}
\usepackage{svg}
\usepackage{algorithm}
\usepackage{algpseudocode}

\ThesisAuthor{Bc. Tomáš Bezděk}

\ThesisSupervisor{Ing. Martin Beseda, Ph.D.}

\CzechThesisTitle{Metody numerické optimalizace v prostředí s kvantovým šumem}

\EnglishThesisTitle{Numerical Optimization Methods in the environment with Quantum Noise}

\SubmissionYear{2025}

\ThesisAssignmentFileName{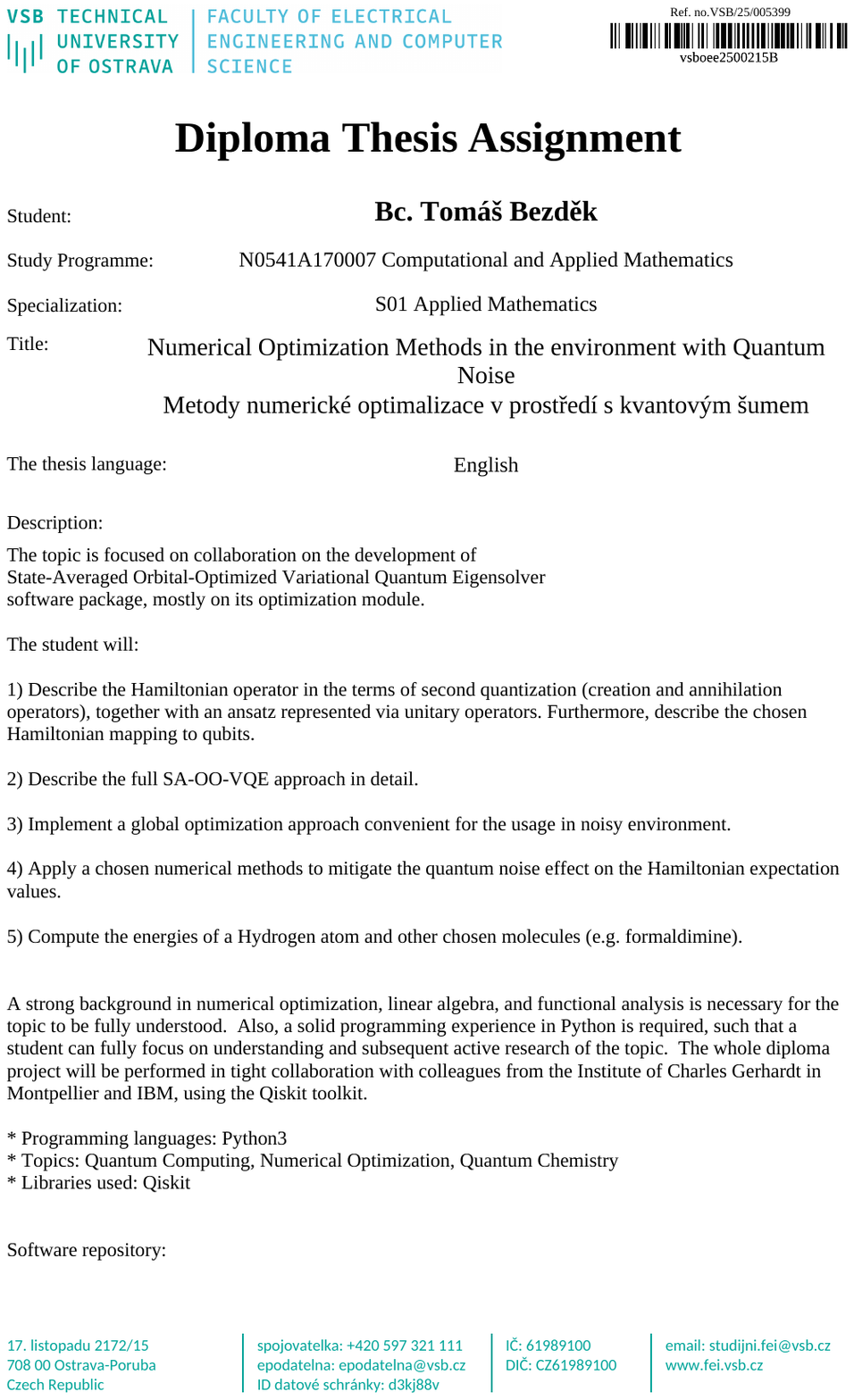}

\Acknowledgement{I would like to express my sincere gratitude to my supervisor, Ing. Martin Beseda, Ph.D., for his invaluable guidance and insightful advice throughout the course of this thesis research. His expertise and encouragement were instrumental in navigating the complexities of this work.

Furthermore, I wish to extend my deepest appreciation to my family for their unwavering support, understanding, and patience during my studies and the writing of this thesis. Their belief in me has been a constant source of motivation.}

\CzechAbstract{Přesný výpočet potenciálových nadploch, zahrnujících jak základní, tak excitované stavy, má zásadní význam v kvantové chemii, zejména pro pochopení fotochemických procesů zprostředkovaných útvary, jako jsou kónická křížení.
Zatímco klasické metody jako Úplná konfigurační interakce poskytují referenční přesnost, jejich exponenciální škálování omezuje jejich použitelnost. Aproximační metody jako je samokonzistentní pole s úplným aktivním prostorem selhávají v blízkosti degenerací. 
Nástup zašuměných kvantových počítačů středního rozsahu nabízí potenciální alternativy, ale algoritmy jako variační kvantový eigensolver jsou omezeny hardwarovými limity a často vyžadují malé aktivní prostory.

Tato práce se zaměřuje na algoritmus stavově průměrovaného variačního kvantového eigensolveru s optimalizovanými orbitaly, hybridní kvantově-klasický přístup navržený tak, aby poskytoval vyvážený popis více elektronových stavů kombinací přípravy kvantového stavu s klasickou, stavově průměrovanou optimalizací orbitalů. 
Práce zahrnuje příspěvek k vývoji softwarového balíčku stavově průměrovaného variačního kvantového eigensolveru s optimalizovanými orbitaly, konkrétně jeho optimalizačního modulu.

Klíčovým přínosem je implementace a vyhodnocení globálního optimalizačního algoritmu diferenciální evoluce v rámci stavově průměrovaného variačního kvantového eigensolveru s optimalizovanými orbitaly, spolu se srovnávací studií oproti jiným klasickým optimalizátorům (metodě největšího spádu, algoritmus Broydena-Fletchera-Goldfarba-Shannoa, optimalizace s omezeními pomocí lineární aproximace, sekvenční programování nejmenších čtverců). 
Výkonnost těchto optimalizátorů je hodnocena při výpočtu energií základního a prvního excitovaného stavu molekul H$_2$, H$_4$ a LiH.

Dále práce demonstruje schopnost stavově průměrovaného variačního kvantového eigensolveru s optimalizovanými orbitaly přesně modelovat potenciálové nadplochy v blízkosti kónických křížení, přičemž jako případovou studii využívá molekulu formaldiminu. 
Výsledky zdůrazňují význam optimalizace orbitalů pro zachycení správné topologie potenciálových nadploch v oblastech silné neadiabatické vazby, což je úkol, ve kterém standardní stavově průměrovaný variační kvantový eigensolver s fixními orbitaly selhává. 
Zjištění naznačují, že zatímco diferenciální evoluce představuje výzvy v oblasti efektivity a robustnosti pro tyto specifické úlohy stavově průměrovaného variačního kvantového eigensolveru s optimalizovanými orbitaly, gradientní metody jako algoritmus Broydena-Fletchera-Goldfarba-Shannoa a sekvenční programování nejmenších čtverců nabízejí vyšší výkonnost a samotný přístup stavově průměrovaného variačního kvantového eigensolveru s optimalizovanými orbitaly je klíčový pro popis komplexních elektronových struktur, jako jsou kónická křížení.}
\CzechKeywords{kvantové počítání; numerická optimalizace; kvantová chemie; variační kvantový eigensolver; stavově průměrovaný variační kvantový eigensolver s optimalizovanými orbitaly; diferenciální evoluce; kónická křížení; zašuměné kvantové počítače středního rozsahu}

\EnglishAbstract{The accurate calculation of electronic potential energy surfaces, encompassing both ground and excited states, is paramount in quantum chemistry, particularly for understanding photochemical processes mediated by features like conical intersections. While classical methods like Full Configuration Interaction provides benchmark accuracy, their exponential scaling limits applicability. 
Approximate methods such as Complete Active Space Self-Consistent Field struggle near degeneracies. 
The advent of noisy intermediate-scale quantum computers offers potential alternatives, but algorithms like the Variational Quantum Eigensolver are constrained by hardware limitations, often requiring small active spaces.

This thesis focuses on the State-Averaged Orbital-Optimized Variational Quantum Eigensolver algorithm, a hybrid quantum-classical approach designed to provide a balanced description of multiple electronic states by combining quantum state preparation with classical state-averaged orbital optimization. 
The work involves contributing to the State-Averaged Orbital-Optimized Variational Quantum Eigensolver software package, specifically its optimization module.

A key contribution is the implementation and evaluation of the Differential Evolution, a global optimization algorithm within the State-Averaged Orbital-Optimized Variational Quantum Eigensolver framework, alongside a comparative study against other classical optimizers (Gradient Descent, Broyden-Fletcher-Goldfarb-Shanno algorithm, Constrained Optimization by Linear Approximation, Sequential Least Squares Programming). The performance of these optimizers is assessed for calculating the ground and first excited state energies of H$_2$, H$_4$, and LiH molecules.

Furthermore, the thesis demonstrates the capability of State-Averaged Orbital-Optimized Variational Quantum Eigensolver to accurately model potential energy surfaces near conical intersections, using the formaldimine molecule as a case study. 
The results highlight the importance of orbital optimization for capturing the correct potential energy surfaces topology in regions of strong non-adiabatic coupling, a task where standard State-Averaged Variational Quantum Eigensolver with fixed orbitals fails. 
The findings indicate that while Differential Evolution presents challenges in efficiency and robustness for these specific State-Averaged Orbital-Optimized Variational Quantum Eigensolver tasks, gradient-based methods like Broyden-Fletcher-Goldfarb-Shanno algorithm and Sequential Least Squares Programming offer superior performance, and the State-Averaged Orbital-Optimized Variational Quantum Eigensolver approach itself is crucial for treating complex electronic structures like conical intersections.
}
\EnglishKeywords{Quantum Computing; Numerical Optimization; Quantum Chemistry; Variational Quantum Eigensolver; State-Averaged Orbital-Optimized Variational Quantum Eigensolver; Differential Evolution; Conical Intersections; Noisy Intermediate-Scale Quantum computer}

\AddAcronym{TISE}{Time-Independent Schrödinger equation}
\AddAcronym{TDSE}{Time-Dependent Schrödinger Equation}
\AddAcronym{BOA}{Born-Oppenheimer approximation}
\AddAcronym{CCR}{Canonical Commutation Relations}
\AddAcronym{CAR}{Canonical Anticommutation Relations}
\AddAcronym{JW}{Jordan-Wigner}

\AddAcronym{SA-OO-VQE}{State-Averaged Orbital-Optimized Variational Quantum Eigensolver}
\AddAcronym{NISQ}{noisy intermediate-scale quantum}
\AddAcronym{VQE}{Variational Quantum Eigensolver}
\AddAcronym{SA-VQE}{State-Averaged Variational Quantum Eigensolver}
\AddAcronym{CASCI}{Complete Active Space Configuration Interaction}
\AddAcronym{CASSCF}{Complete Active Space Self-Consistent Field}
\AddAcronym{SA-CASSCF}{State-Averaged Complete Active Space Self-Consistent Field}
\AddAcronym{SA-OO}{State-Averaged Orbital Optimization}
\AddAcronym{PES}{Potential Energy Surface}
\AddAcronym{FCI}{Full Configuration Interaction}
\AddAcronym{MO}{Molecular Orbital}
\AddAcronym{RDM}{Reduced Density Matrix}
\AddAcronym{HOMO}{Highest Occupied Molecular Orbital}
\AddAcronym{LUMO}{Lowest Occupied Molecular Orbital}
\AddAcronym{UCC}{Unitary Coupled Cluster Ansatz}
\AddAcronym{UCCSD}{Unitary Coupled Cluster Ansatz with Single- and Double-excitation Operators} 

\AddAcronym{DE}{Differential evolution}

\AddAcronym{BFGS}{Broyden-Fletcher-Goldfarb-Shanno}
\AddAcronym{COBYLA}{Constrained Optimization by Linear Approximation}
\AddAcronym{SLSQP}{Sequential Least Squares Programming}
\AddAcronym{QP}{Quadratic Programming}

\AddAcronym{CI}{Conical Intersection}

\newcolumntype{d}[1]{D{.}{.}{#1}}

\begin{document}

\MakeTitlePages

\listoffigures
\clearpage

\listoftables
\clearpage


\chapter{Introduction}
\label{chap:introduction}
Gaining insight into chemical reaction pathways frequently requires precise knowledge of electronic PESs, covering both the lowest-energy (ground) state and relevant higher-energy (excited) states~\cite{TRUHLAR20039, HOU2023295}. 
This need becomes especially pronounced in regions where distinct PESs approach closely or intersect, creating topological features like conical intersections and avoided crossings~\cite{klessinger1995excited, RobbBernardiOlivucci+1995+783+789, CS9962500321, Domcke_Yarkony_Koppel_2004}. 
These features are fundamental to photochemistry and photobiology, acting as critical mediators for important processes such as light-induced isomerization, energy dissipation in photosynthesis, and DNA photostability~\cite{photochemistry_DNA, doi:10.1021/ja904932x, doi:10.1073/pnas.1014982107}. 
Consequently, accurately modeling these intersections is vital for predicting molecular behavior following light absorption~\cite{May_Kuhn_2004}.

However, predicting these intricate PES features reliably faces substantial theoretical and computational obstacles. 
The benchmark for theoretical accuracy, FCI~\cite{TF9524800973}, is computationally infeasible for all but the smallest molecules due to its exponential scaling. 
Practical methods like CASCI and its self-consistent field extension CASSCF~\cite{10.1063/5.0042147} reduce the cost by focusing on a limited active space, but their accuracy depends heavily on this choice, and standard approaches struggle near degeneracies. 
State-specific CASSCF optimizes orbitals for the ground state, potentially degrading the description of excited states and missing conical intersections. 
Optimizing for each state individually can lead to non-orthogonal wavefunctions and convergence issues like root flipping~\cite{Pople19991267}.

The emergence of NISQ computers presents a potentially transformative approach~\cite{Abrams_1999, O_Brien_2019}. 
However, current quantum hardware has significant limitations, including qubit counts and circuit depths prone to noise~\cite{Preskill2018quantumcomputingin}. 
This necessitates the use of hybrid quantum-classical algorithms with shallow circuits, such as the VQE~\cite{Peruzzo_2014, McClean_2016}. 
Standard VQE, typically targeting the ground state within a constrained active space, inherits the limitations of the underlying electronic structure description, potentially failing to capture the essential physics of state interactions~\cite{SEINO2014341, Yu_2021}.

This convergence of difficulties highlights a compelling need for a computational strategy that provides a balanced (democratic) treatment of multiple electronic states concurrently, functions effectively within NISQ constraints, and incorporates robust orbital optimization~\cite{Mao2025_rdm, krompiec2022stronglycontractednelectronvalence, PhysRevResearch.3.033230}. 
Inspired by the classical SA-CASSCF method~\cite{ROOS1980157, Helgaker_Jørgensen_Olsen_2014}, the SA-OO-VQE algorithm was developed~\cite{Yalouz_2021, saoovqe_art_1}. 
SA-OO-VQE synergistically combines the state preparation and measurement capabilities of SA-VQE~\cite{PhysRevResearch.1.033062} on a quantum processor with a classical SA-OO procedure. 
This hybrid approach aims to compute accurate energies and wavefunctions for multiple low-lying states, leverage quantum computation while mitigating active space limitations via orbital optimization, and enable the precise modeling of complex PES features like conical intersections, paving the way for reliable simulations of photochemical processes.

The initial sections of this thesis are dedicated to establishing the necessary theoretical framework. 
This includes describing the Hamiltonian operator using second quantization (Section~\ref{sec:second_quantization}), outlining the mapping of fermionic systems to qubits, exemplified by the Jordan-Wigner transformation (Section~\ref{section:jordan-wigner_mapping}), and providing a detailed explanation of the complete State-Averaged Orbital-Optimized Variational Quantum Eigensolver (SA-OO-VQE) approach (Chapter~\ref{chapter:saoovqe}).

Building upon this theoretical groundwork, the thesis then presents its practical contributions. 
The principles of the Differential Evolution global optimization algorithm~\cite{de2005a, AHMAD20223831, Neri2010, DAS20161, 5601760} are explained theoretically in Chapter~\ref{chapter:DE}.
A key achievement is the implementation and subsequent evaluation of the Differential Evolution global optimization algorithm~\cite{bezdek_2025_15292845}, chosen for its suitability for the potentially noisy optimization landscapes encountered in VQE, integrated within the SA-OO-VQE framework~\cite{joss_saooveq}. 
The effectiveness of this DE implementation is assessed through a comparative study (Section~\ref{sec:comparison_of_optimizers}) against several established classical optimization algorithms (BFGS~\cite{liu1989limited}, COBYLA~\cite{Conn_Scheinberg_Vicente_2009}, SLSQP~\cite{kraft1988software}, Gradient Descent~\cite{ruder2016overview}) applied to the calculation of ground and excited state energies for H$_2$, H$_4$, and LiH molecules using the SA-OO-VQE method~\cite{joss_saooveq}. 
Finally, the practical capability of the SA-OO-VQE algorithm is demonstrated (Section~\ref{sec:formaldimin}) by investigating a conical intersection in the formaldimine molecule, which critically highlights the importance of the orbital optimization step for accurately describing such complex electronic features.

\chapter{Quantum Chemistry}
\label{chapter:quantum_chemistry}
This chapter delves into the fundamental principles of quantum chemistry, exploring the application of quantum mechanics to understand the structure, bonding, and reactivity of chemical systems~\cite{Kauzmann_2013, book:modern_qch}. 
We begin by introducing the time-independent Schrödinger equation as the cornerstone for describing the stationary states crucial for determining electronic structure. 
The discussion progresses from the foundational single-particle case to the complexities of multiparticle systems (Section~\ref{sec:multiparticle_qm}), introducing the molecular Hamiltonian and the vital Born-Oppenheimer approximation, which allows the separation of nuclear and electronic motion~\cite{book:modern_qch}. 
Following this, we transition from the specific wavefunction representation to the more general and abstract Hilbert space formalism, elucidating the concept of quantum states and operators (Section~\ref{sec:wavefunction_to_state})~\cite{Griffiths_QM, Dirac_Principles, VonNeumann_Foundations}. 
The profound consequences of particle indistinguishability are examined, leading to the principles of quantum statistics for fermions and bosons (Section~\ref{sec:identical_particles}), including the Pauli Exclusion Principle~\cite{Sakurai_QM}.
To effectively handle the intricacies of many-electron systems common in chemistry, the powerful formalism of second quantization is presented (Section~\ref{sec:second_quantization}), utilizing creation and annihilation operators defined within Fock space~\cite{Fetter_Walecka, Negele_Orland}. 
Finally, this chapter connects these theoretical foundations to the burgeoning field of quantum computation by detailing the Jordan-Wigner mapping (Section~\ref{section:jordan-wigner_mapping}), a key technique for translating fermionic problems into the qubit representations required by quantum algorithms~\cite{Jordan_Wigner_1928}.

\section{Introduction to Quantum Chemistry}
Quantum chemistry is a branch of chemistry that applies the principles of quantum mechanics to the study of chemical systems. Central to this field is the TISE, which serves as a cornerstone for understanding the behavior of electrons in atoms and molecules.
The TISE is predominantly utilized because it effectively addresses the properties of quantum systems in stationary states, where energy levels are fixed.
This is crucial for studying the electronic structure of atoms and molecules, as it allows for the calculation of energy eigenvalues and wave functions that describe the spatial distribution of electrons.

The TISE simplifies the analysis of chemical systems, enabling researchers to gain insights into molecular geometries, bonding, and reactivity.
Many chemical processes of interest, such as molecular optimization and spectral analysis, are inherently static, making the TISE more appropriate than the TDSE, which accounts for time evolution and dynamic interactions~\cite{Griffiths_QM}.
By focusing on the TISE, quantum chemists can efficiently explore and predict the behavior of complex molecular systems, thereby advancing our understanding of chemical phenomena.

The TISE can be expressed as
\begin{equation}
\label{eq:tise}
\hat{H} \psi(\mathbf{r}) = E \psi(\mathbf{r}),
\end{equation}
where $ \hat{H} $ is the Hamiltonian operator, $ \psi(\mathbf{r}) $ is the wave function of the system, and $ E $ represents the energy eigenvalues.
The Hamiltonian operator is defined as
\begin{equation}
\hat{H} = -\frac{\hbar^2}{2m} \nabla^2 + V(\mathbf{r}),
\end{equation}
where $ \hbar $ is the reduced Planck's constant, $ m $ is the mass of the particle, $ \nabla^2 $ is the Laplacian operator representing the kinetic energy, and $ V(\mathbf{r}) $ is the potential energy as a function of position.

The wave function $ \psi(\mathbf{r}) $ encodes all the information about the quantum state of the system, including the probability density of finding a particle at a given position, given by $ |\psi(\mathbf{r})|^2 $.
The normalization condition requires that the integral of the probability density over all space equals one
\begin{equation}
\int |\psi(\mathbf{r})|^2 \, dV = 1 .
\end{equation}
Solving the TISE for various systems allows chemists to derive important properties such as energy levels, molecular orbitals, and electron distributions.
For example, in the case of the hydrogen atom, the solution yields quantized energy levels given by,
\begin{equation}
E_n = -\frac{m e^4}{2 \hbar^2 n^2},
\end{equation}
where $ n $ is the principal quantum number, $ e $ is the elementary charge, and $ m $ is the mass of the electron.
The corresponding wave functions, known as atomic orbitals, describe the spatial distribution of electrons around the nucleus.

In more complex systems, such as multi-electron atoms or molecules, the Schrödinger equation becomes significantly more complicated due to electron-electron interactions. 
Approximations such as the Hartree-Fock method and Density Functional Theory are often employed to simplify the calculations while still providing valuable insights into the electronic structure and properties of molecules.

In summary, the TISE is a foundational element of quantum chemistry, providing a powerful tool for predicting and understanding the behavior of electrons in atoms and molecules.
Its solutions reveal the underlying quantum mechanical nature of chemical systems, enabling chemists to explore the intricate relationships between structure, reactivity, and properties in the molecular world.

\section{Multiparticle Quantum Mechanics}
\label{sec:multiparticle_qm}
In quantum chemistry, the study of systems with multiple particles introduces additional complexity compared to single-particle systems.
Multiparticle quantum mechanics is essential for understanding the behavior of electrons in atoms and molecules, where interactions between particles significantly influence the overall properties of the system.
The fundamental principle in multiparticle quantum mechanics is that the total wave function of a system containing $ N $ particles must account for the indistinguishability of identical particles and their interactions.
For a system of $ N $ non-relativistic particles, the TISE equation can be generalized to
\begin{equation}
\hat{H} \Psi(\mathbf{r}_1, \mathbf{r}_2, \ldots, \mathbf{r}_N) = E \Psi(\mathbf{r}_1, \mathbf{r}_2, \ldots, \mathbf{r}_N),
\end{equation}
where $ \Psi $ is the total wave function of the system, which depends on the positions of all $ N $ particles and $ \hat{H} $ is the total Hamiltonian operator that includes both kinetic and potential energy terms for all particles.
The total Hamiltonian for a system of $ N $ particles can be expressed as
\begin{equation}
\hat{H} = \sum_{i=1}^{N} \left( -\frac{\hbar^2}{2m_i} \nabla_i^2 \right) + \sum_{i \neq j} V_{ij}(\mathbf{r}_i, \mathbf{r}_j),
\end{equation}
where $ m_i $ is the mass of the $ i $th particle, $ \nabla_i^2 $ is the Laplacian operator acting on the coordinates of the $ i $th particle, and $ V_{ij} $ represents the potential energy due to interactions between particles $ i $ and $ j $.
\begin{example}[Hamiltonian operator of H$_2^+$]
The ion of a hydrogen molecule consists of three particles, two nuclei, and one electron, as can be seen in figure~\ref{fig:h2_ion}.
\begin{figure}[h!]
    \centering
    \begin{tikzpicture}
  \coordinate (HA) at (0,0);
  \coordinate (HB) at (4,0);
  \coordinate (e) at (3.4,2.4);

  \draw (HA) -- (HB) node[midway, below=0.15cm] {$r_{AB}$};
  \draw (HA) -- (e) node[midway, left=0.15cm] {$r_A$};
  \draw (HB) -- (e) node[midway, right=0.1cm] {$r_B$};

  \fill[ball color=blue!80!cyan] (HA) circle (0.3) node[below=0.2cm] {$H_A$};
  \fill[ball color=blue!80!cyan] (HB) circle (0.3) node[below=0.2cm] {$H_B$};
  \fill[ball color=red!80!orange] (e) circle (0.2) node[above=0.1cm] {$e^-$} node[below right=0.1cm] {$m_e$};

\end{tikzpicture}
    \caption{Ion of H$_2^+$.}
    \label{fig:h2_ion}
\end{figure}
Therefore, the Hamiltonian will contain three terms corresponding to the kinetic energy of each particle.
The potential energy is determined by the Coulomb attraction or repulsion~\cite{physic_book} between the particles.
Overall, we obtain the expression
\begin{equation}
\hat{H} = -\frac{\hbar^2}{2m_A} \nabla_A^2 - \frac{\hbar^2}{2m_B} \nabla_B^2 - \frac{\hbar^2}{2m_e} \nabla_e^2 - \frac{1}{4\pi \varepsilon_0} \left( \frac{e^2}{r_{AB}} - \frac{e^2}{r_{A}} - \frac{e^2}{r_{B}} \right) ,
\end{equation}
where, for example, the distance $ r_A $ in Cartesian coordinates is given by
\begin{equation}
r_A = |\mathbf{r}_A - \mathbf{r}_e| = \sqrt{(x_A - x_e)^2 + (y_A - y_e)^2 + (z_A - z_e)^2}.
\end{equation}
\end{example}
In multiparticle systems, the wave function must be symmetrized or antisymmetrized depending on whether the particles are bosons or fermions, respectively. Bosons, which obey Bose-Einstein statistics, have symmetric wave functions, while fermions, which obey Fermi-Dirac statistics, have antisymmetric wave functions due to the Pauli exclusion principle. This leads to significant differences in the statistical behavior of systems composed of identical particles.

The study of multiparticle quantum mechanics is crucial for understanding phenomena such as electron correlation, molecular bonding, and the emergence of collective behaviors in systems like superconductors and superfluids. Techniques such as configuration interaction, coupled cluster methods, and density functional theory are often employed to solve the many-body Schrödinger equation and obtain accurate predictions for the properties of complex molecular systems.

\section{Molecular Hamiltonian}
The description of molecules using quantum theory methods is a central theme in quantum chemistry.
Unlike atoms, molecules are not spherically symmetric, which complicates the calculations of their properties.
Due to the lower symmetry, for example, the conservation of angular momentum is not maintained during the electron motion in molecules.
In the case of molecules, we must also account for the motion of atomic nuclei in addition to the motion of electrons.

It should not be burdensome for us to write down the Hamiltonian for a molecule. It must include all force interactions between the nuclei and electrons. Neglecting relativistic effects, it has the following form
\begin{equation}
\label{eq:molecular_tise}
\hat{H} = \hat{T}_N + \hat{T}_e + \hat{V}_{NN} + \hat{V}_{Ne} + \hat{V}_{ee},
\end{equation}
where $ \hat{T}_N $ is the kinetic energy operator for the nuclei
\begin{equation}
\hat{T}_N = -\sum_{J} \frac{\hbar^2}{2M_J} \nabla_J^2.
\end{equation}
$ \hat{T}_e $ is the kinetic energy operator for the electrons
\begin{equation}
\hat{T}_e = -\sum_{i} \frac{\hbar^2}{2m_e} \nabla_i^2.
\end{equation}
$ \hat{V}_{NN} $ describes the Coulomb repulsion between nuclei
\begin{equation}
\hat{V}_{NN} = \sum_{J > J'} \frac{Z_J Z_{J'} e^2}{4 \pi \varepsilon_0 |\mathbf{R}_J - \mathbf{R}_{J'}|}.
\end{equation}
$ \hat{V}_{Ne} $ describes the attraction between nuclei and electrons
\begin{equation}
\hat{V}_{Ne} = -\sum_{J} \sum_{i} \frac{Z_J e^2}{4 \pi \varepsilon_0 |\mathbf{R}_J - \mathbf{r}_i|},
\end{equation}
and $ \hat{V}_{ee} $ describes the repulsion between electrons
\begin{equation}
\hat{V}_{ee} = \sum_{i > i'} \frac{e^2}{4 \pi \varepsilon_0 |\mathbf{r}_i - \mathbf{r}_{i'}|}.
\end{equation}
In the above expressions, $ \mathbf{R}_J $ and $ \mathbf{r}_i $ are symbols for the position vectors of nucleus $ J $ and electron $ i $, respectively, and $ Z_J $ denotes the atomic number of nucleus $ J $.

Our task, as before, is to solve the equation~\ref{eq:tise}, where the wave function $ \psi $ is a function of both the coordinates of the electrons and the coordinates of the atomic nuclei.

\section{Born-Oppenheimer's Approximation}
Atomic nuclei are significantly heavier than electrons, which means that their quantum-level description is not always essential. 
However, it is crucial to recognize the limitations of the classical perspective on atomic nuclei.
Due to the substantial difference in mass between nuclei and electrons, we can often treat their motions separately. 
This separation forms the foundation of the BOA, which leads to the concept of potential energy hypersurfaces.
While solving the Schrödinger equation with the molecular Hamiltonian in its entirety is quite complex, the BOA simplifies this situation by allowing us to decouple the motions of electrons and atomic nuclei.
We will begin with a brief overview of the BOA before proceeding to a more mathematically rigorous treatment.

We start with the electronic Hamiltonian, which describes the electrons for fixed nuclei in a specific geometry $ \mathbf{R}_i $ given by 
\begin{equation}
\hat{H}_{el} = \hat{T}_e + \hat{V}_{Ne} + \hat{V}_{ee} + \hat{V}_{NN}.
\end{equation}
The electronic Hamiltonian acts only on the functions of the electron coordinates, although this electronic Hamiltonian varies for different nuclear coordinates $ \mathbf{R}_i $.
Let us first solve the electronic TISE for each possible geometry, which is expressed as 
\begin{equation}
\hat{H}_{el} \psi_{el}^{(i)} = E_{el}^{(i)} \psi_{el}^{(i)}.
\end{equation}
The index $ i $ denotes the electronic state. We recall that the solution is a wave function of the electron coordinates that depend parametrically on the nuclear coordinates, represented as 
\begin{equation}
\psi_{el}^{(i)} = \psi_{el}^{(i)}(\mathbf{r}; \mathbf{R}).
\end{equation}
By ``parametric dependence'', we mean that the wave function varies for different nuclear coordinates, while the square of the wave function does not represent the probability density of finding the nuclei in a specific geometry $ \mathbf{R}_i $.

The set of eigenfunctions of the electronic Hamiltonian forms a complete set of functions, allowing us to expand any function of the electron coordinates in terms of the basis of eigenfunctions of the electronic Hamiltonian.
We can also do this for the total wave function, which can be expressed as 
\begin{equation}
\psi(\mathbf{r}, \mathbf{R}) = \sum_i \chi_i(\mathbf{R}) \psi_{el}^{(i)}(\mathbf{r}; \mathbf{R}),
\end{equation}
where $ \chi_i(\mathbf{R}) $ are the expansion coefficients that depend on the positions of the nuclei.
Up to this point, we have made no approximations.

We will substitute the wave function $ \psi(\mathbf{r}, \mathbf{R}) $ into the equation~\ref{eq:molecular_tise}, and we will further manipulate the left-hand side,
\begin{equation}
\label{eq:tise_molecular_complex}
\hat{H} \psi = \sum_i \left( \hat{T}_N + \hat{H}_{el} \right) \chi_i \psi_{el}^{(i)} = \sum_i \hat{T}_N (\chi_i \psi_{el}^{(i)}) + \sum_i \chi_i E_{el}^{(i)} \psi_{el}^{(i)}.
\end{equation}
Now, we will focus on the action of the nuclear kinetic energy operator,
\begin{equation}
\label{eq:kinetic_operator}
\hat{T}_N \chi_i \psi_{el}^{(i)} = -\frac{\hbar^2}{2M} \left( \psi_{el}^{(i)} \nabla_R^2 \chi_i  + 2 \nabla_R \chi_i \nabla_R  \psi_{el}^{(i)} + \chi_i \nabla_R^2 \psi_{el}^{(i)} \right).
\end{equation}

At this point, we will make an approximation – we will neglect the last two terms from equation~\ref{eq:kinetic_operator}, thus we set
\begin{equation}
\hat{T}_N \chi_i \psi_{el} \approx -\frac{\hbar^2}{2M} \psi_{el} \nabla_R^2 \chi_i .
\end{equation}
This neglect is the essence of the Born-Oppenheimer approximation. 
We see that, in general, this will hold better the less the electronic wave function changes with geometry. 
This also means that for rapidly moving nuclei, the Born-Oppenheimer approximation may not work very well.

Let us return to equation~\ref{eq:tise_molecular_complex}.
We first multiply the equation by the complex conjugate electronic wave function $ \psi_{el}^{(j)} $ and then integrate over the electron coordinates $ \mathbf{r} $,
\begin{equation}
    \int \sum_i \left( \hat{T}_N \left(\chi_i \psi_{el}^{(i)}\right) + \chi_i E_{el} \psi_{el}^{(i)} \right) \psi_{el}^{(j)*} \, d\tau = \int E \sum_i \left( \chi_i \psi_{el}^{(i)} \right) \psi_{el}^{(j)*}  \, d\tau
\end{equation}
The TISE will take the form
\begin{equation}
    \int \sum_j  \left( \hat{T}_N \chi_i \psi_{el}^{(i)} \psi_{el}^{(j)*} + \chi_i E_{el} \psi_{el}^{(i)} \psi_{el}^{(j)*} \right) d\tau = \int E \sum_i \left( \chi_i \psi_{el}^{(i)} \psi_{el}^{(j)*} \right) d\tau
\end{equation}
Given that the electronic wave functions are orthonormal, the equation simplifies to
\begin{equation}
    \sum_i \left( \hat{T}_N \chi_i \delta_{ij} + \chi_i E_{el} \delta_{ij} \right) = E \delta_{ij} \chi_i.
\end{equation}
The Kronecker delta $ \delta_{ij} $ will ultimately select only the terms for $ i = j $,
\begin{equation}
\label{eq:nucleons_tise}
    \left( \hat{T}_N + E_{el} \right) \chi_j = E \chi_j.
\end{equation}
Equation~\ref{eq:nucleons_tise} represents the TISE for the motion of atomic nuclei. 
We see that the nuclei move in a potential given by the electronic energy for individual geometries.

\section{From Wavefunction to Quantum State}
\label{sec:wavefunction_to_state}
While the previous sections focused on the TISE and its solutions in the form of wavefunctions, such as 
$\Psi(\mathbf{r}_1, \ldots, \mathbf{r}_N)$ for multiparticle systems (Section~\ref{sec:multiparticle_qm}), it is crucial to understand the deeper and more abstract concept of the quantum state these functions represent. 
The wavefunction, particularly in the position representation $\psi(\mathbf{r})$, provides a description of the system's probability amplitude distribution in space, where $|\psi(\mathbf{r})|^2 dV$ gives the probability of finding the particle within the volume element $dV$ around position $\mathbf{r}$~\cite[e.g.,][]{Griffiths_QM}. 
However, this function is just one specific representation of the underlying quantum state.

Quantum mechanics employs a more abstract and powerful mathematical formalism, largely developed by Paul Dirac~\cite{Dirac_Principles}, where the physical state of a quantum system is represented by a state vector residing in a separable complex vector space known as a separable Hilbert space, denoted $\mathcal{H}$~\cite{VonNeumann_Foundations}. 
Hilbert spaces are complete inner product spaces, meaning they possess structures allowing for notions of distance, orthogonality, and convergence, which are essential for the mathematical rigor of quantum theory~\cite{Reed_Simon_FA}.

The state vector, often denoted using Dirac's bra-ket notation as $|\psi\rangle$, encapsulates all the information about the quantum system. 
This abstract vector exists independently of any specific coordinate system or basis. 
The linearity of the Hilbert space directly reflects the superposition principle, a fundamental tenet of quantum mechanics, allowing valid states to be formed by linear combinations of other states, $|\psi\rangle = \sum_i c_i |\psi_i\rangle$, where $c_i$ are complex coefficients.

The connection between the abstract state vector $|\psi\rangle$ and the familiar position-space wavefunction $\psi(\mathbf{r})$ is established through the concept of basis vectors in the Hilbert space. 
We can consider a basis composed of position eigenstates, denoted $|\mathbf{r}\rangle$, which represent states where the particle is perfectly localized at position $\mathbf{r}$ (though these are idealized states and not strictly normalizable in $\mathcal{H}$). 
The position-space wavefunction is then obtained by projecting the abstract state vector $|\psi\rangle$ onto this position basis using the inner product defined on the Hilbert space~\cite{Sakurai_QM},
\begin{equation}
\psi(\mathbf{r}) = \langle \mathbf{r} | \psi \rangle.
\end{equation}
Here, $\langle \mathbf{r} |$ represents the ``bra'' vector dual to the ``ket'' vector $|\mathbf{r}\rangle$. 
The inner product $\langle \phi | \psi \rangle$ generalizes the concept of the overlap integral $\int \phi^*(\mathbf{r}) \psi(\mathbf{r}) dV$. 
Crucially, the choice of basis is not unique. 
One could equally project the state vector $|\psi\rangle$ onto the momentum basis $\{|\mathbf{p}\rangle\}$ to obtain the momentum-space wavefunction $\phi(\mathbf{p}) = \langle \mathbf{p} | \psi \rangle$. 
The abstract state vector $|\psi\rangle$ contains the information accessible in any valid representation.

Furthermore, physical observables like position, momentum, energy (Hamiltonian $\hat{H}$), and angular momentum are represented by linear Hermitian operators acting on the state vectors within the Hilbert space. 
The TISE (\ref{eq:tise}), in this more general formalism, becomes an eigenvalue equation for the Hamiltonian operator acting on the state vector
\begin{equation}
\hat{H} |\psi\rangle = E |\psi\rangle,
\end{equation}
where $E$ is the energy eigenvalue and $|\psi\rangle$ is the corresponding energy eigenstate (or eigenvector).

This abstract Hilbert space formalism offers significant advantages:
\begin{enumerate}
    \item \textbf{Basis Independence:} Physical predictions (like probabilities and expectation values, $\langle \hat{A} \rangle = \langle \psi | \hat{A} | \psi \rangle / \langle \psi | \psi \rangle$) are independent of the chosen representation (basis).
    \item \textbf{Inclusion of Internal Degrees of Freedom:} It naturally incorporates properties like electron spin, which lack a classical analog and cannot be described solely by a spatial wavefunction $\psi(\mathbf{r})$. 
    The total state vector must include these degrees of freedom, often represented in a tensor product space, e.g., $\mathcal{H}_{\text{spatial}} \otimes \mathcal{H}_{\text{spin}}$.
    \item \textbf{Foundation for Advanced Methods:} It provides the necessary framework for more sophisticated techniques required for many-body systems, such as the formalism of second quantization discussed in Section~\ref{sec:second_quantization}. 
\end{enumerate}

In summary, while the wavefunction $\psi(\mathbf{r})$ is a practical and intuitive tool, particularly for simple systems, the fundamental description of a quantum system lies in its state vector $|\psi\rangle$ within a Hilbert space. 
Recognizing this distinction is essential for comprehending the full structure of quantum mechanics and for tackling the complexities of multi-particle systems and their statistics, which will be explored further.

\section{Identical Particles and Quantum Statistics}
\label{sec:identical_particles}
A fundamental aspect of quantum mechanics arises when dealing with systems containing multiple particles of the same type, such as electrons in an atom or molecule. 
Unlike classical objects, identical quantum particles are truly indistinguishable. 
There is no measurement we can perform to uniquely identify and track an individual particle if identical counterparts are present. 
This indistinguishability is not merely a practical limitation but a fundamental property reflected in the structure of the quantum state itself~\cite{Griffiths_QM, Sakurai_QM}.

Consider a system of $N$ identical particles. As established in Section~\ref{sec:wavefunction_to_state}, the state of such a system is described by a state vector $|\Psi\rangle$ in the appropriate Hilbert space. For $N$ particles, this space is typically constructed as the tensor product of single-particle Hilbert spaces: $\mathcal{H}^{(N)} = \mathcal{H}_1 \otimes \mathcal{H}_2 \otimes \dots \otimes \mathcal{H}_N$, where each $\mathcal{H}_i$ is a copy of the single-particle space~\cite{Shankar_QM}.

To formalize the concept of indistinguishability, we introduce the permutation operator, $\hat{P}_{ij}$, which acts on the $N$-particle state vector $|\Psi\rangle$ by exchanging the roles (all coordinates, including spin) of particle $i$ and particle $j$. For example, in a two-particle system with a state formed from single-particle states $|\phi\rangle$ and $|\chi\rangle$, $\hat{P}_{12} (|\phi\rangle_1 |\chi\rangle_2) = |\chi\rangle_1 |\phi\rangle_2$.

The principle of indistinguishability demands that any physically observable property must remain unchanged if we swap two identical particles. Since observables are represented by Hermitian operators $\hat{A}$, this means that $\hat{A}$ must commute with any permutation operator,
\begin{equation}
[\hat{A}, \hat{P}_{ij}] = 0 \quad \text{for all } i, j.
\end{equation}
Here, the notation $[\hat{A}, \hat{P}_{ij}] \equiv \hat{A}\hat{P}_{ij} - \hat{P}_{ij}\hat{A}$ represents the commutator of the operators $\hat{A}$ and $\hat{P}_{ij}$.
This has profound consequences. 
For instance, the Hamiltonian $\hat{H}$ of a system of identical particles must be symmetric under particle exchange, i.e., $[\hat{H}, \hat{P}_{ij}] = 0$. 
As a result, the energy eigenstates $|\Psi\rangle$ can be chosen to be simultaneous eigenstates of all permutation operators $\hat{P}_{ij}$.

A cornerstone postulate of quantum mechanics, supported by all experimental evidence and deeply connected to relativistic quantum field theory via the spin-statistics theorem~\cite{Streater_Wightman, Peskin_Schroeder_QFT}, dictates that state vectors describing systems of identical particles must possess a specific symmetry under permutation. 
They must be eigenvectors of $\hat{P}_{ij}$ with eigenvalues restricted to only $+1$ or $-1$.
\begin{itemize}
    \item \textbf{Bosons:} Particles whose state vectors are symmetric under the exchange of any pair $(i, j)$:
      \begin{equation}
      \hat{P}_{ij} |\Psi\rangle = +1 |\Psi\rangle \quad (\text{for all } i, j).
      \end{equation}
      Particles with integer intrinsic spin (e.g., photons (spin 1), Helium-4 atoms (spin 0), Higgs boson (spin 0)) are bosons. The subspace of the total Hilbert space $\mathcal{H}^{(N)}$ containing only these symmetric states is denoted $\mathcal{H}^{(N)}_S$.
    \item \textbf{Fermions:} Particles whose state vectors are antisymmetric under the exchange of any pair $(i, j)$:
      \begin{equation}
      \label{eq:fermion_antisymmetry}
      \hat{P}_{ij} |\Psi\rangle = -1 |\Psi\rangle \quad (\text{for all } i, j).
      \end{equation}
      Particles with half-integer intrinsic spin (e.g., electrons (spin 1/2), protons (spin 1/2), neutrons (spin 1/2)) are fermions. The subspace containing only these antisymmetric states is denoted $\mathcal{H}^{(N)}_A$.
\end{itemize}
Nature chooses one symmetry type for each species of elementary particle. 
Composite particles (like atoms) behave as bosons or fermions depending on whether their total spin (sum of spins and orbital angular momenta of constituents) is integer or half-integer.

The antisymmetry requirement (\ref{eq:fermion_antisymmetry}) for fermions leads directly to the \textit{Pauli Exclusion Principle}.
Consider two identical fermions ($i=1, j=2$) and suppose they occupy the same single-particle quantum state $|\phi\rangle$ (this state includes spatial and spin components). 
A naive two-particle state vector might be written as $|\phi\rangle_1 |\phi\rangle_2$. 
Applying the permutation operator, $\hat{P}_{12} (|\phi\rangle_1 |\phi\rangle_2) = |\phi\rangle_2 |\phi\rangle_1$. 
Since the particles are in the same state, $|\phi\rangle_1 |\phi\rangle_2 = |\phi\rangle_2 |\phi\rangle_1$. 
However, for fermions, we must have $\hat{P}_{12} |\Psi\rangle = -|\Psi\rangle$. 
The only vector that satisfies $|\Psi\rangle = -|\Psi\rangle$ is the zero vector, $|\Psi\rangle = 0$. 
This means that a state where two identical fermions occupy the same single-particle quantum state cannot exist. 
In simpler terms, no two identical fermions can occupy the same quantum state simultaneously. 
This principle is fundamental to atomic structure (electron shell filling), molecular bonding, and the stability of matter.

Bosons, being symmetric, face no such restriction. 
Multiple bosons can occupy the same single-particle state, leading to phenomena like Bose-Einstein condensation and lasers.

Finally, let's connect this back to the position-space wavefunction discussed earlier. 
The symmetry properties of the abstract state vector $|\Psi\rangle$ are inherited by its representations in any basis. Specifically, for the position-space wavefunction $\Psi(\mathbf{x}_1, \dots, \mathbf{x}_N) = \langle \mathbf{x}_1, \dots, \mathbf{x}_N | \Psi \rangle$ (where $\mathbf{x}_i$ includes position $\mathbf{r}_i$ and spin coordinate $s_i$), the action of $\hat{P}_{ij}$ on $|\Psi\rangle$ corresponds to swapping the arguments $\mathbf{x}_i$ and $\mathbf{x}_j$ in the wavefunction,
\begin{align}
\Psi(\dots, \mathbf{x}_j, \dots, \mathbf{x}_i, \dots) &= \langle \dots \mathbf{x}_j \dots \mathbf{x}_i \dots | \Psi \rangle \\
&= \langle \dots \mathbf{x}_i \dots \mathbf{x}_j \dots | \hat{P}_{ij} | \Psi \rangle \\
&= \pm \langle \dots \mathbf{x}_i \dots \mathbf{x}_j \dots | \Psi \rangle \\
&= \pm \Psi(\dots, \mathbf{x}_i, \dots, \mathbf{x}_j, \dots),
\end{align}
where the sign is $+$ for bosons and $-$ for fermions. 
This recovers the wavefunction symmetry properties mentioned previously, but now grounded in the fundamental symmetry postulate applied to the abstract state vector in Hilbert space. 
This framework is essential for constructing valid many-body states and forms the basis for methods like second quantization (Section~\ref{sec:second_quantization}). 

\section{Second Quantization}
\label{sec:second_quantization}
The description of quantum systems with many identical particles, as discussed in Section~\ref{sec:identical_particles}, requires state vectors $|\Psi\rangle$ that exhibit specific symmetries under particle exchange. 
For fermions, this involves constructing properly antisymmetrized states, often represented by Slater determinants~\cite{book:modern_qch}. 
While conceptually fundamental, explicitly writing and manipulating these highly complex, (anti)symmetrized N-particle state vectors becomes exceedingly cumbersome as the number of particles $N$ increases~\cite{Fetter_Walecka}. 
Second quantization offers a powerful and elegant alternative formalism that automatically incorporates particle statistics and simplifies the treatment of many-body systems~\cite{book:modern_qch, Negele_Orland}.

The core idea of second quantization is to shift focus from tracking individual particles to tracking the occupation numbers of single-particle states. 
This framework operates within a mathematical structure called Fock space, $\mathcal{F}$~\cite{PhysRev_Foc, BarhoumiAndreani2016FockSpace}. 
For a given type of particle (boson or fermion), Fock space is constructed as the direct sum of Hilbert spaces corresponding to different particle numbers $N$
\begin{equation}
\mathcal{F} = \bigoplus_{N=0}^{\infty} \mathcal{F}_N = \mathcal{F}_0 \oplus \mathcal{F}_1 \oplus \mathcal{F}_2 \oplus \dots.
\end{equation}
Here, $\mathcal{F}_0$ is the space containing only the vacuum state $|\Omega\rangle$ (or $|0\rangle$), representing the absence of any particles. 
$\mathcal{F}_1$ is the single-particle Hilbert space $\mathcal{H}_1$. 
For $N \ge 2$, $\mathcal{F}_N$ is the subspace of the $N$-particle tensor product space $\mathcal{H}_1^{\otimes N}$ containing only the states with the correct symmetry, the totally symmetric subspace $\mathcal{H}^{(N)}_S$ for bosons, or the totally antisymmetric subspace $\mathcal{H}^{(N)}_A$ for fermions~\cite{Sakurai_QM}.

Central to this formalism is the creation ($a_{\lambda}^{\dagger}$) and annihilation ($a_{\lambda}$) operators, defined for each state $|\lambda\rangle$ in a chosen complete orthonormal basis $\{|\lambda\rangle\}$ of the single-particle Hilbert space $\mathcal{H}_1$. 
These operators act on the Fock space $\mathcal{F}$.
The creation operator $a_{\lambda}^{\dagger}$ adds a particle in the single-particle state $|\lambda\rangle$ to the system state, ensuring the resulting state maintains the correct overall symmetry.
The annihilation operator $a_{\lambda}$ removes a particle in the single-particle state $|\lambda\rangle$ from the system state, again preserving the correct symmetry.
By definition, annihilating a particle from the vacuum state yields zero, $a_{\lambda} |\Omega\rangle = 0$ for all $\lambda$. 
Any $N$-particle state in $\mathcal{F}_N$ with the correct symmetry can be constructed by applying $N$ appropriate creation operators to the vacuum state, e.g., $|\lambda_1, \dots, \lambda_N \rangle \propto a_{\lambda_N}^{\dagger} \dots a_{\lambda_1}^{\dagger} |\Omega\rangle$.

The crucial step is that the particle statistics (bosonic or fermionic symmetry) are encoded directly into the algebraic properties of these operators through their \textit{commutation relations}.
\begin{itemize}
    \item \textbf{Bosons} (operators often denoted $b, b^{\dagger}$): Obey the CCR~\cite{Fetter_Walecka}:
      \begin{gather}
      [b_{\lambda}, b_{\mu}] = b_{\lambda} b_{\mu} - b_{\mu} b_{\lambda} = 0 \\
      [b_{\lambda}^{\dagger}, b_{\mu}^{\dagger}] = b_{\lambda}^{\dagger} b_{\mu}^{\dagger} - b_{\mu}^{\dagger} b_{\lambda}^{\dagger} = 0 \\
      [b_{\lambda}, b_{\mu}^{\dagger}] = b_{\lambda} b_{\mu}^{\dagger} - b_{\mu}^{\dagger} b_{\lambda} = \delta_{\lambda\mu}
      \end{gather}
      where $\delta_{\lambda\mu}$ is the Kronecker delta.
    \item \textbf{Fermions} (operators often denoted $c, c^{\dagger}$): Obey the CAR~\cite{Fetter_Walecka, Negele_Orland}:
      \begin{gather}
      \{c_{\lambda}, c_{\mu}\} = c_{\lambda} c_{\mu} + c_{\mu} c_{\lambda} = 0 \label{eq:car_cc} \\
      \{c_{\lambda}^{\dagger}, c_{\mu}^{\dagger}\} = c_{\lambda}^{\dagger} c_{\mu}^{\dagger} + c_{\mu}^{\dagger} c_{\lambda}^{\dagger} = 0 \label{eq:car_cdag_cdag}\\
      \{c_{\lambda}, c_{\mu}^{\dagger}\} = c_{\lambda} c_{\mu}^{\dagger} + c_{\mu}^{\dagger} c_{\lambda} = \delta_{\lambda\mu} \label{eq:car_c_cdag}
      \end{gather}
      where $\{\hat{A}, \hat{B}\} = \hat{A}\hat{B} + \hat{B}\hat{A}$ denotes the anticommutator.
\end{itemize}
Note that the CAR directly enforce the Pauli exclusion principle. 
Setting $\lambda = \mu$ in Eq. (\ref{eq:car_cdag_cdag}) gives $\{c_{\lambda}^{\dagger}, c_{\lambda}^{\dagger}\} = 2 c_{\lambda}^{\dagger} c_{\lambda}^{\dagger} = 0$, which implies $(c_{\lambda}^{\dagger})^2 = 0$. 
This means it is impossible to create two identical fermions in the same single-particle state $|\lambda\rangle$. 
Similarly, from Eq. (\ref{eq:car_cc}), $c_{\lambda}^2 = 0$.

A convenient basis for Fock space is the \textit{occupation number basis}. 
A state in this basis is specified by listing the number of particles $n_\lambda$ occupying each single-particle basis state $|\lambda\rangle$,
\begin{equation}
|n_1, n_2, \dots, n_{\lambda}, \dots \rangle.
\end{equation}
For bosons, each $n_{\lambda}$ can be any non-negative integer ($0, 1, 2, \dots$). For fermions, due to the Pauli principle, each $n_{\lambda}$ can only be $0$ or $1$. The action of the creation and annihilation operators on these basis states are defined as~\cite{Sakurai_QM}:
\begin{align}
\text{Bosons:}\quad & b_{\lambda}^{\dagger} |\dots, n_{\lambda}, \dots \rangle = \sqrt{n_{\lambda}+1} |\dots, n_{\lambda}+1, \dots \rangle \\
& b_{\lambda} |\dots, n_{\lambda}, \dots \rangle = \sqrt{n_{\lambda}} |\dots, n_{\lambda}-1, \dots \rangle \\
\text{Fermions:}\quad & c_{\lambda}^{\dagger} |\dots, n_{\lambda}, \dots \rangle = (-1)^{\sum_{\nu < \lambda} n_{\nu}} \sqrt{1-n_{\lambda}} |\dots, n_{\lambda}+1, \dots \rangle \quad (\text{if } n_{\lambda}=0, \text{ else } 0) \\
& c_{\lambda} |\dots, n_{\lambda}, \dots \rangle = (-1)^{\sum_{\nu < \lambda} n_{\nu}} \sqrt{n_{\lambda}} |\dots, n_{\lambda}-1, \dots \rangle \quad (\text{if } n_{\lambda}=1, \text{ else } 0)
\end{align}
The phase factor $(-1)^{\sum_{\nu < \lambda} n_{\nu}}$ for fermions ensures that the resulting state maintains the correct antisymmetry under permutations defined by the ordering of the basis states $|\lambda\rangle$.

The \textit{number operator} for state $|\lambda\rangle$ is defined as $\hat{n}_{\lambda} = a_{\lambda}^{\dagger} a_{\lambda}$. 
Its eigenvalues are the occupation numbers $n_{\lambda}$. The total number operator $\hat{N} = \sum_{\lambda} \hat{n}_{\lambda}$ has eigenvalues equal to the total number of particles $N$ in the state.

Physical observables, represented by operators in first quantization (acting on $\mathcal{H}^{(N)}$), can be translated into the second quantization formalism (acting on $\mathcal{F}$)~\cite{Fetter_Walecka, book:modern_qch}. Since observables must be symmetric with respect to particle exchange, they typically consist of sums of one-body and two-body terms.
\begin{itemize}
    \item A one-body operator $\hat{\mathcal{O}}_1 = \sum_{i=1}^{N} \hat{o}_i$ (where $\hat{o}_i$ acts only on particle $i$) becomes:
      \begin{equation}
      \hat{\mathcal{O}}_1 = \sum_{\lambda, \mu} \langle \lambda | \hat{o} | \mu \rangle a_{\lambda}^{\dagger} a_{\mu}
      \end{equation}
      where $\langle \lambda | \hat{o} | \mu \rangle = \int \phi_{\lambda}^*(\mathbf{x}) \hat{o}(\mathbf{x}) \phi_{\mu}(\mathbf{x}) d\mathbf{x}$ are the matrix elements of the single-particle operator $\hat{o}$ in the chosen basis $\{|\lambda\rangle\}$.
    \item A two-body operator $\hat{\mathcal{O}}_2 = \frac{1}{2}\sum_{i \neq j}^{N} \hat{o}_{ij}$ (where $\hat{o}_{ij}$ acts on particles $i$ and $j$) becomes:
      \begin{equation}
      \label{eq:sq_two_body}
      \hat{\mathcal{O}}_2 = \frac{1}{2} \sum_{\lambda, \mu, \nu, \rho} \langle \lambda \mu | \hat{o}_{12} | \rho \nu \rangle a_{\lambda}^{\dagger} a_{\mu}^{\dagger} a_{\nu} a_{\rho}
      \end{equation}
      where $\langle \lambda \mu | \hat{o}_{12} | \rho \nu \rangle = \iint \phi_{\lambda}^*(\mathbf{x}_1) \phi_{\mu}^*(\mathbf{x}_2) \hat{o}_{12}(\mathbf{x}_1, \mathbf{x}_2) \phi_{\rho}(\mathbf{x}_1) \phi_{\nu}(\mathbf{x}_2) d\mathbf{x}_1 d\mathbf{x}_2$ are the two-particle matrix elements. Note the conventional order of annihilation operators $a_{\nu} a_{\rho}$, corresponding to destroying particles in states $\nu$ and $\rho$.
\end{itemize}

As a key example, consider the electronic Hamiltonian $\hat{H}_{el}$ within the BOA (Eq.~\ref{eq:molecular_tise}, electronic part only). Let $\{|\phi_p\rangle\}$ be a basis of single-electron spin-orbitals. Using fermion operators $c_p^{\dagger}, c_q$, the Hamiltonian in second quantization is~\cite{book:modern_qch, Helgaker_Jørgensen_Olsen_2014},
\begin{equation}
\hat{H}_{el} = \sum_{p,q} h_{pq} c_{p}^{\dagger} c_{q} + \frac{1}{2} \sum_{p,q,r,s} g_{pqrs} c_{p}^{\dagger} c_{q}^{\dagger} c_{s} c_{r} + V_{NN},
\end{equation}
where.
\begin{itemize}
    \item $h_{pq} = \langle \phi_p | \hat{h} | \phi_q \rangle = \int \phi_p^*(\mathbf{r}) \left( -\frac{\hbar^2}{2m_e}\nabla^2 - \sum_J \frac{Z_J e^2}{4\pi\varepsilon_0 |\mathbf{R}_J - \mathbf{r}|} \right) \phi_q(\mathbf{r}) d\mathbf{r}$ are the one-electron integrals (kinetic energy + nuclear attraction).
    \item $g_{pqrs} = \langle \phi_p \phi_q | \hat{g}_{12} | \phi_r \phi_s \rangle = \iint \phi_p^*(\mathbf{r}_1) \phi_q^*(\mathbf{r}_2) \left( \frac{e^2}{4\pi\varepsilon_0 |\mathbf{r}_1 - \mathbf{r}_2|} \right) \phi_r(\mathbf{r}_1) \phi_s(\mathbf{r}_2) d\mathbf{r}_1 d\mathbf{r}_2$ are the two-electron repulsion integrals. (This uses the common physics notation; chemists often use $\langle pq | rs \rangle$ or $(pr|qs)$ with different index conventions for the same integral).
    \item $V_{NN}$ is the classical nuclear-nuclear repulsion energy, a constant for fixed nuclear geometry.
\end{itemize}

In summary, the second quantization formalism provides a powerful and compact framework for many-body quantum mechanics. By focusing on the occupation of single-particle states within Fock space and using creation/annihilation operators whose algebraic relations encode particle statistics, it automatically handles the symmetry requirements of identical particles and simplifies the representation and manipulation of operators, proving indispensable in quantum chemistry and condensed matter physics.


\section{Jordan-Wigner Mapping}
\label{section:jordan-wigner_mapping}
The second quantization formalism (Section~\ref{sec:second_quantization}) provides a natural language for describing systems of identical fermions, such as the electrons in molecules whose behavior we often wish to simulate. 
However, current quantum computers operate on qubits, which are two-level systems analogous to spin-1/2 particles, described by Pauli operators. 
To leverage quantum algorithms like the VQE (Section~\ref{sec:vqe}) for solving fermionic problems (e.g., finding the ground state energy of a molecule), we first need a way to translate the fermionic Hamiltonian into an equivalent Hamiltonian acting on qubits. 
The Jordan-Wigner mapping~\cite{Jordan_Wigner_1928} provides a fundamental and widely used method to achieve this translation, particularly for systems that can be effectively ordered in one dimension.

\subsection{Mapping Fermions to Qubits}
The core idea is to establish a correspondence between the occupation state of a fermionic mode (orbital) and the state of a qubit. 
Let us consider $N$ fermionic modes (e.g., spin-orbitals resulting from a basis set choice in quantum chemistry), ordered $j = 1, 2, \dots, N$. 
We associate each fermionic mode $j$ with a unique qubit $j$. 
The empty state $|0\rangle_j$ of the fermionic mode is mapped to a specific qubit state, typically the $|0\rangle$ state (spin down, eigenvalue $-1$ for $\sigma^z$), and the occupied state $|1\rangle_j = c_j^\dagger |0\rangle_j$ is mapped to the qubit $|1\rangle$ state (spin up, eigenvalue $+1$ for $\sigma^z$).

The JW transformation provides an explicit mapping between the fermionic creation ($c_j^\dagger$) and annihilation ($c_j$) operators acting on mode $j$ and the Pauli operators ($\sigma_j^x, \sigma_j^y, \sigma_j^z$) acting on qubit $j$.
This mapping, is given as
\begin{align}
c_j &\leftrightarrow \left( \prod_{k=1}^{j-1} \sigma_k^z \right) \sigma_j^- \label{eq:jw_c}, \\
c_j^\dagger &\leftrightarrow \left( \prod_{k=1}^{j-1} \sigma_k^z \right) \sigma_j^+ \label{eq:jw_cdagger},
\end{align}
where $\sigma_j^\pm = \frac{1}{2}(\sigma_j^x \pm i\sigma_j^y)$. 
The crucial \textit{Jordan-Wigner string}, $\prod_{k=1}^{j-1} \sigma_k^z$, involves Pauli Z operators acting on all qubits with indices less than $j$. 
This non-local string is necessary to enforce the fermionic anticommutation relations (Eqs.~\ref{eq:car_cc},~\ref{eq:car_cdag_cdag},~\ref{eq:car_c_cdag}) using qubit operators, which naturally commute between different sites ($[\sigma_j^\alpha, \sigma_k^\beta] = 0$ for $j \neq k$). 
The string effectively accumulates the necessary $(-1)$ phase factors corresponding to anticommuting fermionic operators past occupied modes preceding site $j$. 
The fermion number operator maps locally 
\begin{equation}
    \hat{n}_j = c_j^\dagger c_j \leftrightarrow \frac{1}{2} ( \sigma_j^z + \mathbb{I} ) .
\end{equation}

\subsection{The Qubit Hamiltonian for Quantum Algorithms}
The primary utility of the JW mapping in the context of quantum computation is its ability to transform the fermionic Hamiltonian (e.g., the electronic Hamiltonian $\hat{H}_{el}$ from Section~\ref{sec:second_quantization}) into a qubit Hamiltonian $\hat{H}_q$. 
This qubit Hamiltonian acts on the Hilbert space of $N$ qubits and has the same energy spectrum as the original fermionic Hamiltonian within the corresponding Fock space sector.
The transformation is applied term by term. 
A one-body term $h_{pq} c_p^\dagger c_q$ and a two-body term $g_{pqrs} c_p^\dagger c_q^\dagger c_s c_r$ in the fermionic Hamiltonian are mapped by substituting the JW expressions (\ref{eq:jw_c},~\ref{eq:jw_cdagger}) for each creation and annihilation operator. 
After simplification using Pauli algebra relations (e.g., $(\sigma^z)^2 = \mathbb{I}$, $\sigma^+\sigma^- = (\mathbb{I}+\sigma^z)/2$, $\sigma^-\sigma^+ = (\mathbb{I}-\sigma^z)/2$, etc.), each original fermionic term becomes a sum of products of Pauli operators acting on different qubits. 
The resulting qubit Hamiltonian takes the general form
\begin{equation}
\label{eq:qubit_hamiltonian_pauli}
\hat{H}_q = \sum_{k} w_k \hat{P}_k,
\end{equation}
where each $\hat{P}_k$ is a tensor product of Pauli operators (a ``Pauli string'') acting on the $N$ qubits (e.g., $\hat{P}_k = \sigma_{i_1}^{\alpha_1} \otimes \sigma_{i_2}^{\alpha_2} \otimes \dots \otimes \sigma_{i_m}^{\alpha_m}$, with $\alpha \in \{x, y, z\}$ and identity operators implied on other qubits), and $w_k$ is a real coefficient derived from the original $h_{pq}$ or $g_{pqrs}$ integrals~\cite{seeley2012bravyi}.

\subsection{Relevance for VQE}
The qubit Hamiltonian in the Pauli string form (\ref{eq:qubit_hamiltonian_pauli}) is precisely the input required for algorithms like VQE~\cite{Peruzzo_2014, McClean_2016}. 
These algorithms aim to find the ground state energy (or low-lying excited state energies in the case of SA-VQE~\cite{PhysRevResearch.1.033062}) by variationally optimizing the parameters $\boldsymbol{\theta}$ of a parameterized quantum circuit (ansatz) $|\psi(\boldsymbol{\theta})\rangle$ to minimize the expectation value
\begin{equation}
E(\boldsymbol{\theta}) = \langle \psi(\boldsymbol{\theta}) | \hat{H}_q | \psi(\boldsymbol{\theta}) \rangle = \sum_{k} w_k \langle \psi(\boldsymbol{\theta}) | \hat{P}_k | \psi(\boldsymbol{\theta}) \rangle 
\end{equation}
(Section~\ref{sec:vqe}).
The structure of $\hat{H}_q$ obtained via the JW mapping directly impacts the implementation and resource requirements of VQE/SA-VQE:
\begin{itemize}
    \item \textbf{Measurement Cost:} Evaluating the expectation value requires measuring each Pauli string $\hat{P}_k$. The total number of terms in the sum (\ref{eq:qubit_hamiltonian_pauli}) dictates the number of distinct measurement settings needed (though techniques exist to group compatible terms)~\cite{McClean_2016}.
    \item \textbf{Non-locality Impact:} The JW strings can lead to Pauli strings $\hat{P}_k$ that act non-trivially on many qubits, even if the original fermionic interaction was local. This increases the complexity of the quantum circuits needed to prepare the ansatz state and potentially the circuits needed for measurements, especially on hardware with limited qubit connectivity.
    \item \textbf{Algorithm Performance:} The number of terms and the complexity of the Pauli strings influence the convergence properties of the classical optimization loop within VQE/SA-VQE and the susceptibility to noise in near-term quantum hardware.
\end{itemize}

While the JW mapping is conceptually straightforward and widely implemented, its non-locality has motivated the development of alternative mappings, such as the Bravyi-Kitaev transformation~\cite{bravyi2002fermionic, seeley2012bravyi}, which can sometimes result in qubit Hamiltonians with lower-weight Pauli strings, potentially reducing measurement overhead for certain problems. 
However, the choice of mapping is often problem-dependent and involves trade-offs.

In the context of optimizing molecular simulations or other fermionic problems on quantum computers using variational methods like SA-VQE, the Jordan-Wigner mapping serves as a critical initial step, translating the problem from the language of fermions into the operational language of qubits and Pauli operators, thereby defining the objective function landscape that the quantum-classical optimization seeks to navigate.

\chapter{SA-OO-VQE}
\label{chapter:saoovqe}
Quantum computing has the potential to tackle complex problems that classical computers struggle with~\cite{novak2025predicting, gupta2022quantum, yuan2024quantifying, abdel2025kodama, illesova2025qmetric, illesova2025importance, zhang2025qracle, trovato2025preliminary}, particularly in the field of chemistry.
One of the most promising applications is solving the electronic structure problem, which is crucial for understanding chemical reactions.
While traditional quantum algorithms like quantum phase estimation~\cite{Abrams_1999, O_Brien_2019} can theoretically address this issue, they often require longer circuits, making them less practical for current NISQ devices~\cite{Preskill2018quantumcomputingin, bauer2025efficient}.
Instead, variational hybrid quantum/classical algorithms, such as the VQE~\cite{Peruzzo_2014, McClean_2016}, offer a more feasible approach by using shorter circuits, albeit with more measurements.

In this context, the SA-OO-VQE~\cite{Yalouz_2021, saoovqe_art_1, beseda2024state, illesova2025transformation} was introduced, a new algorithm designed to effectively capture complex features like conical intersections—key points where the energies of different electronic states come together.
This method builds on the principles of the SA-CASSCF approach~\cite{ROOS1980157, Helgaker_Jørgensen_Olsen_2014}, allowing for a more accurate representation of excited states while remaining suitable for NISQ devices.

In the following part of this thesis, Bra-ket notation, also called Dirac notation, will be used~\cite{article_bracket}.

\section{Introduction to SA-OO-VQE}
Gaining insight into chemical reaction pathways frequently requires precise knowledge of electronic PESs~\cite{TRUHLAR20039, HOU2023295}, covering both the lowest-energy (ground) state and relevant higher-energy (excited) states.
This need becomes especially pronounced in regions where distinct PESs approach closely or intersect, creating topological features like conical intersections and avoided crossings~\cite{klessinger1995excited, RobbBernardiOlivucci+1995+783+789, CS9962500321, Domcke_Yarkony_Koppel_2004}.
These features are fundamental to photochemistry and photobiology, acting as critical mediators for important processes.
Well-known examples include the light-induced isomerization of retinal (crucial for vision), pathways for energy dissipation in photosynthesis, mechanisms conferring photostability upon DNA bases~\cite{photochemistry_DNA, doi:10.1021/ja904932x, doi:10.1073/pnas.1014982107}, and various modes of excitation energy transfer within molecular assemblies~\cite{May_Kuhn_2004}.
Consequently, the ability to accurately model these intersections is vital for predicting the behavior of molecules following light absorption.

Predicting these intricate PES features reliably, however, faces substantial theoretical and computational obstacles.
The benchmark for theoretical accuracy, FCI~\cite{TF9524800973}, considers every possible arrangement of electrons within a chosen basis set, yielding an exact solution for that basis.
Unfortunately, FCI's computational demands scale exponentially with the size of the system, making it infeasible for all but the most minimal molecules.
Therefore, practical computational studies typically employ approximations, prominently the CASCI method and its self-consistent field extension CASSCF~\cite{10.1063/5.0042147}.
These approaches lessen the computational load by concentrating on a restricted, chemically significant subset of ``active'' orbitals and electrons, where the primary electronic changes are anticipated. Nevertheless, the accuracy of CASCI/CASSCF is fundamentally linked to the suitability and completeness of this active space, the selection of which often relies on user expertise and lacks a straightforward, universally optimal procedure. Moreover, the physics near conical intersections is further complicated by the breakdown of the BOA, which separates nuclear and electronic motion.

Standard active space techniques encounter inherent difficulties when applied to the near-degenerate conditions found at conical intersections.
A straightforward CASCI calculation using a small active space, often imposed by computational resource limits, might not even qualitatively reproduce these crucial topological elements.
Furthermore, the typical starting molecular orbitals, often derived from a preliminary Hartree-Fock calculation~\cite{Bechstedt2015}, may be inadequate for representing multiple interacting electronic states with the required precision.
While state-specific CASSCF enhances CASCI by optimizing the orbitals variationally, it does so solely to lower the ground state's energy.
This focus on the ground state can unintentionally degrade the description of important excited states, potentially distorting or completely missing conical intersections by artificially inflating the energy separation between the interacting states.
Efforts to address this by optimizing orbitals for each state individually run into different problems, such as generating non-orthogonal wavefunctions and facing practical convergence difficulties known as ``root flipping'' near degeneracies, where the optimization algorithm becomes unstable as the identities of the states interchange between iterations.
Attaining the benchmark of ``chemical accuracy'' (typically defined as errors below $1.6 mHa$ or $1 kcal/mol$) for energy differences becomes exceptionally difficult in these challenging situations~\cite{Pople19991267}.

The emergence of NISQ computers presents a potentially transformative approach to the electronic structure problem, offering a way to potentially bypass the scaling limitations of classical methods.
However, contemporary quantum hardware faces significant restrictions, including limited numbers of qubits and shallow circuit depths prone to decoherence and noise.
These constraints compel quantum algorithms, like the prominent VQE~\cite{Peruzzo_2014}, to operate within comparatively small active spaces.
This limitation intensifies the challenge of active space selection.
Standard VQE algorithms, usually configured to determine the ground state energy, thus directly inherit the intrinsic shortcomings of the underlying CASCI-level description within that constrained space, potentially failing to capture the essential physics of electronic state interactions accurately.

This convergence of difficulties—the steep cost of exact classical methods, the inherent limitations of standard approximations near degeneracies, and the active space restrictions imposed by NISQ technology—highlights a compelling need for a computational strategy that can deliver a balanced and equitable (democratic) treatment of multiple electronic states concurrently.
Such a method must also function effectively within the confined active spaces accessible on current NISQ platforms.
Critically, it requires a robust mechanism for optimizing the molecular orbitals in a manner that collectively benefits all states involved, thereby enabling the faithful representation of complex PES features like conical intersections.
The established classical technique, SA-CASSCF, achieves this balanced orbital optimization by minimizing an averaged energy across the target states.
Drawing inspiration from this successful classical approach, the SA-OO-VQE algorithm was conceived. 
It seeks to harness the potential of state-averaged orbital optimization within the quantum computing framework by synergistically combining the state-preparation and measurement capabilities of VQE (run on a quantum processor) with an SA-OO procedure performed classically.
This hybrid quantum-classical methodology presents a promising pathway for exploring complex photochemical problems involving multiple electronic states using near-term quantum computers.

The SA-OO-VQE is a hybrid quantum-classical algorithm specifically designed to address the challenges outlined previously. 
Its fundamental purpose is to compute accurate energies and wavefunctions for multiple low-lying electronic states (including both ground and excited states) of a molecular system, particularly in situations involving near-degeneracies or conical intersections.
The core goals of the SA-OO-VQE approach are:
\begin{itemize}
    \item To provide a ``democratic'' description, treating all targeted electronic states on an equal footing, avoiding the bias towards the ground state inherent in state-specific methods.
    \item To leverage the capabilities of near-term quantum computers (via the VQE component) for solving the electronic structure problem within a chosen active space, while mitigating the limitations of small active spaces through orbital optimization.
    \item To incorporate the concept of state-averaged orbital optimization (inspired by classical SA-CASSCF and performed on a classical computer) to find a molecular orbital basis that provides a balanced description for all states of interest simultaneously.
    \item To enable the accurate modeling of complex PES features, such as conical intersections, which are crucial for understanding photochemical dynamics but are difficult to capture with standard VQE or state-specific classical methods, especially within the constraints of NISQ hardware.
    \item Ultimately, to pave the way for reliable quantum simulations of photochemical processes that are currently intractable for classical computers due to the complexity or size of the required active space.
\end{itemize}
By iteratively combining quantum state preparation and measurement with classical orbital optimization, SA-OO-VQE aims to achieve a level of accuracy for multiple states that would be difficult to obtain using either component in isolation, particularly for challenging systems relevant to photochemistry and materials science.

\section{Theoretical Framework}
This section delves into the fundamental theoretical concepts underpinning the SA-OO-VQE algorithm. 
To fully appreciate the hybrid nature and capabilities of SA-OO-VQE, it is essential to first understand its core components. 
We begin by outlining the basics of the VQE itself (Section~\ref{sec:vqe}), the foundational hybrid quantum-classical algorithm upon which SA-OO-VQE is built. 
VQE leverages the variational principle to approximate ground state energies using parameterized quantum circuits optimized via classical feedback. 
However, standard VQE is typically state-specific. 
To address the challenge of describing multiple electronic states, particularly near degeneracies or conical intersections, SA-OO-VQE incorporates the principles of State-Averaged methods (Section~\ref{sec:state_averaged_methods}), which aim to provide a balanced, ``democratic'' description of an ensemble of states. 
Finally, to enhance accuracy, especially within the limited active spaces often dictated by near-term quantum hardware, SA-OO-VQE employs Orbital Optimization (Section~\ref{sec:orbital_optimization}), specifically a state-averaged variant, to find a more suitable molecular orbital basis that benefits all targeted states. 
Understanding these three pillars, VQE, state averaging, and orbital optimization, provides the necessary context for the detailed description of the integrated SA-OO-VQE algorithm presented later.

\subsection{VQE Basics}
\label{sec:vqe}
The VQE~\cite{Peruzzo_2014, McClean_2016} represents a prominent hybrid quantum-classical approach tailored for finding eigenvalues, particularly the ground state energy, of a given Hamiltonian, frequently applied within quantum chemistry to tackle the electronic structure problem~\cite{kandala2017hardware}.
It has emerged as a leading candidate algorithm for achieving quantum advantage on near-term quantum processors.

Solving such problems, like finding the electronic ground state of a molecule, on a quantum computer first requires mapping the system's Hamiltonian, often initially expressed in terms of fermionic operators, onto a Hamiltonian composed of qubit operators, $\hat{H}_{q}$. 
This mapping can be achieved through various transformations, for example, Jordan-Wigner mapping (Section~\ref{section:jordan-wigner_mapping}) or Bravyi-Kitaev transformation~\cite{seeley2012bravyi, bravyi2002fermionic}, which maps $N$ spin-orbitals onto $N$ qubits, preserving the essential eigenvalue structure. 
In practice, due to resource limitations, one often works within a chemically motivated active space, leading to an effective Hamiltonian (such as the frozen core Hamiltonian, $\hat{H}^{FC}$~\cite{SEINO2014341, Yu_2021}) acting on a reduced number of qubits.
The VQE algorithm then targets the ground state energy of this effective qubit Hamiltonian, let's denote it $\hat{H}_{target}$ (which could be $\hat{H}_q$ derived from $\hat{H}^{FC}$), whose ground state energy is $E_{0, target}$.

The fundamental principle enabling VQE is the Rayleigh-Ritz variational principle~\cite{fernandez2023rayleighritzvariationalmethod}. 
It states that for any normalized trial wavefunction $|\Psi(\boldsymbol{\theta})\rangle$ parameterized by a set of variables $\boldsymbol{\theta}$, the expectation value of the target Hamiltonian $\hat{H}_{target}$ provides an upper bound to its true ground state energy $E_{0, target}$,
\begin{equation}
    \langle \Psi(\boldsymbol{\theta}) | \hat{H}_{target} | \Psi(\boldsymbol{\theta}) \rangle \ge E_{0, target}.
    \label{eq:vqe_variational_principle}
\end{equation}
VQE systematically searches for the optimal parameters $\boldsymbol{\theta}^*$ that minimize this expectation value, thereby providing the best possible approximation to $E_{0, target}$ achievable with the chosen parameterized wavefunction family.

The trial state $|\Psi(\boldsymbol{\theta})\rangle$ is prepared on a quantum computer using a parameterized quantum circuit, often referred to as the ``ansatz''. 
This circuit corresponds to a unitary operator $\hat{U}(\boldsymbol{\theta})$ applied to a readily preparable initial reference state $|\Phi_{ref}\rangle$ (typically the Hartree-Fock state in quantum chemistry applications),
\begin{equation}
    |\Psi(\boldsymbol{\theta})\rangle = \hat{U}(\boldsymbol{\theta})|\Phi_{ref}\rangle.
    \label{eq:vqe_ansatz_state}
\end{equation}
The parameters $\boldsymbol{\theta}$ are adjustable variables (e.g., rotation angles in quantum gates) within the structure of the ansatz circuit $\hat{U}(\boldsymbol{\theta})$.

The VQE algorithm operates through a hybrid quantum-classical optimization loop:
\begin{enumerate}
    \item \textbf{State Preparation (Quantum):} For a given set of parameters $\boldsymbol{\theta}$, the quantum computer executes the ansatz circuit $\hat{U}(\boldsymbol{\theta})$ acting on $|\Phi_{ref}\rangle$ to generate the trial state $|\Psi(\boldsymbol{\theta})\rangle$.
    \item \textbf{Energy Evaluation (Quantum):} The quantum computer measures the expectation value of the energy, $E(\boldsymbol{\theta}) = \langle \Psi(\boldsymbol{\theta}) | \hat{H}_{target} | \Psi(\boldsymbol{\theta}) \rangle$. Since $\hat{H}_{target}$ is typically a sum of many simple operators (e.g., Pauli strings), this involves measuring the expectation value of each term and summing them classically. Techniques like partial state tomography or specific measurement circuits for each term are employed~\cite{kandala2017hardware, huggins2022unifying}.
    \item \textbf{Parameter Update (Classical):} The computed energy $E(\boldsymbol{\theta})$ is fed as a cost function value into a classical optimization algorithm (e.g., gradient descent, conjugate gradient, Nelder-Mead, BFGS, SLSQP). The optimizer proposes a new set of parameters $\boldsymbol{\theta}'$ intended to lower the energy value.
    \item \textbf{Iteration:} The process repeats from step 1 with the updated parameters $\boldsymbol{\theta}'$. The loop continues until the energy $E(\boldsymbol{\theta})$ converges to a minimum value, ideally approximating the target ground state energy $E_{0, target}$.
\end{enumerate}

The final minimized energy $E(\boldsymbol{\theta}^*)$ represents the VQE approximation to the ground state energy $E_{0, target}$ of the Hamiltonian $\hat{H}_{target}$. It is crucial to recognize that the accuracy of this result compared to the true ground state energy of the original, full physical system depends significantly on the approximations made in deriving $\hat{H}_{target}$ (e.g., the choice of basis set and active space) and the expressiveness of the chosen ansatz $\hat{U}(\boldsymbol{\theta})$. While primarily designed for ground states, extensions to VQE exist for targeting excited states as well~\cite{higgott2019variational, jones2019variational, nause2021tutorial}.

VQE is considered a leading algorithm for near-term quantum computing, particularly in the NISQ era, due to several key advantages.
Compared to algorithms like Quantum Phase Estimation, which theoretically guarantees finding the exact eigenvalue but requires deep, fault-tolerant circuits, VQE typically demands much shallower quantum circuits.
Shorter circuit depth means fewer quantum gates, reducing the computation time and making the algorithm less susceptible to decoherence and gate errors prevalent in NISQ devices.
While VQE does not guarantee finding the exact ground state (the result depends on the expressiveness of the ansatz and the success of classical optimization), its resilience to noise and its hybrid nature, which offloads the optimization task to powerful classical computers, make it a pragmatic choice for leveraging the capabilities of current and near-future quantum hardware for problems like quantum chemistry.

\subsection{State-Averaged Methods}
\label{sec:state_averaged_methods}
State-averaged approaches provide a framework for treating multiple electronic states in a balanced manner, which is crucial for systems exhibiting near-degeneracies.
Instead of optimizing a wavefunction or orbitals for a single state (like the ground state), state-averaged methods optimize based on an ensemble energy functional. 
For a set of $N_s$ states $\{ |\Psi_k\rangle \}$, the state-averaged energy $E^{SA}$ is defined as a weighted sum of the individual state energies:
$$
E^{SA} = \sum_{k=1}^{N_s} w_k E_k = \sum_{k=1}^{N_s} w_k \langle \Psi_k | \hat{\mathcal{H}} | \Psi_k \rangle
$$
where $w_k$ are non-negative weights summing to unity ($\sum_{k=1}^{N_s} w_k = 1$). 
These weights determine the relative importance of each state in the averaging process. 
An equi-ensemble corresponds to the case where all weights are equal ($w_k = 1/N_s$). 
The optimization process then seeks to minimize this $E_{SA}$, leading to wavefunctions or orbitals that represent a compromise, providing a reasonable description for all states included in the average.
This concept is central to classical methods like SA-CASSCF and is adapted for both the VQE energy evaluation (SA-VQE~\cite{PhysRevResearch.1.033062}) and the orbital optimization (SA-OO) steps in the SA-OO-VQE algorithm.

The primary advantage of state-averaging lies in its ability to treat nearly degenerate or degenerate electronic states on an equal footing.
State-specific methods, by focusing on optimizing a single state (usually the ground state), can artificially increase the energy gap between close-lying states, potentially distorting the PES topology and failing to locate or correctly describe features like conical intersections or avoided crossings.
Furthermore, attempting to optimize degenerate states individually often leads to convergence problems known as root flipping, where the optimization algorithm oscillates unstably between the states.
State-averaging inherently avoids these issues by minimizing a single functional that depends on all states of interest.
This democratic treatment ensures that the resulting orbitals and wavefunctions provide a consistent and balanced description across the relevant manifold of states, which is essential for accurately modeling the complex dynamics occurring near PES degeneracies.

\subsection{Orbital Optimization}
\label{sec:orbital_optimization}
In electronic structure theory, MOs form the single-particle basis used to construct many-body wavefunctions (like Slater determinants or configuration interaction expansions).
The initial guess for these MOs, often obtained from a mean-field method like Hartree-Fock, provides a reasonable starting point but is generally not optimal for describing electron correlation effects or multiple electronic states accurately.
Orbital optimization refers to the process of finding a unitary transformation of the initial MO basis that minimizes the total electronic energy (for state-specific methods) or a state-averaged energy (for state-averaged methods) for a given wavefunction form (e.g., a CASCI expansion). 
By optimizing the MOs, one obtains a more compact and accurate wavefunction representation within a given level of theory (like CASCI or CASSCF). 
This process is crucial for capturing static correlation (strong electron correlation effects often present in systems with degeneracies or bond breaking) and improving the overall quality of the calculation, often significantly impacting the computed energies and properties.

The SA-OO procedure aims to minimize the state-averaged energy of a set of pre-determined correlated states, $|\Psi_k\rangle$, by performing an optimal rotation of the system's MOs.
Throughout the SA-OO process, the form of these correlated states is considered fixed.
A unitary operator $\hat{U}_{OO}(\boldsymbol{\kappa})$ is introduced to perform the orbital rotation
\begin{equation}
    \hat{U}_{OO}(\boldsymbol{\kappa}) = e^{-\hat{\kappa}} \quad \text{with} \quad \hat{\kappa} = \sum_{p>q}^{\text{MOs}} \kappa_{pq} (\hat{E}_{pq} - \hat{E}_{qp}),
\end{equation}
where $p$ and $q$ denote spatial orbitals within a chosen set, and $\kappa_{pq}$ are the orbital rotation parameters.
The operators $\hat{E}_{pq}$ are the generators of the unitary transformation. 
This operator transforms the system's full second-quantized Hamiltonian
\begin{equation}
    \hat{H}(\boldsymbol{\kappa}) = \hat{U}^{OO\dagger}(\boldsymbol{\kappa}) \hat{H} \hat{U}^{OO}(\boldsymbol{\kappa}).
\end{equation}
The state-averaged energy with respect to the transformed frozen core Hamiltonian, $\hat{H}_{FC}(\boldsymbol{\kappa})$, is defined as
\begin{equation}
\label{eq:e_saoo}
    E^{SA-OO}(\boldsymbol{\kappa}) = \sum_k w_k \langle\Psi_k|\hat{H}_{FC}(\boldsymbol{\kappa})|\Psi_k\rangle ,
\end{equation}

where $w_k$ is the weight associated with state k. 
The SA-OO procedure seeks to find the optimal parameters $\boldsymbol{\kappa}$ that minimize this state-averaged energy. 
This optimization typically employs a state-averaged version of the classical Newton-Raphson method~\cite{GALLARDOALVARADO202277}, utilizing the one- and two-electron reduced density matrices of the correlated states to analytically compute the orbital gradient and Hessian, which guide the optimization process.

The SA-OO component within the SA-OO-VQE algorithm is directly analogous to the orbital optimization performed in classical SA-CASSCF. 
Both methods aim to find the optimal MO basis by minimizing a state-averaged energy functional. 
They typically employ iterative techniques, often based on the Newton-Raphson method, which require the calculation of the energy gradient and Hessian with respect to orbital rotation parameters. 
These gradients and Hessians depend on the one- and two-electron reduced density matrices of the state(s) involved. 
In classical SA-CASSCF, the wavefunctions and their RDMs are computed using classical diagonalization or iterative techniques within the active space. 
In SA-OO-VQE, the crucial difference is that the required RDMs for the SA-OO step are obtained from measurements performed on the quantum states prepared by the SA-VQE component on the quantum computer~\cite{Mao2025_rdm, krompiec2022stronglycontractednelectronvalence, PhysRevResearch.3.033230}. 
The subsequent computationally intensive task of solving the Newton-Raphson equations to update the orbitals is then performed classically. 
This contrasts with state-specific CASSCF, which optimizes orbitals for a single state's energy, and standard VQE, which typically uses fixed (e.g., Hartree-Fock) orbitals without optimization.

\section{Algorithm Components}
Having established the theoretical underpinnings, this section details the specific components and procedures that constitute the SA-OO-VQE algorithm.

\subsection{State-Averaged VQE Implementation}
\label{sec:sa_vqe}
The core quantum computation within SA-OO-VQE involves a state-averaged variant of the VQE algorithm, which consists of finding two (or more) low-lying eigenstates of a given Hamiltonian.
The quantum circuit implements the chosen wavefunction ansatz, parameterized by $\boldsymbol{\theta}$.
The energy expectation value for each state, $\langle \Psi_k(\boldsymbol{\theta}) | \hat{\mathcal{H}} | \Psi_k(\boldsymbol{\theta}) \rangle$, is obtained by decomposing the qubit Hamiltonian $\hat{\mathcal{H}}_q$ into a sum of Pauli strings and measuring the expectation value of each string on the state prepared by the quantum circuit. 
These measurements are repeated sufficiently often to obtain statistically reliable estimates.
Additionally, measurements required to reconstruct the 1- and 2-RDMs of each state $|\Psi_k(\boldsymbol{\theta})\rangle$ are performed, as these are needed for the subsequent classical orbital optimization step.

\subsection{Wavefunction Ansatz}
\label{section:wafefunction_ansatz}
In VQE-type algorithms, such as the SA-VQE, the selection of a specific, parameterized quantum circuit, referred to as the wavefunction ansatz $\hat{U}(\boldsymbol{\theta})$, is crucial.
This choice dictates the region of the Hilbert space that can be explored during the variational search for the desired quantum state(s) and significantly impacts the algorithm's performance, accuracy, and resource requirements~\cite{cerezo2021variational}.
There are broadly two philosophies for designing ansätze, chemically inspired and hardware-efficient.

One prominent chemically-inspired ansatz class, particularly relevant for quantum chemistry applications, is the UCC method~\cite{romero2018strategies}. 
UCC translates the highly successful classical Coupled Cluster theory into a unitary form suitable for quantum computers. 
A common variant is UCCSD~\cite{Peruzzo_2014, Anand_2022}, which incorporates single and double electron excitations from a reference state (like Hartree-Fock) to approximate the true ground state wavefunction.
\begin{equation}
    |\Psi_{UCC}(\boldsymbol{\theta})\rangle = e^{\hat{T}(\boldsymbol{\theta}) - \hat{T}^\dagger(\boldsymbol{\theta})} |\Phi_{ref}\rangle
\end{equation}
where $\hat{T}(\boldsymbol{\theta}) = \sum_i \theta_i \hat{\tau}_i$ is the cluster operator containing excitation operators $\hat{\tau}_i$ (like single $a^\dagger_i a_j$ and double $a^\dagger_i a^\dagger_j a_k a_l$ excitations after mapping to qubits) weighted by variational parameters $\theta_i$. While UCCSD offers a systematic way to capture electron correlation and can potentially reach high chemical accuracy, implementing the full UCCSD ansatz often results in deep quantum circuits with many gates, posing a challenge for current NISQ devices~\cite{McClean_2016}.

Alternatively, hardware-efficient ansätze are designed with the specific limitations and connectivity of the target quantum hardware in mind~\cite{kandala2017hardware}. 
These ansätze typically consist of repeating layers of parameterized single-qubit rotations (e.g., $R_y(\theta), R_z(\phi)$) and fixed blocks of entangling gates (e.g., CNOTs or ECRs) arranged according to the device's qubit coupling map. 
For example, a layer might look like:
\begin{equation}
    \hat{U}_{layer}(\boldsymbol{\theta}_l) = \hat{U}_{ent} \prod_{i=1}^{N} \hat{R}_x(\theta_{l,i}^{(1)}) \hat{R}_z(\theta_{l,i}^{(2)})
\end{equation}
where $\hat{U}_{ent}$ represents the entangling block for layer $l$. 
Hardware-efficient ansätze often lead to shallower circuits compared to UCC for a similar number of parameters, making them more amenable to NISQ hardware. 
However, they may lack direct chemical intuition, can be harder to initialize effectively, and might be more susceptible to encountering issues like barren plateaus (regions where the cost function gradient vanishes exponentially with the number of qubits), potentially hindering the optimization process~\cite{mcclean2018barren}.

The choice of ansatz thus involves a trade-off between theoretical accuracy guarantees, expressibility within the target Hilbert space, and practical implementability on available quantum hardware.
Selecting an appropriate ansatz remains a critical area of research in the field of variational quantum algorithms~\cite{cerezo2021variational, Anand_2022}.

\subsection{Handling Multiple States}
The SA-VQE procedure starts with building a frozen Hamiltonian $H^{FC}$ and a set of $N$ mutually orthogonal initial states $\{|\Phi_k\rangle\}$.
For instance, for two states, one might use the Hartree-Fock determinant $|\Phi_A\rangle = |HF\rangle$ and a singly excited determinant $|\Phi_B\rangle = \hat{E}_{LH}|HF\rangle$, where $\hat{E}_{LH}$ excites an electron from the HOMO (H) to the LUMO (L). 
Crucially, the same parameterized unitary ansatz $\hat{U}(\boldsymbol{\theta})$ is applied to each initial state to generate the corresponding trial states
\begin{equation}
    |\Psi_k(\boldsymbol{\theta})\rangle = \hat{U}(\boldsymbol{\theta})|\Phi_k\rangle .
\end{equation} 
Since $\hat{U}(\boldsymbol{\theta})$ is unitary, the resulting trial states $\{|\Psi_k(\boldsymbol{\theta})\rangle\}$ remain orthogonal.
Using the same unitary for all the states also simplifies the implementation, even more in the current NISQ era. 
The energy $E_k(\boldsymbol{\theta})$ is measured for each state $|\Psi_k(\boldsymbol{\theta})\rangle$,
\begin{equation}
    E_k = \langle \Psi_k(\boldsymbol{\theta}) | \hat{\mathcal{H}}^{FC} | \Psi_k(\boldsymbol{\theta}) \rangle .
\end{equation}
These individual energies are then combined into the state-averaged cost function 
\begin{equation}
\label{eq:e_savqe}
    E^{SA-VQE}(\boldsymbol{\theta}) = \sum_{k=1}^N w_k E_k(\boldsymbol{\theta}) ,
\end{equation}
which is minimized by the classical optimizer with respect to $\boldsymbol{\theta}$. 
An equi-ensemble ($w_k = 1/N$) is often used to ensure democratic treatment.

\subsection{Integration of Orbital Optimization Procedure with VQE}
The classical part of the hybrid algorithm focuses on optimizing the molecular orbital basis in a state-averaged manner, using information obtained from the quantum component.
The SA-OO step usually employs a state-averaged Newton-Raphson algorithm to find the optimal orbital rotation parameters $\boldsymbol{\kappa}$ that minimize the state-averaged energy $E^{SA-OO}(\boldsymbol{\kappa})$ (Eq.~\ref{eq:e_saoo}), keeping the VQE parameters $\boldsymbol{\theta}$ fixed from the previous SA-VQE step. 
The process is iterated until the orbital rotation parameters $\boldsymbol{\kappa}$ converge according to a predefined threshold (e.g., $10^{-4}$ Ha change in the state-averaged energy for the global SA-OO-VQE loop).

SA-OO-VQE operates through a macro-iteration cycle that alternates between the quantum SA-VQE and classical SA-OO steps, as depicted schematically in Figure~\ref{fig:saoovqe}.
\colorlet{vqeColor}{red!60!black}
\colorlet{ooColor}{blue!50!black!75}
\colorlet{vqeColorDark}{vqeColor!70!black}
\colorlet{ooColorDark}{ooColor!70!black}
\colorlet{arrowColorVQE}{red!70!black}
\colorlet{arrowColorOO}{blue!60!black!80}
\colorlet{arrowColorConnect1}{red!70!black} 
\colorlet{arrowColorConnect2}{gray!70!black} 
\colorlet{labelBgColor}{white!100}
\colorlet{cycleFillColor}{gray!20} 
\colorlet{circuitFillColor}{gray!20} 

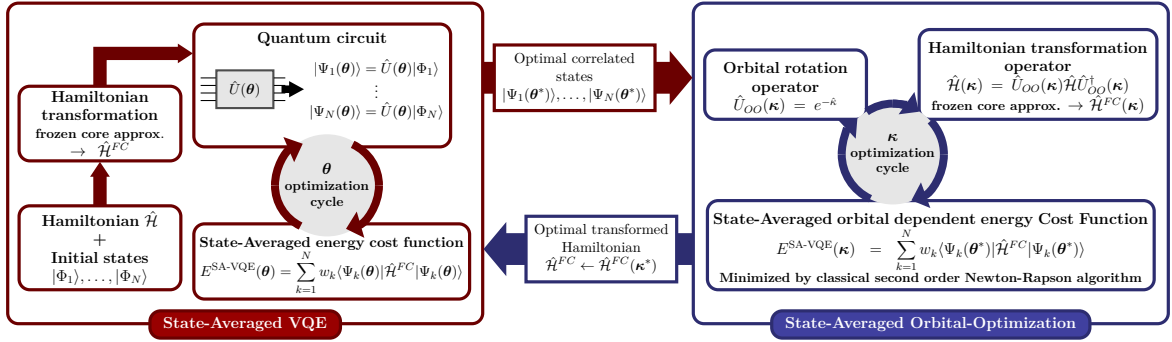
\begin{figure}[!hbpt]
    \centering
\begin{tikzpicture}[
    scale=0.5, transform shape, 
    node distance=15mm and 25mm, 
    arrow/.style={ 
        -{Stealth[length=4mm, width=3mm]}, 
        very thick 
    },
    main_arrow/.style={ 
        -{Triangle[length=3mm, width=20pt]}, 
        line width=12pt 
    },
    sectionlabel/.style={ 
        rectangle,
        draw,
        very thick, 
        rounded corners,
        font=\bfseries, 
        inner ysep=5pt, 
        inner xsep=10pt, 
        text=white 
    },
    mainbox/.style={ 
        draw,
        very thick, 
        rounded corners,
        minimum width=12.5cm,  
        minimum height=8.5cm, 
        inner sep=8mm 
    },
    arrowlabel/.style={ 
        rectangle,
        fill=labelBgColor, 
        inner sep=6pt, 
        text=black, 
        align=center,
        very thick
    },
    block/.style={ 
        rectangle,
        draw=vqeColorDark, 
        fill=labelBgColor,   
        very thick,
        text centered,
        rounded corners,
        minimum height=1.5cm,
        text width=4.5cm, 
        align=center,
        text=black 
    },
    fitBlock/.style={
        rectangle,
        draw=vqeColorDark,
        fill=labelBgColor,
        very thick,
        rounded corners,
        inner sep=10pt 
    },
    internalBlock/.style={
        rectangle,
        draw=none, 
        fill=none, 
        minimum height=1.5cm,
        align=center,
        text=black
    },
    vqe_block/.style={ 
        rectangle,
        draw=vqeColorDark, 
        fill=labelBgColor,   
        very thick,
        rounded corners,
        minimum height=1.5cm, 
        text width=4cm,     
        align=center,
        text=black
    },
    circuitBlock/.style={ 
        rectangle,
        draw=black!70,
        fill=circuitFillColor,
        thick,
        minimum width=1.5cm,
        minimum height=1cm
    },
    circ/.style={ 
        fill=cycleFillColor,
        circle,
        inner sep=2pt,
        font=\small,
        align=center,
        thick
    },
    vqe_arrow/.style={ 
        -{Triangle[length=2mm, width=7pt]}, 
        line width=5pt,
        draw=vqeColorDark
    },
    circuitArrow/.style={ 
        -{Triangle[length=2mm, width=2mm]},
        line width=1mm,
        draw=black
    },
    vqe_cycle_arrow/.style={ 
        -{Triangle[length=2mm, width=3.5mm]},
        line width=3pt,
        draw=vqeColorDark 
    },
    vqe_cycle_line/.style={ 
        -,
        line width=3pt,
        draw=vqeColorDark 
    },
    oo_cycle_arrow/.style={ 
        -{Triangle[length=2mm, width=3.5mm]},
        line width=3pt,
        draw=ooColorDark 
    },
    oo_cycle_line/.style={ 
        -,
        line width=3pt,
        draw=ooColorDark 
    },
    oo_arrow/.style={ 
        -{Triangle[length=2mm, width=7pt]}, 
        line width=5pt,
        draw=ooColorDark
    },
    oo_block/.style={ 
        rectangle,
        draw=ooColorDark, 
        fill=labelBgColor,   
        very thick,
        rounded corners,
        minimum height=1.5cm, 
        text width=4cm,     
        align=center,
        text=black
    },
]


\node[mainbox, draw=vqeColorDark] (vqe_box) {}; 
\node[sectionlabel, fill=vqeColor, draw=vqeColorDark, anchor=center, at=(vqe_box.south)] {\large State-Averaged VQE};

\node[mainbox, draw=ooColorDark, right=55mm of vqe_box] (oo_box) {}; 
\node[sectionlabel, fill=ooColor, draw=ooColorDark, anchor=center, at=(oo_box.south)] {\large State-Averaged Orbital-Optimization};

\coordinate (oo_initial_pos) at ($(oo_box.south) + (0, 0.9cm)$);
\node[oo_block, anchor=south, at=(oo_initial_pos), text width=11.5cm] (oo_initial) {\large \textbf{\\State-Averaged orbital dependent energy Cost Function} \\
$\displaystyle E^{\text{SA-VQE}}(\boldsymbol{\kappa}) = \sum_{k=1}^N w_k \langle \Psi_k(\boldsymbol{\theta}^*) | \hat{\mathcal{H}}^{FC} | \Psi_k(\boldsymbol{\theta}^*) \rangle$\\
{\small \textbf{Minimized by classical second order Newton-Rapson algorithm}\\}
};

\coordinate (oo_ht_cords) at ($(oo_box.north east) + (-0.3, -0.7cm)$);
\node[oo_block, anchor=north east, at=(oo_ht_cords), align=center, text width=6cm] (oo_ht) {\large\newline \textbf{Hamiltonian transformation} \\
\textbf{operator}\\
$\hat{\mathcal{H}}(\boldsymbol{\kappa}) =  \hat{U}_{OO} (\boldsymbol{\kappa}) \hat{\mathcal{H}} \hat{U}^\dagger_{OO}(\boldsymbol{\kappa})$\\
{\small \textbf{frozen core approx.}} $\rightarrow \hat{\mathcal{H}}^{FC}(\boldsymbol{\kappa}) $
};

\coordinate (oo_oro_cords_side) at ($(oo_box.north west) + (0.3, -0.7cm)$);
\coordinate (oo_oro_cords_top) at (oo_ht.south west);
\coordinate (oo_oro_cords) at ( oo_oro_cords_top -| oo_oro_cords_side);
\node[oo_block, anchor=south west, at=(oo_oro_cords), align=center] (oo_oro) {\large\newline \textbf{Orbital rotation} \\
\textbf{operator}\\
$\hat{U}_{OO} (\boldsymbol{\kappa}) = e^{-\hat{\kappa}}$\\
};

\coordinate (initial_pos) at ($(vqe_box.south west) + (0.3, 0.9cm)$);
\node[vqe_block, anchor=south west, at=(initial_pos)] (initial) {\large \textbf{Hamiltonian} $\hat{\mathcal{H}}$ \\ \textbf{+} \\ \textbf{Initial states} \\ $|\Phi_1\rangle, \ldots, |\Phi_N\rangle$};

\coordinate (h_transform_pos) at ($(initial.north west) + (0cm, 1.15cm)$);
\node[vqe_block, anchor=south west, at=(h_transform_pos)] (h_transform) {\textbf{\large Hamiltonian transformation} \\ \textbf{\small frozen core approx}. \\ \large$\rightarrow \hat{\mathcal{H}}^{FC}$};

\coordinate (qc_outer_pos) at ($(h_transform.north east) + (0.3cm, 0)$);
\node[vqe_block, text width=6.5cm, text height=3cm, align=center, anchor=west, at=(qc_outer_pos)] (quantum_circuit_outer) {};

\coordinate (qc_schema) at ($(quantum_circuit_outer.west) + (0.6cm, 0)$);
\node[circuitBlock, below=1mm of qc_schema, anchor=west] (circuit_schematic) {$\hat{U}(\boldsymbol{\theta})$};

\foreach \y in {-0.3, -0.1, 0.1, 0.3} {
    \draw[thin] ($(circuit_schematic.west) + (0, \y)$) -- ++(-0.4, 0); 
    \draw[thin] ($(circuit_schematic.east) + (0, \y)$) -- ++(0.3, 0);
}
\coordinate (circ_arr_cord) at ($(circuit_schematic.east) + (0.2cm, 0)$);
\draw[circuitArrow] (circ_arr_cord) -- ++(0.6, 0); 

\node[above=-1mm of quantum_circuit_outer.north, anchor=north, align=center] {
    \large\textbf{Quantum circuit}};

\node[right=8mm of circuit_schematic, anchor=west, align=center] {
    $|\Psi_1(\boldsymbol{\theta})\rangle = \hat{U}(\boldsymbol{\theta})|\Phi_1\rangle$ \\
    $\vdots$\\
    $|\Psi_N(\boldsymbol{\theta})\rangle = \hat{U}(\boldsymbol{\theta})|\Phi_N\rangle$
};

\coordinate (cost_vqe_pos) at ($(quantum_circuit_outer.south west) + (0, -2cm)$);
\node[vqe_block, text width=7cm, anchor=north west, at=(cost_vqe_pos)] (cost_vqe) {\\\textbf{State-Averaged energy cost function} \\
$\displaystyle E^{\text{SA-VQE}}(\boldsymbol{\theta}) = \sum_{k=1}^N w_k \langle \Psi_k(\boldsymbol{\theta}) | \hat{\mathcal{H}}^{FC} | \Psi_k(\boldsymbol{\theta}) \rangle$};

\draw[vqe_arrow] (initial) -- (h_transform); 

\coordinate (qc_input) at ($(quantum_circuit_outer.north west) + (0cm, -0.8)$);
\draw[vqe_arrow] (h_transform.north) |- (qc_input);


\coordinate (cycle_center) at ($(quantum_circuit_outer.south)!0.5!(cost_vqe.north)$); 

\node[circ, at=(cycle_center)] (theta_cycle) {\large $\boldsymbol{\theta}$\\\textbf{optimization}\\\textbf{cycle}};

\coordinate (arc_center_right) at (theta_cycle.east);
\coordinate (arc_center_left) at (theta_cycle.west);
\draw[vqe_cycle_arrow] (arc_center_left) arc (-180:-247:1.2cm);
\draw[vqe_cycle_line] (arc_center_left) arc (-180:-116:1.2cm);
\draw[vqe_cycle_line] (arc_center_right) arc (0:65:1.2cm);
\draw[vqe_cycle_arrow] (arc_center_right) arc (0:-74:1.2cm);

\coordinate (oo_cycle_center_side) at ($(oo_ht.south west)!0.5!(oo_oro.south east)$);
\coordinate (oo_cycle_center_top) at ($(oo_ht.south)!0.5!(oo_initial.north) + (0,0.15cm)$);
\coordinate (oo_cycle_center) at (oo_cycle_center_side |- oo_cycle_center_top);
\node[circ, at=(oo_cycle_center)] (kappa_cycle) {\large $\boldsymbol{\kappa}$\\\textbf{optimization}\\\textbf{cycle}};

\coordinate (oo_arc_center_right) at (kappa_cycle.east);
\coordinate (oo_arc_center_left) at (kappa_cycle.west);
\coordinate (oo_arc_center_top) at (kappa_cycle.north);
\draw[oo_cycle_arrow] (oo_arc_center_left) arc (-180:-227:1.3cm);
\draw[oo_cycle_line] (oo_arc_center_left) arc (-180:-115:1.3cm);
\draw[oo_cycle_line] (oo_arc_center_right) arc (0:43:1.3cm);
\draw[oo_cycle_arrow] (oo_arc_center_right) arc (0:-70:1.3cm);
\draw[oo_cycle_line] (oo_arc_center_top) arc (90:125:1.2cm);
\draw[oo_cycle_arrow] (oo_arc_center_top) arc (90:55:1.2cm);
\draw[oo_cycle_line] (oo_arc_center_top) arc (90:66:1.2cm);

\coordinate (vqe_out) at ($(vqe_box.north east) + (0cm, -2cm)$); 
\coordinate (oo_in) at ($(oo_box.north west) + (0cm, -2cm)$);    
\coordinate (oo_out) at ($(oo_box.south west) + (0cm, 2cm)$);    
\coordinate (vqe_in) at ($(vqe_box.south east) + (0cm, 2cm)$);   

\draw[main_arrow, draw=vqeColorDark] (vqe_out) -- (oo_in) node[arrowlabel, draw=vqeColorDark, midway, anchor=center, xshift=-10pt] {Optimal correlated\\ states\\ $|\Psi_1(\boldsymbol{\theta}^*)\rangle, \ldots, |\Psi_N(\boldsymbol{\theta}^*)\rangle $};

\draw[main_arrow, draw=ooColorDark] (oo_out) --  (vqe_in) node[arrowlabel, draw=ooColorDark, midway, anchor=center, xshift=+10pt] {Optimal transformed\\ Hamiltonian\\
$\displaystyle
    \hat{\mathcal{H}}^{FC} \leftarrow \hat{\mathcal{H}}^{FC} (\boldsymbol{\kappa}^*)
$
};

\end{tikzpicture}

\caption[SA-OO-VQE scheme.]{The SA-OO-VQE method, depicted schematically based on the diagram in~\cite{Yalouz_2021}, treats multiple states (e.g., ground and several excited states) on an equal footing through an iterative cycle. 
It combines: (Left) A hybrid quantum-classical SA-VQE step, where the parameters of a single, shared quantum circuit are optimized to reduce the average energy of the resulting correlated wavefunctions for the selected states.
(Right) A classical SA-orbital-optimization (SA-OO) step, which takes the correlated states from SA-VQE and performs an optimal molecular orbital rotation on the Hamiltonian $\hat{H}$, further minimizing the state-averaged energy.
The updated Hamiltonian is returned to the SA-VQE step, completing the loop.
}
    \label{fig:saoovqe}
\end{figure}
\begin{enumerate}
    \item 
    \label{step:saoovqe_1}
    \textbf{SA-VQE Step:} For a given set of MOs defined by parameters $\boldsymbol{\kappa}_i$, the SA-VQE algorithm optimizes the ansatz parameters $\boldsymbol{\theta}$ to minimize $E^{SA-VQE}(\boldsymbol{\theta})$ (Eq.~\ref{eq:e_savqe}).
    \item \textbf{Information Transfer:} Once converged, the optimal states $|\Psi_k(\boldsymbol{\theta}^*)\rangle$ are used to measure the necessary 1- and 2-RDMs on the quantum computer. These RDMs are passed to the classical computer.
    \item \textbf{SA-OO Step:} Using the RDMs, the classical computer performs the state-averaged orbital optimization (e.g., Newton-Raphson) to find the optimal orbital rotation parameters $\boldsymbol{\kappa}_{i+1}$ that minimize $E^{SA-OO}(\boldsymbol{\kappa})$ (Eq.~\ref{eq:e_saoo}).
    \item \textbf{Hamiltonian Update:} The MO coefficients and consequently the one- and two-electron integrals defining the Hamiltonian are updated based on the new rotation $\boldsymbol{\kappa}_{i+1}$. 
    This yields the updated Hamiltonian $\hat{\mathcal{H}}^{FC}(\boldsymbol{\kappa}_{i+1})$ for the next cycle.
    \item \textbf{Iteration:} The updated Hamiltonian is fed back into the SA-VQE step (Step~\ref{step:saoovqe_1}), and the cycle repeats.
\end{enumerate}
This entire process continues until a global convergence criterion on the state-averaged energy is met.

\chapter{Differential Evolution}
\label{chapter:DE}
In the realm of optimization, local and global methods serve as two fundamental approaches to finding optimal solutions.
Local optimization techniques, such as gradient descent, focus on refining solutions within a limited neighborhood, often leading to quick convergence but risking entrapment in local minima.
In contrast, global optimization methods aim to explore the entire solution space, seeking the best possible outcome regardless of initial conditions.
Among these global strategies, DE~\cite{de2005a, AHMAD20223831, Neri2010, DAS20161, 5601760} stands out as a powerful biological algorithm inspired by the principles of natural selection and evolution.
By mimicking the processes of mutation, crossover, and selection, DE effectively navigates complex landscapes, adapting and evolving candidate solutions over generations to converge on optimal or near-optimal results.
This chapter delves into the mechanics of DE, exploring its unique features and the advantages it offers in tackling challenging optimization problems.

DE is a robust optimization algorithm that operates on a population of candidate solutions, evolving them over successive generations to find optimal solutions in complex search spaces.
The process begins with the initialization of a diverse population of D-dimensional vectors, where each individual represents a potential solution composed of real numbers.
In each iteration, DE employs a mutation strategy to create new candidate solutions by combining existing ones, introducing variability that allows exploration of the solution space.
This is followed by a crossover operation, where the mutated candidates are blended with the original solutions to generate offspring that inherit characteristics from both parents.
The selection phase then determines which individuals will survive to the next generation, favoring those with better fitness values while maintaining diversity within 
the population.
The algorithm continues to iterate through these steps until a predefined termination criterion is met, such as reaching a maximum number of generations or achieving a satisfactory fitness level.
Each of these components plays a crucial role in the effectiveness of DE, which will be examined in greater detail in the following sections.

\section{Population Structure}
In its most flexible implementation, DE operates with two populations of vectors, each containing $ N_p $ D-dimensional vectors characterized by real-valued parameters.
The current population in generation $g$, referred to as $ P^{(g)}_x $, consists of vectors $ x^{(g)}_{i} $ that have been recognized as viable candidates, either as initial solutions or through comparative evaluations with other vectors
\begin{equation}
P^{(g)}_{x} = \left(x^{(g)}_{i}\right), \quad i = 0, 1, \ldots, N_p - 1, \quad g = 0, 1, \ldots, g_{\text{max}},
\end{equation}
where
\begin{equation}
x^{(g)}_{i} = (x^{(g)}_{i,j}), \quad j = 0, 1, \ldots, D - 1.
\end{equation}
The indexing begins at 0 to simplify the handling of arrays and modular arithmetic. 
The generation index $ g = 0, 1, \ldots, g_{\text{max}} $ indicates the generation to which each vector belongs, while the population index $ i $ ranges from 0 to $ N_p - 1 $. 
The parameters within each vector are indexed by $ j $, which spans from $0$ to $ D - 1 $.

Once the initialization is complete, DE proceeds to create an intermediary population $ P^{(g)}_{v} $ consisting of $ N_p $ mutant vectors $ v^{(g)}_{i} $ through the mutation of randomly selected vectors,
\begin{equation}
P^{(g)}_{v} = \left(v^{(g)}_{i}\right), \quad i = 0, 1, \ldots, N_p - 1, \quad g = 0, 1, \ldots, g_{\text{max}},
\end{equation}
where
\begin{equation}
v^{(g)}_{i} = \left(v^{(g)}_{i,j}\right), \quad j = 0, 1, \ldots, D - 1.
\end{equation}
Following this, each vector from the current population is recombined with a mutant vector to form a trial population $ P^{(g)}_{u} $ comprising $ N_p $ trial vectors $ u^{(g)}_{i} $,
\begin{equation}
P^{(g)}_{u} = \left(u^{(g)}_{i}\right), \quad i = 0, 1, \ldots, N_p - 1, \quad g = 0, 1, \ldots, g_{\text{max}},
\end{equation}
where
\begin{equation}
u^{(g)}_{i} = \left(u^{(g)}_{i,j}\right), \quad j = 0, 1, \ldots, D - 1.
\end{equation}
During the recombination phase, the trial vectors replace the mutant population, allowing both populations to be stored within a single array.

\section{Initialization}
Multiple different initialization methods exist for evolutionary algorithms~\cite{BAJER2016294, inproceedings, POIKOLAINEN2015216}.
For the purpose of this thesis, only one of the simplest initializations based on uniform or normal random distribution will be explained, together with the idea of how to deal with population initialization in the case where no boundaries are set.

Before initializing the population, it is essential to define both the upper and lower bounds for each parameter.
These two sets of values can be organized into two D-dimensional initialization vectors, $ b_L $ and $ b_U $, where the subscripts $ L $ and $ U $ denote the lower and upper bounds, respectively.
If no lower bound is specified, the algorithm defaults to using the smallest possible floating-point value as the lower bound.
Similarly, if no upper bound is provided, the largest possible floating-point value is used as the upper bound.
This ensures that the initialization process remains feasible even in the absence of explicit bounds.

Once the initialization bounds are established, a random number generator assigns a value to each parameter of every vector within the specified range.
The generator can utilize either a uniform distribution or a normal distribution, depending on the desired characteristics of the initial population.
For instance, if a uniform distribution is employed, the initial value (when $ g = 0 $) of the $ j $-th parameter of the $ i $-th vector can be expressed as
\begin{equation}
x^{(0)}_{i,j} = \text{rand}_j(0, 1) \cdot (b_{U,j} - b_{L,j}) + b_{L,j}.
\end{equation}
In this case, the random number generator, $ \text{rand}_j(0,1) $, produces a uniformly distributed random number within the interval $[0,1)$, meaning $ 0 \leq \text{rand}_j(0,1) < 1 $.
Alternatively, if a normal distribution is used, the initial values can be generated based on a specified mean and standard deviation, ensuring that the values are centered around the mean while still adhering to the defined bounds.
The subscript $ j $ indicates that a new random value is generated for each parameter.

\section{Mutation}
Mutation is a fundamental operation in DE that enhances the genetic diversity of the population, thereby facilitating effective exploration of the solution space. 
The core principle of mutation involves the generation of mutant vectors by perturbing existing candidate solutions, which introduces variability that aids in escaping local optima. 
In DE, mutation is typically executed by selecting a target vector $x^{(g)}_{i} $ from the current population and creating a mutant vector $v^{(g)}_{i} $ through the addition of scaled differences between randomly selected vectors.
Several mutation strategies, including \textit{DE/rand/1}, \textit{DE/best/1}, and \textit{DE/current-to-random/1}, are employed within the DE framework, each offering distinct advantages tailored to various optimization scenarios~\cite{de_art_1, de_art_2, OPARA201853}. 
These strategies will be examined in greater detail in the following sections, elucidating their specific mechanisms and applications in enhancing the performance of the DE algorithm.

\subsection{DE/Rand/1}
\begin{figure}[!ht]
    \centering
    \begin{tikzpicture}[scale=0.9]

  \draw[->] (0,0) -- (9,0) node[below] {$x_0$};
  \draw[->] (0,0) -- (0,6) node[left] {$x_1$};

  \coordinate (x0g) at (2,2.5);
  \coordinate (xr1g) at (3,3.2);
  \coordinate (fdiff) at (3.65, 2.5);
  \coordinate (xr2g) at (5,1);
  \coordinate (vig) at (1,4); 

  \filldraw (x0g) circle (2.5pt) node[left] {$\mathbf{x}^{(g)}_{r_0}$};
  \filldraw (xr1g) circle (2.5pt) node[above right] {$\mathbf{x}^{(g)}_{r_1}$}; 
  \filldraw (xr2g) circle (2.5pt) node[below right] {$\mathbf{x}^{(g)}_{r_2}$}; 
  \filldraw (vig) node[above right] {$\mathbf{v}^{(g)}_{i}= \mathbf{x}^{(g)}_{r_0} + F \cdot \left(\mathbf{x}^{(g)}_{r_1} - \mathbf{x}^{(g)}_{r_2}\right)$};

  \draw[-stealth, line width = 1.4] (x0g) -- (vig);
  \draw[-, line width = 0.6] (fdiff) -- (xr1g);
  \draw[-stealth, line width = 1.4] (xr2g) -- (fdiff); 

  \draw[dashed] (fdiff) -- (vig);
  \draw[dashed] (xr2g) -- (x0g);

  \node[right] at (4.4, 1.8) {$F \cdot \left(\mathbf{x}^{(g)}_{r_1} - \mathbf{x}^{(g)}_{r_2}\right)$}; 

\end{tikzpicture}
    \caption[DE mutation strategy rand/1.]{Differential mutation rand/1: the weighted differential, $F \cdot \left(\mathbf{x}^{(g)}_{r1} - \mathbf{x}^{(g)}_{r2}\right)$, is added to the base vector, $\mathbf{x}^{(g)}_{r0}$, to produce a mutant, $\mathbf{v}^{(g)}_{i}$.}
    \label{fig:de_mutation}
\end{figure}
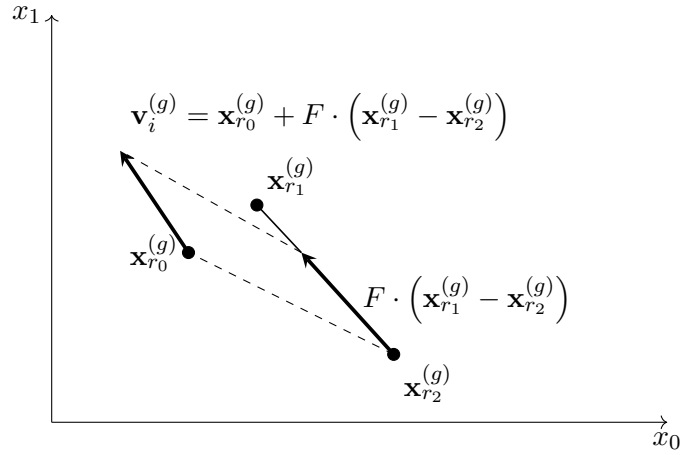

The \textit{rand/1} mutation strategy is a fundamental component of DE, which enhances the diversity of the population by introducing variability through the combination of randomly selected vectors. 
In this approach, a mutant vector $ v^{(g)}_{i} $ is generated by adding a scaled difference between two randomly chosen vectors to a third randomly selected base vector. 
The mathematical representation of this process is given by Equation
\begin{equation}
v^{(g)}_{i} = x^{(g)}_{r_0} + F \cdot \left(x^{(g)}_{r_1} - x^{(g)}_{r_2}\right).
\label{eq:mutation}
\end{equation}
In this equation, $ x^{(g)}_{r_0} $ serves as the base vector, while $ x^{(g)}_{r_1} $ and $ x^{(g)}_{r_2} $ are the difference vectors.
The scale factor $ F $, which lies within the range $ (0, 1) $, plays a crucial role in determining the extent of the mutation.
It is a positive real number that influences the rate of evolution within the population.
Although there is no theoretical upper limit for $ F $, practical applications typically utilize values that do not exceed 1 to maintain effective mutation rates.
Graphical demonstration of this mutation strategy for two-dimensional vectors can be seen in Figure~\ref{fig:de_mutation}.

The selection of the base vector index $ r_0 $ is performed randomly, ensuring that it differs from the target vector index $ i $.
Similarly, the indices for the difference vectors $ r_1 $ and $ r_2 $ are also chosen randomly, with the stipulation that they must be distinct from one another and the base and target vector indices.
This randomness in vector selection is essential for fostering exploration in the solution space, thereby enhancing the algorithm's ability to converge toward optimal solutions.
The \textit{rand/1} strategy exemplifies the balance between exploration and exploitation in DE, making it a widely used method in various optimization problems~\cite{Neri2010}.

\subsection{DE/Best/1}
The \textit{best/1} mutation strategy is a variant of DE that focuses on enhancing the convergence of the population towards optimal solutions by utilizing the best vector in the current generation.
In this approach, a mutant vector $ v_{i,g} $ is generated by adding a scaled difference between a randomly selected vector and the best vector to a third randomly selected base vector.
The mathematical representation of this process can be expressed as follows,
\begin{equation}
v^{(g)}_{i} = x^{(g)}_{best} + F \cdot \left(x^{(g)}_{r_1} - x^{(g)}_{r_2}\right).
\label{eq:ed_best_1}
\end{equation}
In this equation, $ x^{(g)}_{best} $ represents the best vector in the current generation, while $ x^{(g)}_{r_1} $ and $ x^{(g)}_{r_2} $ are randomly chosen difference vectors.
The scale factor $F$, typically in $(0, 1)$, controls the magnitude of the mutation.

The selection of the difference vector indices $r_1$ and $r_2$ is performed randomly, ensuring that they are distinct from each other and the index of the target vector $i$. 
By incorporating the best vector into the mutation process, the \textit{best/1} strategy introduces a strong exploitation pressure, aiming to accelerate convergence towards the best-found region~\cite{Neri2010}. 
However, this greediness can sometimes lead to premature convergence, especially on multimodal problems, as it reduces exploration compared to \textit{rand/1}.

\subsection{DE/Current-to-Rand/1}
The \textit{current-to-rand/1} mutation strategy is a widely used approach in DE that combines information from the current target vector with a randomly selected vector to create a mutant vector.
This method aims to maintain diversity in the population while leveraging the current state of the optimization process.
The mathematical representation of this strategy can be expressed as follows,
\begin{equation}
v^{(g)}_{i} = x^{(g)}_{i} + F \cdot \left(x^{(g)}_{r_1} - x^{(g)}_{r_2}\right).
\label{eq:current_random_1}
\end{equation}
In this equation, $ x^{(g)}_{i} $ is the current target vector, while $ x^{(g)}_{r_1} $ and $ x^{(g)}_{r_2} $ are two randomly selected difference vectors.
The scale factor $F$, typically in $(0, 1)$, controls the magnitude of the mutation.

The indices for the difference vectors $r_1$ and $r_2$ are chosen randomly, ensuring that they are distinct from each other and the index of the target vector $i$. 
By using the target vector itself as the base, this strategy explores the vicinity of the current solution. 
Its properties regarding exploration and exploitation are different from \textit{rand/1} and \textit{best/1}, and it's less commonly cited as a standalone high-performing strategy compared to others like \textit{current-to-best/1} or \textit{rand/1}~\cite{Neri2010}.

\subsection{DE/Current-to-Best/1}
The \textit{current-to-best/1} mutation strategy is an effective variant of DE that enhances convergence by utilizing both the current target vector and the best vector in the population.
This approach aims to exploit the best-known solution while still allowing for exploration of the solution space.
The mathematical representation of this strategy can be expressed as follows,
\begin{equation}
v^{(g)}_{i} = x^{(g)}_{i} + F \cdot \left(x^{(g)}_{best} - x^{(g)}_{i}\right) + F \cdot \left(x^{(g)}_{r_1} - x^{(g)}_{r_2}\right).
\label{eq:current_to_best_1}
\end{equation}
In this equation, $ x^{(g)}_{i} $ is the current target vector, $ x^{(g)}_{best} $ represents the best vector in the current generation, and $ x^{(g)}_{r_1} $ and $ x^{(g)}_{r_2} $ are two randomly selected difference vectors.
The scale factor $F$, typically in $(0, 1)$, controls the magnitude of the mutation.

The indices for the difference vectors $r_1$ and $r_2$ are chosen randomly, ensuring that they are distinct from each other and the index $i$. 
By incorporating both the current target vector's position and the direction towards the best vector, along with random perturbations, the \textit{current-to-best/1} strategy effectively balances exploitation and exploration~\cite{Neri2010}. 
This makes it a powerful method often associated with good performance on many optimization problems.

\subsection{DE/Current-to-pBest/1}
A popular and related strategy, often used in high-performing DE variants like JADE~\cite{zhang2009jade} and SHADE~\cite{tanabe2013success}, is \textit{current-to-pbest/1}. 
This strategy replaces the single global best vector $x^{(g)}_{best}$ with a vector $x^{(g)}_{pbest}$ chosen randomly from the top $p \times 100\%$ of the individuals in the current population. 
This aims to reduce the greediness of \textit{current-to-best/1} while still guiding the search towards good regions~\cite{Bilal2020review}. 
The formulation is
\begin{equation}
 v^{(g)}_{i} =  x^{(g)}_{i} + F \cdot \left( x^{(g)}_{pbest} -  x^{(g)}_{\neq i}\right) + F \cdot \left( x^{(g)}_{r_1} -  x^{(g)}_{r_{\neq i}}\right).
\label{eq:current_to_pbest_1}
\end{equation}
Here, $ x^{(g)}_{pbest}$ is randomly selected from the top $p \times N_P$ individuals (where $N_P$ is the population size), $ x^{(g)}_{r_1}$ is chosen randomly from the entire population, and $ x^{(g)}_{r_{\neq i}}$ is chosen randomly from the population excluding the current individual $i$. 
The indices $i, r_1, r_{\neq i}, pbest$ must all be distinct. 
The scale factor $F$, typically in $(0, 1)$, controls the magnitude of the mutation.
This approach is known for its strong performance in modern adaptive DE algorithms.

\subsection{DE/Rand-to-Best/1}
The \textit{rand-to-best/1} mutation strategy is a variant of DE that combines the exploration capabilities of randomly selected vectors with the exploitation of the best vector in the population.
This approach aims to enhance convergence towards optimal solutions while maintaining diversity within the population.
The mathematical representation of this strategy can be expressed as follows,
\begin{equation}
v^{(g)}_{i} = x^{(g)}_{r_1} + F\cdot \left( x^{(g)}_{best} - x^{(g)}_{r_1}\right) + F\cdot \left( x^{(g)}_{r_2} - x^{(g)}_{r_3}\right).
\label{eq:random_to_best_1}
\end{equation}
In this equation, $x^{(g)}_{best}$ represents the best vector in the current generation, while ${x}^{(g)}_{r_1}$, ${x}^{(g)}_{r_2}$, and ${x}^{(g)}_{r_3}$ are three distinct, randomly selected vectors from the population, also distinct from the target index $i$. 
The scale factor $F$, typically in $(0, 1)$, controls the magnitude of the mutation.

By incorporating the best vector into the mutation process while relying on randomly selected vectors for the base and differences, the \textit{rand-to-best/1} strategy attempts to balance exploration and exploitation effectively~\cite{Neri2010}. 
It introduces guidance from the best solution without being as purely exploitative as \textit{best/1}.

\section{Boundary Handling Strategies}
DE algorithms are frequently employed to solve optimization problems where the decision variables are constrained within specific lower and upper bounds, defining the feasible search space. 
During the DE process, the mutation operation generates new candidate solutions, known as trial vectors. 
However, this operation can produce a trial vector where one or more of its components lie outside the predefined boundaries~\cite{Price2005Differential}. 
Since evaluating solutions outside the feasible region is often undesirable or nonsensical, effective boundary handling strategies are required to reposition the violating parameters back within their allowed range. 
Several techniques have been developed to address this issue, with some of the most commonly used methods including the clamping (or projection), toroidal, and re-initialization strategies~\cite{Neri2010, storn1997differential, BIEDRZYCKI2019100453}.

\subsection{Clamping}
\label{subsubsec:boundary_clamping}
The clamping (or projection) strategy is perhaps the most intuitive method used for handling boundary violations in DE. 
If a component $v_{i,j}$ of the trial vector $v_i$ falls outside the allowed range $[b_{L,j}, b_{U,j}]$ for the $j$-th dimension after mutation and crossover, it is simply set to the value of the violated boundary. 
Specifically, if the component $v_{i,j}$ is less than the lower bound $b_{L,j}$, it is assigned the value $b_{L,j}$. 
Conversely, if $v_{i,j}$ exceeds the upper bound $b_{U,j}$, it is assigned the value $b_{U,j}$.
This can be expressed concisely as
\begin{equation}
    v_{i,j} = 
    \begin{cases} 
        b_{L,j} & \text{if } v_{i,j} < b_{L,j}, \\
        b_{U,j} & \text{if } v_{i,j} > b_{U,j} ,\\
        v_{i,j} & \text{otherwise} ,
    \end{cases}
\end{equation}
or using functions as $v_{i,j} = \max(b_{L,j}, \min(v_{i,j}, b_{U,j}))$. While simple and effective at ensuring feasibility, this strategy can lead to an accumulation of parameter values precisely at the boundaries, which might sometimes hinder exploration near the edge of the search space if the optimum lies there or slightly beyond where the unconstrained step would have landed~\cite{Neri2010}.

\subsection{Toroidal}
\label{subsubsec:boundary_toroidal}
The toroidal strategy treats the search space for each dimension as if it were periodic. 
If a component $v_{i,j}$ moves beyond an upper or lower bound, it ``wraps around'' and re-enters the feasible range from the opposite side, maintaining the magnitude of the intended change relative to the boundary it crossed. 
The amount by which the parameter exceeds the boundary is effectively added to the opposite boundary. 
For a range $\Delta_j = b_{U,j} - b_{L,j}$, the repositioning can be implemented as follows,
\begin{itemize}
    \item If $v_{i,j} < b_{L,j}$: The component has gone below the lower bound. 
    Calculate the underflow $b_{L,j} - v_{i,j}$. 
    The new value is $v_{i,j} = b_{U,j} - \text{fmod}(b_{L,j} - v_{i,j}, \Delta_j)$, where $\text{fmod}$ is the floating-point remainder function ensuring the result wraps correctly within the range.
    \item If $v_{i,j} > b_{U,j}$: The component has gone above the upper bound. 
    Calculate the overflow $v_{i,j} - b_{U,j}$. 
    The new value is $v_{i,j} = b_{L,j} + \text{fmod}(v_{i,j} - b_{U,j}, \Delta_j)$.
\end{itemize}
A careful implementation of the modulo operation is needed, especially with floating-point numbers. 
This strategy might be suitable for problems with inherent periodicity but can feel unnatural for others. 
It preserves the step size suggested by DE's operations but displaces the location significantly, which could be disruptive or beneficial depending on the landscape~\cite{Price2005Differential}. 

\subsection{Re-initialization}
\label{subsubsec:boundary_randomize}
The re-initialization strategy takes a different approach. 
If a component $v_{i,j}$ of the trial vector violates either the lower bound $b_{L,j}$ or the upper bound $b_{U,j}$, it is discarded and replaced with a new value generated randomly and uniformly within the feasible range $[b_{L,j}, b_{U,j}]$. 
The process is straightforward,
\begin{equation}
    v_{i,j} = 
    \begin{cases} 
        \text{rand}(b_{L,j}, b_{U,j}) & \text{if } v_{i,j} < b_{L,j} \text{ or } v_{i,j} > b_{U,j} \\
        v_{i,j} & \text{otherwise} 
    \end{cases}
\end{equation}
where $\text{rand}(b_{L,j}, b_{U,j})$ denotes a uniformly distributed random number drawn from the interval $[b_{L,j}, b_{U,j}]$. 
This method is simple to implement and ensures feasibility while potentially introducing diversity that might help escape local optima. 
However, it completely disregards the information contained in the direction and magnitude of the move that led to the boundary violation, which could otherwise have been useful exploration. 
Frequent re-initializations might also slow down the convergence process~\cite{Price2005Differential, Neri2010}.

\section{Crossover}
The crossover operation in DE plays a crucial role in generating trial vectors by combining parameter values from the target and mutant vectors, thereby maintaining population diversity and facilitating effective exploration of the solution space.
Two primary crossover strategies are commonly employed: Uniform (Binomial) Crossover and Exponential crossover~\cite{Lin2011, ZAHARIE20091126, 10.1007/978-3-030-51859-2_35}.

\subsection{Uniform (Binomial) Crossover}
The Uniform (Binomial) Crossover, also known as discrete recombination, is a widely used strategy in DE, having been part of the original algorithm proposal~\cite{storn1997differential}.
The Uniform (Binomial) crossover, also known as discrete recombination, is a widely used strategy in DE.
It generates trial vectors by merging parameter values from two distinct vectors, the target vector and the mutant vector.
The formation of the trial vector $ u^{(g)}_{i} $ is governed by the following rule,
\begin{equation}
u^{(g)}_{i,j} =
\begin{cases}
v^{(g)}_{i,j} & \text{if } \text{rand}_j(0,1) \leq Cr \text{ or } j = j_{rand}, \\
x^{(g)}_{i,j} & \text{otherwise}.
\end{cases}
\label{eq:binomial_crossover}
\end{equation}
In this equation, $ Cr \in [0,1]$ represents the crossover rate(crossover probability), a user-defined parameter that specifies the fraction of parameters to be inherited from the mutant vector $ v^{(g)}_{i} $.
The decision on which source to draw a particular parameter from is made by comparing $ Cr $ with a value generated by a uniform random number generator, $ \text{rand}_j(0,1) $.
If the generated random number is less than or equal to $ Cr $, the parameter is taken from the mutant vector; otherwise, it is sourced from the target vector $ x^{(g)}_{i} $.

To ensure that the trial vector does not exactly replicate the target vector, one parameter is always selected from the mutant vector at a randomly chosen index $ j_{rand} $.
This requirement introduces a slight deviation from the expected crossover probability, which may not perfectly align with $ Cr $ due to this additional constraint.

\subsection{Exponential Crossover}
Another common crossover strategy is the Exponential crossover~\cite{Price2005Differential}.
In the Exponential crossover, the trial vector $ u^{(g)}_{i} $ is formed by copying parameters from the mutant vector $ v^{(g)}_{i} $ until a randomly determined length $ L $ is reached.
The starting point of this copying process is a randomly chosen index $ j_{rand} $.
The copying wraps around the parameter vector, effectively forming a circular process.
The formation of the trial vector $ u^{(g)}_{i} $ using Exponential crossover is defined as
\begin{equation}
u^{(g)}_{i,j} =
\begin{cases}
v^{(g)}_{i,j} & \text{if } j \in \langle j_{rand}, (j_{rand} + L - 1) \rangle_{\text{mod } D}, \\
x^{(g)}_{i,j} & \text{otherwise},
\end{cases}
\label{eq:exponential_crossover}
\end{equation}
where $ D $ is the dimension of the problem and $ [ a, b ]_{\text{mod } D} $ represents the indices from $ a $ to $ b $ modulo $ D $.
The length $ L $ is determined by repeatedly comparing a random number $ \text{rand}_j(0,1) $ with the crossover probability $ Cr $ until $ \text{rand}_j(0,1) > Cr $ or $ L $ reaches $ D $.

The Exponential crossover can be beneficial in problems where parameters exhibit linkage or epistasis (i.e., they are correlated or interacting), as it tends to inherit blocks of related parameters together from the mutant vector~\cite{Lin2011}. However, it might also lead to premature convergence compared to binomial crossover if the problem does not have such structure or if $Cr$ is high, leading to consistently large blocks being transferred~\cite{ZAHARIE20091126}.\\

Both the Uniform and Exponential crossover operations play a vital role in maintaining diversity within the population, allowing DE to effectively explore the solution space while converging towards optimal solutions.
The choice between these crossover strategies often depends on the specific characteristics of the optimization problem.

\section{Selection}
\label{sec:de_selection}
If the trial vector $ u^{(g)}_{i} $ achieves an objective function value that is equal to or better than that of its corresponding target vector $ x^{(g)}_{i} $, it will replace the target vector in the subsequent generation.
Conversely, if the trial vector does not meet this criterion, the target vector will remain in the population for at least one more generation, as described by the following equation,
\begin{equation}
x^{(g+1)}_{i} = 
\begin{cases} 
u^{(g)}_{i} & \text{if } f\left(u^{(g)}_{i}\right) \leq f\left(x^{(g)}_{i}\right) ,\\ 
x^{(g)}_{i} & \text{otherwise} .
\end{cases}
\label{eq:selection}
\end{equation}
This mechanism represents one of the simpler selection strategies employed in DE.
It allows DE to effectively integrate the processes of recombination and selection compared to other evolutionary algorithms. 
In addition to this approach, there are various other selection strategies available, such as tournament selection, roulette wheel selection, and rank-based selection, each with its own advantages and applications~\cite{AHMAD20223831, de_art_selection, 350042}.

Once the new population is established, the cycle of mutation, recombination, and selection is repeated until an optimal solution is found or a predefined termination condition is met.

\section{Termination Criteria}
Termination criteria play a crucial role in DE, dictating when the iterative optimization process should cease.
Given the stochastic nature of DE, determining an appropriate stopping point is essential to balance solution quality with computational efficiency.
A variety of criteria exist, ranging from simple limits on the number of function evaluations or generations to more sophisticated measures of convergence and tolerance.
Among the most frequently employed are maximum function evaluations, maximum generations, and tolerance-based criteria that monitor fitness improvement~\cite{de_art_termination, RAVBER2022109478, 10.1145/3205455.3205466}.
The subsequent sections will delve deeper into the nuances of these and other termination strategies, exploring their strengths, weaknesses, and suitability for different problem domains.

\subsection{Maximum Evaluations or Generations}

The most straightforward and widely implemented termination criteria in DE are based on limiting the computational effort, either by setting a maximum number of fitness function evaluations ($FE_{max}$) or a maximum number of generations ($G_{max}$)~\cite{storn1997differential, Price2005Differential}.

Maximum Function Evaluations criterion halts the algorithm when the total number of fitness function evaluations reaches a predefined limit, denoted as $FE_{max}$. 
Since the fitness function evaluation is often the most computationally expensive operation in DE, this criterion directly controls the overall computational cost. 
Mathematically, the termination condition is expressed as
$$ FE \geq FE_{max}, $$
where $FE$ represents the current number of fitness function evaluations. 
While simple to implement, this criterion does not guarantee a specific solution quality, as the algorithm might terminate prematurely before reaching a satisfactory solution or continue unnecessarily if a good solution is found early on~\cite{Neri2010}.

Analogous to maximum function evaluations, the Maximum Generations/Iterations criterion terminates the algorithm after a predetermined number of generations, denoted as $G_{max}$.
Each generation represents a complete iteration of the DE algorithm, including mutation, crossover, and selection.
The termination condition is
$$ G \geq G_{max}, $$
where $G$ represents the current generation number.
This criterion is also easy to implement and control, but like $FE_{max}$, it does not ensure a certain solution quality.
The choice of $G_{max}$ is often determined empirically, based on the problem's complexity and the available computational resources~\cite{Price2005Differential}.
Both $FE_{max}$ and $G_{max}$ provide a deterministic stopping point, but they lack adaptive behavior based on the algorithm's progress.

\subsection{Fitness Improvement Tolerance}
Tolerance-based criteria offer a more adaptive approach to terminating the DE algorithm, focusing on the convergence of the solution rather than a fixed computational limit~\cite{Price2005Differential, Neri2010}.
These criteria monitor the improvement in the best-found fitness value over successive generations and terminate the algorithm when the improvement falls below a predefined tolerance threshold.

The Absolute Fitness Improvement Tolerance criterion assesses the change in the best fitness value between consecutive generations.
Let $f_{best}(G)$ represent the best fitness value at generation $G$.
The algorithm terminates when the absolute change in the best fitness value over a specified number of consecutive generations, $N_{tol}$, is less than a tolerance value, $\epsilon_{tol}$. Mathematically, this can be expressed as
$$ |f_{best}(G) - f_{best}(G-1)| < \epsilon_{tol} \quad \text{for } N_{tol} \text{ consecutive generations}, $$
where $\epsilon_{tol}$ is a small positive value representing the tolerance threshold.
This criterion aims to stop the algorithm when the improvement in the solution quality becomes negligible, indicating that the algorithm has likely converged~\cite{Neri2010}. 
The choice of $\epsilon_{tol}$ and $N_{tol}$ is crucial; inadequate settings may lead to premature termination or unnecessary computation.

The Relative Fitness Improvement Tolerance criterion is an alternative approach that uses a relative tolerance, which considers the relative change in fitness rather than the absolute change.
This is particularly useful when dealing with fitness functions that have a wide range of values.
The termination condition is
$$ \frac{|f_{best}(G) - f_{best}(G-1)|}{|f_{best}(G)| + \delta} < \epsilon_{rtol} \quad \text{for } N_{tol} \text{ consecutive generations} ,$$
where $\epsilon_{rtol}$ is the relative tolerance threshold, and $\delta$ is a small constant (e.g., machine epsilon or a small positive number) added to the denominator to prevent division by zero or issues when $f_{best}(G)$ is near zero~\cite{Price2005Differential}. 
This relative criterion provides a potentially more robust measure of convergence across problems with different fitness scales.

Tolerance-based criteria offer a more intelligent termination strategy by focusing on the convergence of the solution, but they require careful selection of the tolerance parameters to ensure accurate and efficient termination.

\subsection{Running Mean Tolerance}
A variation of tolerance-based criteria employs a running mean of the fitness improvement to provide a smoother and more robust measure of convergence, particularly useful in noisy environments~\cite{Price2005Differential}. 
This approach aims to mitigate the effects of occasional fluctuations in fitness values, which can trigger premature termination in standard tolerance-based methods. 
Instead of considering the instantaneous fitness improvement between consecutive generations, this criterion calculates the running mean of the absolute fitness improvement over a window of $N_{mean}$ recent generations. 
Let $f_{best}(G)$ represent the best fitness value at generation $G$. 
The running mean of the fitness improvement, $\mu_{improv}(G)$, is defined as
$$ \mu_{improv}(G) = \frac{1}{N_{mean}} \sum_{k=0}^{N_{mean}-1} |f_{best}(G-k) - f_{best}(G-k-1)| .$$
The algorithm terminates when the running mean of the fitness improvement falls below a tolerance value, $\epsilon_{mean}$, potentially sustained over $N_{tol}$ generations (though often checked just at the current generation). 
A common condition is
$$ \mu_{improv}(G) < \epsilon_{mean}  \quad \text{for } N_{tol} \text{ consecutive generations}$$
where $\epsilon_{mean}$ is the tolerance threshold for the running mean. 
Using a running mean provides a more stable measure of convergence, as it averages out short-term fluctuations~\cite{Neri2010}. 
The choice of $N_{mean}$ and $\epsilon_{mean}$ affects the sensitivity and robustness of the termination criterion. 
A larger $N_{mean}$ smooths out fluctuations more effectively but may delay termination.

This approach is particularly useful in noisy optimization problems where fitness values can exhibit significant variations between generations.

\subsection{Population Fitness Convergence}
Another approach to termination focuses on the convergence of the population itself by monitoring the difference between the best and worst fitness values within the current generation~\cite{Price2005Differential, Neri2010}. 
This criterion aims to stop the algorithm when the population has sufficiently converged, indicating that the individuals have become relatively similar in terms of fitness, suggesting limited potential for further significant improvement through recombination. 
This criterion calculates the difference between the best fitness value, $f_{best}(G)$, and the worst fitness value, $f_{worst}(G)$, within generation $G$.
The algorithm terminates when this difference falls below a predefined tolerance value, $\epsilon_{bw}$, potentially checked over $N_{tol}$ consecutive generations
$$ |f_{worst}(G) - f_{best}(G)| < \epsilon_{bw} \quad \text{for } N_{tol} \text{ consecutive generations}, $$
where $\epsilon_{bw}$ is the tolerance threshold for the best-worst fitness difference. 
This criterion indicates that the population has converged to a state where individuals have similar fitness values. 
The choice of $\epsilon_{bw}$ and $N_{tol}$ is critical; a small $\epsilon_{bw}$ implies a high degree of convergence, while a large $\epsilon_{bw}$ allows for more diversity~\cite{Neri2010}.

This approach is informative about the population's state but may lead to premature termination on flat landscapes or very rugged multimodal landscapes where diverse solutions with similar fitness might still be valuable. It's often used in conjunction with other criteria.

\section{Algorithm of Differential Evolution}
Building upon the theoretical components detailed in the preceding sections (mutation, crossover, selection, boundary handling, termination criteria), the general structure of the Differential Evolution algorithm can be formalized. 
The procedural steps common to most DE variants are presented in Algorithm~\ref{alg:general_de}. The algorithm iteratively refines a population of candidate solutions by applying differential mutation, crossover, and selection operators until a specified termination criterion is satisfied.
\begin{algorithm}[!ht]
\caption{General algorithm of DE}
\label{alg:general_de}
\begin{algorithmic}[1]
\Require Population size $N_p$, Dimension $D$, Mutation Factor $F$, Crossover rate $CR$
\Ensure Best solution $x_{\text{best}}$

\State Initialize population $P^{(0)}_{x} = \left(x^{(0)}_{i}\right)$ for $i = 0, 1, \ldots, N_p - 1$, where $x^{(0)}_{i} = \left(x^{(0)}_{i,j}\right)$ for $j = 0, 1, \ldots, D - 1$
\State Evaluate fitness of each $x^{(0)}_{i}$
\State $g \leftarrow 0$
\While{Stop condition not met}
    \For{$i = 0$ to $N_p - 1$}
        \State $v^{(g)}_i \leftarrow \text{differential mutation}\left(F, x^{(g)}_{i}, \ldots,x^{(g)}_{N_p} \right)$ \label{line:mutation_step}
        \State $u^{(g)}_i \leftarrow \text{crossover} \left( CR, x^{(g)}_i, v^{(g)}_i \right)$\label{line:crossover_step}
    \EndFor
    \State $P^{(g+1)}_x \leftarrow \text{selection}\left( P^{(g)}_x , P^{(g)}_u \right) $ \label{line:selection_step}
    \State $g \leftarrow g + 1$
\EndWhile
\State $x_{\text{best}} \leftarrow \text{best solution in } P^{(g)}_{x}$
\Return $x_{\text{best}}$
\end{algorithmic}
\end{algorithm}
This general framework serves as a blueprint for various DE implementations, which differ primarily in the specific mutation (line~\ref{line:mutation_step}) and crossover (line~\ref{line:crossover_step}) strategies employed, along with potential variations in parameter control and selection mechanisms (line~\ref{line:selection_step}). 
Detailed pseudocode for two specific common DE variants is provided in Appendix~\ref{ap:de_pseudocode}. 
Algorithm~\ref{alg:de_rand_1_bin} illustrates the widely used DE/rand/1/bin strategy, while Algorithm~\ref{alg:de_ctb_1_exp} demonstrates the DE/current-to-best/1/exp variant. 
Both referenced algorithms utilize the selection mechanism detailed in Section~\ref{sec:de_selection}.

\section{Implementation}
An implementation of the DE algorithm\footnote{https://gitlab.com/bezdektom/de-for-sa-oo-vqe-qiskit}~\cite{bezdek_2025_15292845} was developed in Python~\cite{python}, designed explicitly for compatibility with the SA-OO-VQE framework~\cite{joss_saooveq}. Adherence to the Qiskit Algorithms Optimizer interface \footnote{https://docs.quantum.ibm.com/api/qiskit/0.39/qiskit.algorithms.optimizers.Optimizer}~\cite{javadiabhari2024quantumcomputingqiskit} ensures its applicability as a classical optimization component for the SA-VQE part of the solver (Section~\ref{sec:sa_vqe}). 
For improved computational efficiency, the implementation leverages the numerical processing capabilities of the NumPy package~\cite{numpy}.

\chapter{Results of SA-OO-VQE Optimizer Comparison and Conical Intersection Analysis}
This chapter initially presents a comparative analysis of various optimization algorithms and their efficacy as energy minimizers within the SA-OO-VQE framework~\cite{joss_saooveq}. 
To this end, we compute the ground state energy and first excited state energy of three molecular systems, hydrogen molecule (H$_2$), lithium hydride (LiH), and a linear tetratomic hydrogen chain (H$_4$). 
Subsequently, we will demonstrate the principal advantage of the SA-OO-VQE method using the formaldimine molecule as a case study, illustrating its capability to address problems involving conical intersections.

\section{Comparison of Numerical Optimizers}
\label{sec:comparison_of_optimizers}
This section aimed to evaluate and compare the performance of several optimization algorithms when integrated with the SA-OO-VQE framework~\cite{joss_saooveq}. 
Specifically, we investigated the efficacy of Gradient Descent, the BFGS algorithm, the COBYLA method, the SLSQP algorithm, and a suite of DE strategies, including DE/Rand1/bin, DE/Rand2/bin, DE/Best1/bin, DE/Best2/bin, DE/Current-to-Best1/bin, DE/Current-to-Random1/bin, and DE/Random-to-Best1/bin.
The comparative analysis for each optimizer encompassed an assessment of the quality of the identified energy minimum, as well as the required number of iterations and function evaluations to achieve convergence.

\subsection{Used methodology}
\label{sec:experiments_methodology}
In this thesis, we utilize the SA-OO-VQE framework, as detailed in Chapter~\ref{chapter:saoovqe}. 
The primary goal here is not to perform a detailed spectroscopic study of the selected molecules, but rather to assess the performance, efficiency, and robustness of various classical optimizers when tasked with minimizing the cost function defined within the SA-OO-VQE context for each of the molecules.

The SA-OO-VQE algorithm simultaneously optimizes both the parameters of the quantum circuit ansatz (representing the wave function's correlation part) and the molecular orbital basis itself. 
The optimization process aims to minimize the state-averaged energy, defined as
$$ E_{SA} = 0.5 E_0 + 0.5 E_1 $$
where $E_0$ and $E_1$ are the energies of the ground and excited states, respectively, computed from the VQE.
The initial solution for the optimization was obtained using the Hartree-Fock method~\cite{Bechstedt2015}, providing a reasonable starting point for all optimizers.

Consequently, when evaluating the performance of the different classical optimizers used within the SA-OO-VQE loop, we will primarily focus on metrics related to the minimization of this state-average energy ($E_{SA}$). 
This includes the final $E_{SA}$ value achieved (accuracy), the number of function evaluations required to reach convergence (efficiency), and the consistency of the results across multiple independent runs (robustness). 
While the individual ground state ($E_0$) and excited state ($E_1$) energies are also obtained and reported, the convergence behavior and efficiency metrics are most directly linked to the optimizer's ability to navigate the landscape defined by $E_{SA}$. 

To provide insight into the optimization dynamics for each optimizer-molecule pair, we will present convergence plots illustrating the progression of the state-average energy ($E_{SA}$). 
We analyze this convergence from three distinct perspectives by plotting $E_{SA}$ against different metrics of progress:
\begin{enumerate}
    \item \textbf{SA-OO-VQE Iteration Number:} Tracks the energy improvement after each complete cycle of the outer SA-OO-VQE loop (orbital optimization plus VQE optimization).
    \item \textbf{Cumulative Evaluations per SA-OO-VQE Iteration:} Plots the energy achieved at the end of each SA-OO-VQE iteration against the total number of function evaluations consumed up to that point. This reflects the cost associated with each outer loop step.
    \item \textbf{Cumulative Evaluations per Optimizer Step:} Shows the energy after each internal step taken by the classical optimizer algorithm itself, plotted against the total number of function evaluations consumed up to that internal step. This offers the most granular view of the optimization trajectory.
\end{enumerate}
For the Differential Evolution configurations and the Gradient Descent method, where monitoring the progress after internal optimizer steps was implemented, all three types of convergence plots will be displayed. 
However, for the BFGS, COBYLA, and SLSQP optimizers, which were used via the standard SciPy library implementation\footnote{\url{https://docs.scipy.org/doc/scipy/reference/optimize.html}}~\cite{2020SciPy-NMeth}, accessing the required information after each internal optimizer step is not straightforwardly supported. 
Therefore, for these specific optimizers, only the first two types of plots (energy vs. SA-OO-VQE iteration number, and energy vs. cumulative evaluations per SA-OO-VQE iteration) will be provided.

\subsection{Optimization Methods Employed}
This study investigated the performance of several gradient-based and derivative-free optimization algorithms within the SA-OO-VQE framework. 
The principles of DE were already covered in Chapter~\ref{chapter:DE}, let us now look at the rest of the optimizers.
\begin{itemize}
    \item \textbf{Gradient Descent}~\cite{ruder2016overview} is a first-order iterative optimization algorithm for finding the local minimum of a differentiable function. 
    It takes steps proportional to the negative of the gradient (or approximate gradient) of the function at the current point.
    While simple to implement, its convergence can be slow, especially in poorly conditioned landscapes or near saddle points.
    \item \textbf{BFGS} algorithm~\cite{liu1989limited} is a quasi-Newton method that approximates the Hessian matrix of the objective function using information from previous iterations.
    This approximation of the second-order curvature allows for more efficient updates to the parameter vector compared to first-order methods.
    BFGS generally exhibits faster convergence than Gradient Descent and is well-suited for problems with a moderate number of variables.
    \item \textbf{COBYLA}~\cite{Conn_Scheinberg_Vicente_2009} is a derivative-free optimization algorithm designed for constrained nonlinear optimization problems.
    It iteratively builds a linear approximation of the objective function and the constraints based on function evaluations at a simplex of points.
    In each iteration, COBYLA solves a linear programming subproblem within a shrinking trust region to determine the next step.
    This method is effective when analytical derivatives are not available or are computationally expensive to obtain.
    \item \textbf{SLSQP}~\cite{kraft1988software} is an iterative method for constrained nonlinear optimization. 
    At each major iteration, it solves a QP subproblem which is a quadratic approximation of the Lagrangian function and a linear approximation of the constraints. 
    By sequentially solving these QP subproblems, SLSQP aims to find a solution that satisfies the problem's constraints and minimizes the objective function. 
    It typically requires gradient information for both the objective and constraint functions.
\end{itemize}
All the optimizers except DE, which were implemented by the author~\cite{bezdek_2025_15292845}, were used from the SciPy package~\cite{2020SciPy-NMeth}.

\subsection{Molecule H$_2$}
\label{sec:molecule_h2}
The first molecular system investigated for the comparison of different optimization algorithms is the hydrogen molecule (H$_2$). 
As the simplest neutral diatomic molecule, H$_2$ serves as a fundamental benchmark system in quantum chemistry and is frequently employed in testing novel quantum computing algorithms due to its small size and well-understood electronic structure~\cite{Helgaker_Jørgensen_Olsen_2014}.

The calculations were performed using the cc-pVDZ basis~\cite{10.1063/1.456153}. 
The SA-OO-VQE procedure targeted an equal-weighted average of the singlet ground state ($S_0$) and the first singlet excited state ($S_1$). 

\begin{figure}[!htbp] 
    \centering
    
    \begin{subfigure}[t]{0.49\textwidth} 
        \centering
        \includegraphics[width=\linewidth]{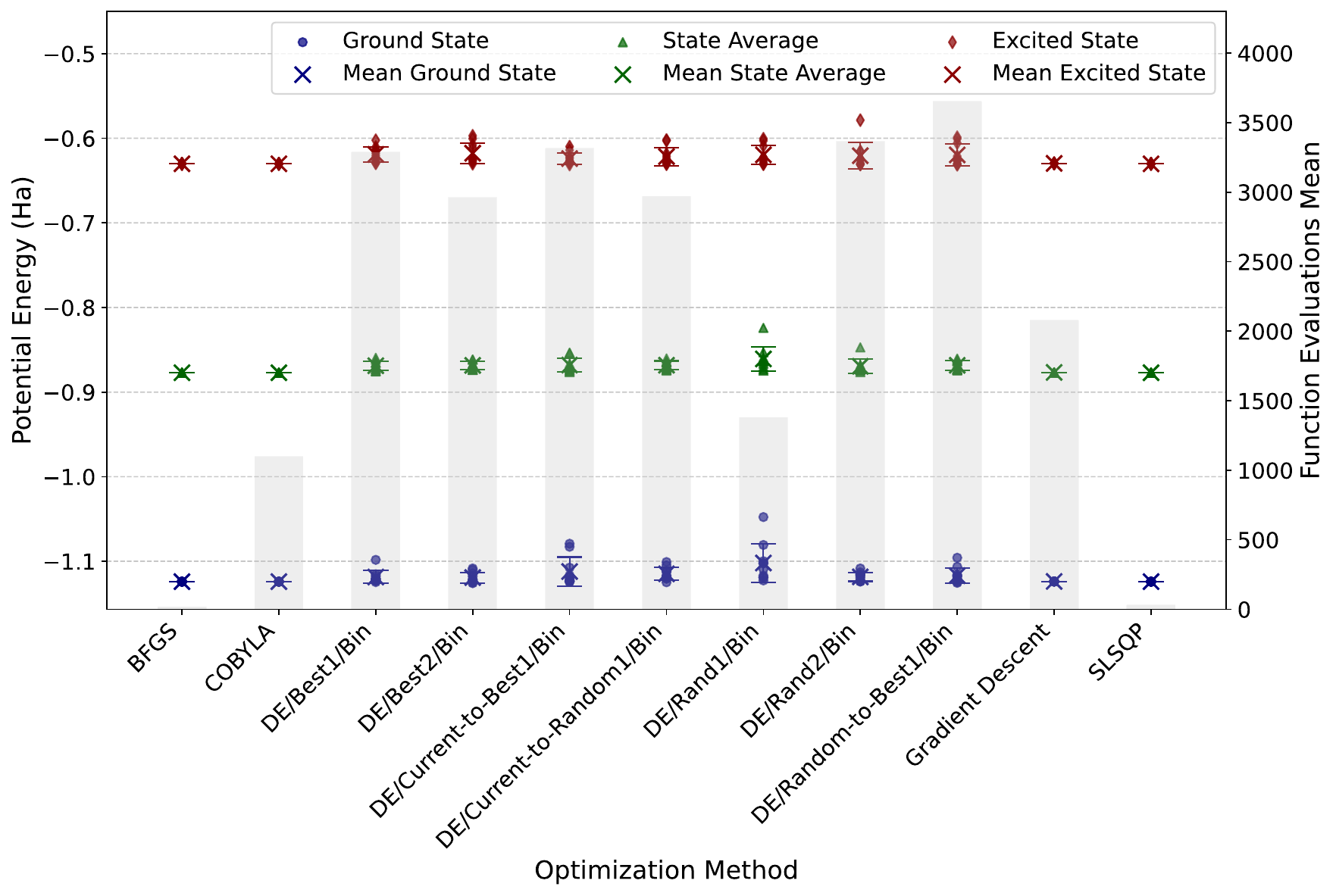}
        \caption{All calculated energies and function evaluations.}
        \label{fig:h2_saoovqe_all_energies} 
    \end{subfigure}
    \hfill 
    \begin{subfigure}[t]{0.49\textwidth} 
        \centering
        \includegraphics[width=\linewidth]{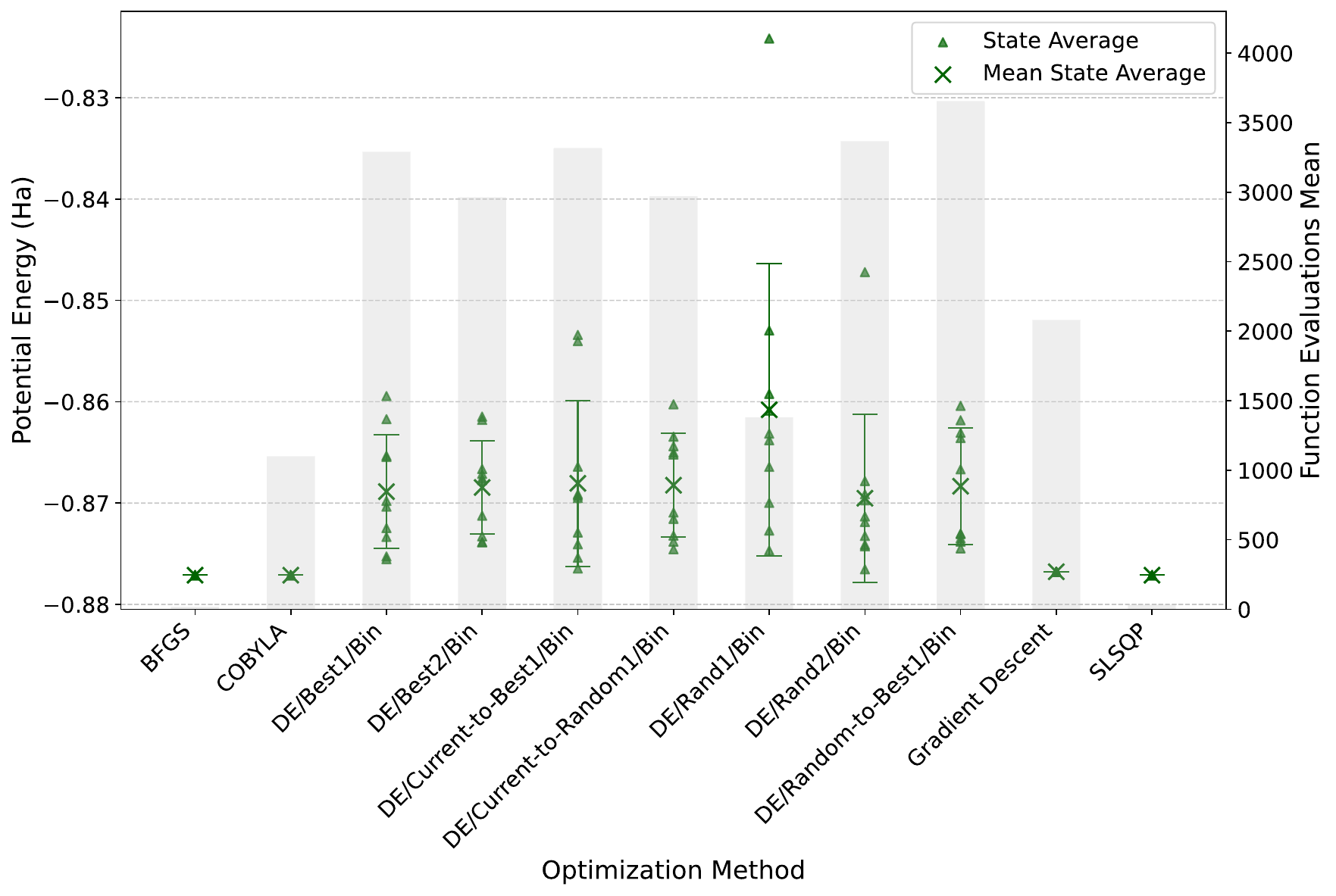}
        \caption{State-average energy detail and function evaluations.}
        \label{fig:h2_saoovqe_sa_detail}
    \end{subfigure}

    \caption[Comparative optimizer results from SA-OO-VQE for H$_2$ molecule.]{Performance of different optimizers within the SA-OO-VQE framework for the H$_2$ molecule. 
    Both plots show results derived from 10 independent optimization runs for each method. Individual run results are shown as markers, means as crosses ($\times$), standard deviations across runs are shown as error bars (left y-axis of each plot, Hartrees), and mean function evaluations shown as gray bars on the right y-axis of each plot.    (\subref{fig:h2_saoovqe_all_energies}) Displays ground state (blue circles), excited state (red diamond), and state-average (green triangle) energies. 
    (\subref{fig:h2_saoovqe_sa_detail}) Provides a detailed view of the state-average energy.}
    \label{fig:h2_saoovqe_comparison} 
\end{figure}

The subsequent parts (e.g., Figure~\ref{fig:h2_saoovqe_comparison} and Table~\ref{tab:calculation_summary_h2}) present these comparative results.
As we can see in Figure~\ref{fig:h2_saoovqe_comparison} and Table~\ref{tab:calculation_summary_h2}, there are significant differences between the used optimizers in terms of efficiency (number of function evaluations) and accuracy and consistency of the final state-average energy achieved for the H$_2$ molecule.

\FloatBarrier
\begin{table}[!htbp]
\centering
\small
\begin{tabular}{l c c c c c c}
\toprule
Method & $evals_{min}$ & $evals_{max}$ & $evals_{mean}$ & $E_{min}$ & $E_{max}$ & $E_{mean}$ \\
\midrule
BFGS & 18 & 18 & 18 & -0.877120 & -0.877120 & -0.877120 \\
COBYLA & 1022 & 1185 & 1100 & -0.877120 & -0.877120 & -0.877120 \\
DE/Best1/Bin & 1380 & 8490 & 3288 & -0.875536 & -0.859444 & -0.868871 \\
DE/Best2/Bin & 1980 & 4860 & 2964 & -0.873872 & -0.861468 & -0.868461 \\
DE/Current-to-Best1/Bin & 1440 & 7800 & 3315 & -0.876457 & -0.853409 & -0.868056 \\
DE/Current-to-Random1/Bin & 1290 & 5610 & 2970 & -0.874561 & -0.860257 & -0.868240 \\
DE/Rand1/Bin & 630 & 2130 & 1380 & -0.874680 & -0.824160 & -0.860801 \\
DE/Rand2/Bin & 2070 & 6570 & 3366 & -0.876537 & -0.847211 & -0.869526 \\
DE/Random-to-Best1/Bin & 1620 & 5610 & 3654 & -0.874475 & -0.860410 & -0.868342 \\
Gradient Descent & 2080 & 2080 & 2080 & -0.876777 & -0.876777 & -0.876777 \\
SLSQP & 30 & 30 & 30 & -0.877120 & -0.877120 & -0.877120 \\
\bottomrule
\end{tabular}
\caption[Comparative optimizer results from SA-OO-VQE for H$_2$ molecule.]{Comparison of optimizer performance for the H$_2$ molecule simulation. 
The table reports statistics based on 10 independent runs: minimum ($min$), maximum ($max$), and mean ($mean$) number of function evaluations ($evals$) required for convergence, and the corresponding final state-average energies ($E$ in Hartrees).} 
\label{tab:calculation_summary_h2}
\end{table}
\FloatBarrier

The gradient-based methods BFGS and SLSQP demonstrate exceptional performance for this problem. They consistently converge to the lowest observed state-average energy ($-0.877120$ Ha) with remarkable efficiency, requiring only 18 and 30 function evaluations, respectively, across all runs. 
Their deterministic nature ensures high robustness, as indicated by the identical minimum, maximum, and mean values for both evaluations and energy.

The gradient-free local optimizer COBYLA also consistently reaches the optimal energy value. However, it requires substantially more function evaluations (mean of 1100) compared to BFGS and SLSQP, highlighting a trade-off between not needing gradient information and computational cost for this particular system. The standard Gradient Descent method, while also deterministic, converged to a slightly higher energy ($-0.876777$ Ha) and required a relatively large number of evaluations (2080).

The standard Gradient Descent method, while also deterministic (unlike the DE variants), proved less effective than the other gradient-based methods tested. 
It converged to a slightly higher energy ($-0.876777$ Ha) compared to BFGS/COBYLA/SLSQP and required a significantly larger number of evaluations (2080) than BFGS and SLSQP. 
This suggests that for this SA-OO-VQE problem, the basic Gradient Descent approach, potentially due to factors like step size selection or sensitivity to the landscape curvature, is less suited than the quasi-Newton (BFGS) or sequential quadratic programming (SLSQP) techniques which implicitly or explicitly incorporate second-order information.

In stark contrast, the stochastic DE variants exhibit considerably different behavior. 
Generally, they required significantly more function evaluations on average (ranging from 1380 for DE/Rand1/Bin to 3654 for DE/Random-to-Best1/Bin) compared to BFGS/SLSQP, and also showed large variability between runs (e.g., $evals_{max}$ reaching up to 8490 for DE/Best1/Bin). 
More importantly, the DE methods struggled to consistently find the optimal energy. 
Their mean final energies were noticeably higher (worse) than those obtained by BFGS, SLSQP, and COBYLA, typically around $-0.868$ to $-0.870$ Ha. 
Furthermore, the large difference between $E_{min}$ and $E_{max}$ for most DE variants (e.g., DE/Rand1/Bin ranging from $-0.874680$ Ha down to $-0.824160$ Ha) underscores their stochastic nature and lack of robustness for this specific problem; multiple runs yielded quite poor results. 
Even the best runs ($E_{min}$) across the 10 trials for the DE optimizers did not consistently match the energy found reliably by the best local methods.

In summary, for the SA-OO-VQE optimization of the H$_2$ molecule under these conditions, the local optimizers, particularly the gradient-based methods BFGS and SLSQP, offered the best combination of accuracy, efficiency, and robustness. The stochastic DE approaches, while potentially useful for more complex, non-convex landscapes, proved less suitable here, exhibiting higher computational costs and lower reliability in finding the optimal state-average energy.

\begin{figure}[!htbp] 
    \centering
    \begin{subfigure}[t]{0.49\textwidth} 
        \centering
        \includegraphics[width=\linewidth]{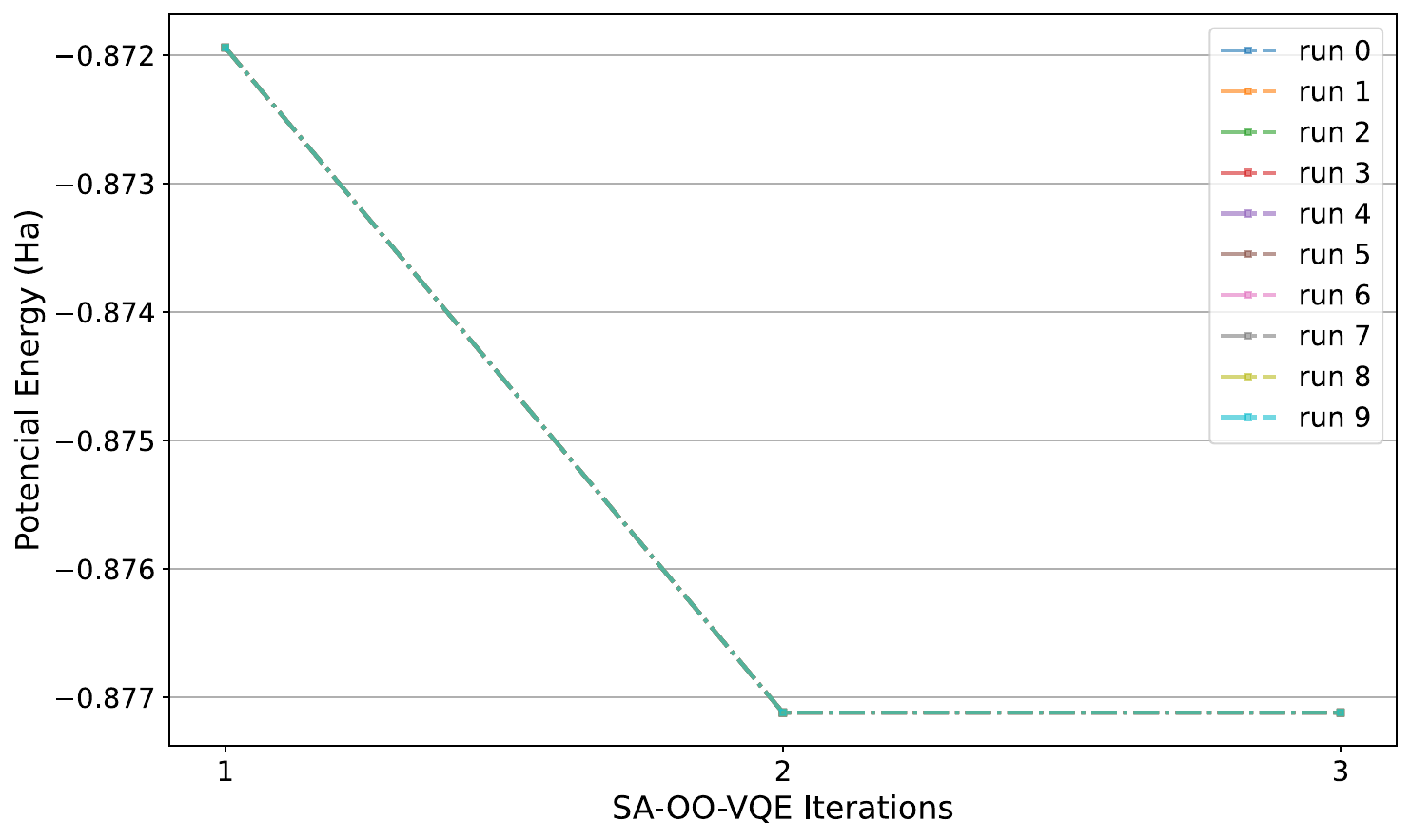}
        \caption{Energy vs. SA-OO-VQE iteration number.} 
        \label{fig:h2_bfgs_conv_iters} 
    \end{subfigure}
    \hfill 
    \begin{subfigure}[t]{0.49\textwidth} 
        \centering
        \includegraphics[width=\linewidth]{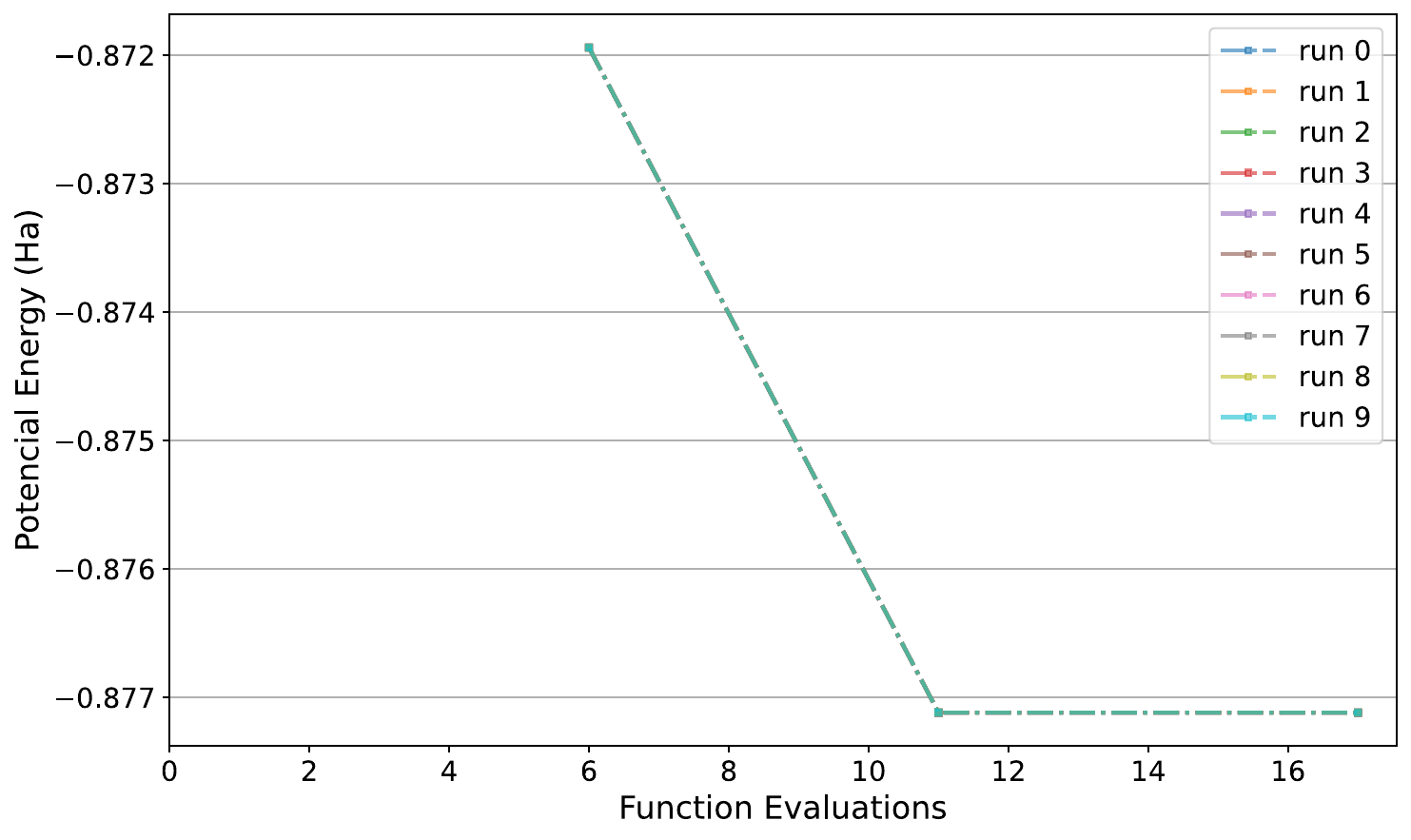}
        \caption{Energy after full SA-OO-VQE iterations vs. cumulative evaluations.} 
        \label{fig:h2_bfgs_conv_evals}
    \end{subfigure}

    \caption[Convergence plots of the BFGS optimizer within the SA-OO-VQE framework for the
H$_2$ molecule]{Convergence analysis of the BFGS optimizer within the SA-OO-VQE framework for the H$_2$ molecule, based on 10 independent runs (shown in different colors/styles, see legend in plots). The plots display the state-average energy (Hartrees) progression viewed against different metrics: 
    (\subref{fig:h2_bfgs_conv_iters}) Energy plotted at the end of each completed SA-OO-VQE iteration against the iteration number. 
    (\subref{fig:h2_bfgs_conv_evals}) Energy plotted at the end of each completed SA-OO-VQE iteration against the cumulative number of function evaluations consumed up to that iteration.}
\label{fig:h2_bfgs} 
\end{figure}

\begin{figure}[!htbp] 
    \centering
    \begin{subfigure}[t]{0.7\textwidth} 
        \centering
        \includegraphics[width=\linewidth]{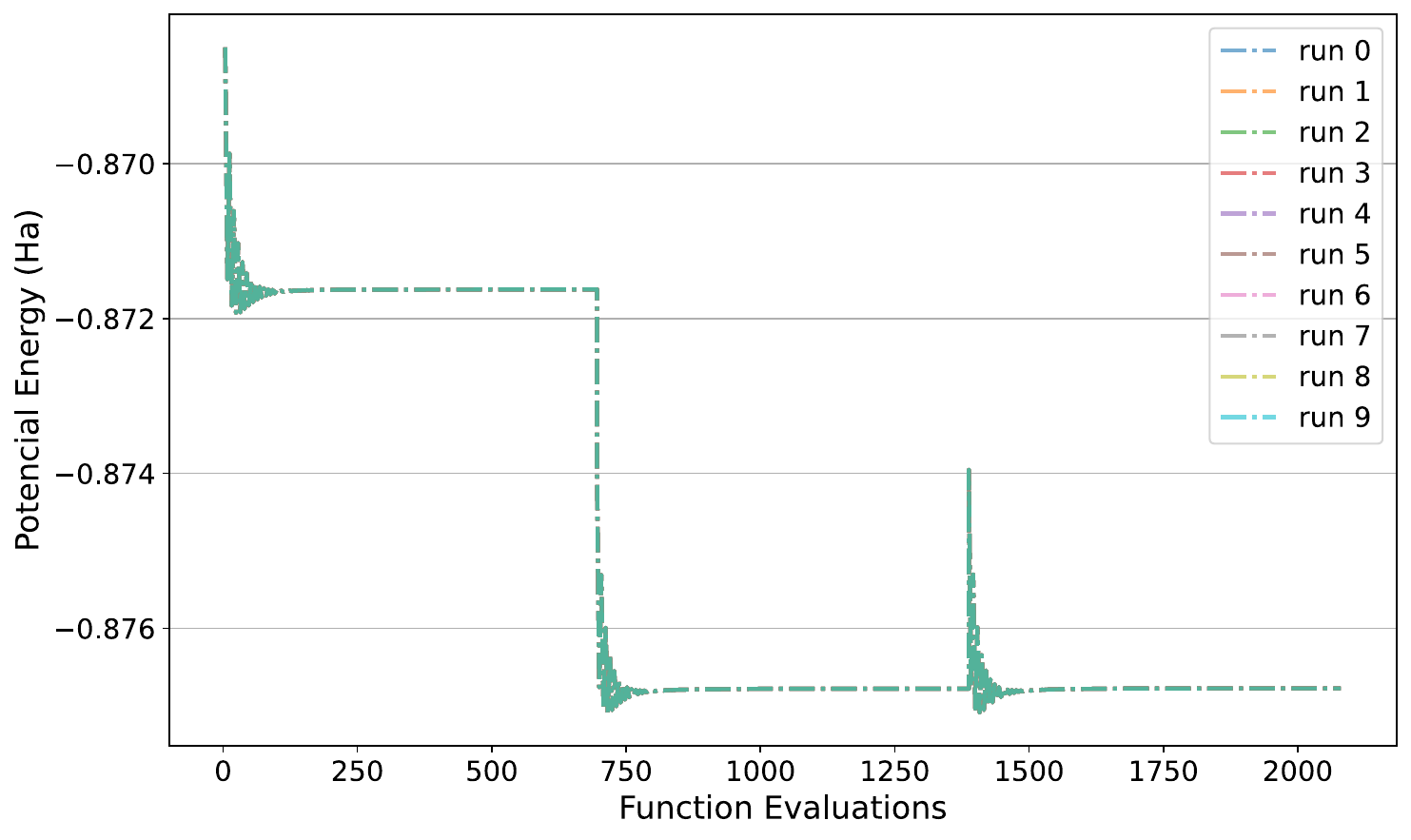}
        \caption{Energy after optimizer iterations vs. cumulative evaluations.} 
        \label{fig:h2_gd_conv} 
    \end{subfigure}
    \begin{subfigure}[t]{0.49\textwidth} 
        \centering
        \includegraphics[width=\linewidth]{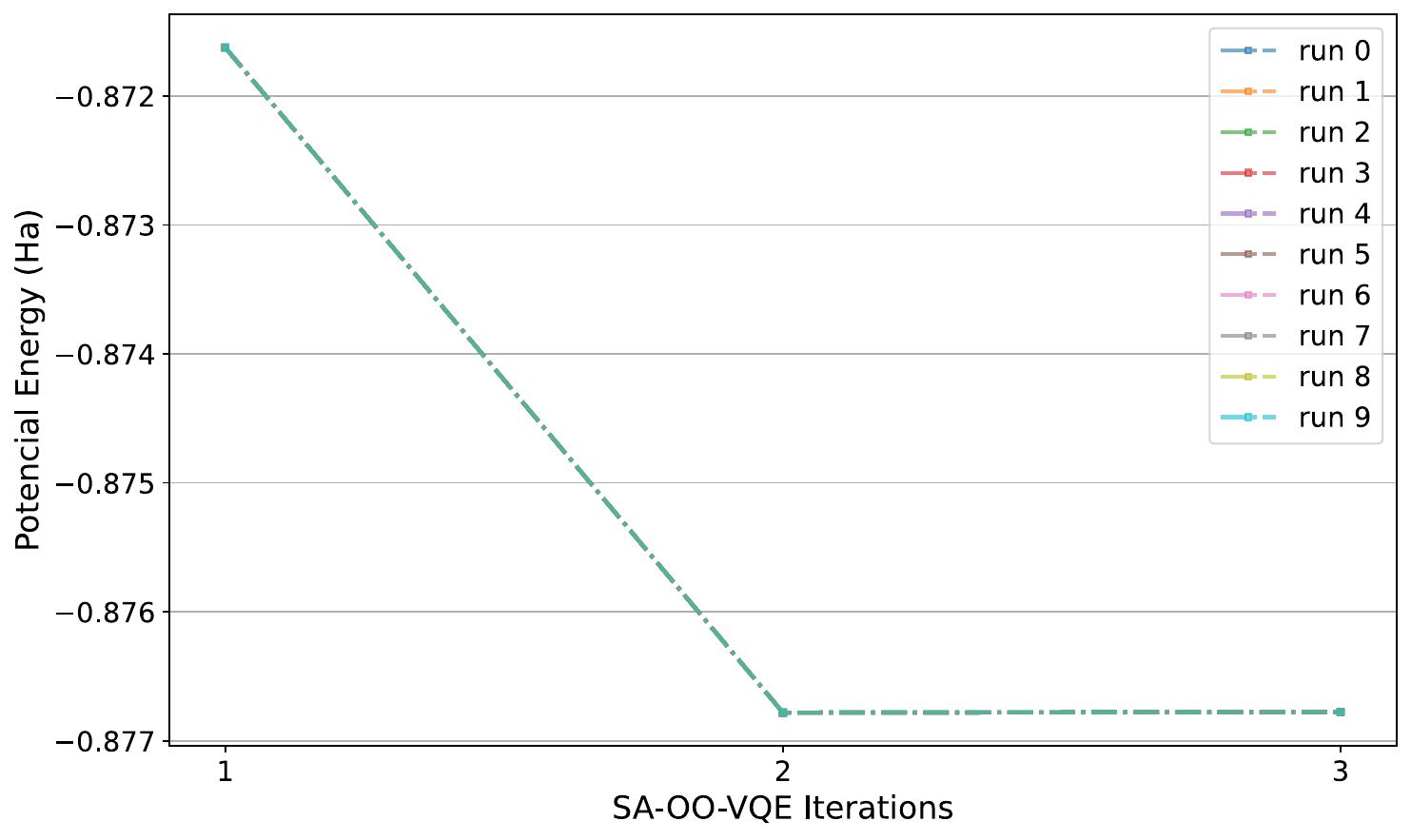}
        \caption{Energy vs. SA-OO-VQE iteration number.} 
        \label{fig:h2_gd_conv_iters} 
    \end{subfigure}
    \hfill 
    \begin{subfigure}[t]{0.49\textwidth} 
        \centering
        \includegraphics[width=\linewidth]{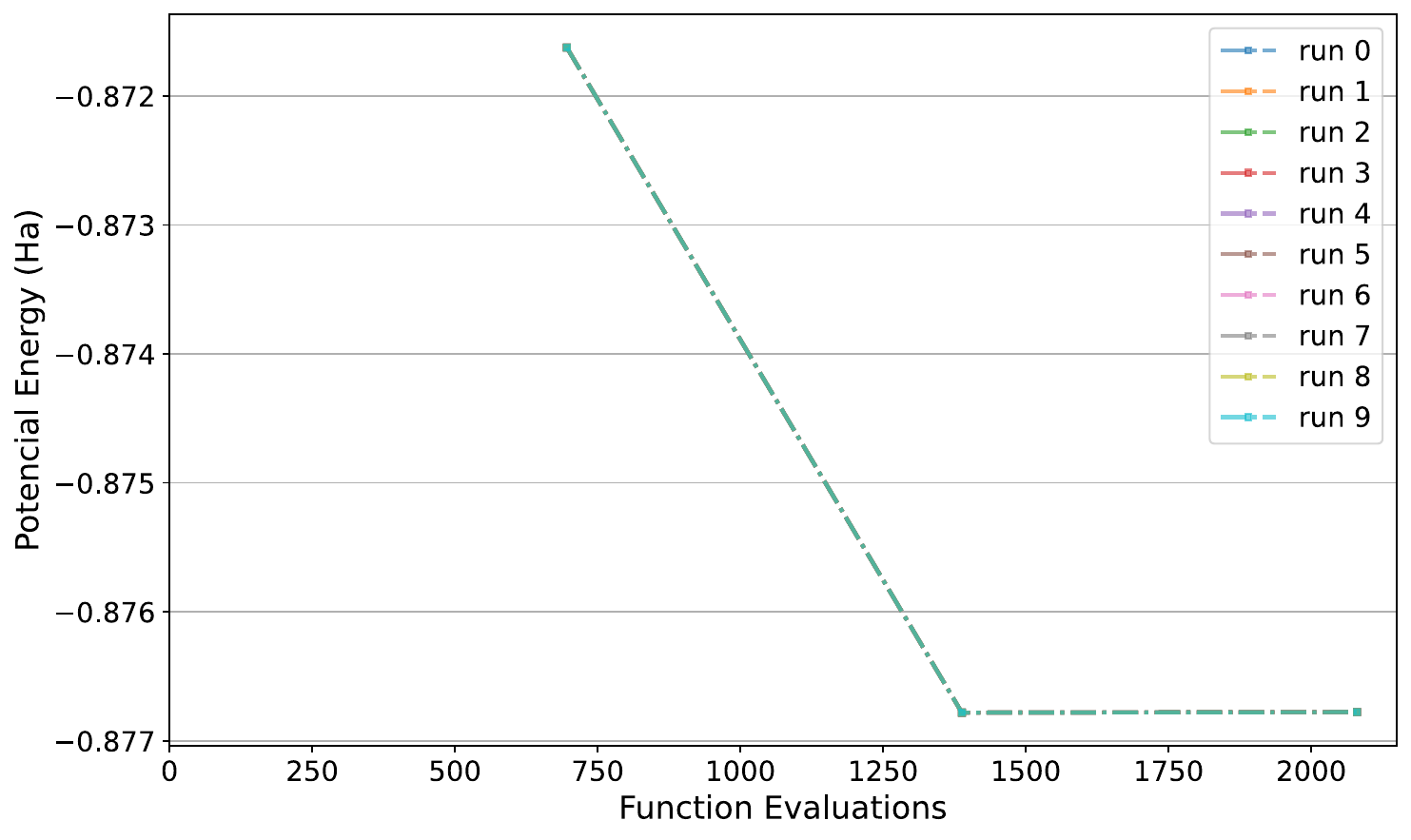}
        \caption{Energy after full SA-OO-VQE iterations vs. cumulative evaluations.} 
        \label{fig:h2_gd_conv_evals}
    \end{subfigure}

    \caption[Convergence plots of the Gradient Descent optimizer within the SA-OO-VQE framework for the
H$_2$ molecule]{Convergence analysis of the Gradient Descent optimizer within the SA-OO-VQE framework for the H$_2$ molecule, based on 10 independent runs (shown in different colors/styles, see legend in plots). The plots display the state-average energy (Hartrees) progression viewed against different metrics: 
    (\subref{fig:h2_gd_conv}) Energy evaluated at the end of each internal Gradient Descent optimizer iteration, plotted against the cumulative number of function evaluations consumed up to that iteration point
    (\subref{fig:h2_gd_conv_iters}) Energy plotted at the end of each completed SA-OO-VQE iteration against the iteration number. 
    (\subref{fig:h2_gd_conv_evals}) Energy plotted at the end of each completed SA-OO-VQE iteration against the cumulative number of function evaluations consumed up to that iteration.}
\label{fig:h2_grad_desc} 
\end{figure}

Now, as was said in section~\ref{sec:experiments_methodology}, we present the convergence plots for each optimizer to demonstrate optimization dynamics.
Just convergence plots for BFGS (Figure~\ref{fig:h2_bfgs}), Gradient Descent (Figure~\ref{fig:h2_grad_desc}) and DE/Rand/2/bin (Figure~\ref{fig:h2_de_r2}) are presented there; the rest of them are located in the appendix in section~\ref{ap_sec:h2_conv}, which contains figures for COBYLA (Figure~\ref{fig:h2_cobyla}), SLSQP (Figure~\ref{fig:h2_slsqp}), and the rest of the DE configurations (Figures~\ref{fig:h2_de_b1}, \ref{fig:h2_de_b2}, \ref{fig:h2_de_ctb}, \ref{fig:h2_de_ctr}, \ref{fig:h2_de_r1}, and \ref{fig:h2_de_rtb}).
Let's first take a look at the convergence plot of BFGS (Figure~\ref{fig:h2_bfgs}). 
Analysis of the iterative SA-OO-VQE procedure reveals two notable characteristics regarding its convergence and computational cost. 
Firstly, the number of function evaluations required per SA-OO-VQE iteration remains relatively consistent throughout the process. 
Secondly, the energy reduction achieved diminishes with successive iterations; the most significant energy improvement occurs between the first and second iteration, compared to a smaller decrease between the second and third. 
This pattern of convergence, characterized by initially large energy drops followed by smaller refinements alongside stable per-iteration cost, is consistently observed across different optimization methods, as exemplified by the results for COBYLA~(Figure~\ref{fig:h2_cobyla}), Gradient Descent~(Figure~\ref{fig:h2_grad_desc}), and SLSQP~(Figure~\ref{fig:h2_slsqp}).

Let's now look in more detail at the Gradient Descent convergence plot (Figure~\ref{fig:h2_grad_desc}). 
Particularly informative is the visualization where the energy, evaluated upon completion of each internal Gradient Descent optimizer iteration, is plotted against the cumulative number of function evaluations consumed up to that iteration. 
Within this plot, distinct peaks are observable. 
These peaks correspond to the main cycles of the SA-OO-VQE procedure and arise immediately after the molecular orbital optimization step. 
They signify a transient increase in the evaluated energy occurring when the system's energy is calculated using the newly optimized orbitals before those parameters are subsequently refined.
Finally, the convergence behavior of the DE optimizers is examined. 
While the DE/Rand/2/Bin strategy (Figure~\ref{fig:h2_de_r2}) is selected as a representative example, the key observations presented here generally apply to all tested \texttt{DE} variants (Figures~\ref{fig:h2_de_b1},~\ref{fig:h2_de_b2},~\ref{fig:h2_de_ctb},~\ref{fig:h2_de_ctr},~\ref{fig:h2_de_r1}, and~\ref{fig:h2_de_rtb}).

Similar to the Gradient Descent results (Figure~\ref{fig:h2_grad_desc}), the DE convergence plots exhibit peaks coinciding with the main SA-OO-VQE cycles, occurring immediately after the molecular orbital optimization step. 
However, a distinct characteristic apparent in the DE results is a frequently non-ideal convergence trajectory between these main cycles. 
In contrast to the more consistent energy reduction typically sought during optimization, many independent DE runs display non-monotonic behavior; specifically, the final energy obtained at the end of some SA-OO-VQE iterations is higher than the energy achieved at the end of the preceding iteration. 
Indeed, only a minority of the DE runs exhibit a consistently decreasing energy profile across the entire multi-iteration SA-OO-VQE procedure.

\clearpage

\begin{figure}[!htbp] 
    \centering
    \begin{subfigure}[t]{0.7\textwidth} 
        \centering
        \includegraphics[width=\linewidth]{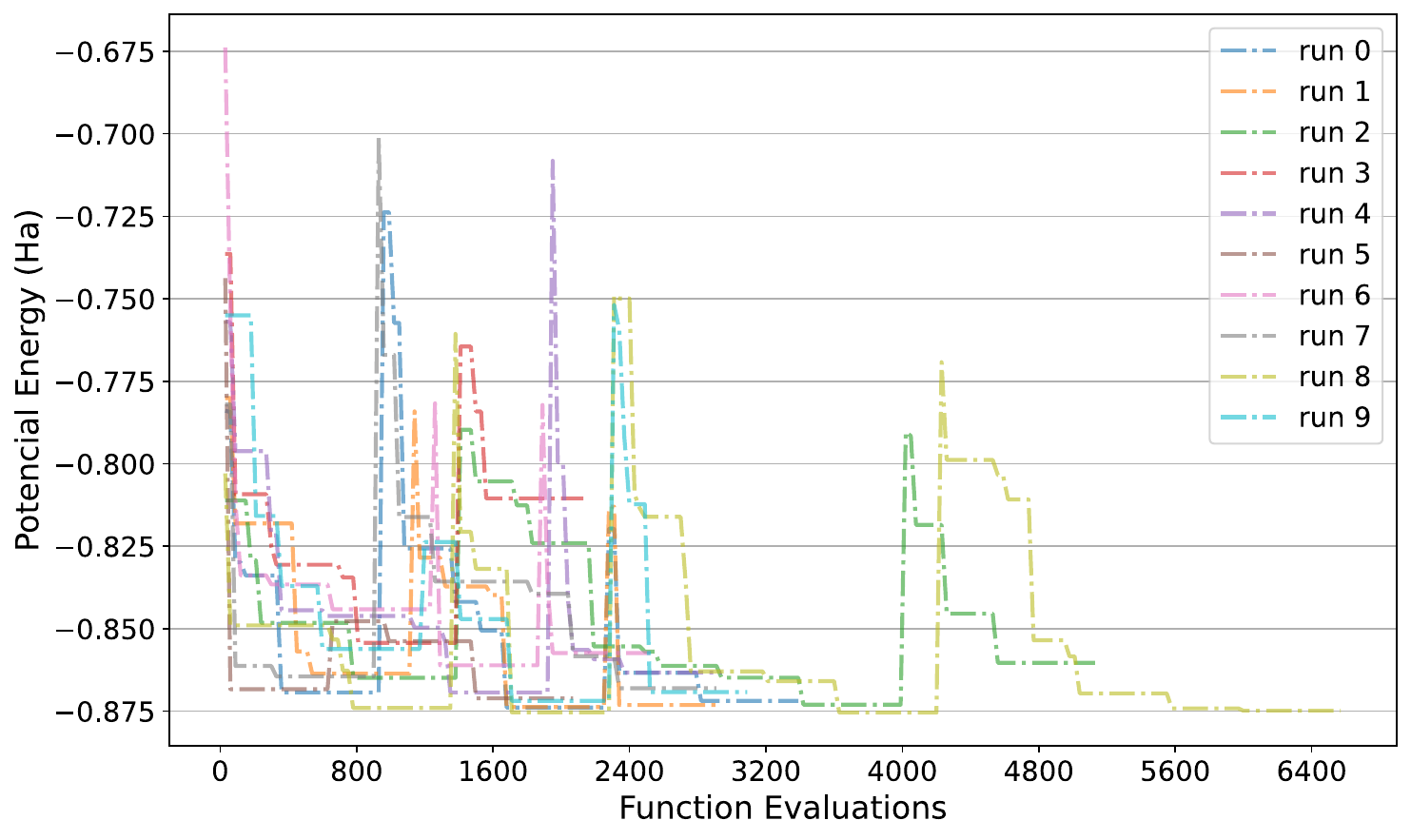}
        \caption{Energy after optimizer iterations vs. cumulative evaluations.} 
        \label{fig:h2_de_r2_conv} 
    \end{subfigure}
    \begin{subfigure}[t]{0.49\textwidth} 
        \centering
        \includegraphics[width=\linewidth]{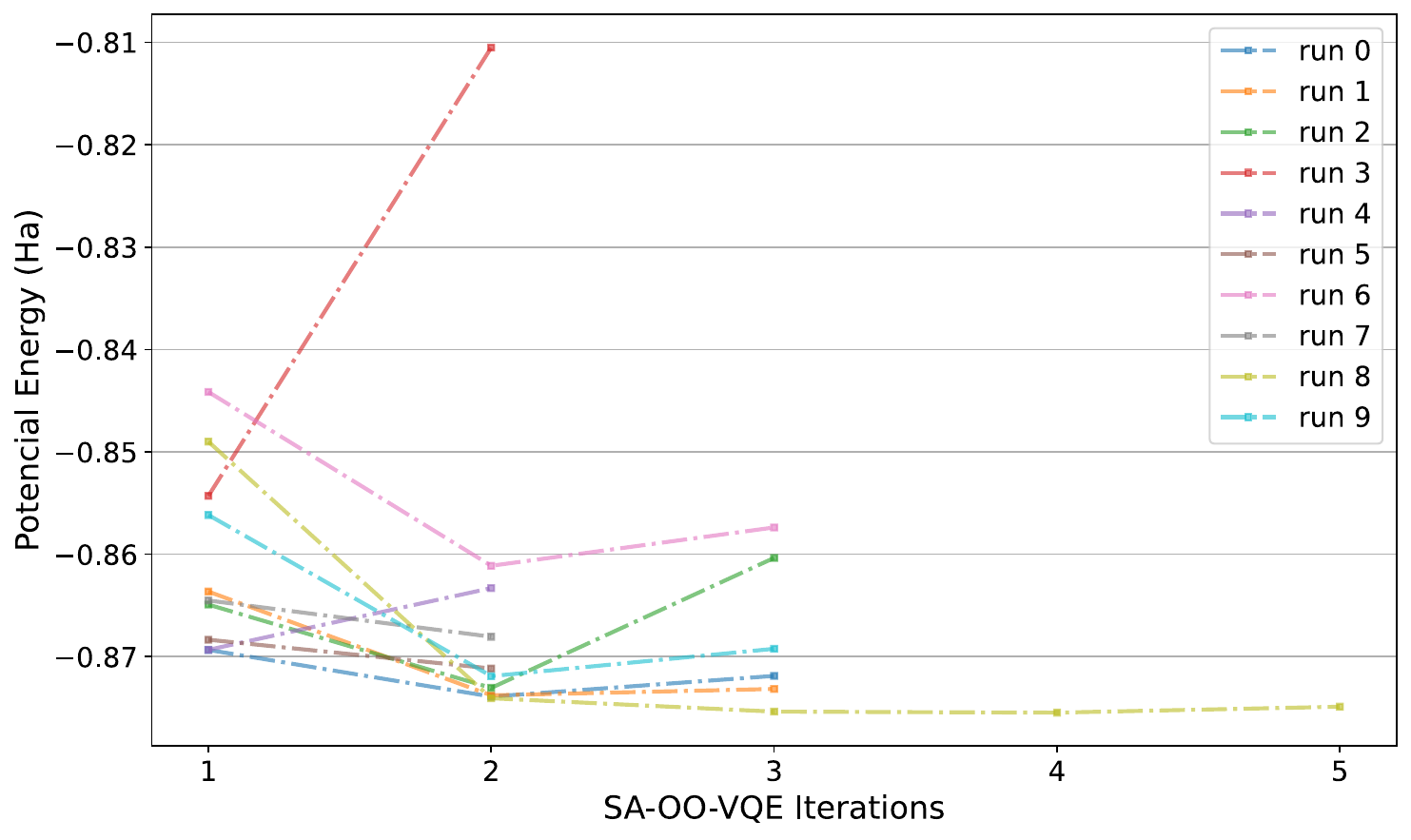}
        \caption{Energy vs. SA-OO-VQE iteration number.} 
        \label{fig:h2_de_r2_conv_iters} 
    \end{subfigure}
    \hfill 
    \begin{subfigure}[t]{0.49\textwidth} 
        \centering
        \includegraphics[width=\linewidth]{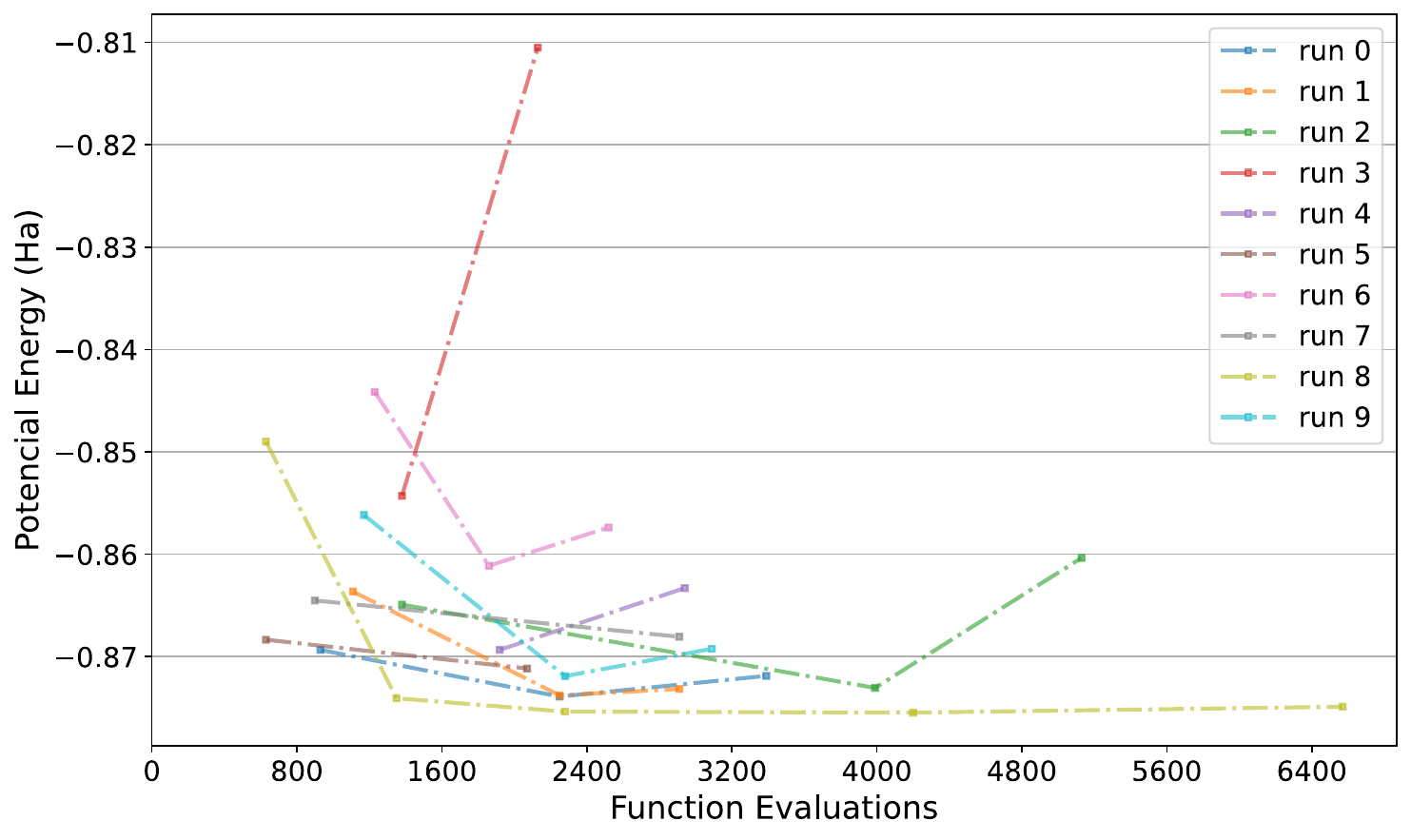}
        \caption{Energy after full SA-OO-VQE iterations vs. cumulative evaluations.} 
        \label{fig:h2_de_r2_conv_evals}
    \end{subfigure}

    \caption[Convergence plots of the DE/Rand/2/bin optimizer within the SA-OO-VQE framework for the
H$_2$ molecule]{Convergence analysis of the DE/Rand/2/bin optimizer within the SA-OO-VQE framework for the H$_2$ molecule, based on 10 independent runs (shown in different colors/styles, see legend in plots). The plots display the state-average energy (Hartrees) progression viewed against different metrics: 
    (\subref{fig:h2_de_r2_conv}) Energy evaluated at the end of each internal Gradient Descent optimizer iteration, plotted against the cumulative number of function evaluations consumed up to that iteration point
    (\subref{fig:h2_de_r2_conv_iters}) Energy plotted at the end of each completed SA-OO-VQE iteration against the iteration number. 
    (\subref{fig:h2_de_r2_conv_evals}) Energy plotted at the end of each completed SA-OO-VQE iteration against the cumulative number of function evaluations consumed up to that iteration.}
\label{fig:h2_de_r2} 
\end{figure}

In summary, the convergence plots presented in this section for selected optimizers (with others available in Appendix~\ref{ap_sec:h2_conv}) illustrate key dynamics of the SA-OO-VQE procedure for H$_2$. 
A general trend, observed across methods like BFGS, COBYLA, Gradient Descent, and SLSQP, is that energy reduction diminishes over successive main SA-OO-VQE iterations, even as the function evaluations per iteration remain relatively constant. 
More detailed plots, particularly for Gradient Descent and DE, reveal transient energy peaks occurring immediately after the molecular orbital optimization step, resulting from the evaluation using updated orbitals before VQE parameter refinement.
Notably distinct is the behavior of the DE optimizers; in addition to the transient peaks, they frequently exhibit non-monotonic convergence between SA-OO-VQE iterations, where the energy minimum may increase relative to the previous iteration, a behavior rarely seen in the smoother convergence profiles of methods like BFGS and indicative of DEs convergence difficulties in this context.

\subsection{Molecule H$_4$}
\label{sec:molecule_h4}
\begin{figure}[!htbp] 
    \centering
    
    \begin{subfigure}[t]{0.49\textwidth} 
        \centering
        \includegraphics[width=\linewidth]{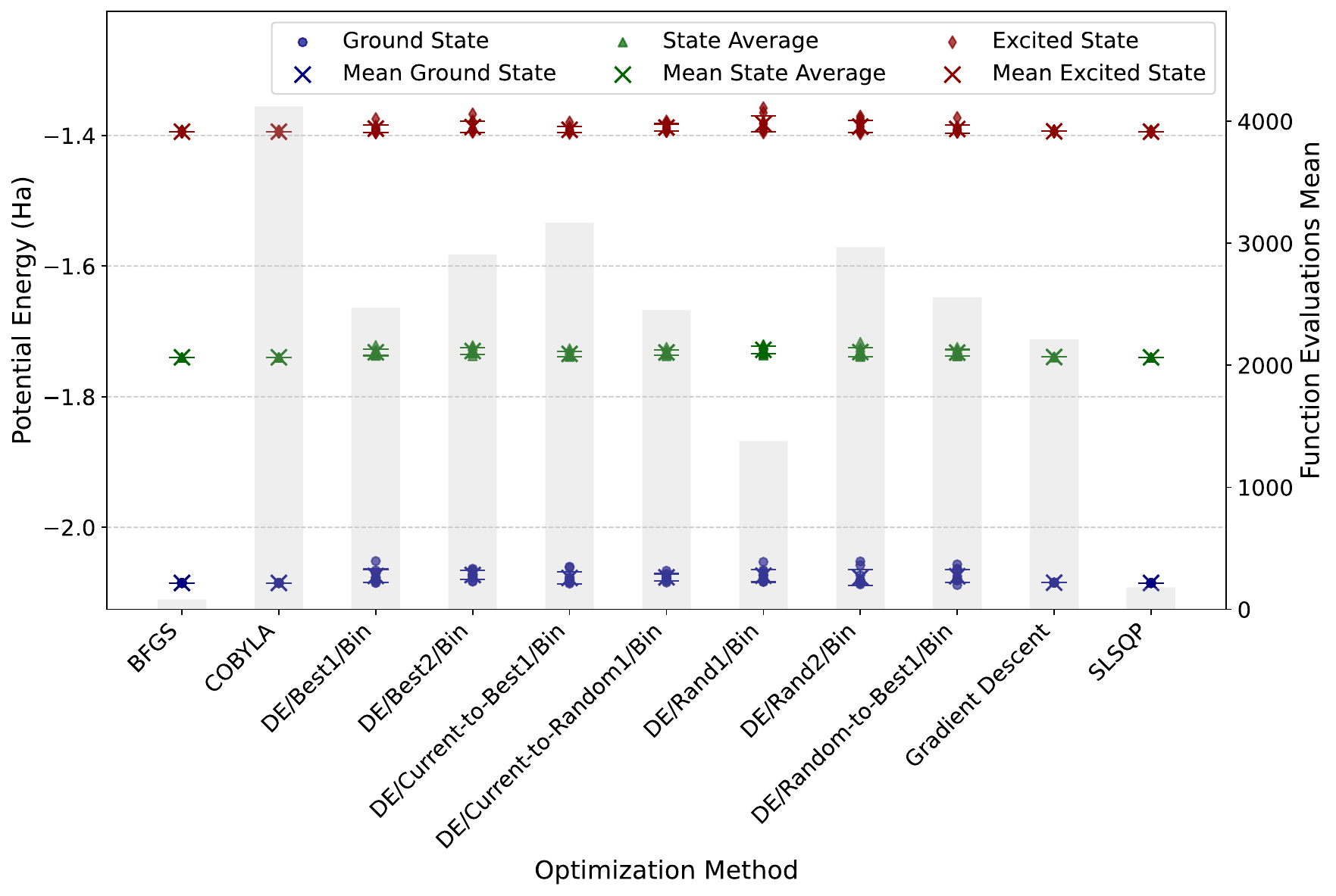}
        \caption{All calculated energies and function evaluations.}
        \label{fig:h4_saoovqe_all_energies} 
    \end{subfigure}
    \hfill 
    \begin{subfigure}[t]{0.49\textwidth} 
        \centering
        \includegraphics[width=\linewidth]{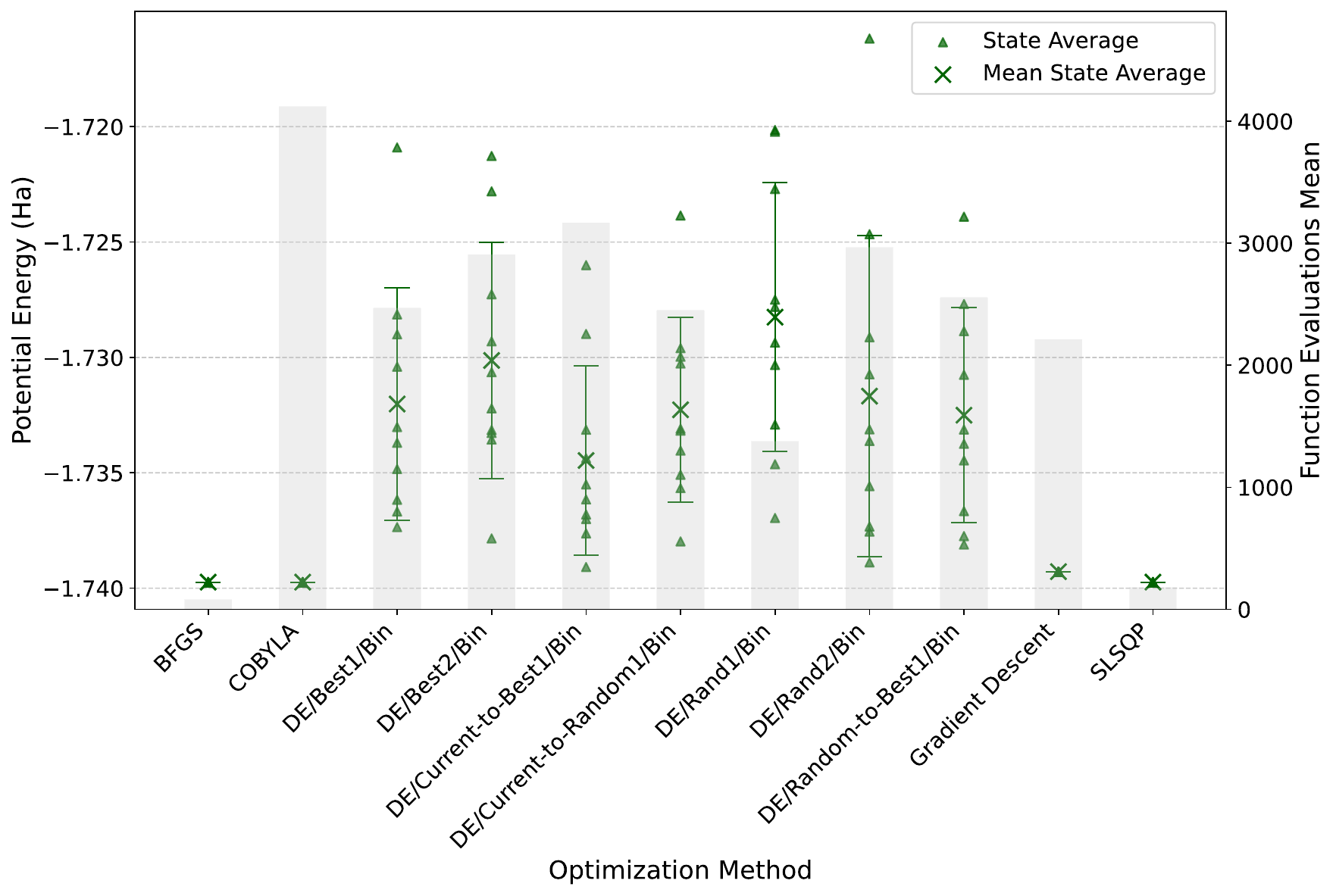}
        \caption{State-average energy detail and function evaluations.}
        \label{fig:h4_saoovqe_sa_detail}
    \end{subfigure}

    \caption[Comparative optimizer results from SA-OO-VQE for H$_4$ molecule.]{Performance of different optimizers within the SA-OO-VQE framework for the H$_4$ molecule. 
    Both plots show results derived from 10 independent optimization runs for each method. Individual run results are shown as markers, means as crosses ($\times$), standard deviations across runs are shown as error bars (left y-axis of each plot, Hartrees), and mean function evaluations shown as gray bars on the right y-axis of each plot. 
    (\subref{fig:h4_saoovqe_all_energies}) Displays ground state (blue circles), excited state (red diamond), and state-average (green triangle) energies. 
    (\subref{fig:h4_saoovqe_sa_detail}) Provides a detailed view of the state-average energy.}
    \label{fig:h4_saoovqe_comparison} 
\end{figure}

This section presents the comparative performance of the different optimization algorithms for the H$_4$ chain molecule simulation, mirroring the analysis performed previously for H$_2$ (Section~\ref{sec:molecule_h2}). 
The calculations were performed using the sto-3g basis~\cite{book_basis}. 
Table~\ref{tab:calculation_summary_h4} details the key statistics derived from 10 independent runs for each optimizer, including the minimum, maximum, and mean function evaluations (evals) required for convergence, alongside the corresponding final state-average energies ($E$ in Hartrees). These findings are also illustrated graphically in Figure~\ref{fig:h4_saoovqe_comparison}.
\begin{table}[!htbp]
\centering
\small
\begin{tabular}{l c c c c c c}
\toprule
Method & $evals_{min}$ & $evals_{max}$ & $evals_{mean}$ & $E_{min}$ & $E_{max}$ & $E_{mean}$ \\
\midrule
BFGS & 79 & 79 & 79 & -1.739735 & -1.739735 & -1.739735 \\
COBYLA & 4067 & 4252 & 4121 & -1.739735 & -1.739735 & -1.739735 \\
DE/Best1/Bin & 1500 & 4020 & 2469 & -1.737352 & -1.720899 & -1.732016 \\
DE/Best2/Bin & 1710 & 4620 & 2907 & -1.737840 & -1.721269 & -1.730129 \\
DE/Current-to-Best1/Bin & 2040 & 4800 & 3168 & -1.739082 & -1.725997 & -1.734465 \\
DE/Current-to-Random1/Bin & 1290 & 4350 & 2451 & -1.737971 & -1.723848 & -1.732268 \\
DE/Rand1/Bin & 690 & 2850 & 1377 & -1.736956 & -1.720139 & -1.728253 \\
DE/Rand2/Bin & 1680 & 4530 & 2967 & -1.738881 & -1.716178 & -1.731676 \\
DE/Random-to-Best1/Bin & 1410 & 4050 & 2556 & -1.738107 & -1.723898 & -1.732502 \\
Gradient Descent & 2212 & 2212 & 2212 & -1.739282 & -1.739282 & -1.739282 \\
SLSQP & 177 & 177 & 177 & -1.739735 & -1.739735 & -1.739735 \\
\bottomrule
\end{tabular}
\caption[Comparative optimizer results from SA-OO-VQE for H$_4$ molecule.]{Comparison of optimizer performance for the H$_4$ molecule simulation. 
The table reports statistics based on 10 independent runs: minimum ($min$), maximum ($max$), and mean ($mean$) number of function evaluations ($evals$) required for convergence, and the corresponding final state-average energies ($E$ in Hartrees).} 
\label{tab:calculation_summary_h4}
\end{table}
Similar to the findings for the H$_2$ molecule, the results for H$_4$ reveal significant differences in efficiency and reliability among the tested optimizers.

BFGS, SLSQP, and COBYLA again demonstrate the ability to consistently locate the same lowest state-average energy ($-1.739735$~Ha). However, their efficiency varies dramatically. BFGS remains highly efficient (79 evaluations), followed by SLSQP (177 evaluations). COBYLA, while accurate, requires substantially more computational effort (mean of 4121 evaluations).

Gradient Descent, while deterministic, converges to a slightly higher energy ($-1.739282$~Ha) compared to the best methods and requires a considerable number of evaluations (2212).

The DE variants continue the trend observed for H$_2$, generally requiring a large number of function evaluations (mean values ranging from 1377 to 3168) with significant variability between runs. 
Crucially, they struggle with robustness and accuracy for H$_4$ as well. Their mean final energies (around $-1.728$ to $-1.734$~Ha) are notably higher than the optimal value found by BFGS/SLSQP/COBYLA. 
The large spread between minimum and maximum achieved energies ($E_{min}$ and $E_{max}$) further highlights their stochastic nature and difficulty in reliably converging to the best solution for this system, with even the best runs ($E_{min}$) failing to match the energy achieved consistently by the top local methods.

The data in Table~\ref{tab:calculation_summary_h4}, visually supported by Figure~\ref{fig:h4_saoovqe_comparison}, therefore reinforces the observations from the H$_2$ study. 
It suggests that for the SA-OO-VQE optimization on these small molecular systems, gradient-based local optimizers like BFGS and SLSQP offer superior performance in terms of finding the optimal energy with high efficiency and consistency compared to standard Gradient Descent and the stochastic DE approaches tested.

\begin{figure}[!htb] 
    \centering
    \begin{subfigure}[t]{0.7\textwidth} 
        \centering
        \includegraphics[width=\linewidth]{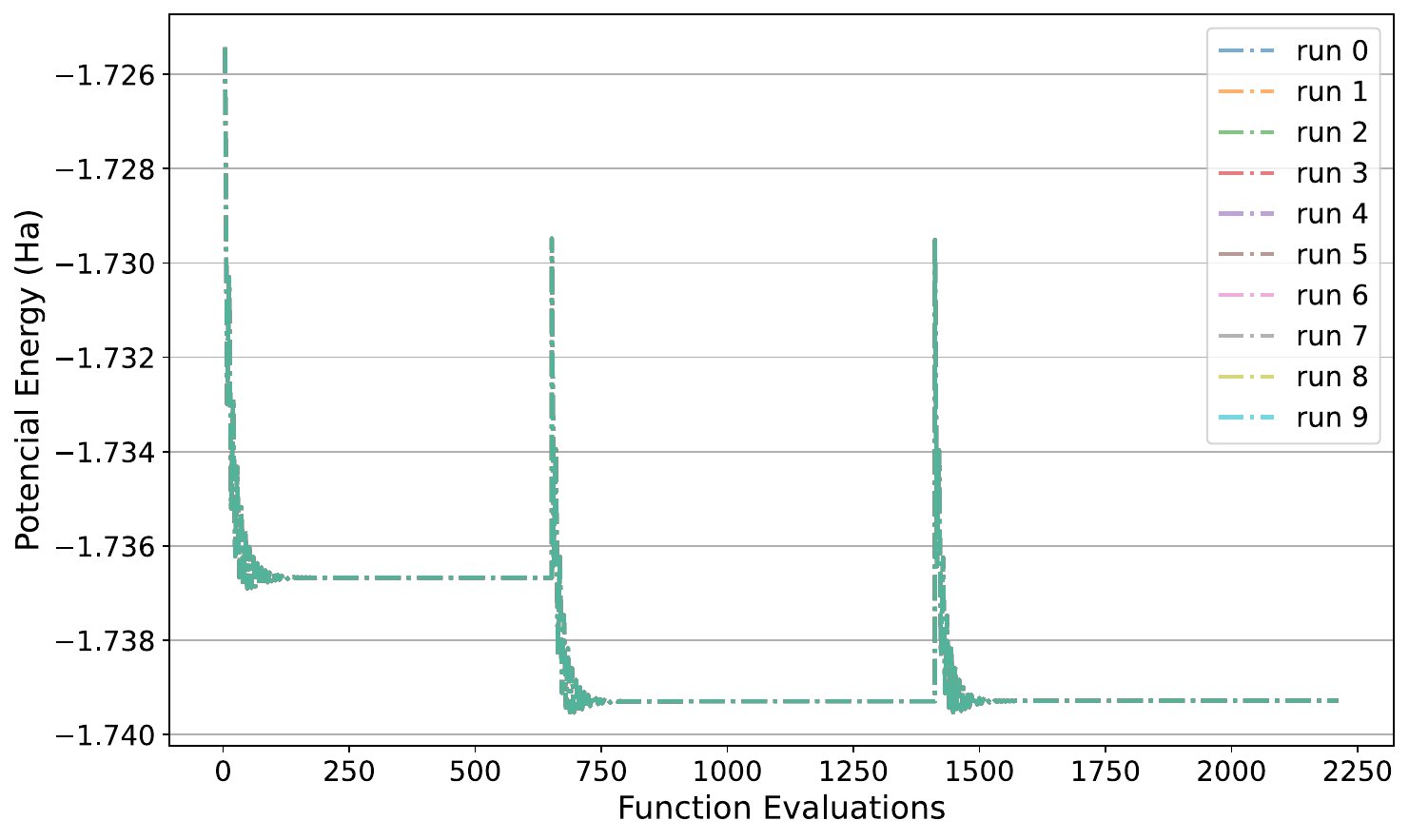}
        \caption{Energy after optimizer iterations vs. cumulative evaluations.} 
        \label{fig:h4_gd_conv} 
    \end{subfigure}
    \begin{subfigure}[t]{0.49\textwidth} 
        \centering
        \includegraphics[width=\linewidth]{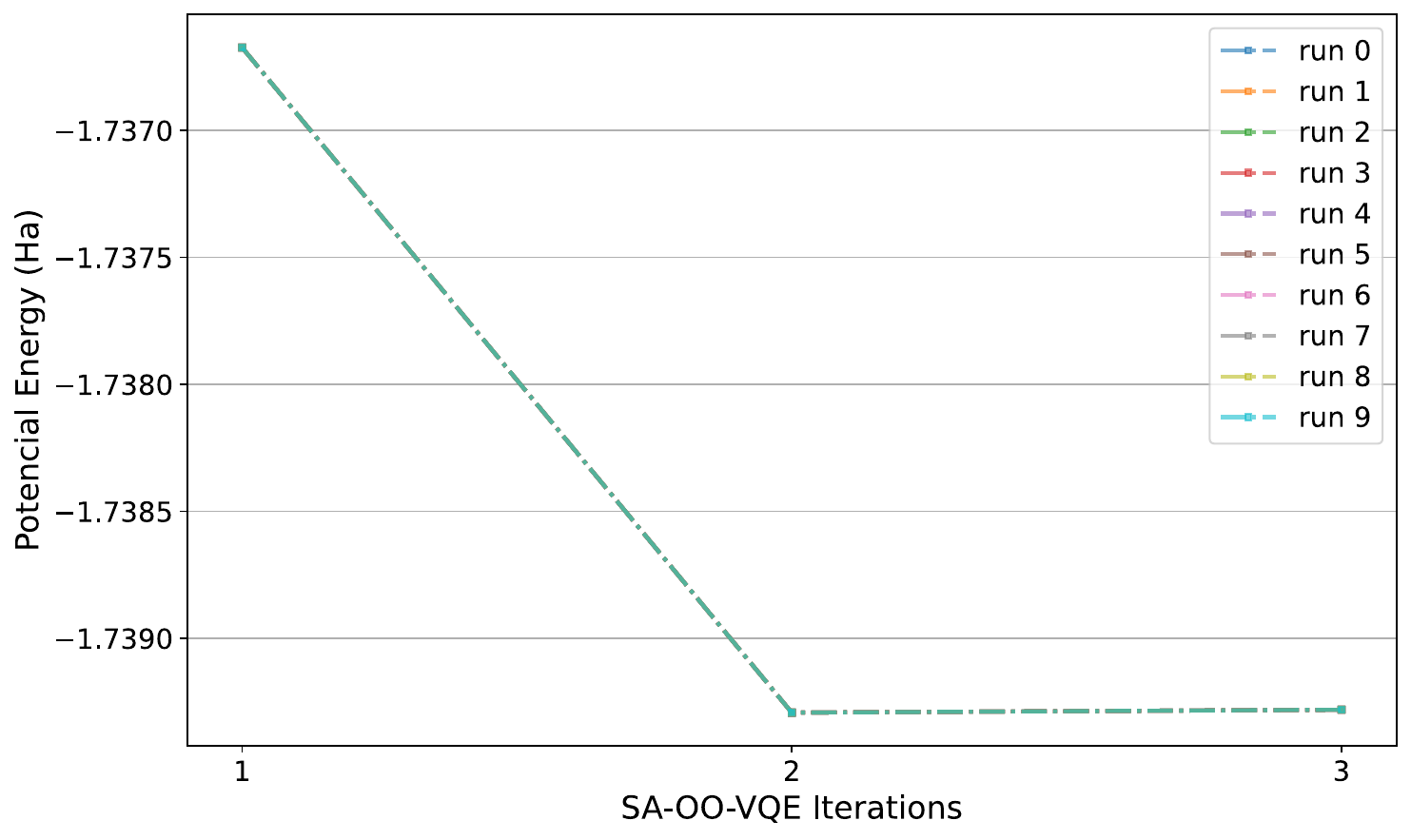}
        \caption{Energy vs. SA-OO-VQE iteration number.} 
        \label{fig:h4_gd_conv_iters} 
    \end{subfigure}
    \hfill 
    \begin{subfigure}[t]{0.49\textwidth} 
        \centering
        \includegraphics[width=\linewidth]{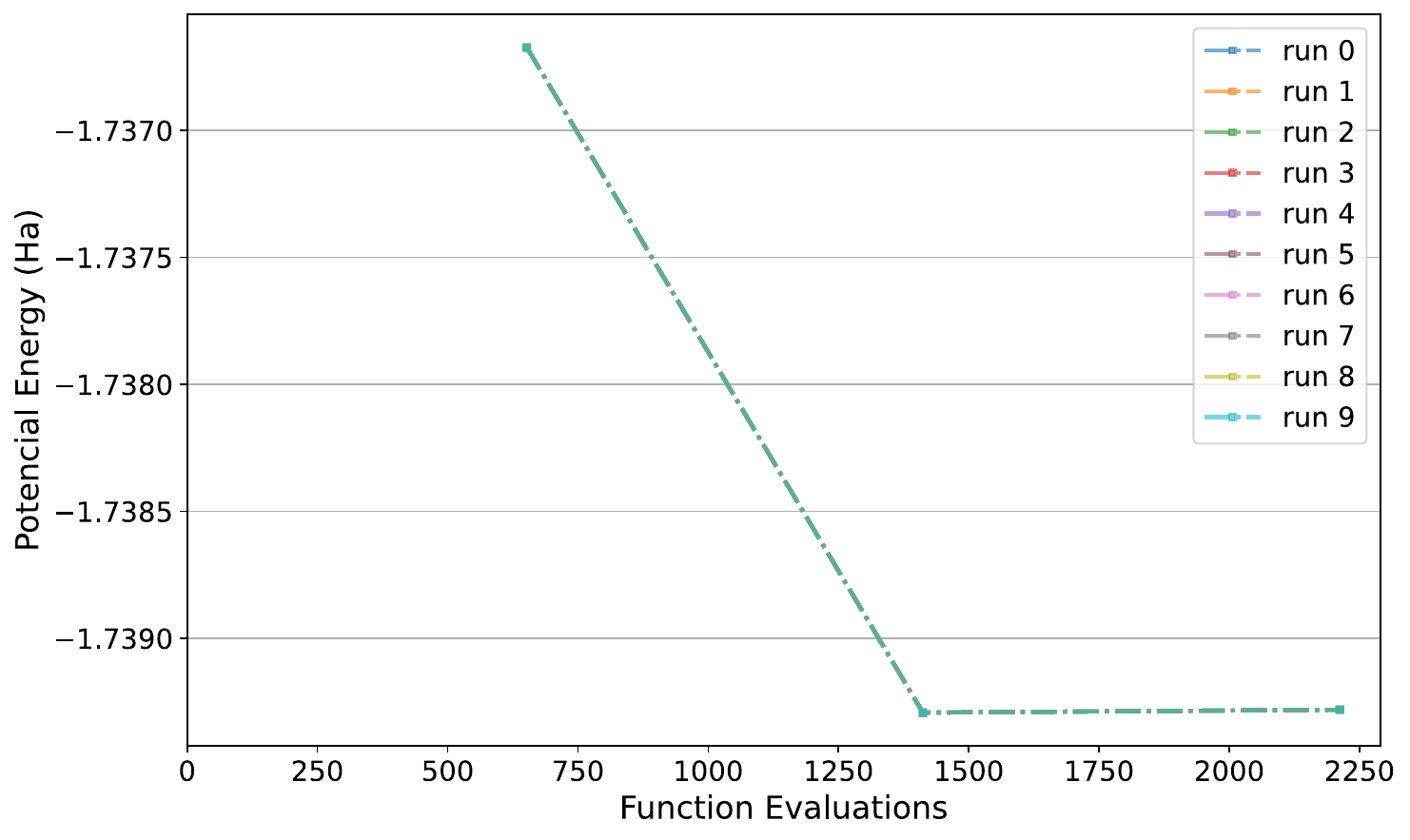}
        \caption{Energy after full SA-OO-VQE iterations vs. cumulative evaluations.}
        \label{fig:h4_gd_conv_evals}
    \end{subfigure}

    \caption[Convergence plots of the Gradient Descent optimizer within the SA-OO-VQE framework for the
H$_4$ molecule.]{Convergence analysis of the Gradient Descent optimizer within the SA-OO-VQE framework for the H$_4$ molecule, based on 10 independent runs (shown in different colors/styles, see legend in plots). The plots display the state-average energy (Hartrees) progression viewed against different metrics: 
    (\subref{fig:h4_gd_conv}) Energy evaluated at the end of each internal Gradient Descent optimizer iteration, plotted against the cumulative number of function evaluations consumed up to that iteration point
    (\subref{fig:h4_gd_conv_iters}) Energy plotted at the end of each completed SA-OO-VQE iteration against the iteration number. 
    (\subref{fig:h4_gd_conv_evals}) Energy plotted at the end of each completed SA-OO-VQE iteration against the cumulative number of function evaluations consumed up to that iteration.}
\label{fig:h4_grad_desc} 
\end{figure}

Now, as was said in section~\ref{sec:experiments_methodology}, we present the convergence plots for each optimizer to demonstrate optimization dynamics.
Because the convergence plots for the H$_4$ molecule share most of the properties with plots for molecule H$_2$, only the convergence plot of Gradient Descent~(Figure~\ref{fig:h4_grad_desc}) is shown in this section.
Specifically, we observe the characteristic decrease in the state-average energy as the optimization progresses, both in terms of individual optimizer steps (Figure~\ref{fig:h4_gd_conv}) and full SA-OO-VQE iterations (Figs.~\ref{fig:h4_gd_conv_iters} and~\ref{fig:h4_gd_conv_evals}). 
The variability between independent runs, reflecting different optimization trajectories potentially influenced by initialization or noise, is also apparent, similar to the H$_2$ case. 
Furthermore, the specific properties previously noted for individual optimizers, such as the convergence characteristics DE observed for H$_2$, remain consistent for the H$_4$ system.

For detailed convergence plots of the other optimizers, please refer to Appendix~\ref{ap_sec:h4_conv}, which contains figures for BFGS (Figure~\ref{fig:h4_bfgs}), COBYLA (Figure~\ref{fig:h4_cobyla}), SLSQP (Figure~\ref{fig:h4_slsqp}), and all DE configurations (Figures~\ref{fig:h4_de_b1}, \ref{fig:h4_de_b2}, \ref{fig:h4_de_ctb}, \ref{fig:h4_de_ctr}, \ref{fig:h4_de_r1}, \ref{fig:h4_de_r2}, and \ref{fig:h4_de_rtb}).

However, a slight difference arises in the computational cost required to reach convergence for BFGS, COBYLA, Gradient Descent, and SLSQP. 
Compared to the H$_2$ molecule, the optimization for H$_4$ required a different number of SA-OO-VQE iterations (visible on the x-axis of Figure~\ref{fig:h4_gd_conv_iters}) and consequently, for some optimizers a  different total number of function evaluations (visible on the x-axes of Figs.~\ref{fig:h4_gd_conv} and~\ref{fig:h4_gd_conv_evals}). 
Despite this, the fundamental convergence patterns and the relative performance characteristics observed among different optimizers (as detailed for H$_2$ and shown for all optimizers for H$_4$ in Appendix~\ref{ap_sec:h4_conv}) remain consistent.

\subsection{Molecule LiH}
\label{sec:molecule_lih}
\begin{figure}[!htb] 
    \centering
    
    \begin{subfigure}[t]{0.49\textwidth} 
        \centering
        \includegraphics[width=\linewidth]{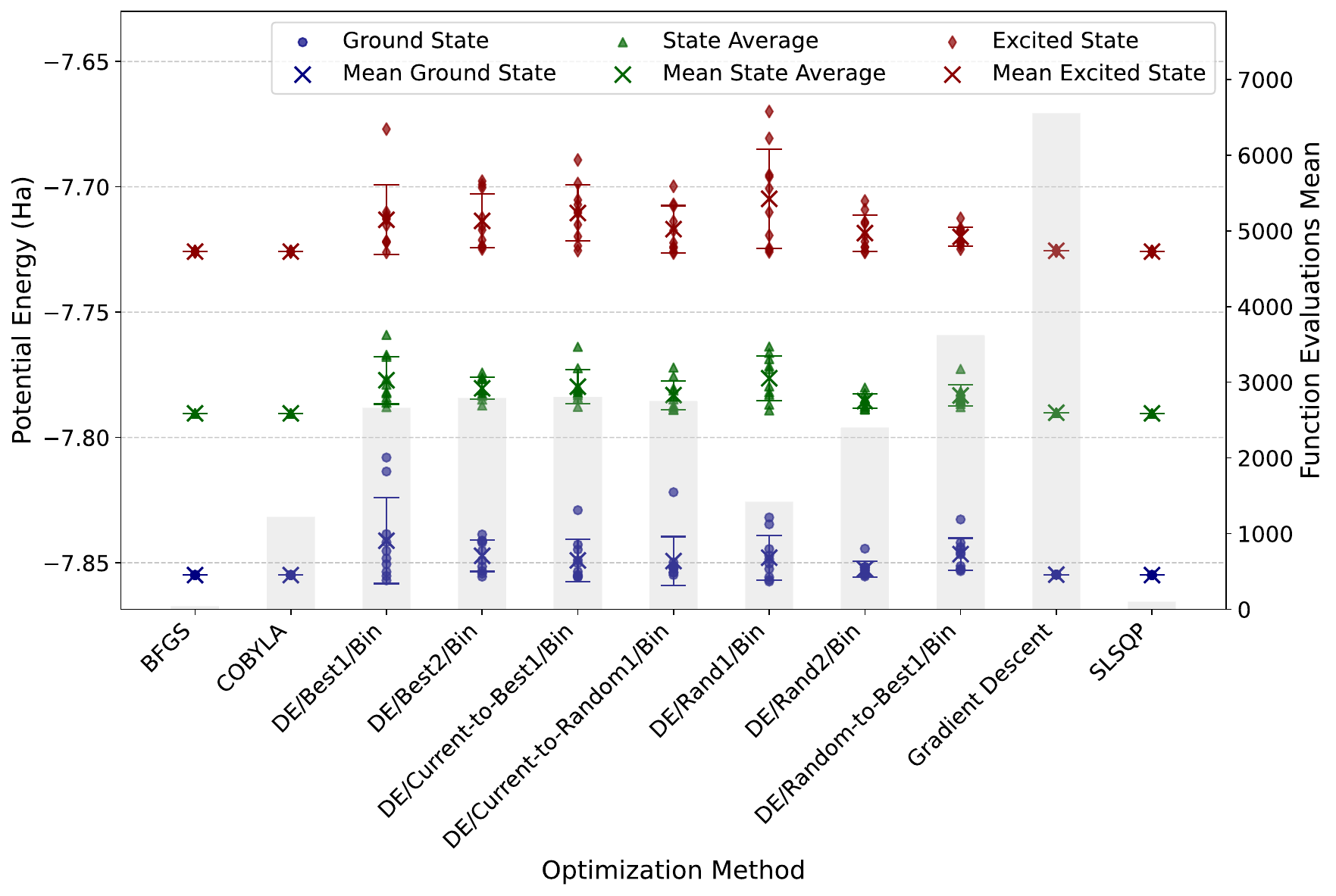}
        \caption{All calculated energies and function evaluations.}
        \label{fig:lih_saoovqe_all_energies} 
    \end{subfigure}
    \hfill 
    \begin{subfigure}[t]{0.49\textwidth} 
        \centering
        \includegraphics[width=\linewidth]{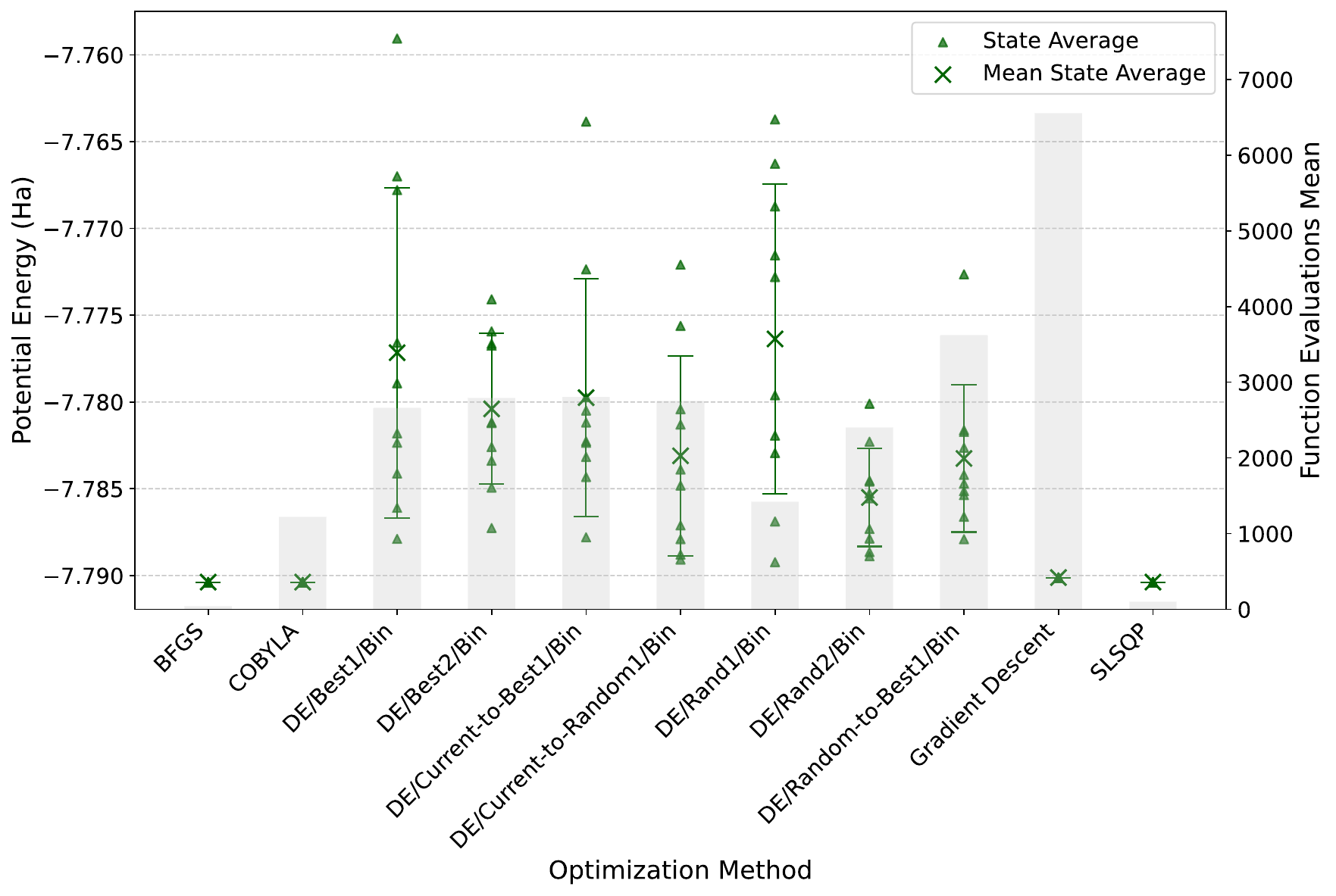}
        \caption{State-average energy detail and function evaluations.}
        \label{fig:lih_saoovqe_sa_detail}
    \end{subfigure}

    \caption[Comparative optimizer results from SA-OO-VQE for LiH molecule.]{Performance of different optimizers within the SA-OO-VQE framework for the LiH molecule. 
    Both plots show results derived from 10 independent optimization runs for each method. Individual run results are shown as markers, means as crosses ($\times$), standard deviations across runs are shown as error bars (left y-axis of each plot, Hartrees), and mean function evaluations shown as gray bars on the right y-axis of each plot. 
    (\subref{fig:lih_saoovqe_all_energies}) Displays ground state (blue circles), excited state (red diamond), and state-average (green triangle) energies. 
    (\subref{fig:lih_saoovqe_sa_detail}) Provides a detailed view of the state-average energy.}
    \label{fig:lih_saoovqe_comparison} 
\end{figure}

Continuing the investigation, the Lithium Hydride (LiH) molecule serves as the third system for comparing optimizer performance within the SA-OO-VQE framework. 
The calculations were performed using the sto-3g basis~\cite{book_basis}. 
The results, obtained using the same methodology as for H$_2$ and H$_4$, are presented graphically in Figure~\ref{fig:lih_saoovqe_comparison} and numerically in Table~\ref{tab:calculation_summary_lih}.

\begin{table}[!htb]
\centering
\small
\begin{tabular}{l c c c c c c}
\toprule
Method & $evals_{min}$ & $evals_{max}$ & $evals_{mean}$ & $E_{min}$ & $E_{max}$ & $E_{mean}$ \\
\midrule
BFGS & 39 & 40 & 40 & -7.790381 & -7.790381 & -7.790381 \\
COBYLA & 1102 & 1293 & 1223 & -7.790381 & -7.790381 & -7.790381 \\
DE/Best1/Bin & 1350 & 4410 & 2658 & -7.787890 & -7.759062 & -7.777164 \\
DE/Best2/Bin & 1440 & 4860 & 2793 & -7.787264 & -7.774088 & -7.780401 \\
DE/Current-to-Best1/Bin & 1470 & 4980 & 2805 & -7.787800 & -7.763849 & -7.779753 \\
DE/Current-to-Random1/Bin & 1620 & 4710 & 2748 & -7.789081 & -7.772093 & -7.783103 \\
DE/Rand1/Bin & 930 & 2310 & 1419 & -7.789232 & -7.763723 & -7.776370 \\
DE/Rand2/Bin & 1410 & 3630 & 2397 & -7.788888 & -7.780103 & -7.785505 \\
DE/Random-to-Best1/Bin & 2070 & 6300 & 3621 & -7.787908 & -7.772646 & -7.783256 \\
Gradient Descent & 6556 & 6556 & 6556 & -7.790118 & -7.790118 & -7.790118 \\
SLSQP & 100 & 100 & 100 & -7.790381 & -7.790381 & -7.790381 \\
\bottomrule
\end{tabular}
\caption[Comparative optimizer results from SA-OO-VQE for LiH molecule.]{Comparison of optimizer performance for the LiH molecule simulation. 
The table reports statistics based on 10 independent runs: minimum ($min$), maximum ($max$), and mean ($mean$) number of function evaluations ($evals$) required for convergence, and the corresponding final state-average energies ($E$ in Hartrees).} 
\label{tab:calculation_summary_lih}
\end{table}
The performance trends observed for LiH largely mirror those seen for H$_2$ and H$_4$. 
As detailed in Table~\ref{tab:calculation_summary_lih} and visualized in Figure~\ref{fig:lih_saoovqe_comparison}.

The local optimizers BFGS, SLSQP, and COBYLA consistently converge to the same lowest state-average energy ($-7.790381$~Ha). 
BFGS demonstrates remarkable efficiency, requiring only 40 function evaluations on average. 
SLSQP is also efficient (100 evaluations). 
COBYLA, while achieving the optimal energy, remains computationally expensive (mean 1223 evaluations).
    
The standard Gradient Descent method again converges to a slightly higher energy ($-7.790118$~Ha) and proves particularly inefficient for LiH, requiring a very large number of function evaluations~(6556).
    
The Differential Evolution variants continue to exhibit significant variability in the number of required function evaluations and struggle to consistently find the optimal energy. 
Their mean achieved energies (ranging from approximately $-7.776$~Ha to $-7.786$~Ha) are notably worse than the value reached by BFGS/SLSQP/COBYLA. 
Furthermore, the substantial difference between the minimum ($E_{min}$) and maximum ($E_{max}$) energies obtained across the 10 runs for each DE method underscores their lack of robustness for this problem, with none of the best runs matching the optimal energy found by the top local methods.

In summary, the LiH results further reinforce the conclusions drawn from the H$_2$ and H$_4$ simulations. 
For the specific SA-OO-VQE task under the conditions studied, the gradient-based local optimizers BFGS and SLSQP consistently provide the best combination of accuracy, efficiency, and robustness, significantly outperforming standard Gradient Descent and the various stochastic DE approaches tested across all three molecular systems.

\begin{figure}[!htb] 
    \centering
    \begin{subfigure}[t]{0.7\textwidth} 
        \centering
        \includegraphics[width=\linewidth]{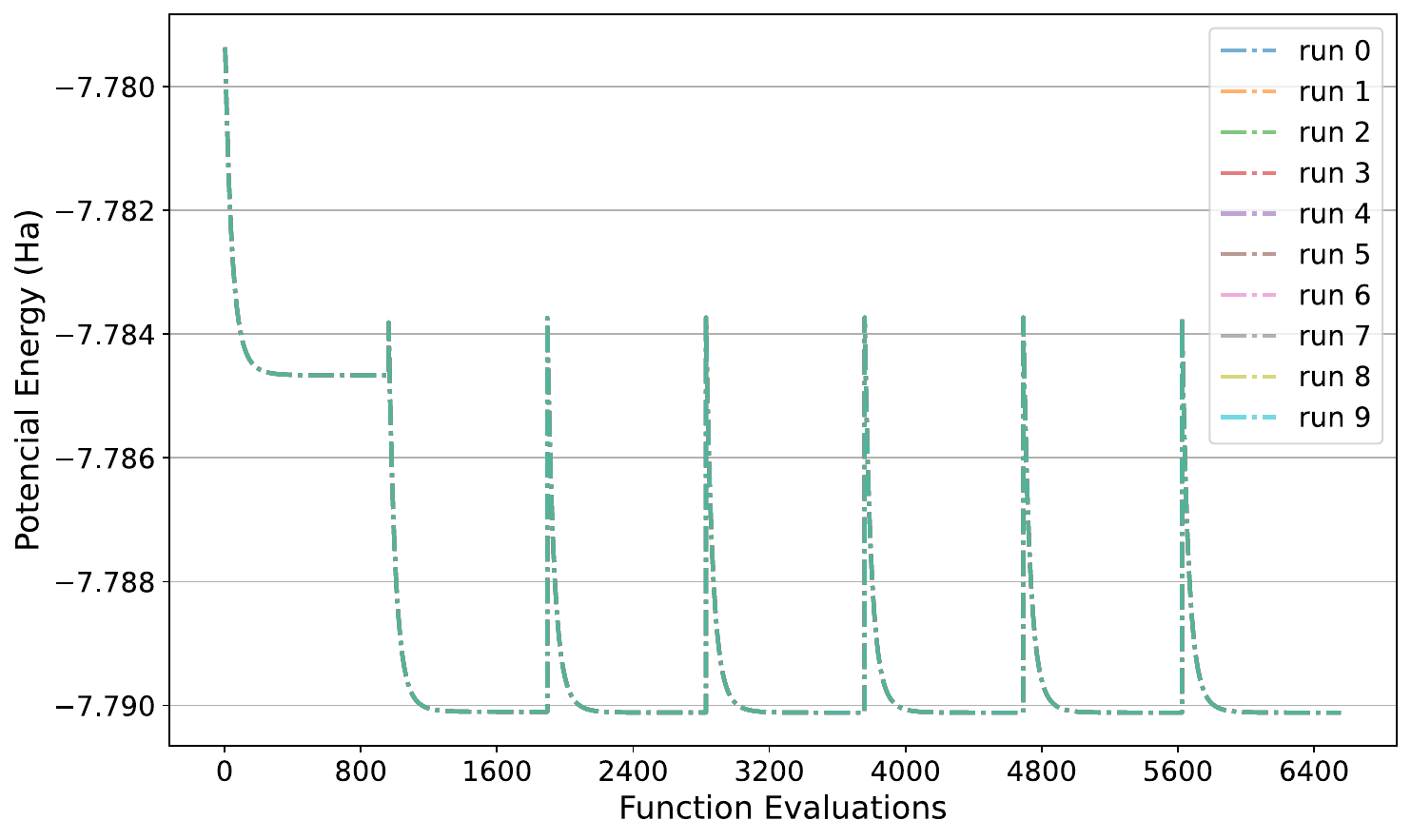}
        \caption{Energy after optimizer iterations vs. cumulative evaluations.} 
        \label{fig:lih_gd_conv} 
    \end{subfigure}
    \begin{subfigure}[t]{0.49\textwidth} 
        \centering
        \includegraphics[width=\linewidth]{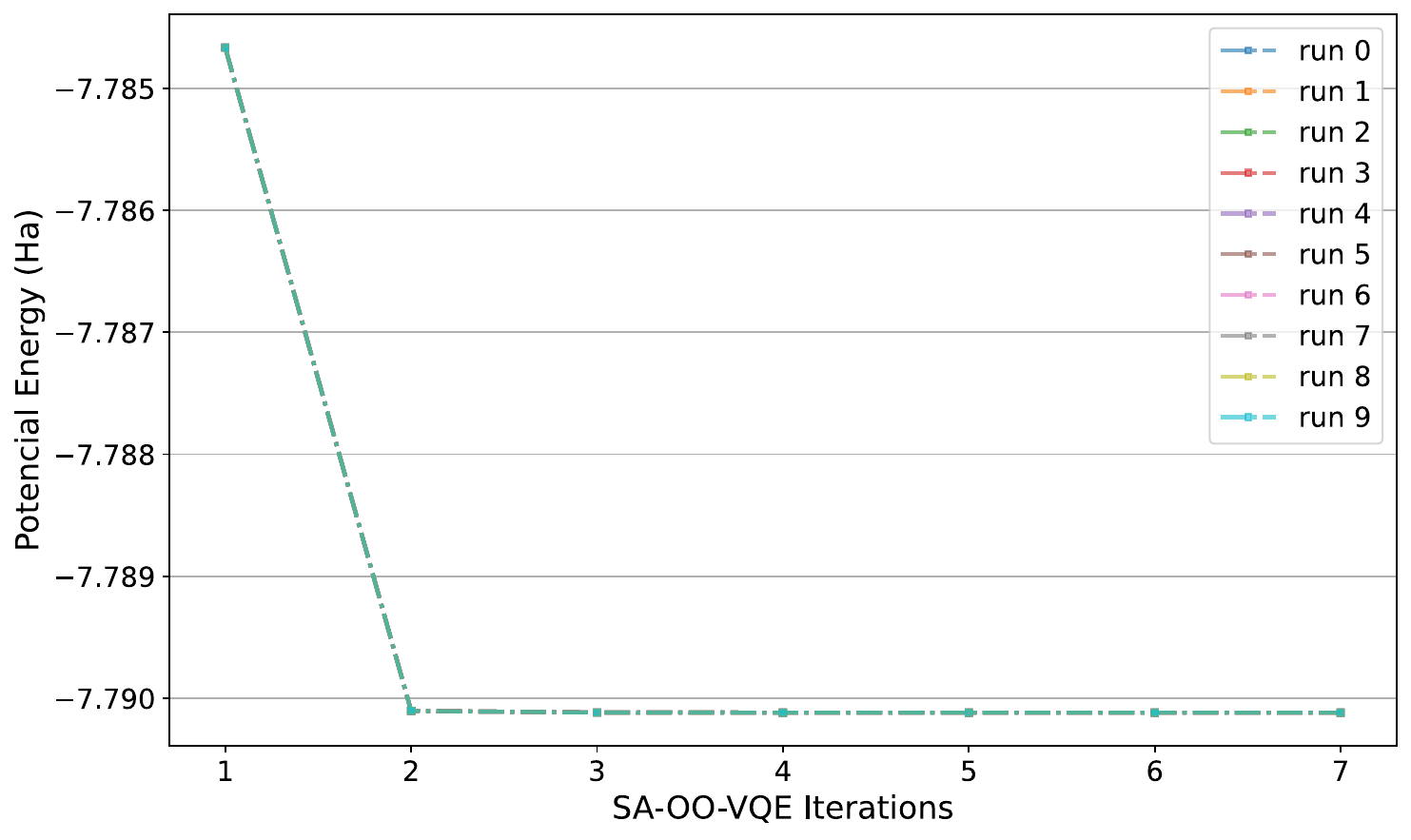}
        \caption{Energy vs. SA-OO-VQE iteration number.} 
        \label{fig:lih_gd_conv_iters} 
    \end{subfigure}
    \hfill 
    \begin{subfigure}[t]{0.49\textwidth} 
        \centering
        \includegraphics[width=\linewidth]{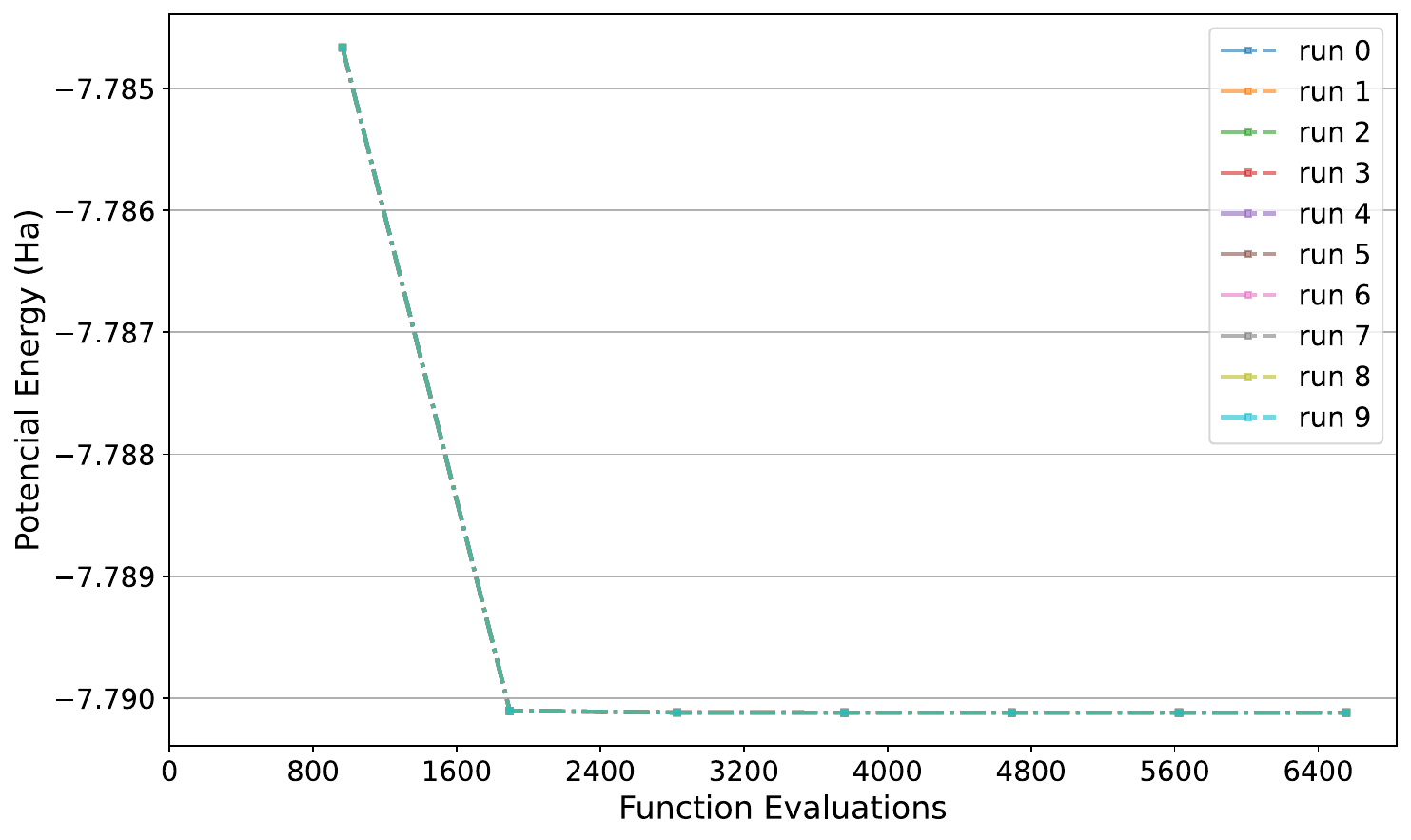}
        \caption{Energy after full SA-OO-VQE iterations vs. cumulative evaluations.} 
        \label{fig:lih_gd_conv_evals}
    \end{subfigure}

    \caption[Convergence plots of the Gradient Descent optimizer within the SA-OO-VQE framework for the
LiH molecule.]{Convergence analysis of the Gradient Descent optimizer within the SA-OO-VQE framework for the LiH molecule, based on 10 independent runs (shown in different colors/styles, see legend in plots). The plots display the state-average energy (Hartrees) progression viewed against different metrics: 
    (\subref{fig:lih_gd_conv}) Energy evaluated at the end of each internal Gradient Descent optimizer iteration, plotted against the cumulative number of function evaluations consumed up to that iteration point
    (\subref{fig:lih_gd_conv_iters}) Energy plotted at the end of each completed SA-OO-VQE iteration against the iteration number. 
    (\subref{fig:lih_gd_conv_evals}) Energy plotted at the end of each completed SA-OO-VQE iteration against the cumulative number of function evaluations consumed up to that iteration.}
\label{fig:lih_grad_desc} 
\end{figure}

Now, as was said in section~\ref{sec:experiments_methodology}, we present the convergence plots for each optimizer to demonstrate optimization dynamics.
Because the convergence plots for the LiH molecule share most of the properties with plots for molecule H$_2$ or H$_4$~(Section~\ref{sec:molecule_h2} and~\ref{sec:molecule_h4}), only the convergence plot of Gradient Descent~(Figure~\ref{fig:lih_grad_desc}) is shown in this section.
For detailed convergence plots of the other optimizers, please refer to Appendix~\ref{ap_sec:lih_conv}, which contains figures for BFGS (Figure~\ref{fig:lih_bfgs}), COBYLA (Figure~\ref{fig:lih_cobyla}), SLSQP (Figure~\ref{fig:lih_slsqp}), and all DE configurations (Figures~\ref{fig:lih_de_b1}, \ref{fig:lih_de_b2}, \ref{fig:lih_de_ctb}, \ref{fig:lih_de_ctr}, \ref{fig:lih_de_r1}, \ref{fig:lih_de_r2}, and \ref{fig:lih_de_rtb}).

The convergence behavior observed for the LiH molecule, exemplified by the Gradient Descent optimizer in Figure~\ref{fig:lih_grad_desc}, mirrors the general trends seen for the H$_2$ and H$_4$ molecules. 
Specifically, we observe the characteristic decrease in the state-average energy as the optimization progresses, both in terms of individual optimizer steps (Figure~\ref{fig:lih_gd_conv}) and full SA-OO-VQE iterations (Figs.~\ref{fig:lih_gd_conv_iters} and~\ref{fig:lih_gd_conv_evals}). 
The variability between independent runs, reflecting different optimization trajectories potentially influenced by initialization or noise, is also apparent, similar to the previous molecules. 
Furthermore, the specific properties previously noted for individual optimizers, such as the convergence characteristics DE observed for H$_2$ and H$_4$, remain consistent for the LiH system.

However, a slight difference arises in the computational cost required to reach convergence for BFGS, COBYLA, Gradient Descent, and SLSQP. 
Compared to the H$_2$ and H$_4$ molecules, the optimization for LiH required a different number of SA-OO-VQE iterations (visible on the x-axis of Figure~\ref{fig:lih_gd_conv_iters}) and consequently, for some optimizers a  different total number of function evaluations (visible on the x-axes of Figs.~\ref{fig:lih_gd_conv} and~\ref{fig:lih_gd_conv_evals}). 
Despite this, the fundamental convergence patterns and the relative performance characteristics observed among different optimizers (as detailed for H$_2$ and shown for all optimizers for LiH in Appendix~\ref{ap_sec:lih_conv}) remain consistent.

\FloatBarrier

\subsection{Overall Summary of Optimizer Performance} 
The comparative studies conducted on the H$_2$, H$_4$, and LiH molecules reveal consistent trends in the performance of various optimization algorithms applied to the SA-OO-VQE problem under the investigated conditions. 

Across all three systems, the gradient-based local optimizers BFGS and SLSQP consistently delivered the best results. 
They reliably located the lowest state-average energies with high efficiency (requiring relatively few function evaluations) and demonstrated excellent robustness across multiple independent runs. 

The gradient-free local optimizer COBYLA also consistently found the same optimal energy but required significantly more function evaluations, highlighting a trade-off between gradient-free operation and computational cost for these problems. 

Standard Gradient Descent, while deterministic, consistently converged to slightly higher energies than BFGS/SLSQP/COBYLA and demanded a substantial number of evaluations.

In contrast, the stochastic Differential Evolution variants proved less suitable for these specific SA-OO-VQE tasks. They generally incurred higher computational costs (more function evaluations) with considerable run-to-run variability, and more importantly, they struggled to reliably converge to the optimal energy, often yielding significantly worse results compared to the best-performing local methods.

These combined findings strongly indicate that for optimizing state-averaged energies in small molecules using the SA-OO-VQE approach as implemented here, gradient-based methods like BFGS and SLSQP offer the most advantageous balance of accuracy, efficiency, and reliability.

\section{Conical intersection for Formaldimin molecule}
\label{sec:formaldimin}
Conical intersections are points or seams of degeneracy between two or more electronic PESs in polyatomic molecules~\cite{klessinger1995excited, RobbBernardiOlivucci+1995+783+789, CS9962500321, Domcke_Yarkony_Koppel_2004}. 
These features play a crucial role in photochemistry and photophysics, mediating ultra-fast non-radiative decay pathways from electronically excited states back to the ground state or to other lower-lying states. 
Accurate theoretical characterization of PESs in the vicinity of CIs is challenging due to the breakdown of the BOA and the often multi-configurational nature of the electronic wave functions involved~\cite{Domcke_Yarkony_Koppel_2004}.
One of the significant capabilities of advanced VQE methods, specifically the SA-OO-VQE, is the potential to compute the energies of both ground and excited states accurately, even in regions where PESs approach or exhibit conical intersections~\cite{PhysRevA.111.022437}. 
To demonstrate this capability, we investigate the formaldimine molecule (H$_2$CNH), a well-studied system known to possess accessible CIs relevant to its photochemistry.

The geometry of formaldimine in our study is parameterized by two key coordinates, the dihedral angle $\phi$ (defining the twist around the C=N bond) and the H-N-C bending angle $\alpha$, as illustrated in Figure~\ref{fig:formaldimin_angles}.
\begin{figure}[!hbpt]
\centering
\begin{subfigure}[b]{0.4\textwidth} 
    \centering

\begin{tikzpicture}[
    atom/.style 2 args={circle, shading=ball, ball color=#1, minimum size=#2},
    bond/.style={line width=4pt, line cap=round}, 
    doublebond/.style={line width=2pt, line cap=round}, 
    lbl/.style={font=\large}, 
    scale=1
  ]

  \coordinate (C)  at (0,0);
  \coordinate (N)  at (2.0, 0.4);          
  \coordinate (H_topC) at (-0.95, 1.1);     
  \coordinate (H_botC) at (-0.1, -1.6);     
  \coordinate (H_N)    at (2.5, 1.5);      

  \draw[bond, draw=green!50!gray] (C) -- (H_topC);
  \draw[bond, draw=green!50!gray] (C) -- (H_botC);
  \draw[bond, draw=blue!50!gray]  (N) -- (H_N);
  \coordinate (CshiftN)    at ($(C)!0.1cm! 90:(N)$); 
  \coordinate (NshiftC)    at ($(N)!0.1cm!-90:(C)$);
  \coordinate (CshiftNneg) at ($(C)!-0.1cm! 90:(N)$);
  \coordinate (NshiftCneg) at ($(N)!-0.1cm!-90:(C)$);
  \draw[doublebond, draw=green!40!blue!70!gray] (CshiftN) -- (NshiftC);
  \draw[doublebond, draw=green!40!blue!70!gray] (CshiftNneg) -- (NshiftCneg);

  \node[atom={green!70!black}{0.9cm}, label={[white, font=\bfseries]center:C}] at (C) {}; 
  \node[atom={blue!70!cyan}{0.8cm},  label={[white, font=\bfseries]center:N}] at (N) {}; 
  \node[atom={gray!50!white}{0.7cm}, label={[black, font=\bfseries]center:H}] at (H_topC) {}; 
  \node[atom={gray!50!white}{0.7cm}, label={[black, font=\bfseries]center:H}] at (H_botC) {};
  \node[atom={gray!50!white}{0.7cm}, label={[black, font=\bfseries]center:H}] at (H_N) {};


  \coordinate (wC1) at ($(C)!1!(H_topC)$); 
  \coordinate (wC2) at ($(C)!0.9!(N)$);      
  \coordinate (wC3) at ($(C)!1!(H_botC)$); 
  \coordinate (wN1) at ($(N)!1!(H_N)$);    
  \coordinate (wN2) at ($(N)!1!(C)$);      
  Fill the wedges
  \fill[red!60!black, fill opacity=0.2] (C) -- (wC1) -- (wC2) -- cycle;
  \fill[red!60!black, fill opacity=0.2] (C) -- (wC3) -- (wC2) -- cycle;
  \fill[blue!60!black, fill opacity=0.3] (N) -- (wN1) -- (wN2) -- cycle;
  \coordinate (CNmid) at ($(C)!0.5!(N)$);
  \draw[draw=black, thick, ->, >=Latex] ($(CNmid) + (-0.4, 0.3)$) arc (140:50:0.6cm); 
  \node[lbl, black] at ($(CNmid)+(0.1,0.8)$) {$\phi$}; 
  \node[lbl, text width=1.8cm, align=center, below right] at ($(N)+(-2,2)$) {dihedral angle}; 

  \node[atom={green!70!black}{0.9cm}, label={[white, font=\bfseries]center:C}] at (C) {}; 
  \node[atom={blue!70!cyan}{0.8cm},  label={[white, font=\bfseries]center:N}] at (N) {}; 
  \node[atom={gray!50!white}{0.7cm}, label={[black, font=\bfseries]center:H}] at (H_topC) {}; 
  \node[atom={gray!50!white}{0.7cm}, label={[black, font=\bfseries]center:H}] at (H_botC) {};
  \node[atom={gray!50!white}{0.7cm}, label={[black, font=\bfseries]center:H}] at (H_N) {};

\end{tikzpicture}
\caption{Dihedral angle $\phi$.} 
\label{subfig:formaldimin_dih}
\end{subfigure}
\begin{subfigure}[b]{0.4\textwidth}
    \centering
\begin{tikzpicture}[
        atom/.style 2 args={circle, shading=ball, ball color=#1, minimum size=#2},
        bond/.style={line width=4pt, line cap=round},
        doublebond/.style={line width=2pt, line cap=round},
        lbl/.style={font=\large},
        scale=1
    ]
    \coordinate (C)  at (0,0);
    \coordinate (N)  at (2.0, 0.4);
    \coordinate (H_topC) at (-0.95, 1.1);
    \coordinate (H_botC) at (-0.1, -1.6);
    \coordinate (H_N)    at (2.5, 1.5);

    \draw[bond, draw=green!50!gray] (C) -- (H_topC);
    \draw[bond, draw=green!50!gray] (C) -- (H_botC);
    \draw[bond, draw=blue!50!gray]  (N) -- (H_N);
    \coordinate (CshiftN)    at ($(C)!0.1cm! 90:(N)$);
    \coordinate (NshiftC)    at ($(N)!0.1cm!-90:(C)$);
    \coordinate (CshiftNneg) at ($(C)!-0.1cm! 90:(N)$);
    \coordinate (NshiftCneg) at ($(N)!-0.1cm!-90:(C)$);
    \draw[doublebond, draw=green!40!blue!70!gray] (CshiftN) -- (NshiftC);
    \draw[doublebond, draw=green!40!blue!70!gray] (CshiftNneg) -- (NshiftCneg);

    \node[atom={green!70!black}{0.9cm}, label={[white, font=\bfseries]center:C}] at (C) {};
    \node[atom={blue!70!cyan}{0.8cm},  label={[white, font=\bfseries]center:N}] at (N) {};
    \node[atom={gray!50!white}{0.7cm}, label={[black, font=\bfseries]center:H}] at (H_topC) {};
    \node[atom={gray!50!white}{0.7cm}, label={[black, font=\bfseries]center:H}] at (H_botC) {};
    \node[atom={gray!50!white}{0.7cm}, label={[black, font=\bfseries]center:H}] at (H_N) {};

    \coordinate (angle_start) at ($(C) + (0.7, 0.1)$); 
    \coordinate (angle_end) at ($(H_N) + (-0.3,0)$);   
    \draw[red, thick, ->, >=Latex] (angle_start) to[bend right=-30] node[midway, above left, black] {$\alpha$} (angle_end);
    \node[lbl, black, below left] at ($(C) + (2.3, 2.2)$) {bending angle};

    \node[atom={green!70!black}{0.9cm}, label={[white, font=\bfseries]center:C}] at (C) {};
    \node[atom={blue!70!cyan}{0.8cm},  label={[white, font=\bfseries]center:N}] at (N) {};
    \node[atom={gray!50!white}{0.7cm}, label={[black, font=\bfseries]center:H}] at (H_topC) {};
    \node[atom={gray!50!white}{0.7cm}, label={[black, font=\bfseries]center:H}] at (H_botC) {};
    \node[atom={gray!50!white}{0.7cm}, label={[black, font=\bfseries]center:H}] at (H_N) {};

    \end{tikzpicture}
    \caption{Bending angle $\alpha$.} 
    \label{subfig:formaldimin_bending}
\end{subfigure}

\caption[Geometry parameters of formaldimine molecule.]{Geometric parameters used to define the structure of the formaldimine molecule (H$_2$CNH). (\subref{subfig:formaldimin_dih}) Illustration of the dihedral angle $\phi$, representing the twist around the C=N bond. 
(\subref{subfig:formaldimin_bending}) Illustration of the H-N-C bending angle $\alpha$.
}
\label{fig:formaldimin_angles}
\end{figure}

To probe the region near a known CI, we perform a one-dimensional PES scan. We fix the dihedral angle at $\phi = 90^{\circ}$, a geometry known to lower the energy gap between the ground state ($S_0$) and the first singlet excited state ($S_1$). We then compute the energies of these two states as a function of the bending angle $\alpha$, varying it within the range $[90^{\circ}, 150^{\circ}]$.

\begin{figure}[!hbpt]
    \centering
    \includegraphics[width=0.9\linewidth]{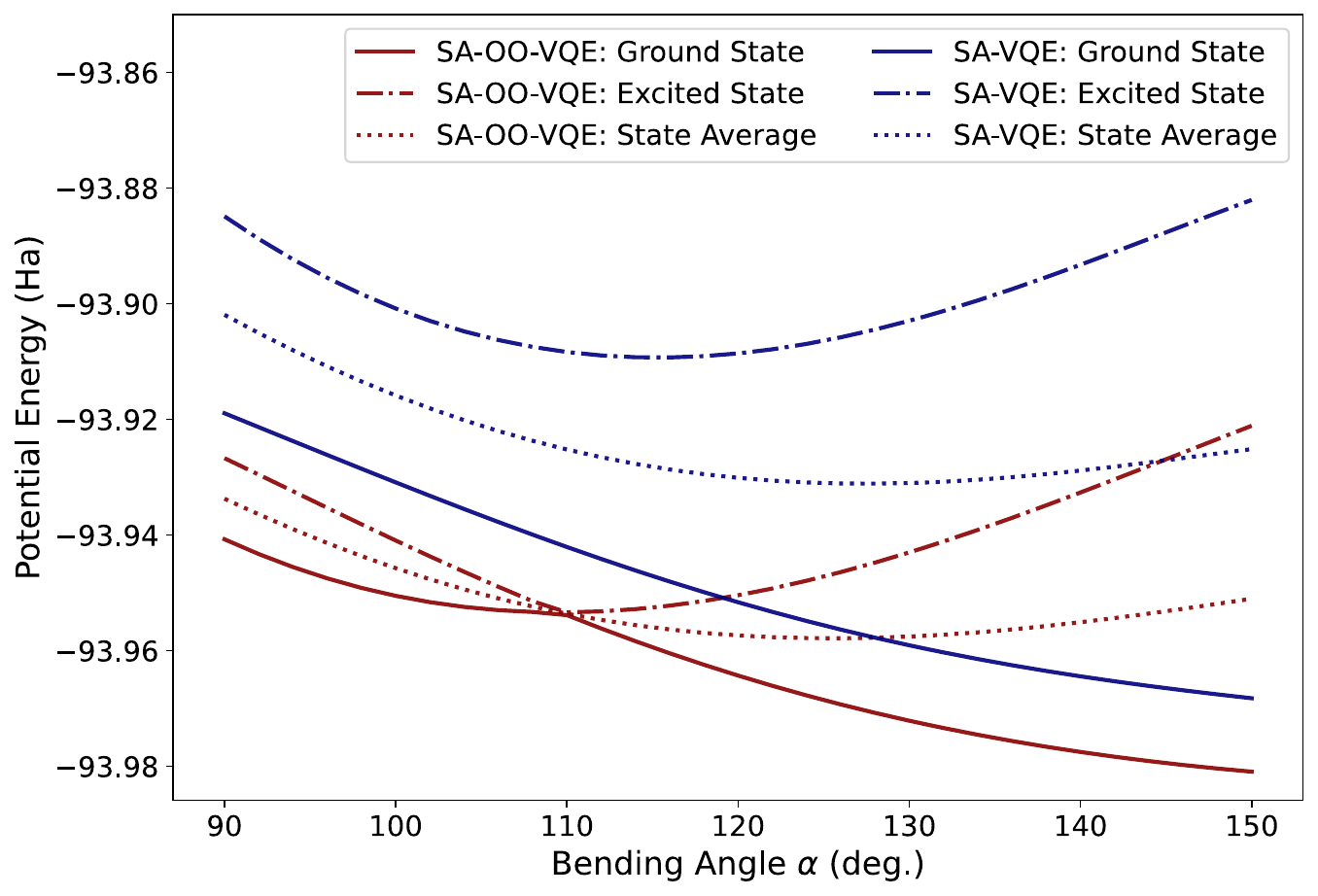}
    \caption[One dimensional PES of formaldimine molecule.]{One dimensional PES of formaldimine as a function of the bending angle $\alpha$ at a fixed dihedral angle $\phi = 90^\circ$. Results are shown for the ground state (solid lines), first excited state (dash-dotted lines), and the state average (dotted lines), obtained using the SA-OO-VQE method (red) and SA-VQE with a Hartree-Fock restricted Hamiltonian (blue).
    The BFGS optimizer was used in both SA-OO-VQE and SA-VQE.}
    \label{fig:avoided_crossing_formaldimine}
\end{figure}

We employ two quantum computing approaches for this task, SA-OO-VQE and SA-VQE (Chapter~\ref{chapter:saoovqe}).
For both methodologies, the SA-OO-VQE framework~\cite{joss_saooveq} was used, with or without molecular optimization, and the optimization of the VQE parameters was performed using the BFGS algorithm.
The calculations were performed using the cc-pVDZ basis sets~\cite{10.1063/1.456153}. 
The computed PESs for the $S_0$, the $S_1$, and the state average energy are presented in Figure~\ref{fig:avoided_crossing_formaldimine}. 
The results obtained using SA-OO-VQE are shown in red, while those from SA-VQE with the fixed Hartree-Fock orbital basis are shown in blue.
The results depicted in Figure~\ref{fig:avoided_crossing_formaldimine} highlight the advantage of orbital optimization in SA-OO-VQE for describing PESs near conical intersections.
\begin{itemize}
    \item \textbf{SA-OO-VQE (Red Curves):} The PESs for the $S_0$ and $S_1$ states computed with the SA-OO-VQE approach track each other closely and appear to intersect near $\alpha \approx 110^{\circ}$. 
    This behavior is characteristic of traversing a conical intersection seam along the chosen coordinate. 
    The method successfully captures the degeneracy point because the orbital optimization allows the electronic basis to adapt, providing a balanced description of both the ground and excited states even in this challenging region of strong non-adiabatic coupling.
    \item \textbf{SA-VQE with Fixed Orbitals (Blue Curves):} In contrast, the SA-VQE calculation using orbitals fixed from a ground-state Hartree-Fock calculation fails to capture the degeneracy. 
    Instead, it predicts an \textit{avoided crossing}, where the energy curves approach each other but then diverge, maintaining a non-zero energy gap. 
    This artifact arises because the fixed Hartree-Fock orbitals, optimized for the ground state at a reference geometry, become increasingly inadequate for describing both the $S_0$ and $S_1$ states, especially the excited state and the mixed-state character near the CI. 
    The lack of orbital relaxation prevents the wave function from correctly representing the degenerate electronic structure.
    \item \textbf{Energy Accuracy:} Furthermore, it is noticeable that the potential energies computed using the standard SA-VQE approach with fixed HF orbitals are generally higher than those obtained with SA-OO-VQE. 
    This suggests that the lack of orbital relaxation not only fails to capture the correct topology near the CI but also leads to less accurate absolute energy predictions across the scanned coordinate range.
\end{itemize}

In summary, this investigation on the formaldimine molecule demonstrates that the SA-OO-VQE method, by incorporating simultaneous optimization of molecular orbitals and VQE circuit parameters, is capable of correctly describing the topology of potential energy surfaces around conical intersections. 
This capability is crucial for the accurate simulation of photochemical dynamics and represents a significant advantage over standard VQE or SA-VQE approaches that rely on a fixed, pre-determined orbital basis. 
The failure of the fixed-orbital SA-VQE to find the intersection underscores the importance of orbital relaxation effects in electronically challenging regions of the PES.

\chapter{Conclusion}
This thesis explored the application and performance of numerical optimization methods within the SA-OO-VQE framework, a hybrid quantum-classical algorithm designed for the accurate description of ground and excited electronic states (Chapter~\ref{chapter:saoovqe}).

A primary focus was the comparative analysis of various classical optimization algorithms tasked with minimizing the state-averaged energy cost function within the SA-OO-VQE loop (Section~\ref{sec:comparison_of_optimizers}). 
Studies conducted on the H$_2$, H$_4$, and LiH molecular systems revealed consistent performance trends. 
The gradient-based local optimizers BFGS and SLSQP consistently demonstrated superior performance, reliably locating the lowest state-average energies (e.g., -0.877120 Ha for H$_2$, -1.739735 Ha for H$_4$, -7.790381 Ha for LiH) with remarkable computational efficiency, often requiring only tens of function evaluations (e.g., 18 for BFGS on H$_2$, 79 for BFGS on H$_4$, 40 for BFGS on LiH). 
Their deterministic nature also ensured excellent robustness across multiple runs. 
The gradient-free method COBYLA also consistently achieved the optimal energy but at a significantly higher computational cost, typically requiring over a thousand function evaluations (e.g., mean of 1100 for H$_2$, 4121 for H$_4$, 1223 for LiH). 
Standard Gradient Descent proved less effective, consistently converging to slightly higher energies (e.g., -0.876777 Ha for H$_2$) and demanding a substantial number of function evaluations (e.g., 2080 for H$_2$, 6556 for LiH).

In contrast, the implemented Differential Evolution strategies (Chapter~\ref{chapter:DE}), while representing a global optimization approach potentially suited for complex landscapes, proved less advantageous for these specific SA-OO-VQE tasks. They generally incurred higher computational costs, often several thousand function evaluations on average (e.g., means ranging from 1380 to 3654 across variants for H$_2$), and exhibited significant run-to-run variability (e.g., max evaluations reaching 8490 for DE/Best1/Bin on H$_2$). 
More importantly, they struggled to reliably converge to the optimal state-average energies found by the best local methods, often yielding significantly worse mean energies and showing a large spread between the best (E$_{min}$) and worst (E$_{max}$) results in repeated runs. 
The convergence plots for DE frequently displayed non-monotonic behavior between the main SA-OO-VQE cycles, indicating difficulties in efficiently navigating the optimization landscape in this context. 
These findings strongly suggest that for the systems and SA-OO-VQE implementation studied here, efficient gradient-based local optimizers like BFGS and SLSQP offer the most practical and reliable choice.

Furthermore, the thesis successfully demonstrated a key strength of the SA-OO-VQE methodology, its ability to accurately model potential energy surfaces in challenging regions near conical intersections (Section~\ref{sec:formaldimin}). 
The investigation of the formaldimine molecule along a specific reaction coordinate (varying the H-N-C bending angle $\alpha$ at a fixed dihedral angle $\phi=90^\circ$) showed that SA-OO-VQE, through its simultaneous optimization of molecular orbitals and VQE parameters, correctly captures the degeneracy point characteristic of a conical intersection, located near $\alpha \approx 110^{\circ}$. 
In stark contrast, a standard SA-VQE calculation using fixed Hartree-Fock orbitals failed to describe the intersection, instead predicting an artificial avoided crossing with a persistent energy gap. 
Additionally, the energies calculated by the fixed-orbital SA-VQE were consistently higher than those from SA-OO-VQE across the scan. 
This comparison explicitly underscores the critical importance of orbital relaxation, facilitated by the SA-OO component, for obtaining a qualitatively correct PES topology and quantitatively more accurate energies in non-adiabatically coupled systems.

In conclusion, this work provides valuable, detailed insights into the choice of classical optimizers for hybrid quantum algorithms like SA-OO-VQE, highlighting the efficiency and reliability of gradient-based methods for the systems studied. 
It also empirically reaffirms the significance and necessity of orbital optimization within the SA-OO-VQE approach for tackling complex problems in quantum chemistry, particularly the description of excited states and conical intersections, using near-term quantum computing techniques.

\chapter{Future Work}
The research presented in this thesis, focusing on the performance of numerical optimizers within the SA-OO-VQE framework and demonstrating its application to conical intersections, naturally leads to several promising avenues for future investigation.
These directions primarily involve enhancing the optimization strategies, extending the capabilities and applicability of the SA-OO-VQE algorithm itself, addressing the challenges posed by quantum noise, applying the methodology to more complex chemical systems and phenomena, and connecting to related research efforts.

Regarding optimization strategies, while the comparative study indicated that gradient-based local optimizers like BFGS and SLSQP performed optimally for the relatively small H$_2$, H$_4$, and LiH systems investigated here, the implemented standard 
Differential Evolution variants were found to be less robust. 
This suggests a clear avenue for future work in enhancing global optimization approaches for SA-OO-VQE. 
Exploring more advanced, adaptive DE strategies, such as JADE~\cite{zhang2009jade} or SHADE~\cite{tanabe2013success}, which dynamically adjust control parameters, could potentially overcome the limitations observed. 
Further investigation into the specific reasons for the standard DE's struggles in this context, alongside evaluating other metaheuristic, noise-aware, or hybrid global-local optimization algorithms, would also be valuable~\cite{novak2025optimization}. 
Such improvements are particularly pertinent as VQE landscapes may become more complex with larger active spaces or different ansätze~\cite{mcclean2018barren}. 
Crucially, the relative performance of global versus local optimizers warrants re-evaluation on larger molecular systems, where it is plausible that the global search capabilities of methods like DE might become more advantageous for navigating potentially rugged energy landscapes, despite their higher per-iteration cost observed in this work.

Addressing the impact of noise on current quantum hardware is another critical step. 
Systematic studies running SA-OO-VQE on real devices or using realistic noise models are needed to understand how noise affects both VQE convergence and the stability of the orbital optimization.

Furthermore, participation in related research provides broader context and potential synergies. 
Ongoing work contributes to a paper further investigating the performance of numerical optimizers, specifically within the context of the Variational Hamiltonian Ansatz, which represents another significant class of ansatz used in variational quantum algorithms~\cite{illesova2025numerical}. 
This research directly complements the findings in Section~\ref{sec:comparison_of_optimizers} by exploring the crucial role and behavior of classical optimization across different types of ansatz of VQE-like methods. 
Understanding how optimizers perform on both the state-preparation ansatz of SA-OO-VQE and the Hamiltonian-based ansatz of Variational Hamiltonian Ansatz contributes to a more comprehensive picture of the classical optimization bottleneck in near-term quantum algorithms and helps identify robust strategies applicable across different problem formulations.

Additionally, participation in another paper in preparation, titled ``Sequentially-Paralleled Quantum Networks for Discrimination of Quantum Channels'', delves into the fundamental limits and practical implementation challenges of different quantum circuit structures under NISQ constraints~\cite{bilek2025experimental}. 
This work analyzes the trade-offs between purely sequential circuit execution (prone to decoherence over time) and purely parallel execution (requiring many qubits) by studying intermediate ``sequentially-paralleled'' network structures. 
Although the application focus is quantum channel discrimination and benchmarking of gate-based quantum devices~\cite{lewandowska2025benchmarking}, the insights gained are highly relevant to SA-OO-VQE. Understanding how circuit depth, width, qubit connectivity, and overall structure impact susceptibility to noise and overall performance on near-term hardware is crucial for designing efficient SA-OO-VQE implementations. 
This knowledge could inform strategies for decomposing the SA-OO-VQE computation, scheduling circuit execution, or choosing hardware topologies to maximize fidelity and minimize resource requirements.

Pursuing these avenues will contribute significantly to advancing the capabilities of hybrid quantum-classical algorithms for tackling complex and important problems in quantum chemistry and materials science.

\printbibliography[heading=bibintoc]
\clearpage

\appendix
\chapter{DE Pseudocode}
\label{ap:de_pseudocode}
\begin{algorithm}[!ht]
\caption{DE/rand/1/bin with simple selection}
\label{alg:de_rand_1_bin}
\begin{algorithmic}[1]
\Require Population size $N_p$, Dimension $D$, Maximum generations $g_{\text{max}}$, Mutation factor $F$, Crossover rate $CR$
\Ensure Best solution $x_{\text{best}}$

\State Initialize population $P^{(0)}_{x} = \left(x^{(0)}_{i}\right)$ for $i = 0, 1, \ldots, N_p - 1$, where $x^{(0)}_{i} = \left(x^{(0)}_{i,j}\right)$ for $j = 0, 1, \ldots, D - 1$
\State Evaluate fitness of each $x^{(0)}_{i}$
\State $g \leftarrow 0$
\While{$g < g_{\text{max}}$\text{ and Stop condition not met}}
    \For{$i = 0$ to $N_p - 1$}
        \State Randomly select $r_1, r_2, r_3 \in \{0, 1, \ldots, N_p - 1\}$, such that $r_1 \neq r_2 \neq r_3 \neq i$
        \For{$j = 0$ to $D - 1$}
            \State $rand \leftarrow \text{random}(0, 1)$
            \If{$rand < CR$ or $j = \text{random}(0, D - 1)$}
                \State $v^{(g)}_{i,j} \leftarrow x^{(g)}_{r_1,j} + F \cdot \left(x^{(g)}_{r_2,j} - x^{(g)}_{r_3,j}\right)$ \Comment{Mutation: rand/1}
                \State $u^{(g)}_{i,j} \leftarrow v^{(g)}_{i,j}$ \Comment{Crossover: binomial}
            \Else
                \State $u^{(g)}_{i,j} \leftarrow x^{(g)}_{i,j}$ \Comment{Crossover: binomial}
            \EndIf
        \EndFor
        \State Evaluate fitness of $u^{(g)}_{i}$
        \If{fitness$\left(u^{(g)}_{i}\right) <$ fitness$\left(x^{(g)}_{i}\right)$} \Comment{Selection}
            \State $x^{(g+1)}_{i} \leftarrow u^{(g)}_{i}$
        \Else
            \State $x^{(g+1)}_{i} \leftarrow x^{(g)}_{i}$
        \EndIf
    \EndFor
    \State $g \leftarrow g + 1$
\EndWhile
\State $x_{\text{best}} \leftarrow \text{best solution in } P^{(g)}_{x}$
\Return $x_{\text{best}}$
\end{algorithmic}
\end{algorithm}

\begin{algorithm}[!ht]
\caption{DE/current-to-best/1/exp with simple selection}
\label{alg:de_ctb_1_exp}
\begin{algorithmic}[1]
\Require Population size $N_p$, Dimension $D$, Maximum generations $g_{\text{max}}$, Mutation factor $F$, Crossover rate $CR$
\Ensure Best solution $x_{\text{best}}$

\State Initialize population $P^{(0)}_{x} = \left(x^{(0)}_{i}\right)$ for $i = 0, 1, \ldots, N_p - 1$, where $x^{(0)}_{i} = \left(x^{(0)}_{i,j}\right)$ for $j = 0, 1, \ldots, D - 1$
\State Evaluate fitness of each $x^{(0)}_{i}$
\State $g \leftarrow 0$
\While{$g < g_{\text{max}} \text{ and Stop condition not met}$}
    \State $x^{(g)}_{\text{best}} \leftarrow \text{best solution in } P^{(g)}_{x}$
    \For{$i = 0$ to $N_p - 1$}
        \State Randomly select $r_1, r_2 \in \{0, 1, \ldots, N_p - 1\}$, such that $r_1 \neq r_2 \neq i$
        \State $L \leftarrow \text{random integer in } [1, D]$
        \State $j \leftarrow \text{random integer in } [0, D-1]$
        \State $k \leftarrow 0$
        \While{$k < L$ or $\text{random}(0,1) < CR$}
            \State $v^{(g)}_{i,j} \leftarrow x^{(g)}_{i,j} + F \cdot \left(x^{(g)}_{\text{best},j} - x^{(g)}_{i,j}\right) + F \cdot \left(x^{(g)}_{r_1,j} - x^{(g)}_{r_2,j}\right)$ \Comment{Mutation: current-to-best}
            \State $u^{(g)}_{i,j} \leftarrow v^{(g)}_{i,j}$ \Comment{Crossover: exponential}
            \State $j \leftarrow (j+1) \bmod D$
            \State $k \leftarrow k+1$
        \EndWhile
        \For{$m \leftarrow k$ to $D-1$}
            \State $u^{(g)}_{i,m} \leftarrow x^{(g)}_{i,m}$ \Comment{Crossover: exponential (rest of the dimensions)}
        \EndFor
        \State Evaluate fitness of $u^{(g)}_{i}$
        \If{fitness$\left(u^{(g)}_{i}\right) <$ fitness$\left(x^{(g)}_{i}\right)$} \Comment{Selection}
            \State $x^{(g+1)}_{i} \leftarrow u^{(g)}_{i}$
        \Else
            \State $x^{(g+1)}_{i} \leftarrow x^{(g)}_{i}$
        \EndIf
    \EndFor
    \State $g \leftarrow g + 1$
\EndWhile
\State $x_{\text{best}} \leftarrow \text{best solution in } P^{(g)}_{x}$
\Return $x_{\text{best}}$
\end{algorithmic}
\end{algorithm}

\clearpage

\chapter{Convergence plots of Different optimizers}
\label{ap:convergence_plots}
\section{Convergence plots for H$_2$ molecule}
\label{ap_sec:h2_conv}
\begin{figure}[!htbp] 
    \centering
    \begin{subfigure}[t]{0.49\textwidth} 
        \centering
        \includegraphics[width=\linewidth]{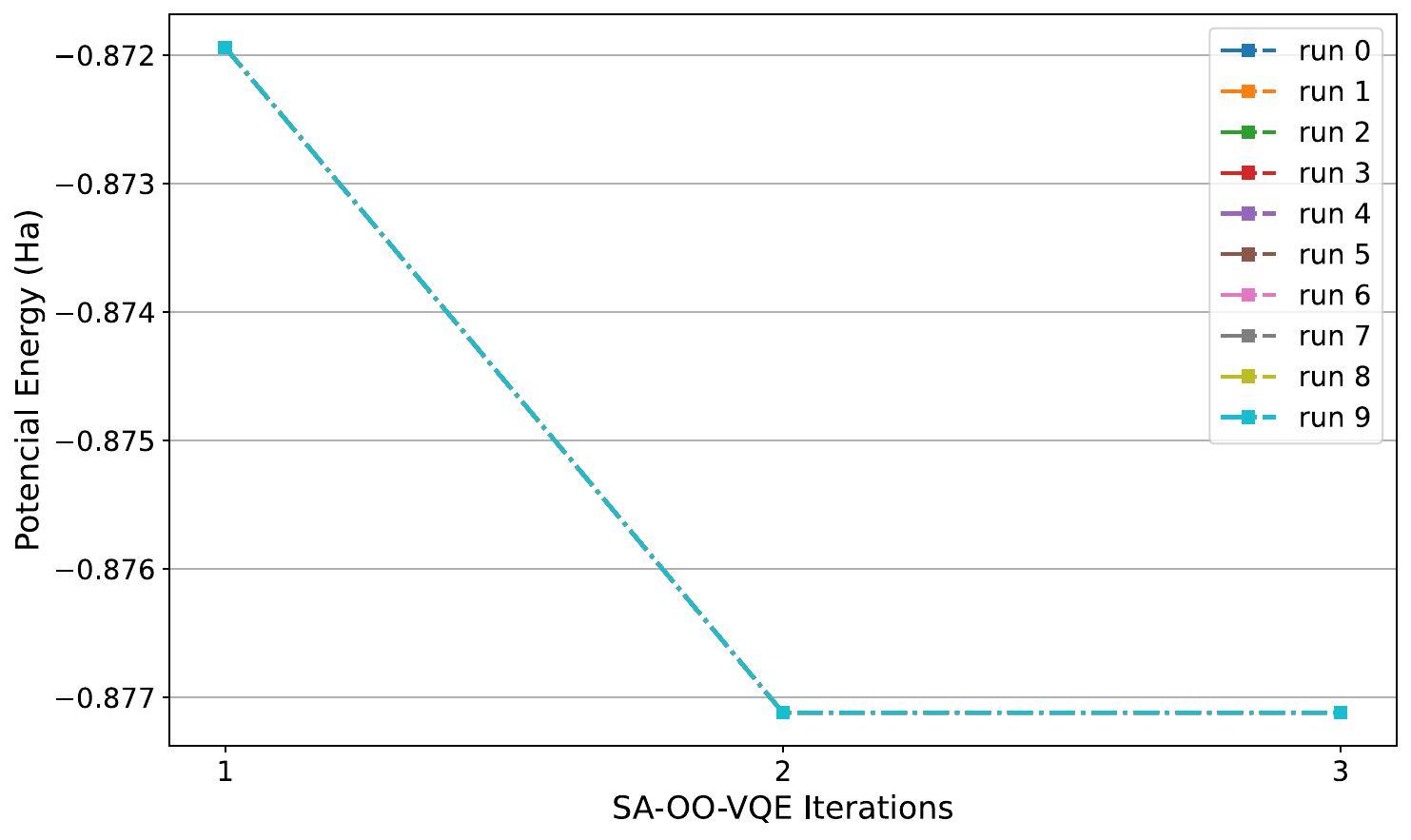}
        \caption{Energy vs. SA-OO-VQE iteration number.} 
        \label{fig:h2_cobyla_conv_iters} 
    \end{subfigure}
    \hfill 
    \begin{subfigure}[t]{0.49\textwidth} 
        \centering
        \includegraphics[width=\linewidth]{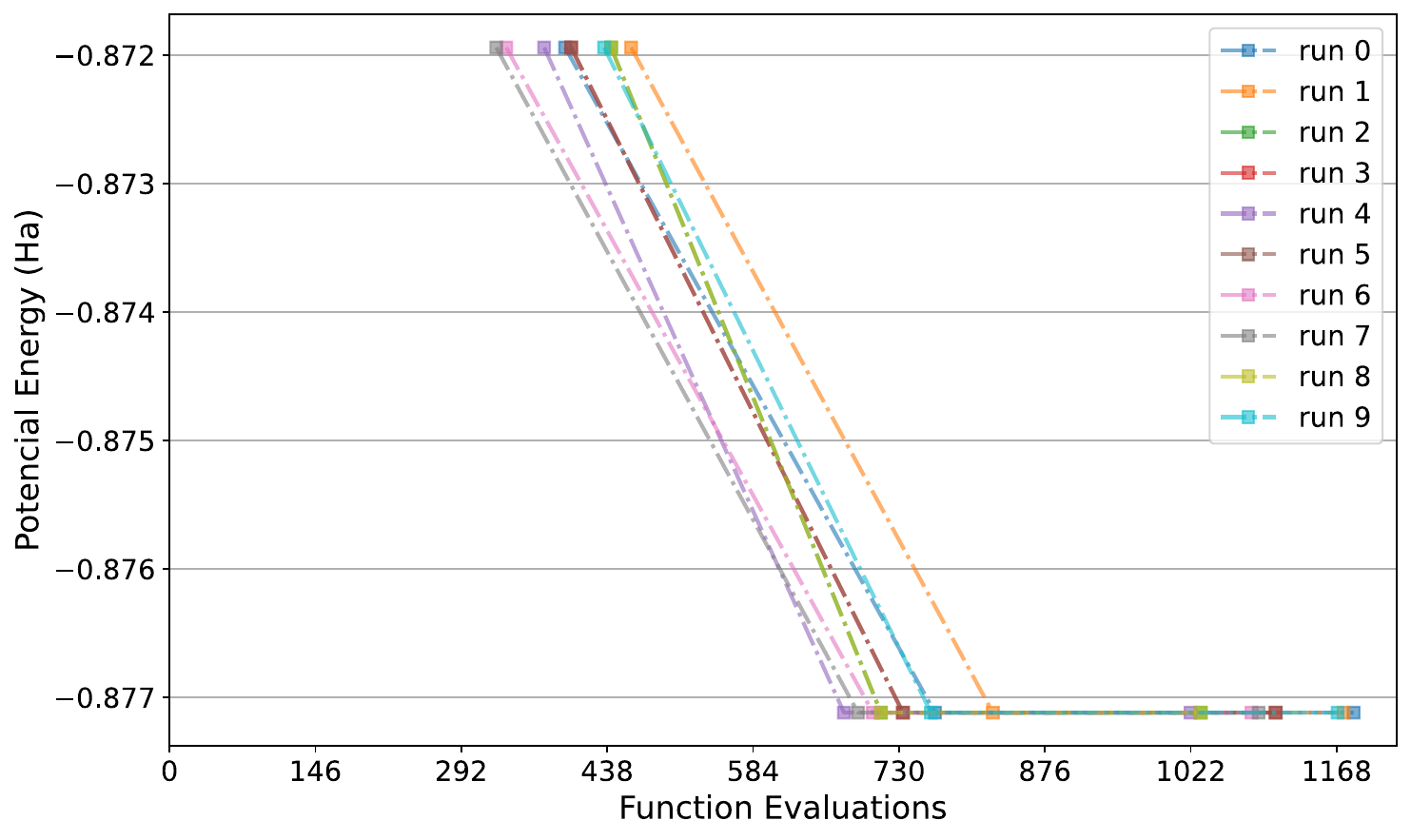}
        \caption{Energy after full SA-OO-VQE iterations vs. cumulative evaluations.} 
        \label{fig:h2_cobyla_conv_evals}
    \end{subfigure}

    \caption[Convergence plots of the COBYLA optimizer within the SA-OO-VQE framework for the
H$_2$ molecule.]{Convergence analysis of the COBYLA optimizer within the SA-OO-VQE framework for the H$_2$ molecule, based on 10 independent runs (shown in different colors/styles, see legend in plots). The plots display the state-average energy (Hartrees) progression viewed against different metrics: 
    (\subref{fig:h2_cobyla_conv_iters}) Energy plotted at the end of each completed SA-OO-VQE iteration against the iteration number. 
    (\subref{fig:h2_cobyla_conv_evals}) Energy plotted at the end of each completed SA-OO-VQE iteration against the cumulative number of function evaluations consumed up to that iteration.}
\label{fig:h2_cobyla} 
\end{figure}

\begin{figure}[!htbp] 
    \centering
    \begin{subfigure}[t]{0.49\textwidth} 
        \centering
        \includegraphics[width=\linewidth]{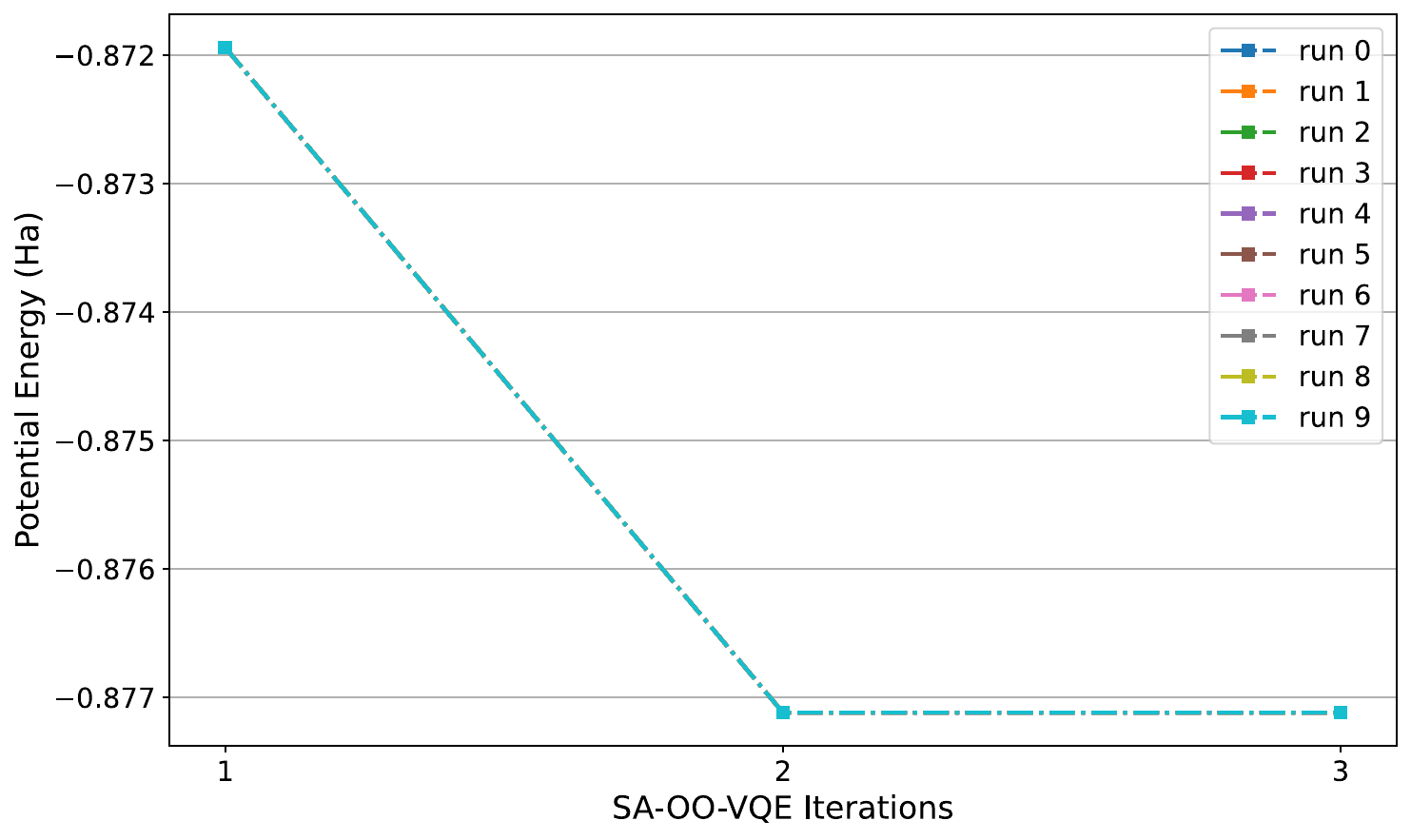}
        \caption{Energy vs. SA-OO-VQE iteration number.} 
        \label{fig:h2_slsqp_conv_iters} 
    \end{subfigure}
    \hfill 
    \begin{subfigure}[t]{0.49\textwidth} 
        \centering
        \includegraphics[width=\linewidth]{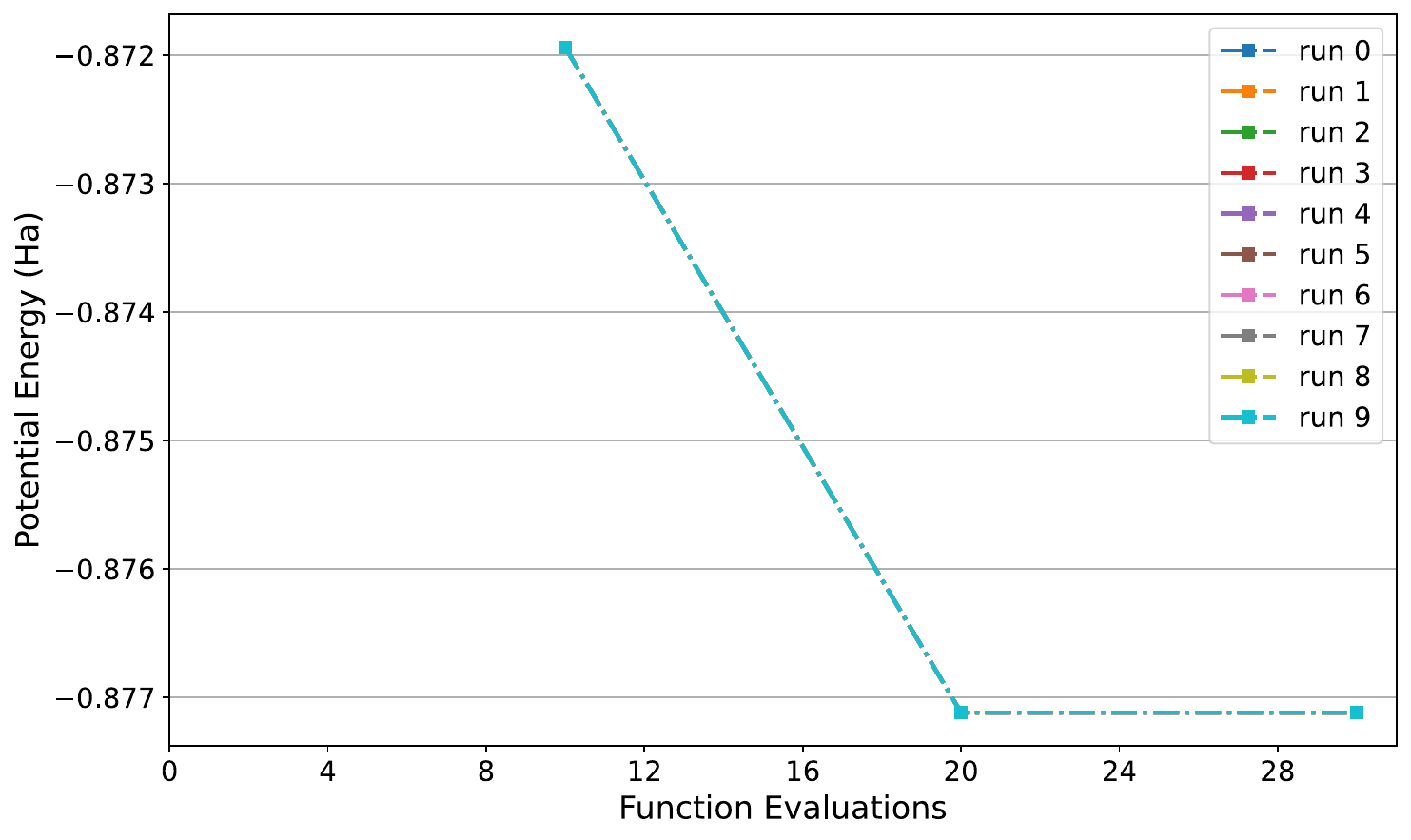}
        \caption{Energy after full SA-OO-VQE iterations vs. cumulative evaluations.} 
        \label{fig:h2_slsqp_conv_evals}
    \end{subfigure}

    \caption[Convergence plots of the SLSQP optimizer within the SA-OO-VQE framework for the
H$_2$ molecule.]{Convergence analysis of the SLSQP optimizer within the SA-OO-VQE framework for the H$_2$ molecule, based on 10 independent runs (shown in different colors/styles, see legend in plots). The plots display the state-average energy (Hartrees) progression viewed against different metrics: 
    (\subref{fig:h2_slsqp_conv_iters}) Energy plotted at the end of each completed SA-OO-VQE iteration against the iteration number. 
    (\subref{fig:h2_slsqp_conv_evals}) Energy plotted at the end of each completed SA-OO-VQE iteration against the cumulative number of function evaluations consumed up to that iteration.}
\label{fig:h2_slsqp} 
\end{figure}

\begin{figure}[!htbp] 
    \centering
    \begin{subfigure}[t]{0.7\textwidth} 
        \centering
        \includegraphics[width=\linewidth]{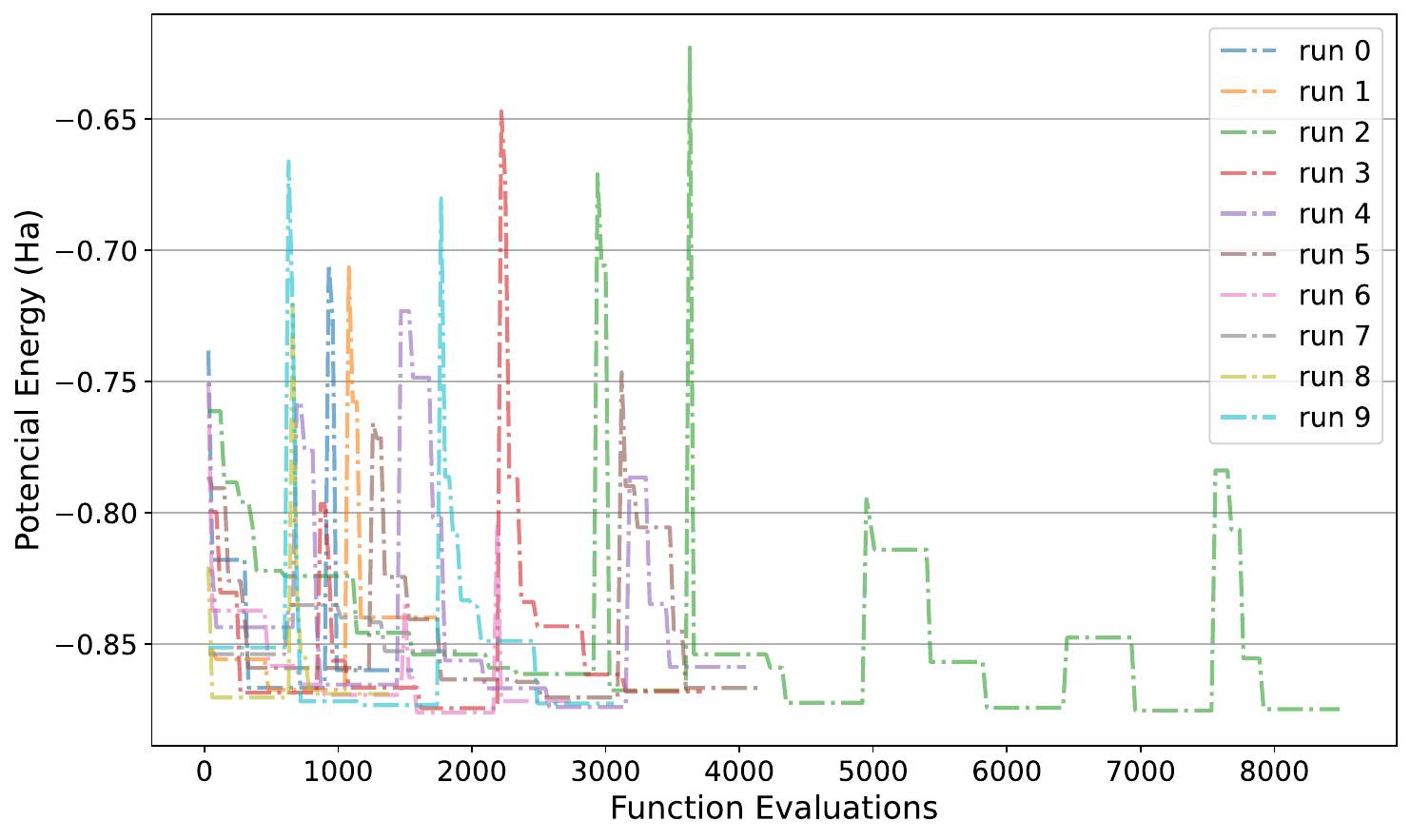}
        \caption{Energy after optimizer iterations vs. cumulative evaluations.} 
        \label{fig:h2_de_b1_conv} 
    \end{subfigure}
    \begin{subfigure}[t]{0.49\textwidth} 
        \centering
        \includegraphics[width=\linewidth]{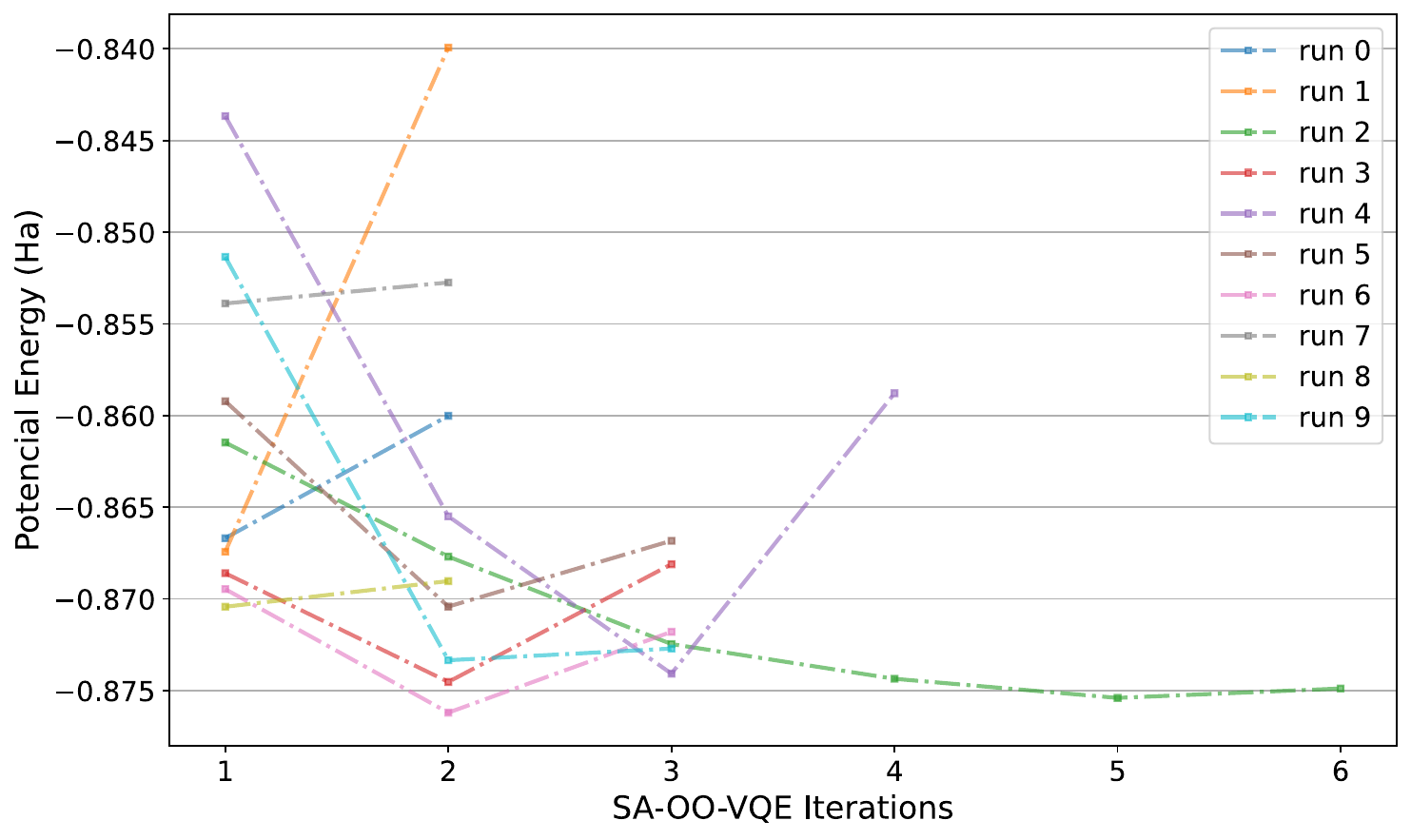}
        \caption{Energy vs. SA-OO-VQE iteration number.} 
        \label{fig:h2_de_b1_conv_iters} 
    \end{subfigure}
    \hfill 
    \begin{subfigure}[t]{0.49\textwidth} 
        \centering
        \includegraphics[width=\linewidth]{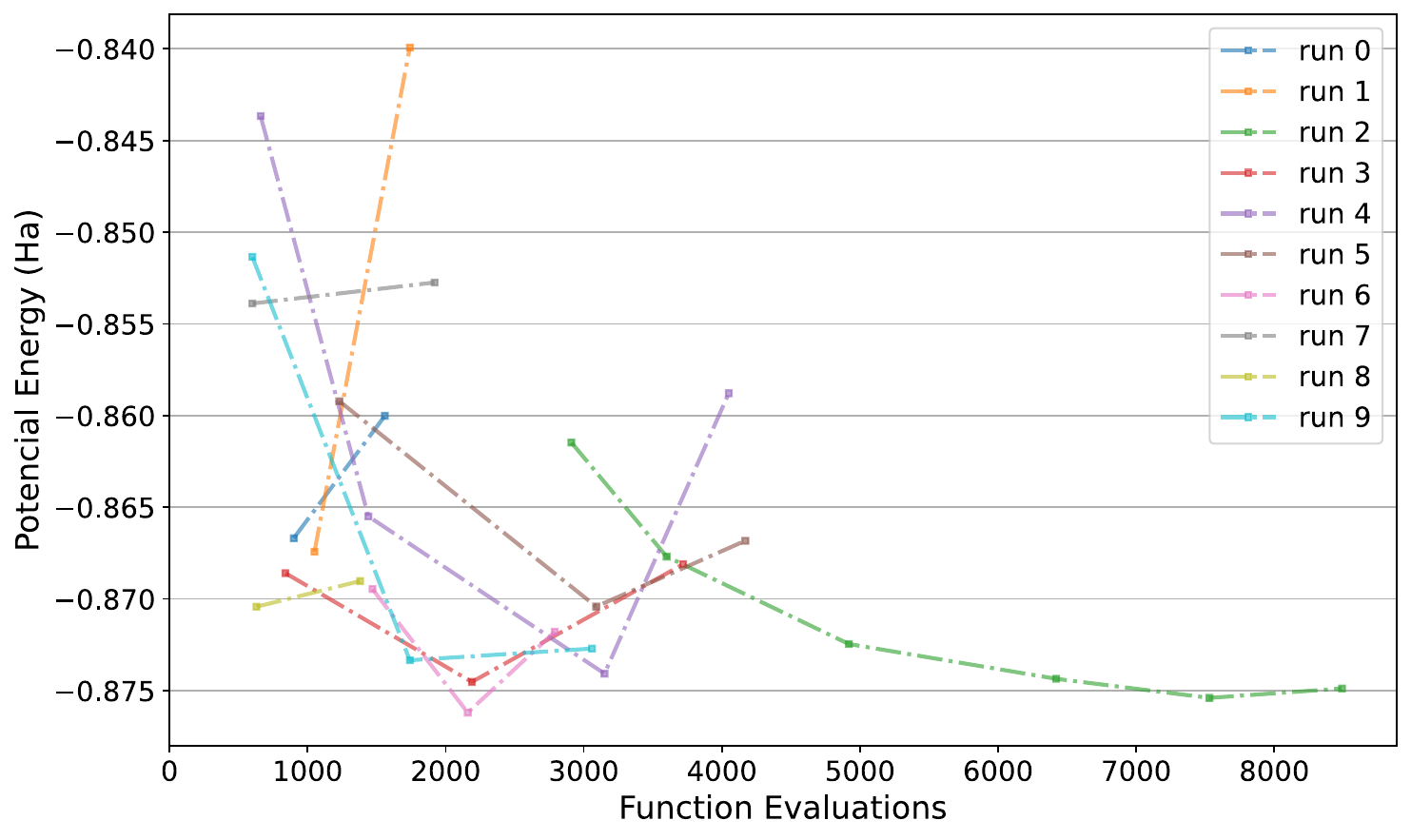}
        \caption{Energy after full SA-OO-VQE iterations vs. cumulative evaluations.} 
        \label{fig:h2_de_b1_conv_evals}
    \end{subfigure}

    \caption[Convergence plots of the DE/Best/1/bin optimizer within the SA-OO-VQE framework for the
H$_2$ molecule.]{Convergence analysis of the DE/Best/1/bin optimizer within the SA-OO-VQE framework for the H$_2$ molecule, based on 10 independent runs (shown in different colors/styles, see legend in plots). The plots display the state-average energy (Hartrees) progression viewed against different metrics: 
    (\subref{fig:h2_de_b1_conv}) Energy evaluated at the end of each internal Gradient Descent optimizer iteration, plotted against the cumulative number of function evaluations consumed up to that iteration point
    (\subref{fig:h2_de_b1_conv_iters}) Energy plotted at the end of each completed SA-OO-VQE iteration against the iteration number. 
    (\subref{fig:h2_de_b1_conv_evals}) Energy plotted at the end of each completed SA-OO-VQE iteration against the cumulative number of function evaluations consumed up to that iteration.}
\label{fig:h2_de_b1} 
\end{figure}

\begin{figure}[!htbp] 
    \centering
    \begin{subfigure}[t]{0.7\textwidth} 
        \centering
        \includegraphics[width=\linewidth]{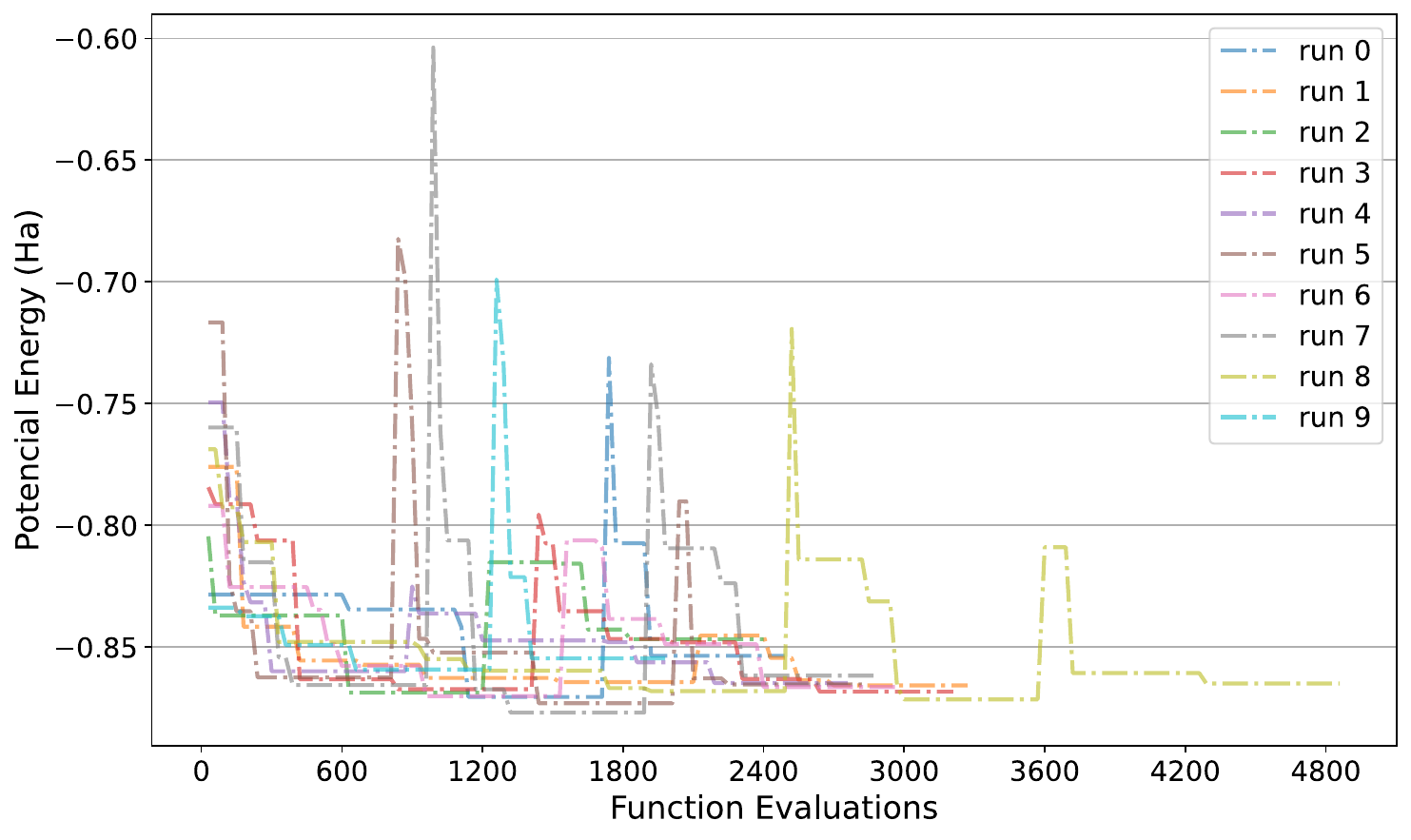}
        \caption{Energy after optimizer iterations vs. cumulative evaluations.} 
        \label{fig:h2_de_b2_conv} 
    \end{subfigure}
    \begin{subfigure}[t]{0.49\textwidth} 
        \centering
        \includegraphics[width=\linewidth]{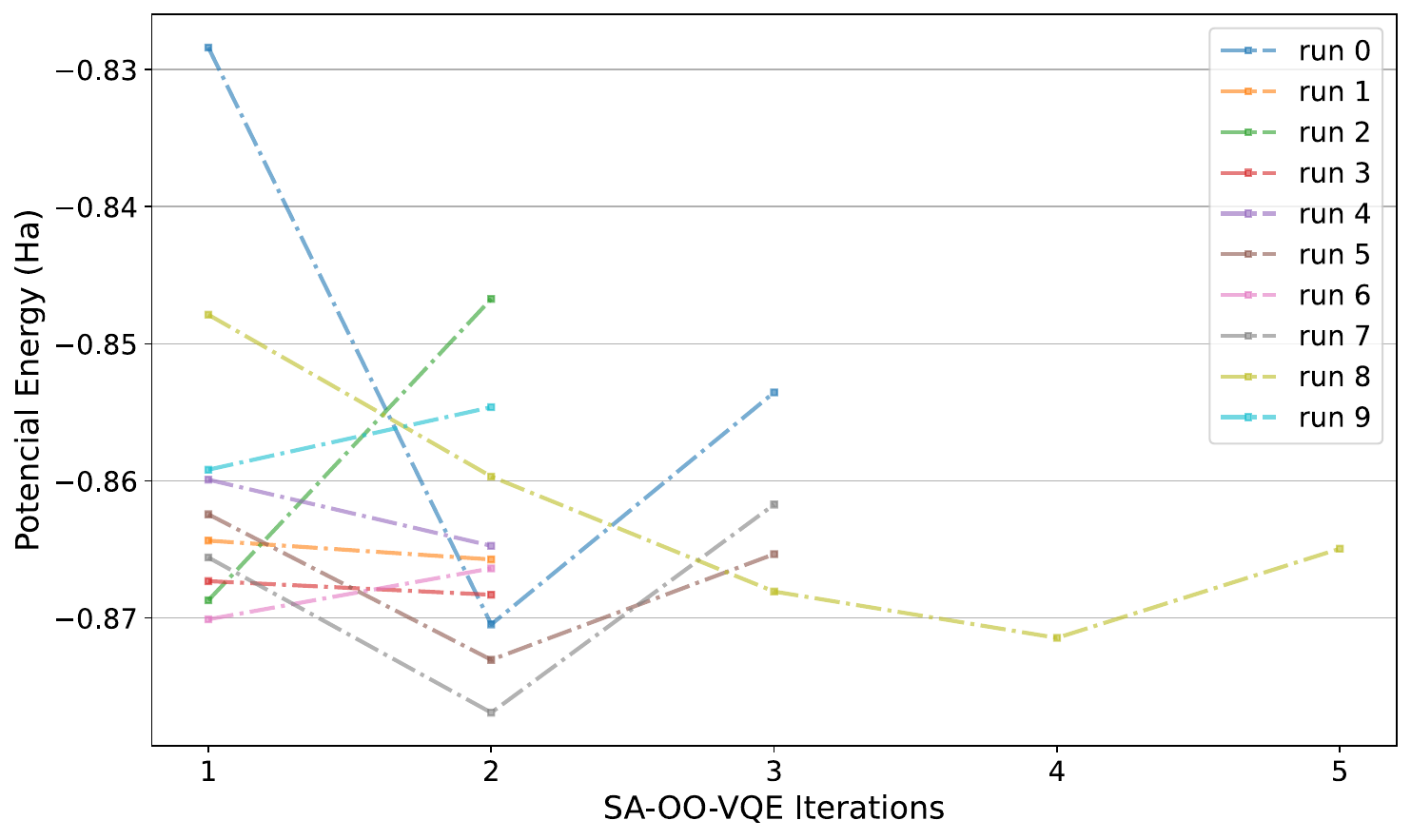}
        \caption{Energy vs. SA-OO-VQE iteration number.} 
        \label{fig:h2_de_b2_conv_iters} 
    \end{subfigure}
    \hfill 
    \begin{subfigure}[t]{0.49\textwidth} 
        \centering
        \includegraphics[width=\linewidth]{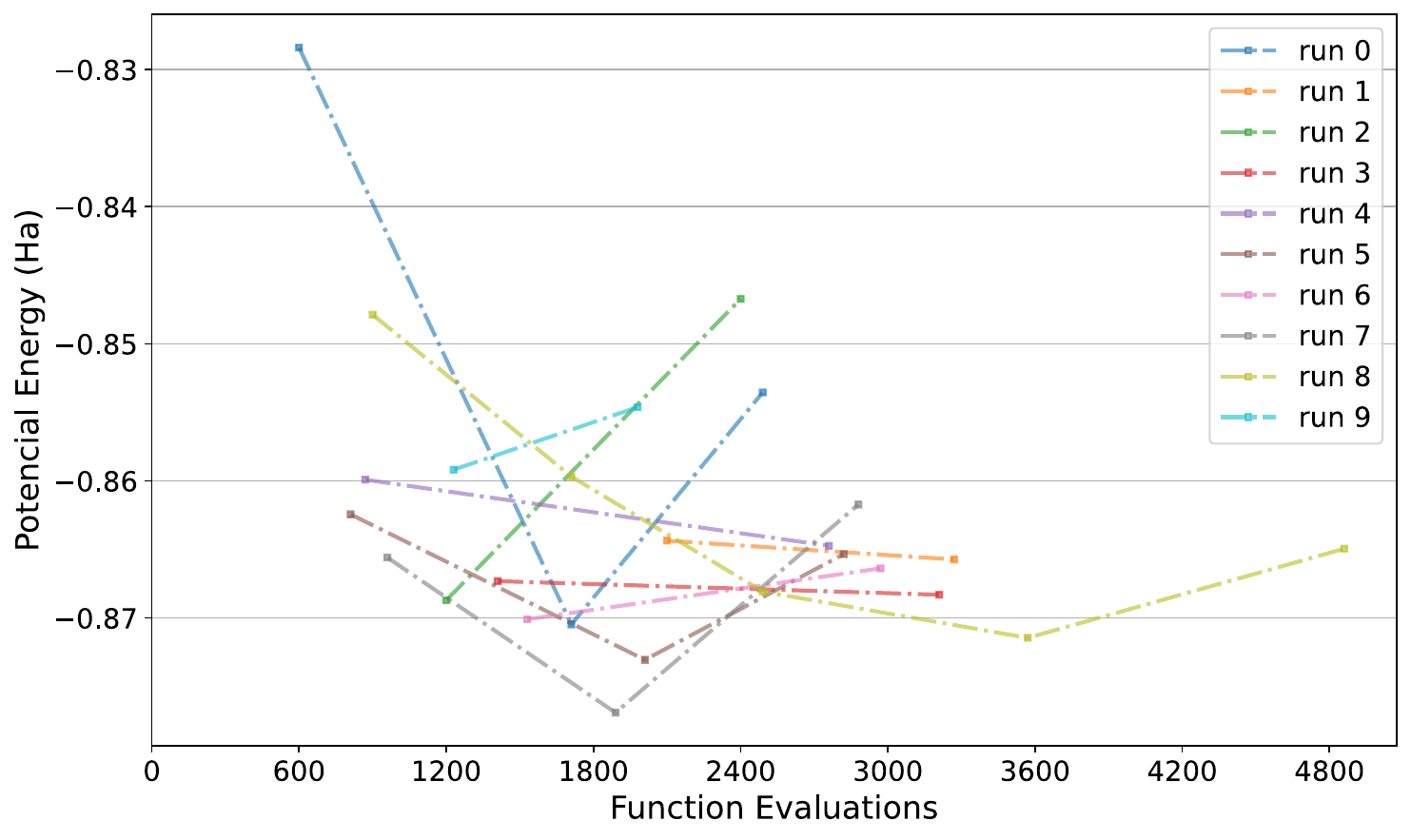}
        \caption{Energy after full SA-OO-VQE iterations vs. cumulative evaluations.} 
        \label{fig:h2_de_b2_conv_evals}
    \end{subfigure}

    \caption[Convergence plots of the DE/Best/2/bin optimizer within the SA-OO-VQE framework for the
H$_2$ molecule.]{Convergence analysis of the DE/Best/2/bin optimizer within the SA-OO-VQE framework for the H$_2$ molecule, based on 10 independent runs (shown in different colors/styles, see legend in plots). The plots display the state-average energy (Hartrees) progression viewed against different metrics: 
    (\subref{fig:h2_de_b2_conv}) Energy evaluated at the end of each internal Gradient Descent optimizer iteration, plotted against the cumulative number of function evaluations consumed up to that iteration point
    (\subref{fig:h2_de_b1_conv_iters}) Energy plotted at the end of each completed SA-OO-VQE iteration against the iteration number. 
    (\subref{fig:h2_de_b2_conv_evals}) Energy plotted at the end of each completed SA-OO-VQE iteration against the cumulative number of function evaluations consumed up to that iteration.}
\label{fig:h2_de_b2} 
\end{figure}

\begin{figure}[!htbp] 
    \centering
    \begin{subfigure}[t]{0.7\textwidth} 
        \centering
        \includegraphics[width=\linewidth]{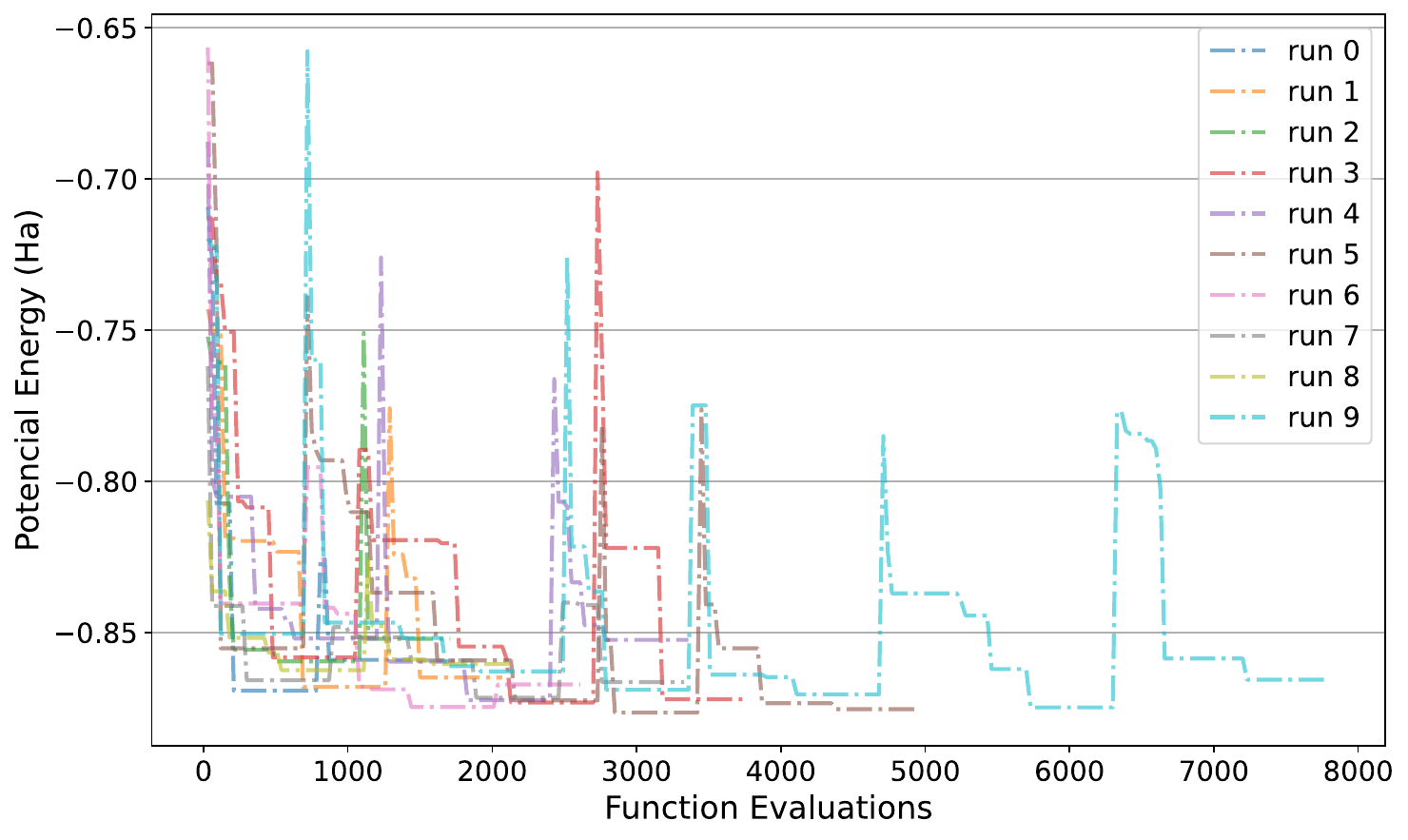}
        \caption{Energy after optimizer iterations vs. cumulative evaluations.} 
        \label{fig:h2_de_ctb_conv} 
    \end{subfigure}
    \begin{subfigure}[t]{0.49\textwidth} 
        \centering
        \includegraphics[width=\linewidth]{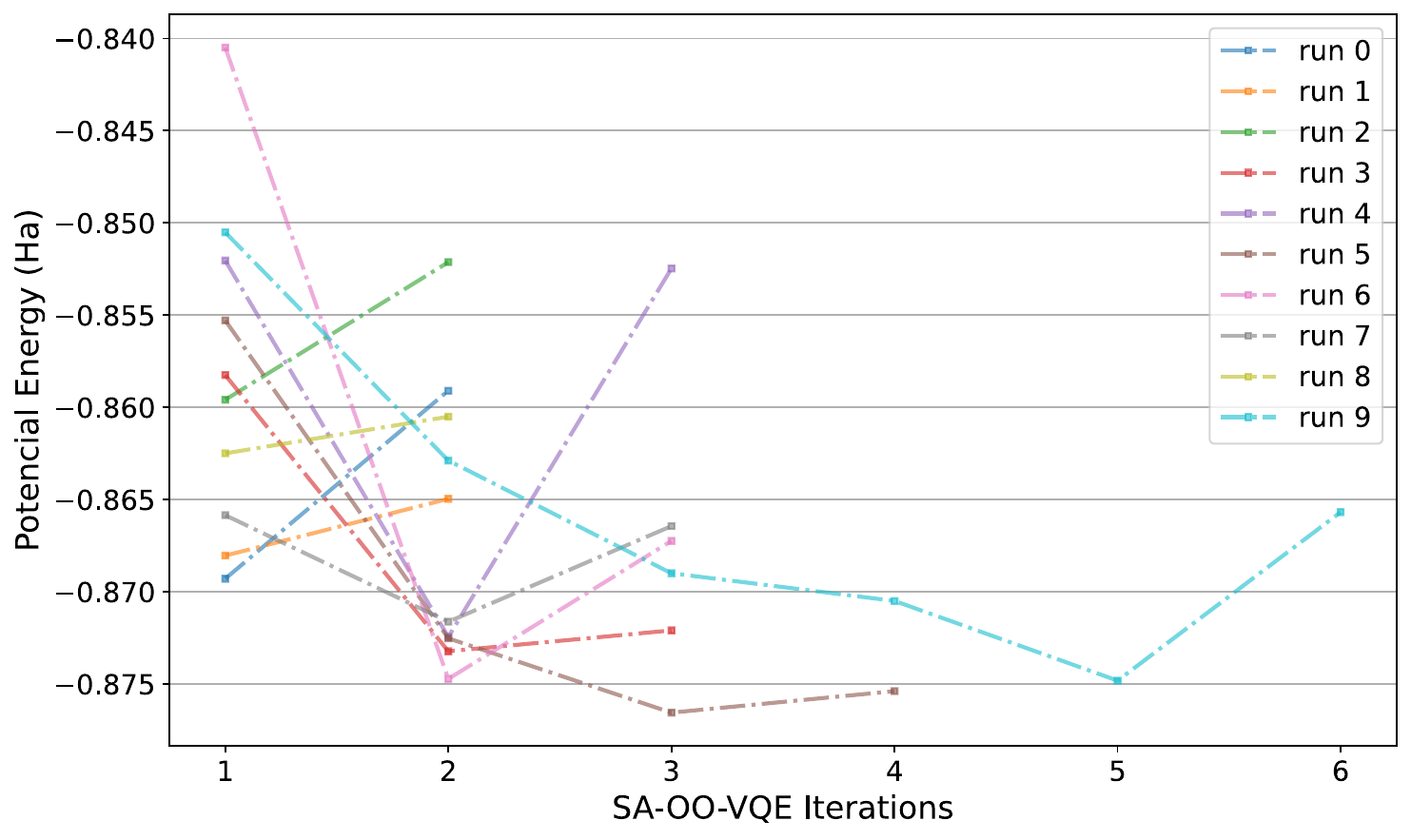}
        \caption{Energy vs. SA-OO-VQE iteration number.} 
        \label{fig:h2_de_ctb_conv_iters} 
    \end{subfigure}
    \hfill 
    \begin{subfigure}[t]{0.49\textwidth} 
        \centering
        \includegraphics[width=\linewidth]{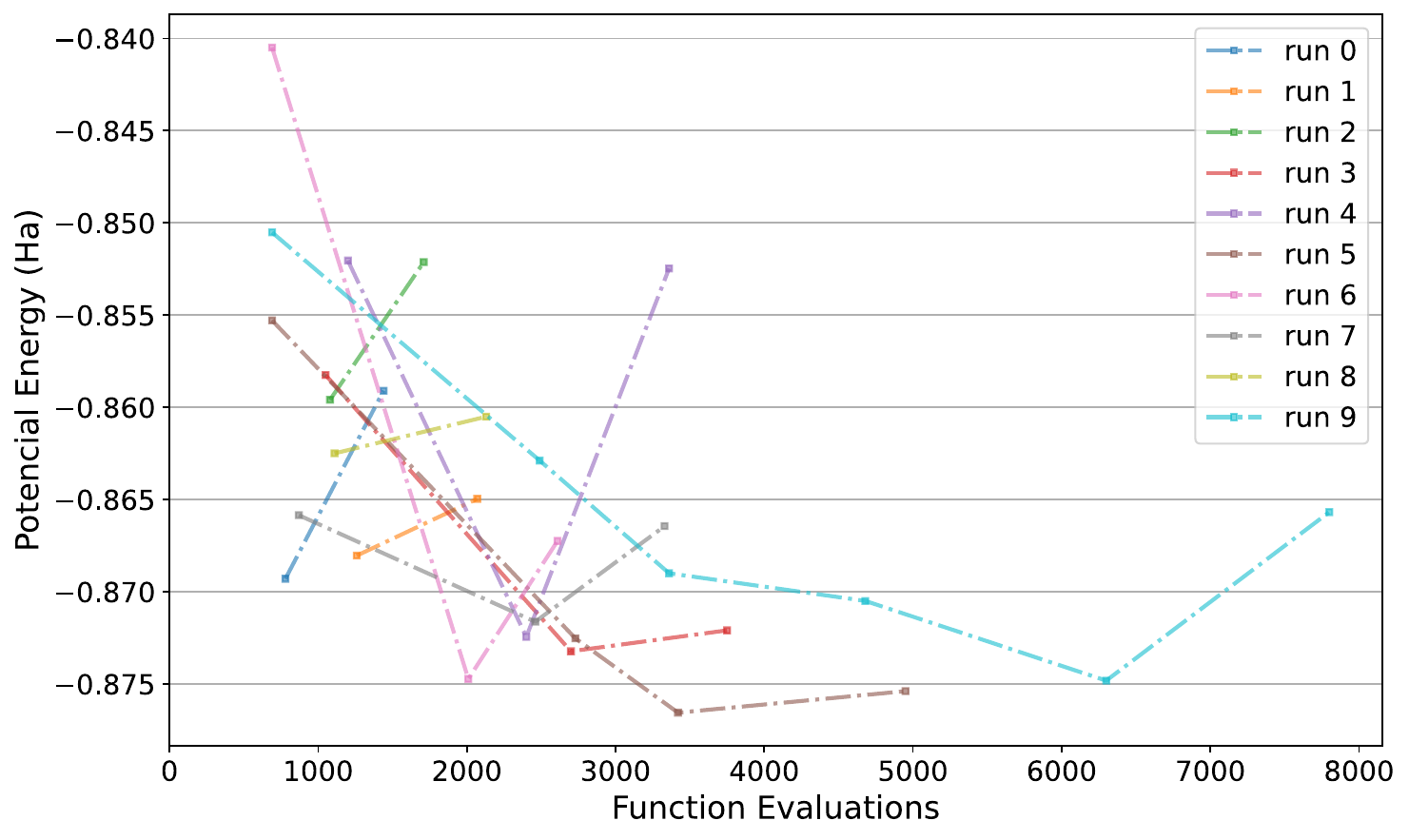}
        \caption{Energy after full SA-OO-VQE iterations vs. cumulative evaluations.} 
        \label{fig:h2_de_ctb_conv_evals}
    \end{subfigure}

    \caption[Convergence plots of the DE/Current-to-Best/1/bin optimizer within the SA-OO-VQE framework for the
H$_2$ molecule.]{Convergence analysis of the DE/Current-to-Best/1/bin optimizer within the SA-OO-VQE framework for the H$_2$ molecule, based on 10 independent runs (shown in different colors/styles, see legend in plots). The plots display the state-average energy (Hartrees) progression viewed against different metrics: 
    (\subref{fig:h2_de_ctb_conv}) Energy evaluated at the end of each internal Gradient Descent optimizer iteration, plotted against the cumulative number of function evaluations consumed up to that iteration point
    (\subref{fig:h2_de_ctb_conv_iters}) Energy plotted at the end of each completed SA-OO-VQE iteration against the iteration number. 
    (\subref{fig:h2_de_ctb_conv_evals}) Energy plotted at the end of each completed SA-OO-VQE iteration against the cumulative number of function evaluations consumed up to that iteration.}
\label{fig:h2_de_ctb} 
\end{figure}

\begin{figure}[!htbp] 
    \centering
    \begin{subfigure}[t]{0.7\textwidth} 
        \centering
        \includegraphics[width=\linewidth]{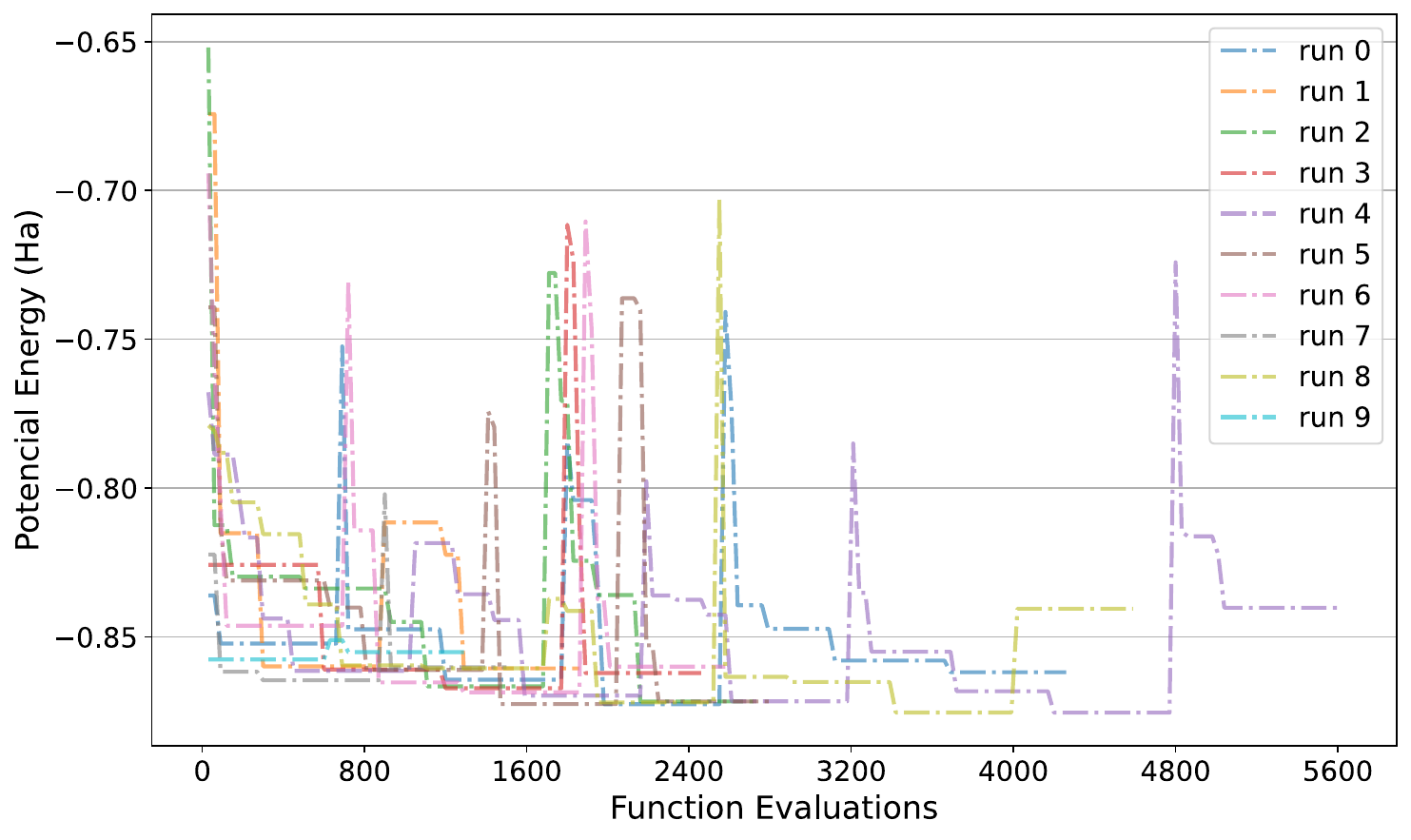}
        \caption{Energy after optimizer iterations vs. cumulative evaluations.} 
        \label{fig:h2_de_ctr_conv} 
    \end{subfigure}
    \begin{subfigure}[t]{0.49\textwidth} 
        \centering
        \includegraphics[width=\linewidth]{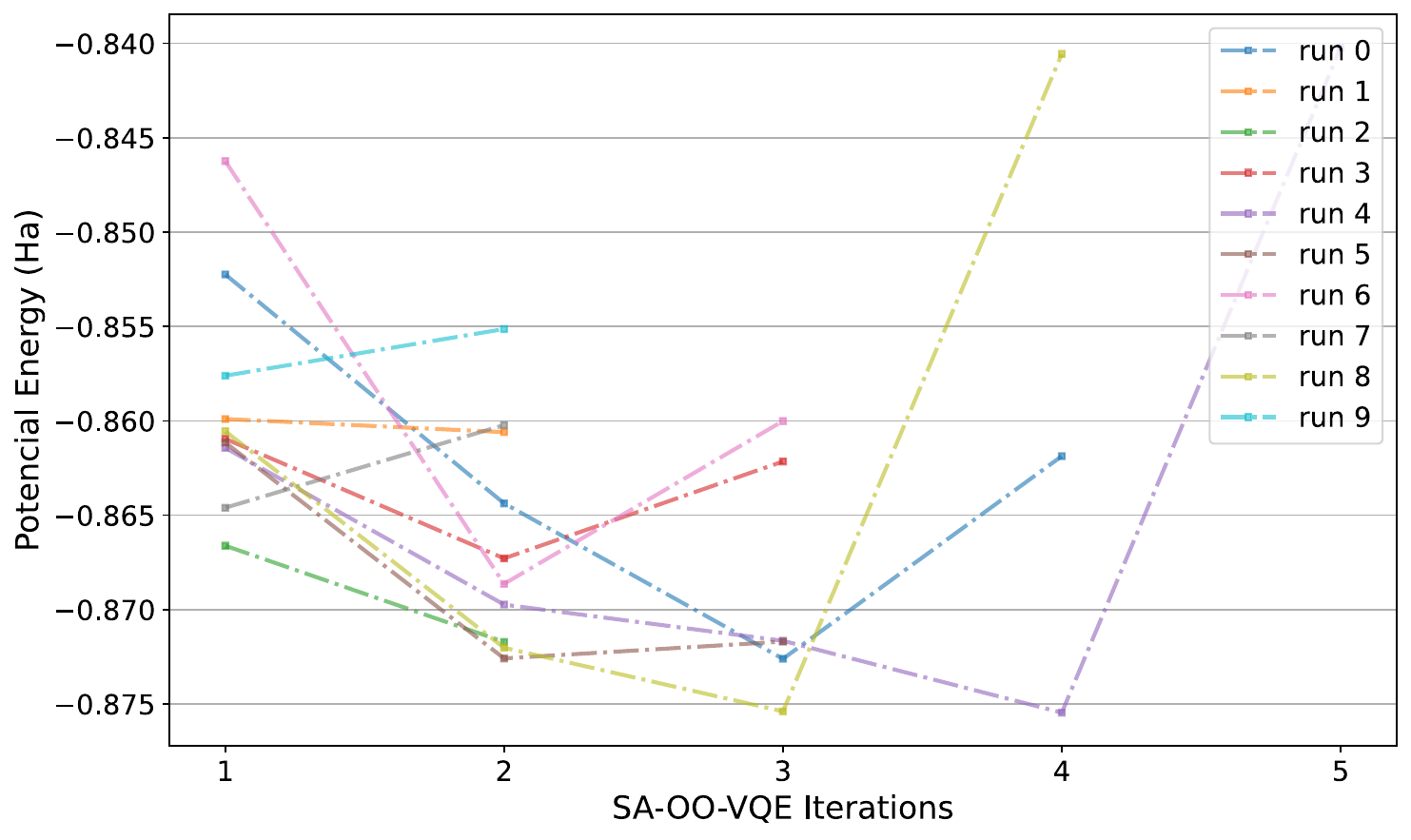}
        \caption{Energy vs. SA-OO-VQE iteration number.} 
        \label{fig:h2_de_ctr_conv_iters} 
    \end{subfigure}
    \hfill 
    \begin{subfigure}[t]{0.49\textwidth} 
        \centering
        \includegraphics[width=\linewidth]{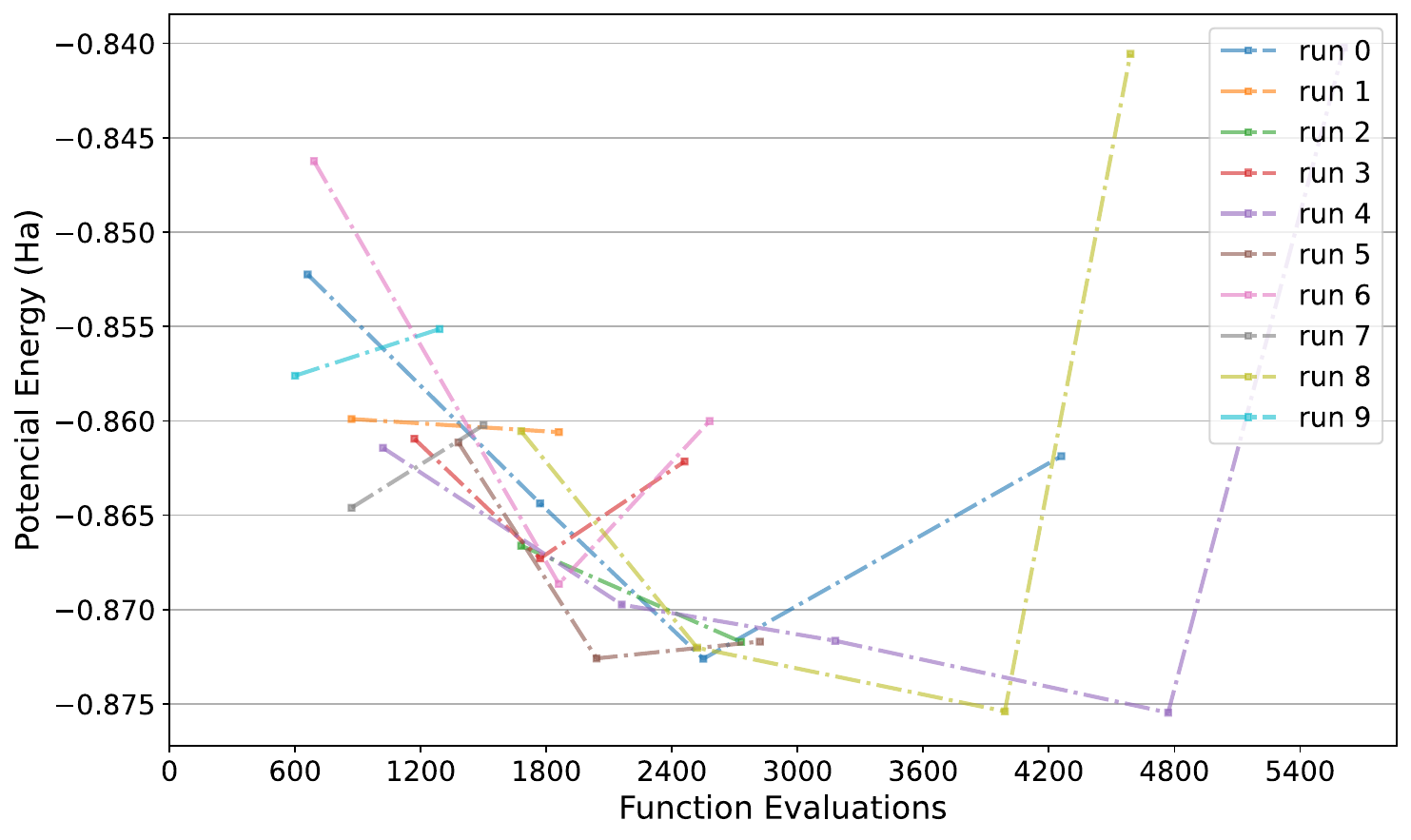}
        \caption{Energy after full SA-OO-VQE iterations vs. cumulative evaluations.} 
        \label{fig:h2_de_ctr_conv_evals}
    \end{subfigure}

    \caption[Convergence plots of the DE/Current-to-Random/1/bin optimizer within the SA-OO-VQE framework for the
H$_2$ molecule.]{Convergence analysis of the DE/Current-to-Random/1/bin optimizer within the SA-OO-VQE framework for the H$_2$ molecule, based on 10 independent runs (shown in different colors/styles, see legend in plots). The plots display the state-average energy (Hartrees) progression viewed against different metrics: 
    (\subref{fig:h2_de_ctr_conv}) Energy evaluated at the end of each internal Gradient Descent optimizer iteration, plotted against the cumulative number of function evaluations consumed up to that iteration point
    (\subref{fig:h2_de_ctr_conv_iters}) Energy plotted at the end of each completed SA-OO-VQE iteration against the iteration number. 
    (\subref{fig:h2_de_ctr_conv_evals}) Energy plotted at the end of each completed SA-OO-VQE iteration against the cumulative number of function evaluations consumed up to that iteration.}
\label{fig:h2_de_ctr} 
\end{figure}

\begin{figure}[!htbp] 
    \centering
    \begin{subfigure}[t]{0.7\textwidth} 
        \centering
        \includegraphics[width=\linewidth]{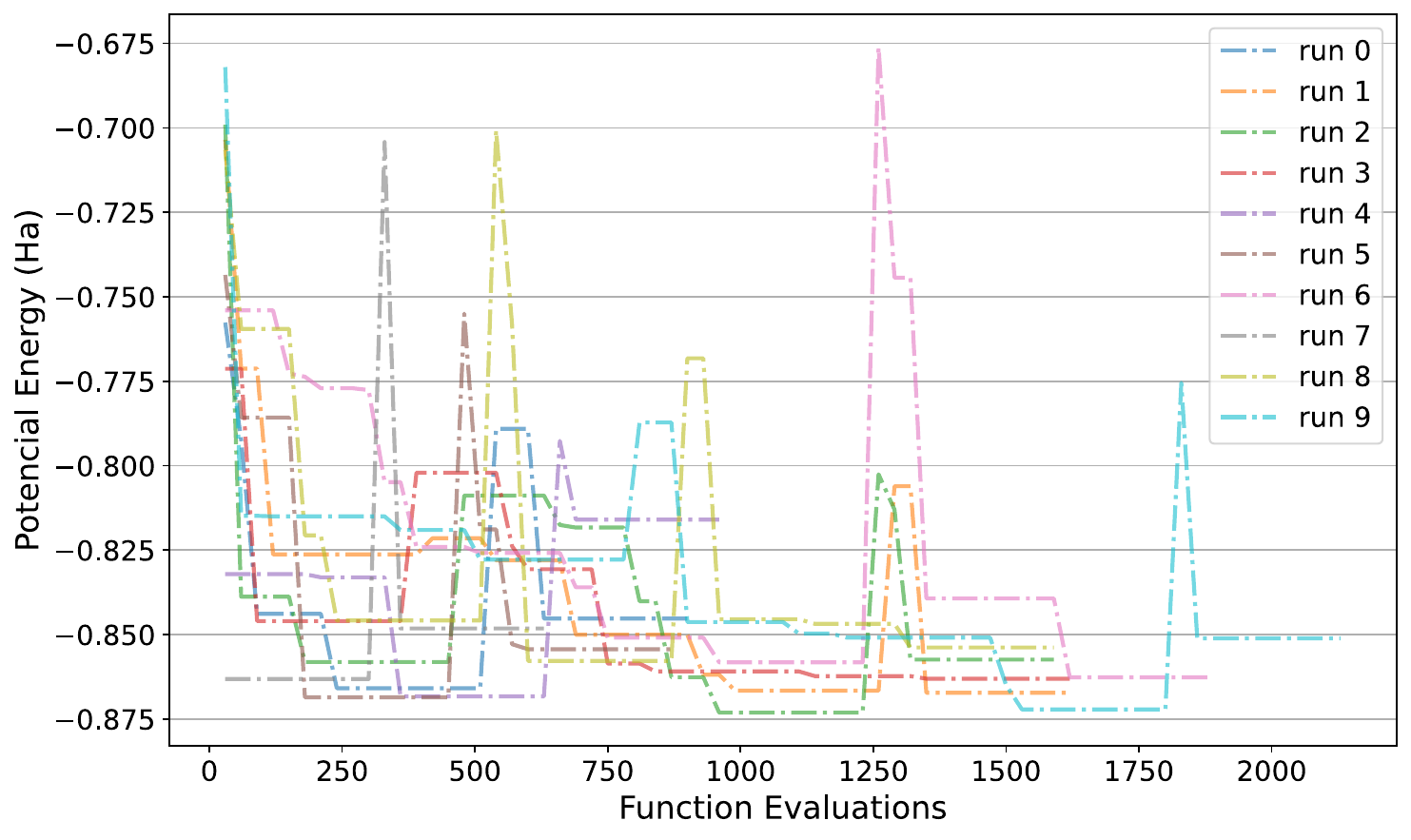}
        \caption{Energy after optimizer iterations vs. cumulative evaluations.} 
        \label{fig:h2_de_r1_conv} 
    \end{subfigure}
    \begin{subfigure}[t]{0.49\textwidth} 
        \centering
        \includegraphics[width=\linewidth]{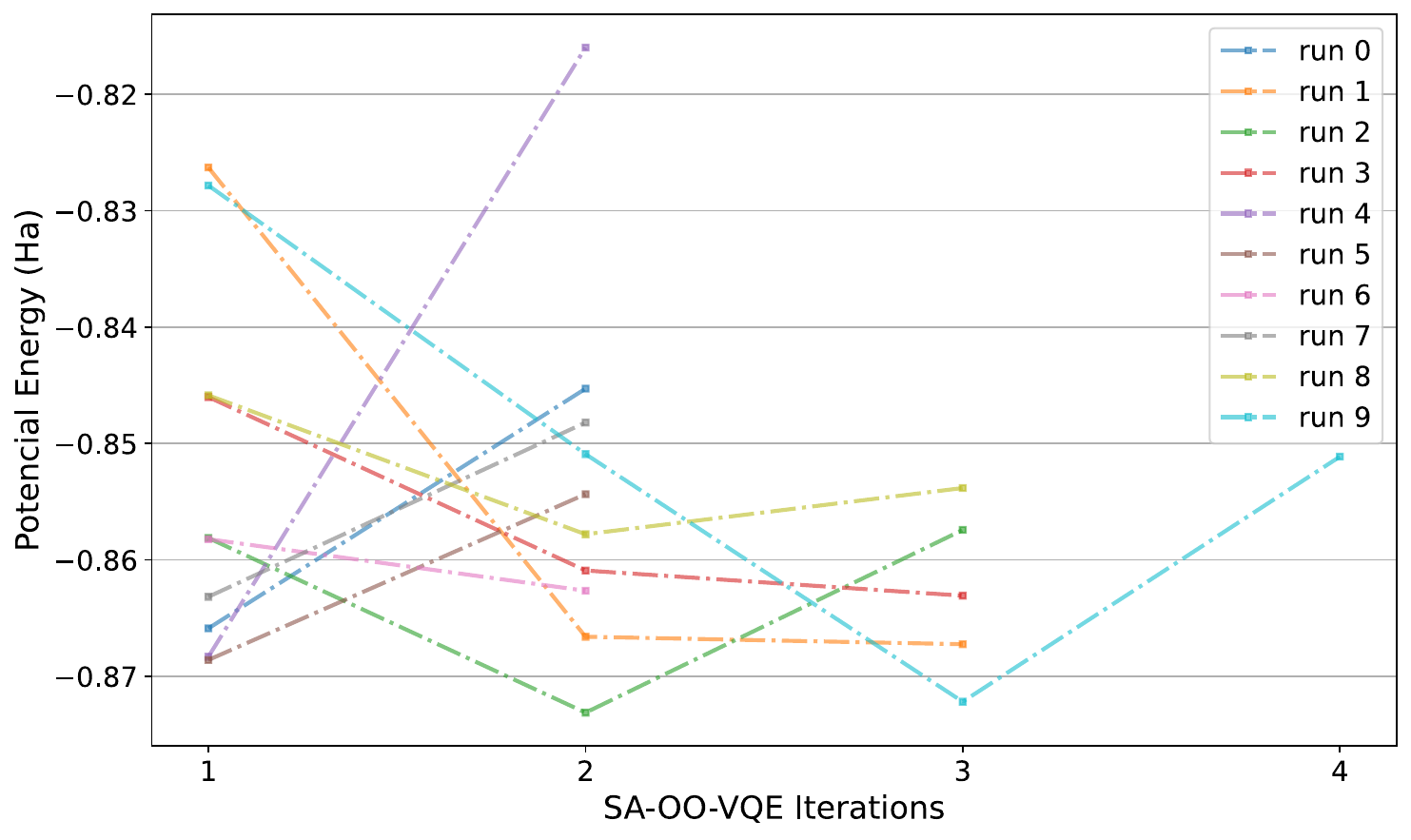}
        \caption{Energy vs. SA-OO-VQE iteration number.} 
        \label{fig:h2_de_r1_conv_iters} 
    \end{subfigure}
    \hfill 
    \begin{subfigure}[t]{0.49\textwidth} 
        \centering
        \includegraphics[width=\linewidth]{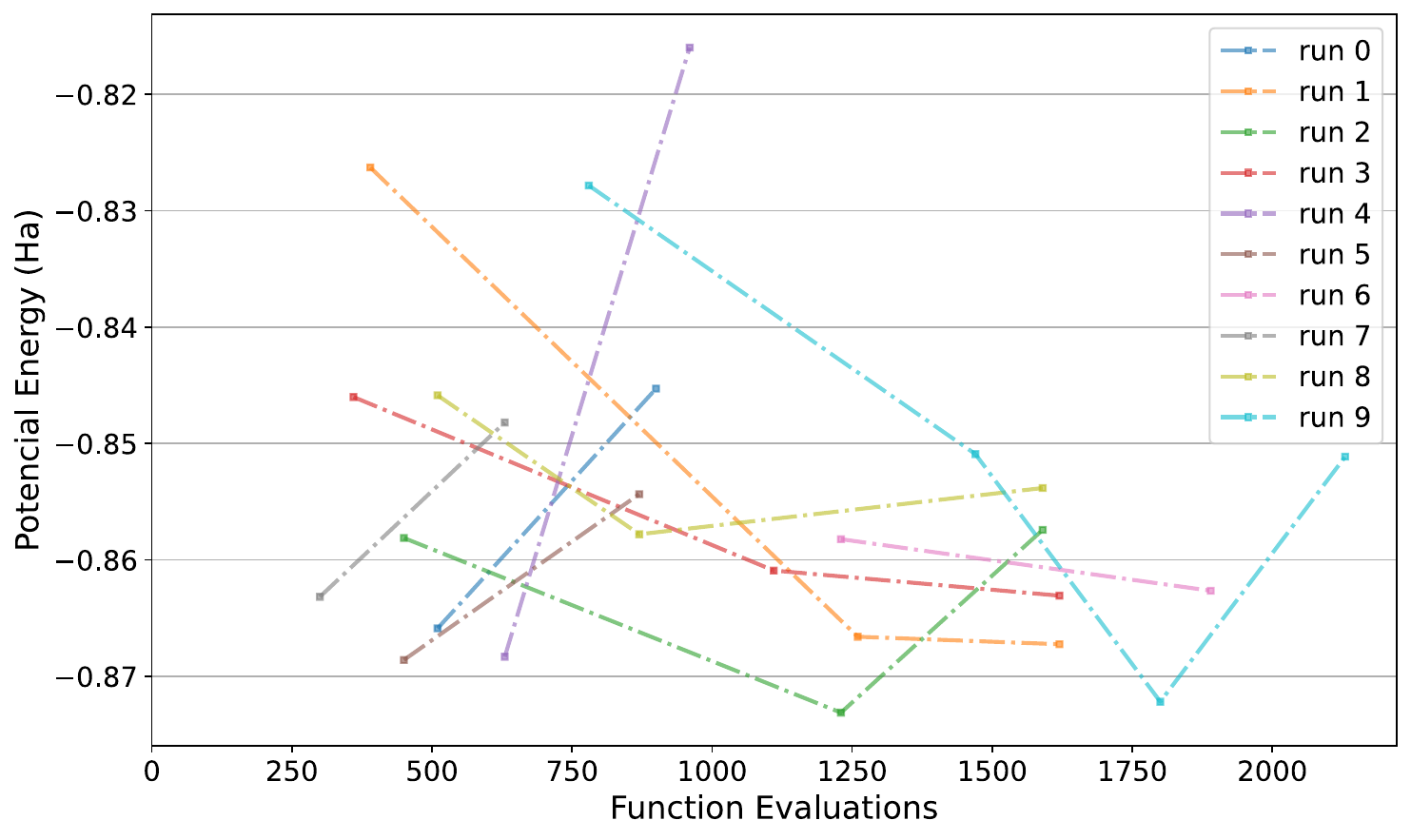}
        \caption{Energy after full SA-OO-VQE iterations vs. cumulative evaluations.} 
        \label{fig:h2_de_r1_conv_evals}
    \end{subfigure}

    \caption[Convergence plots of the DE/Rand/1/bin optimizer within the SA-OO-VQE framework for the
H$_2$ molecule.]{Convergence analysis of the DE/Rand/1/bin optimizer within the SA-OO-VQE framework for the H$_2$ molecule, based on 10 independent runs (shown in different colors/styles, see legend in plots). The plots display the state-average energy (Hartrees) progression viewed against different metrics: 
    (\subref{fig:h2_de_r1_conv}) Energy evaluated at the end of each internal Gradient Descent optimizer iteration, plotted against the cumulative number of function evaluations consumed up to that iteration point
    (\subref{fig:h2_de_r1_conv_iters}) Energy plotted at the end of each completed SA-OO-VQE iteration against the iteration number. 
    (\subref{fig:h2_de_r1_conv_evals}) Energy plotted at the end of each completed SA-OO-VQE iteration against the cumulative number of function evaluations consumed up to that iteration.}
\label{fig:h2_de_r1} 
\end{figure}

\begin{figure}[!htbp] 
    \centering
    \begin{subfigure}[t]{0.7\textwidth} 
        \centering
        \includegraphics[width=\linewidth]{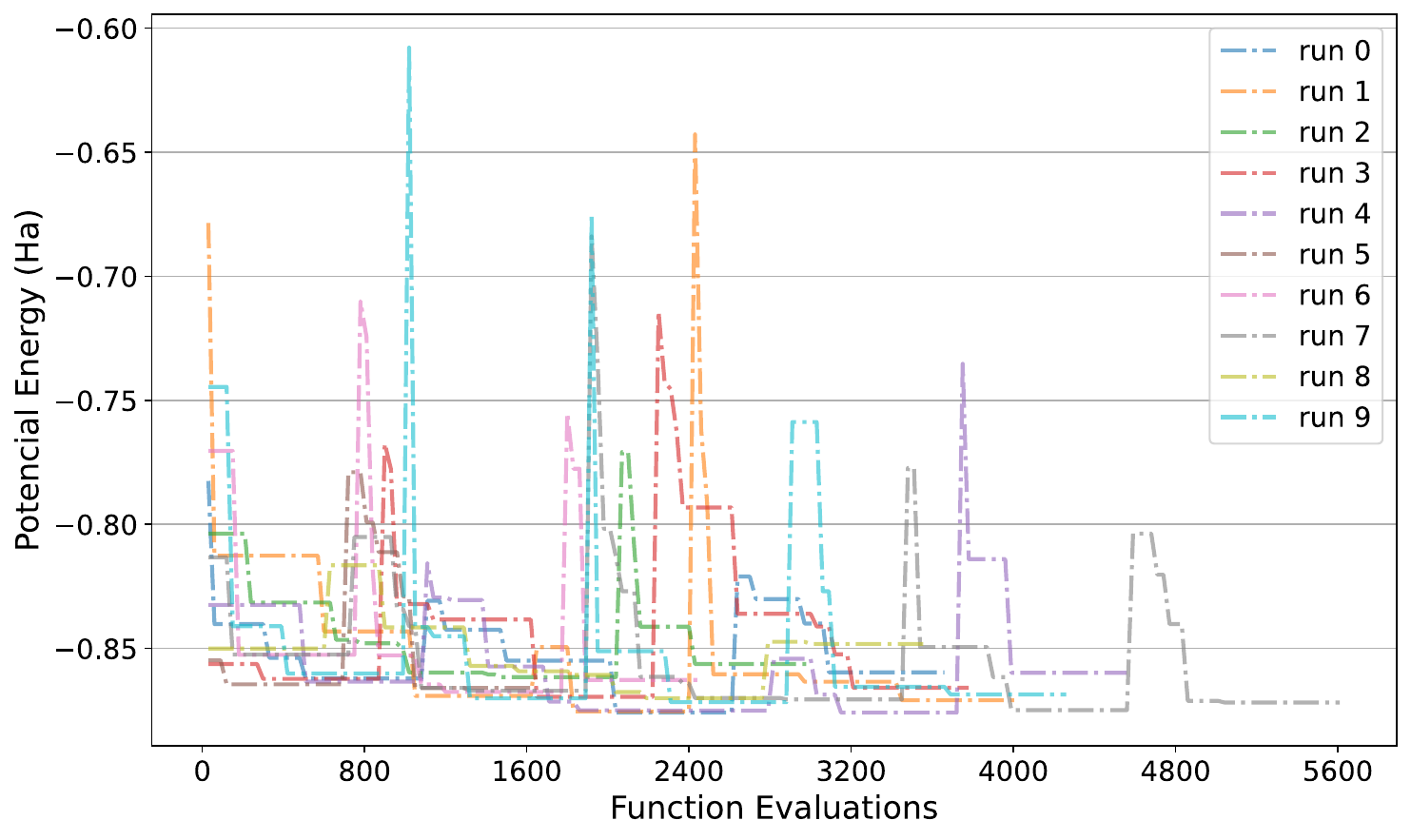}
        \caption{Energy after optimizer iterations vs. cumulative evaluations.} 
        \label{fig:h2_de_rtb_conv} 
    \end{subfigure}
    \begin{subfigure}[t]{0.49\textwidth} 
        \centering
        \includegraphics[width=\linewidth]{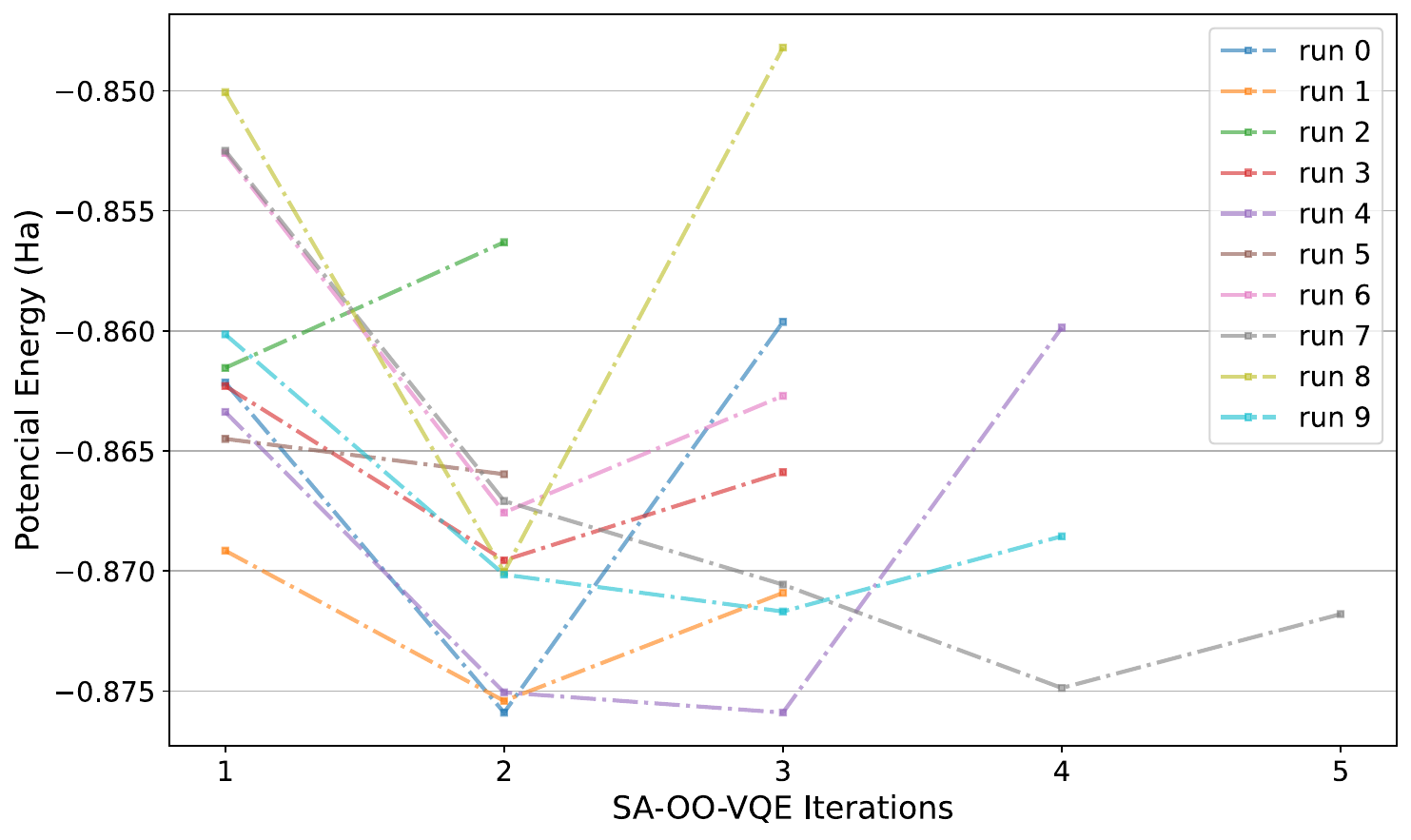}
        \caption{Energy vs. SA-OO-VQE iteration number.} 
        \label{fig:h2_de_rtb_conv_iters} 
    \end{subfigure}
    \hfill 
    \begin{subfigure}[t]{0.49\textwidth} 
        \centering
        \includegraphics[width=\linewidth]{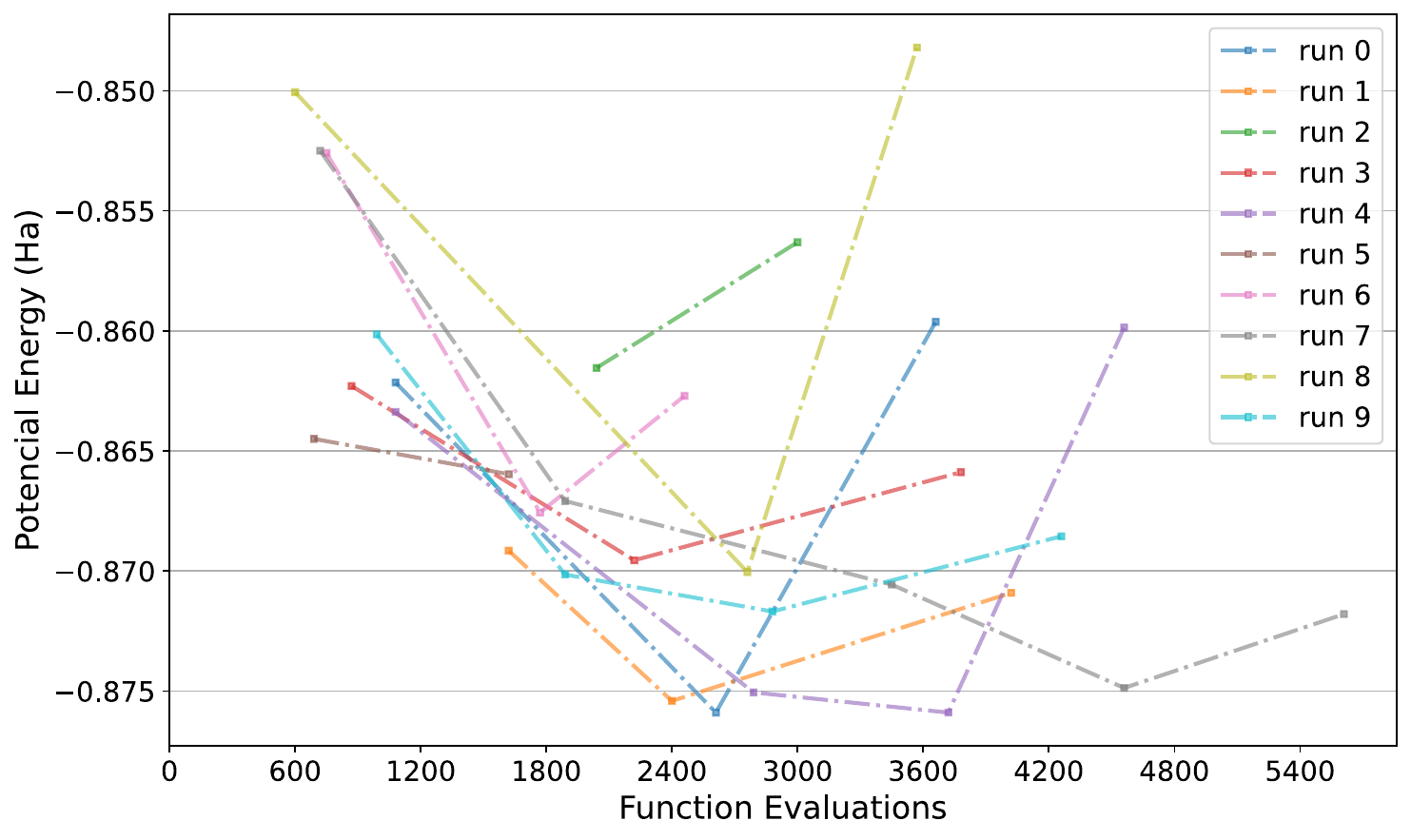}
        \caption{Energy after full SA-OO-VQE iterations vs. cumulative evaluations.} 
        \label{fig:h2_de_rtb_conv_evals}
    \end{subfigure}

    \caption[Convergence plots of the DE/Random-to-Best/1/bin optimizer within the SA-OO-VQE framework for the
H$_2$ molecule.]{Convergence analysis of the DE/Random-to-Best/1/bin optimizer within the SA-OO-VQE framework for the H$_2$ molecule, based on 10 independent runs (shown in different colors/styles, see legend in plots). The plots display the state-average energy (Hartrees) progression viewed against different metrics: 
    (\subref{fig:h2_de_rtb_conv}) Energy evaluated at the end of each internal Gradient Descent optimizer iteration, plotted against the cumulative number of function evaluations consumed up to that iteration point
    (\subref{fig:h2_de_rtb_conv_iters}) Energy plotted at the end of each completed SA-OO-VQE iteration against the iteration number. 
    (\subref{fig:h2_de_rtb_conv_evals}) Energy plotted at the end of each completed SA-OO-VQE iteration against the cumulative number of function evaluations consumed up to that iteration.}
\label{fig:h2_de_rtb} 
\end{figure}

\clearpage
 
\section{Convergence plots for H$_4$ molecule}
\label{ap_sec:h4_conv}
\begin{figure}[!htbp] 
    \centering
    \begin{subfigure}[t]{0.49\textwidth} 
        \centering
        \includegraphics[width=\linewidth]{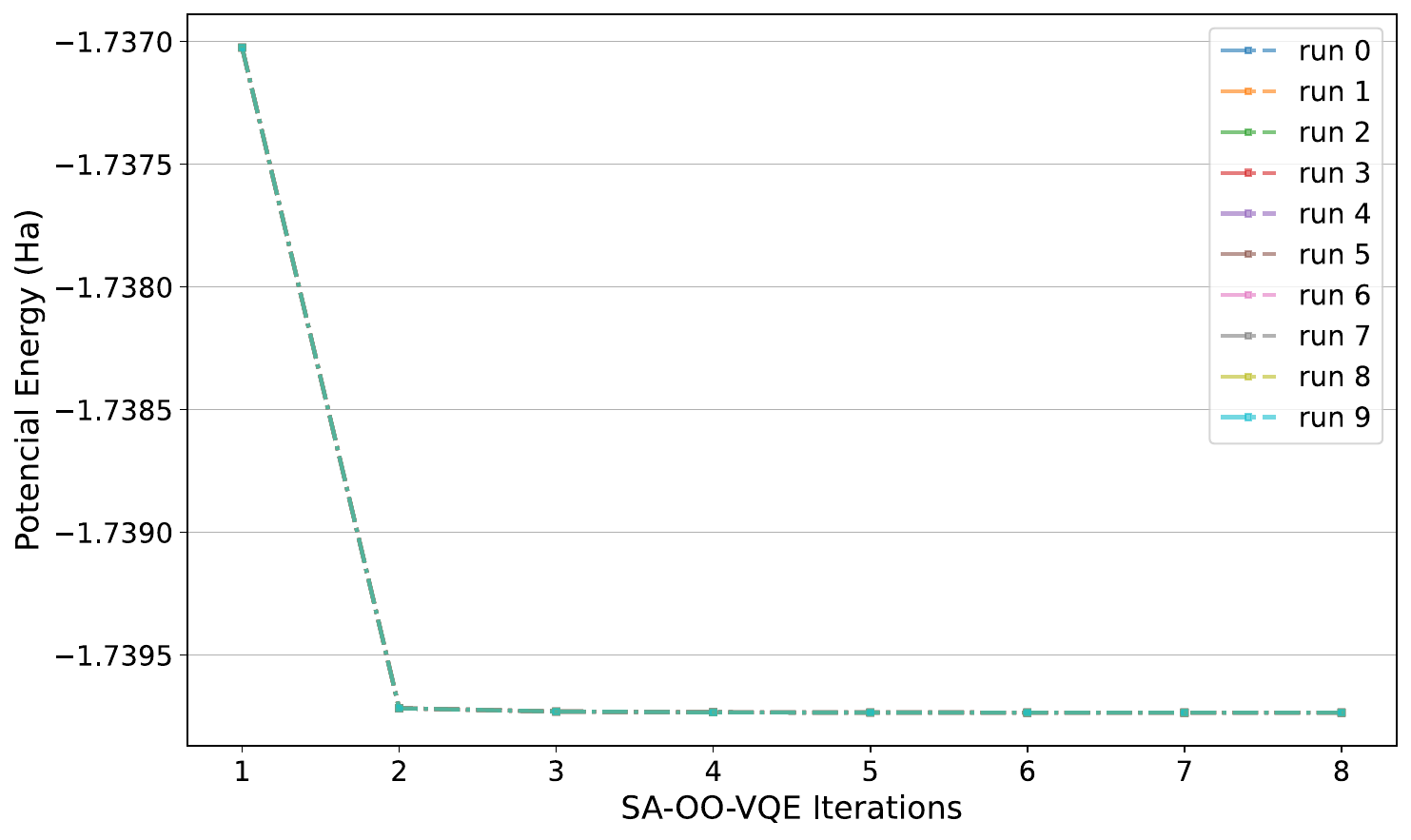}
        \caption{Energy vs. SA-OO-VQE iteration number.} 
        \label{fig:h4_bfgs_conv_iters} 
    \end{subfigure}
    \hfill 
    \begin{subfigure}[t]{0.49\textwidth} 
        \centering
        \includegraphics[width=\linewidth]{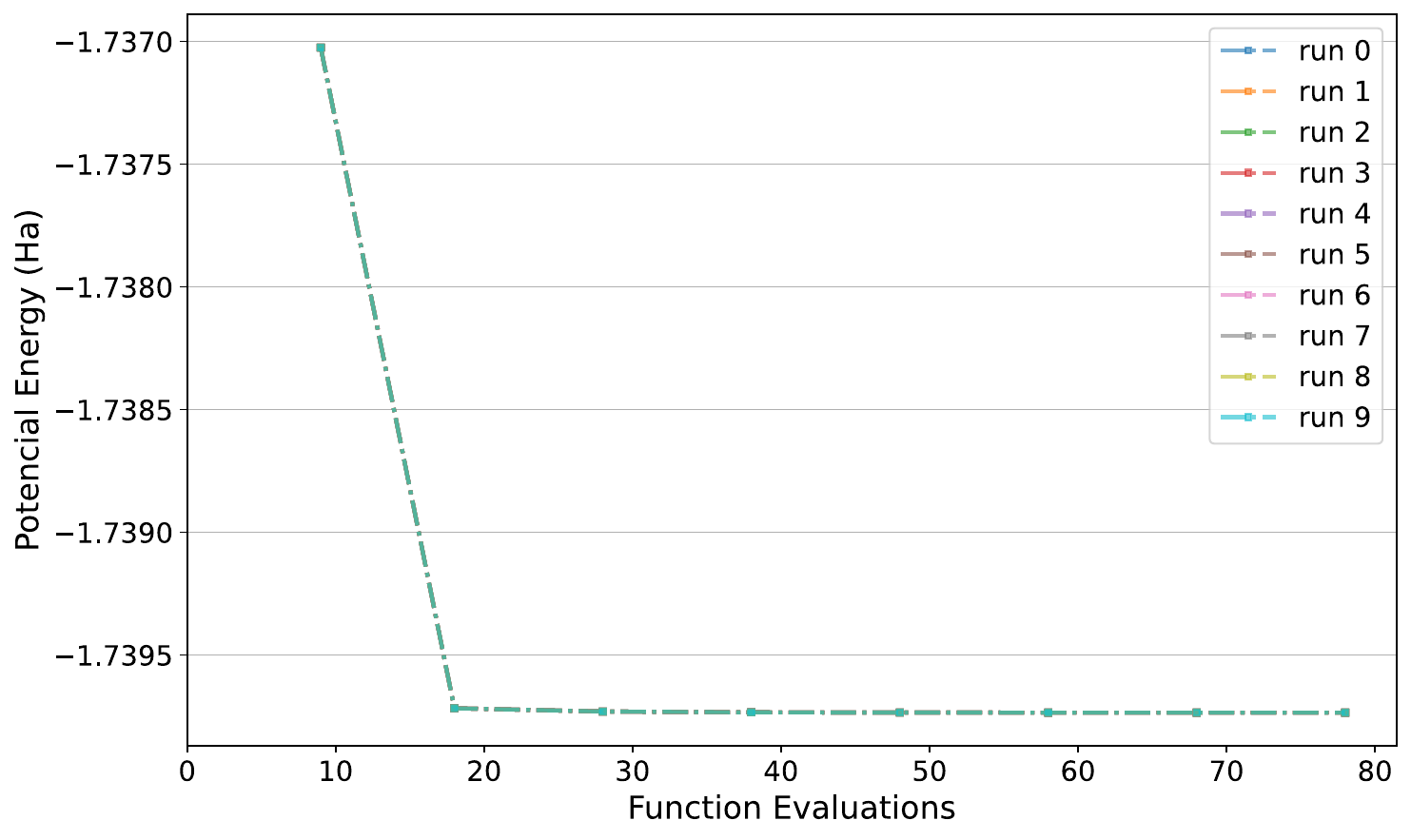}
        \caption{Energy after full SA-OO-VQE iterations vs. cumulative evaluations.} 
        \label{fig:h4_bfgs_conv_evals}
    \end{subfigure}

    \caption[Convergence plots of the BFGS optimizer within the SA-OO-VQE framework for the
H$_4$ molecule.]{Convergence analysis of the BFGS optimizer within the SA-OO-VQE framework for the H$_4$ molecule, based on 10 independent runs (shown in different colors/styles, see legend in plots). The plots display the state-average energy (Hartrees) progression viewed against different metrics: 
    (\subref{fig:h4_bfgs_conv_iters}) Energy plotted at the end of each completed SA-OO-VQE iteration against the iteration number. 
    (\subref{fig:h4_bfgs_conv_evals}) Energy plotted at the end of each completed SA-OO-VQE iteration against the cumulative number of function evaluations consumed up to that iteration.}
\label{fig:h4_bfgs} 
\end{figure}

\begin{figure}[!htbp] 
    \centering
    \begin{subfigure}[t]{0.49\textwidth} 
        \centering
        \includegraphics[width=\linewidth]{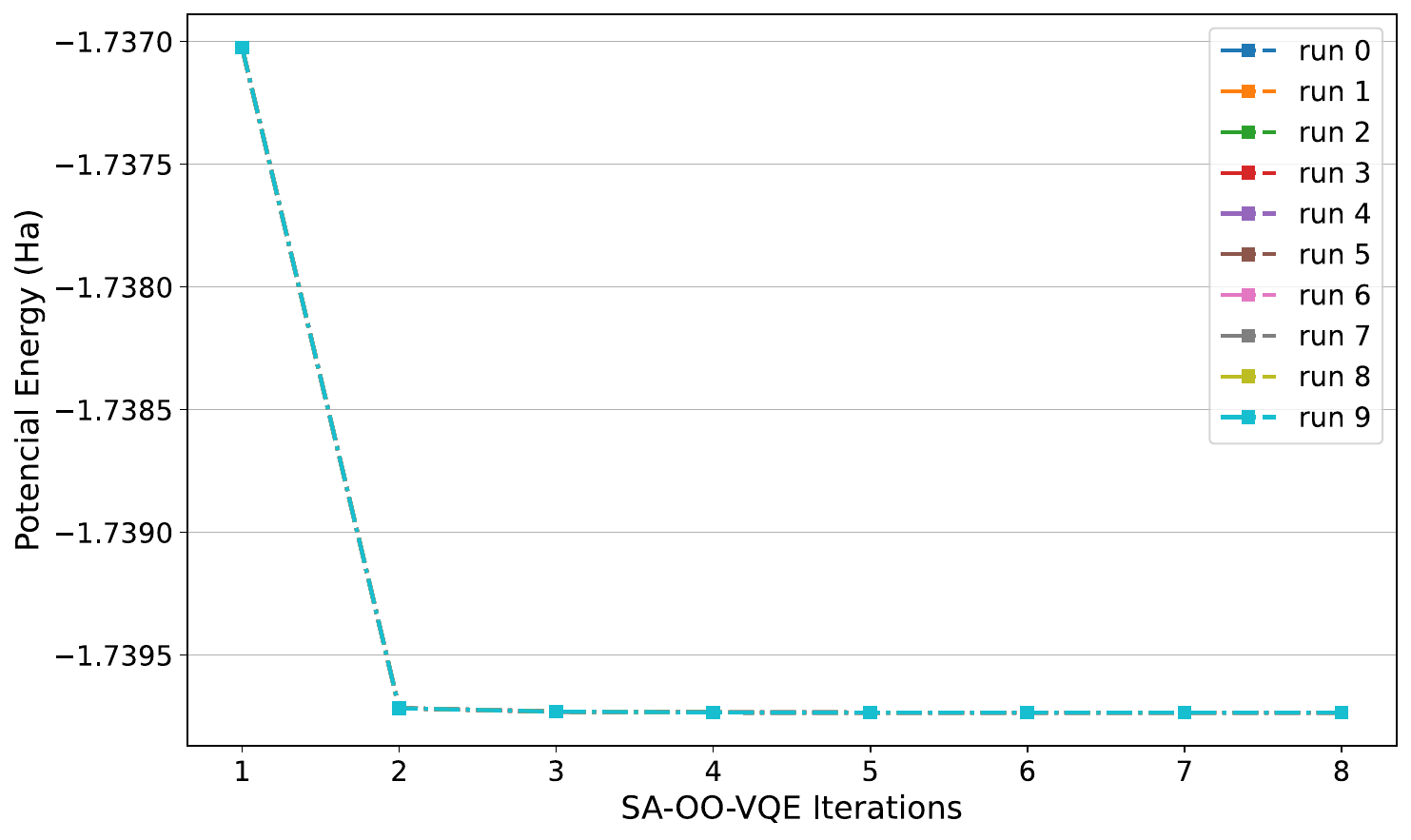}
        \caption{Energy vs. SA-OO-VQE iteration number.} 
        \label{fig:h4_cobyla_conv_iters} 
    \end{subfigure}
    \hfill 
    \begin{subfigure}[t]{0.49\textwidth} 
        \centering
        \includegraphics[width=\linewidth]{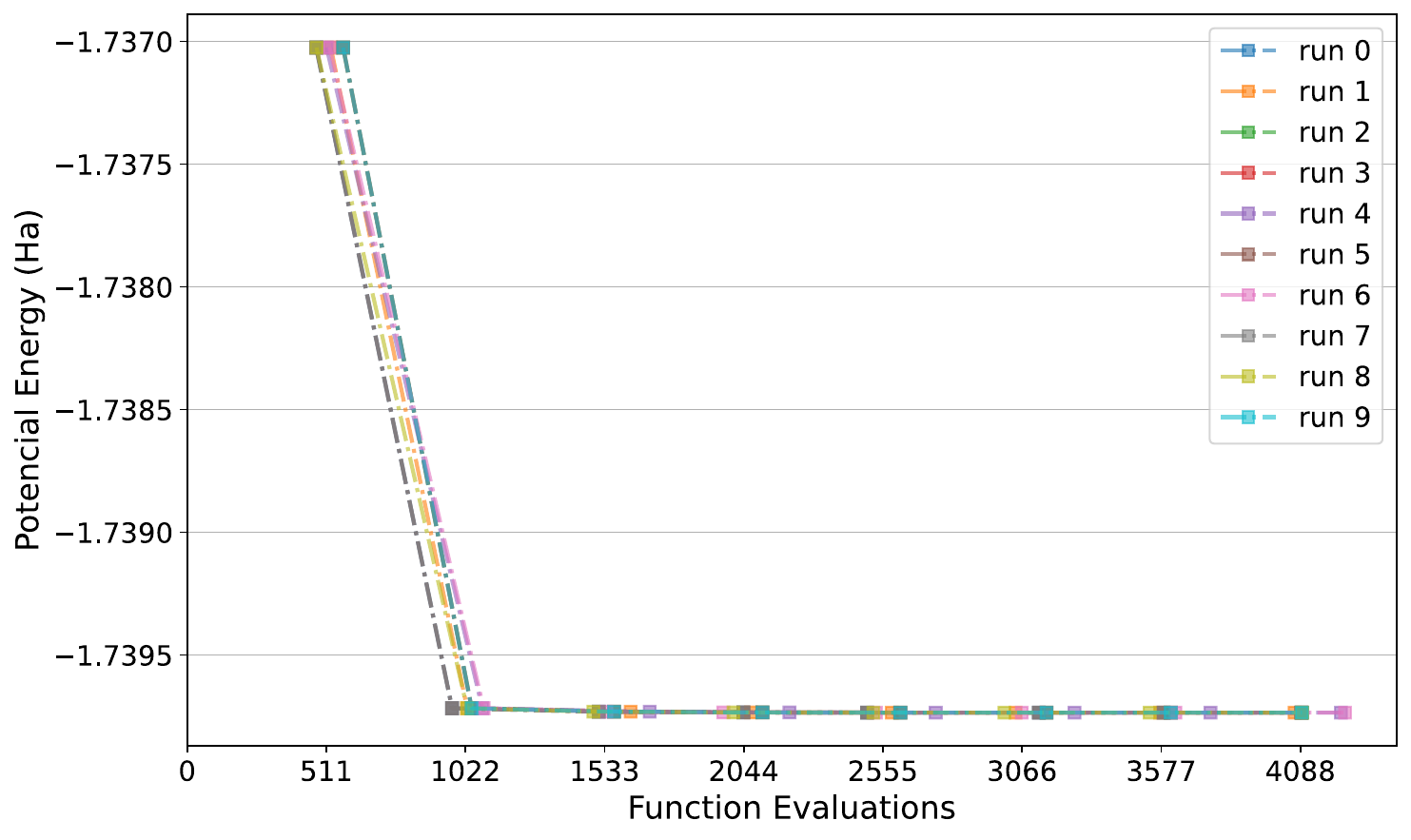}
        \caption{Energy after full SA-OO-VQE iterations vs. cumulative evaluations.} 
        \label{fig:h4_cobyla_conv_evals}
    \end{subfigure}

    \caption[Convergence plots of the COBYLA optimizer within the SA-OO-VQE framework for the
H$_4$ molecule.]{Convergence analysis of the COBYLA optimizer within the SA-OO-VQE framework for the H$_4$ molecule, based on 10 independent runs (shown in different colors/styles, see legend in plots). The plots display the state-average energy (Hartrees) progression viewed against different metrics: 
    (\subref{fig:h4_cobyla_conv_iters}) Energy plotted at the end of each completed SA-OO-VQE iteration against the iteration number. 
    (\subref{fig:h4_cobyla_conv_evals}) Energy plotted at the end of each completed SA-OO-VQE iteration against the cumulative number of function evaluations consumed up to that iteration.}
\label{fig:h4_cobyla} 
\end{figure}

\begin{figure}[!htbp] 
    \centering
    \begin{subfigure}[t]{0.7\textwidth} 
        \centering
        \includegraphics[width=\linewidth]{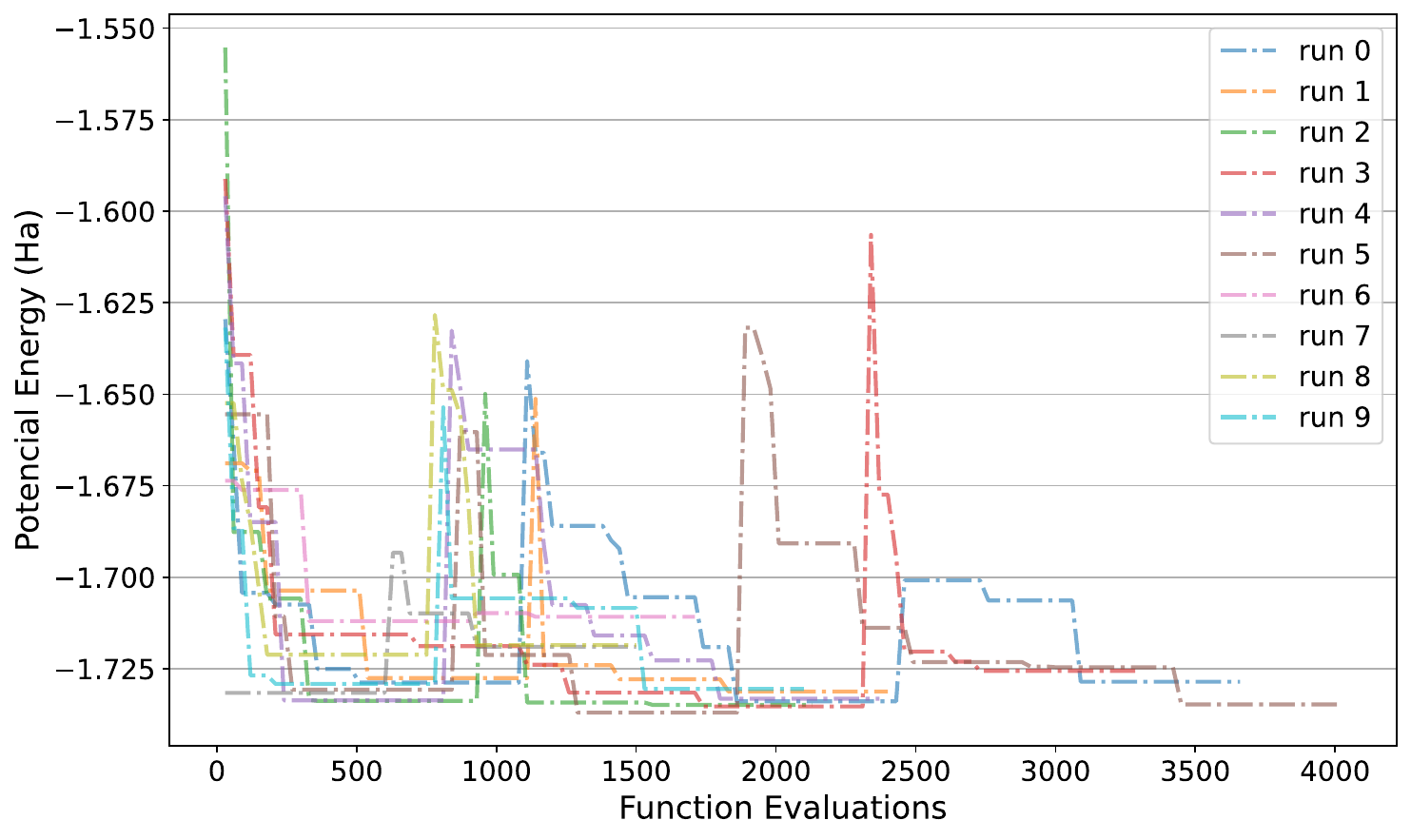}
        \caption{Energy after optimizer iterations vs. cumulative evaluations.} 
        \label{fig:h4_de_b1_conv} 
    \end{subfigure}
    \begin{subfigure}[t]{0.49\textwidth} 
        \centering
        \includegraphics[width=\linewidth]{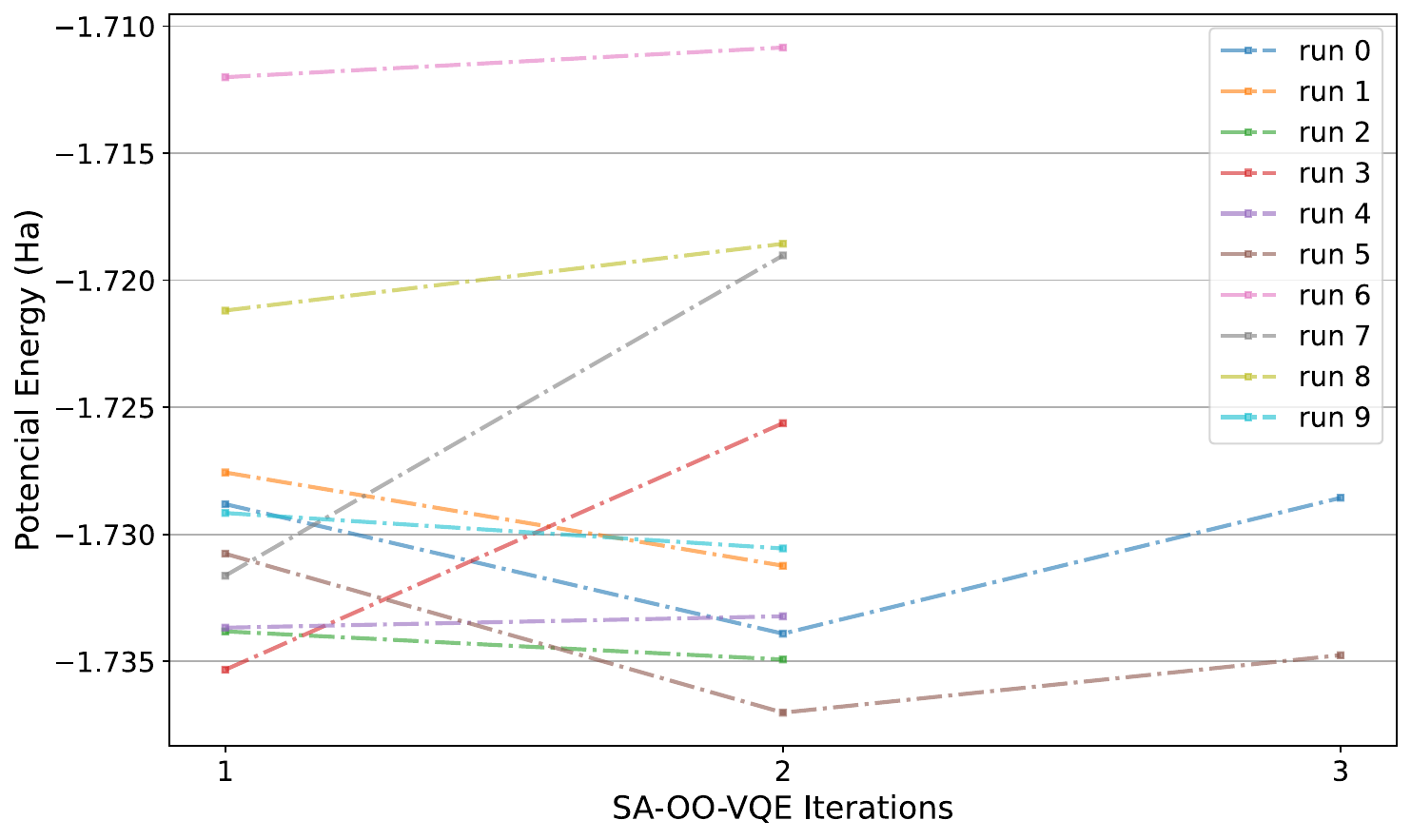}
        \caption{Energy vs. SA-OO-VQE iteration number.} 
        \label{fig:h4_de_b1_conv_iters} 
    \end{subfigure}
    \hfill 
    \begin{subfigure}[t]{0.49\textwidth} 
        \centering
        \includegraphics[width=\linewidth]{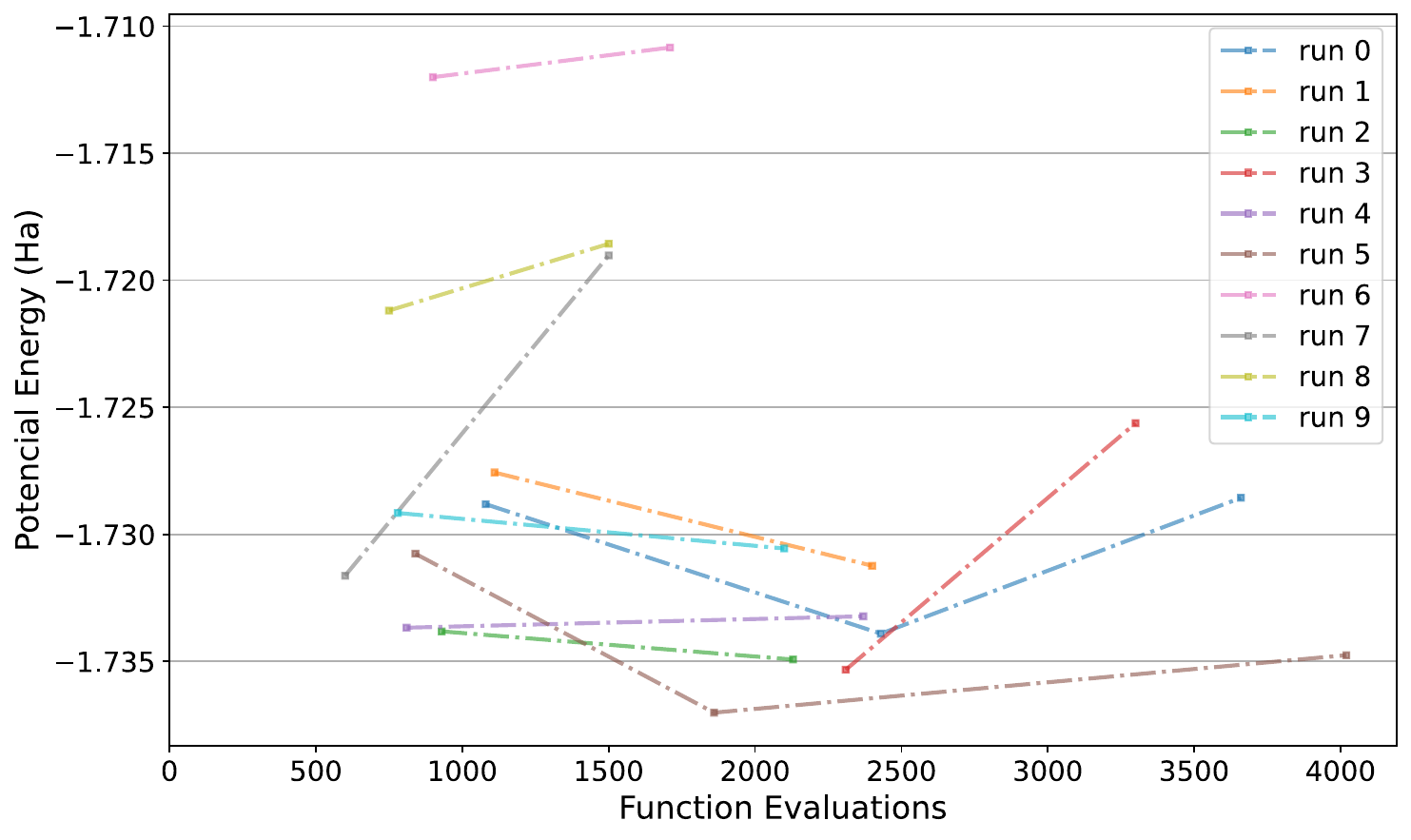}
        \caption{Energy after full SA-OO-VQE iterations vs. cumulative evaluations.} 
        \label{fig:h4_de_b1_conv_evals}
    \end{subfigure}

    \caption[Convergence plots of the DE/Best/1/bin optimizer within the SA-OO-VQE framework for the
H$_4$ molecule.]{Convergence analysis of the DE/Best/1/bin optimizer within the SA-OO-VQE framework for the H$_4$ molecule, based on 10 independent runs (shown in different colors/styles, see legend in plots). The plots display the state-average energy (Hartrees) progression viewed against different metrics: 
    (\subref{fig:h4_de_b1_conv}) Energy evaluated at the end of each internal Gradient Descent optimizer iteration, plotted against the cumulative number of function evaluations consumed up to that iteration point
    (\subref{fig:h4_de_b1_conv_iters}) Energy plotted at the end of each completed SA-OO-VQE iteration against the iteration number. 
    (\subref{fig:h4_de_b1_conv_evals}) Energy plotted at the end of each completed SA-OO-VQE iteration against the cumulative number of function evaluations consumed up to that iteration.}
\label{fig:h4_de_b1} 
\end{figure}

\begin{figure}[!htbp] 
    \centering
    \begin{subfigure}[t]{0.7\textwidth} 
        \centering
        \includegraphics[width=\linewidth]{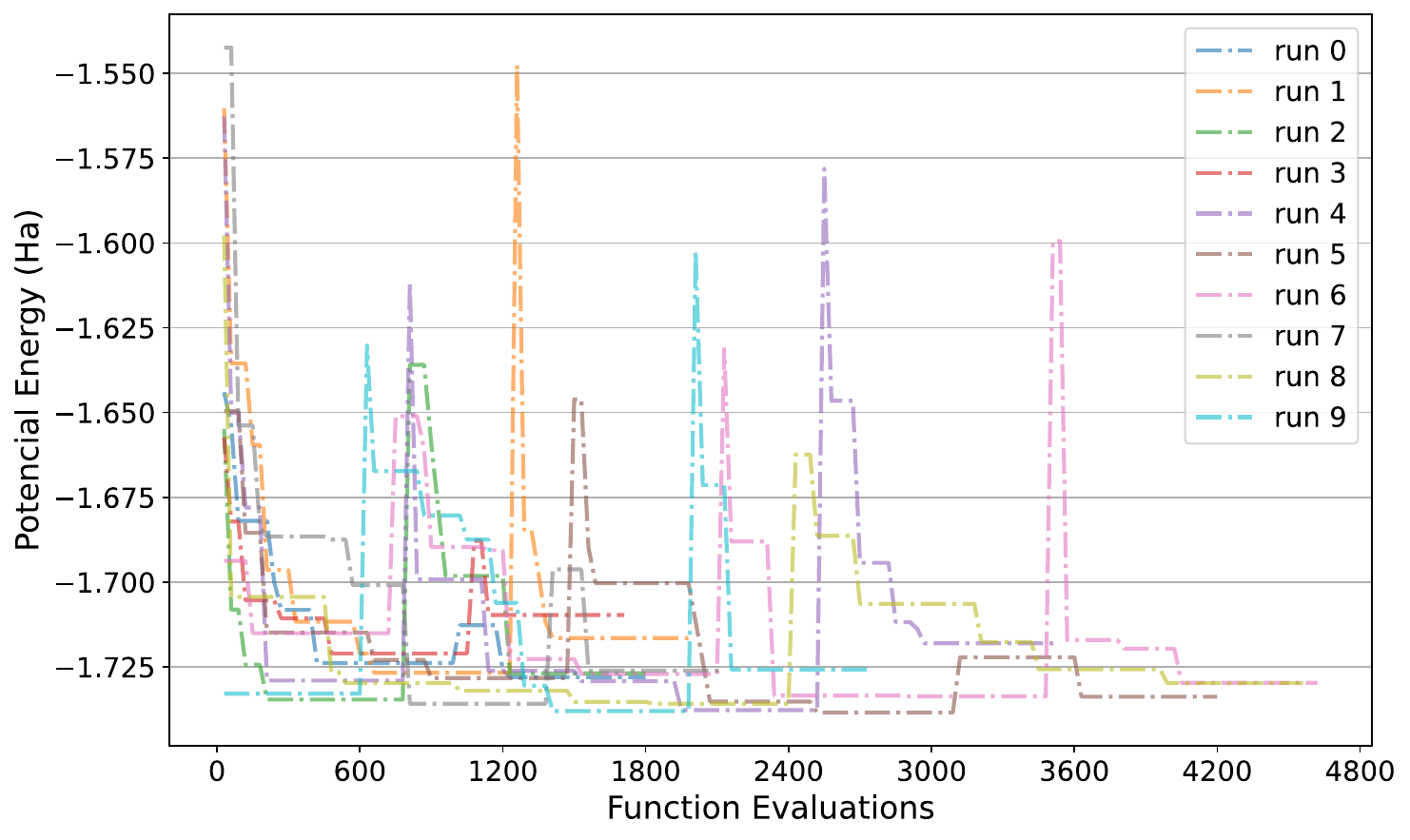}
        \caption{Energy after optimizer iterations vs. cumulative evaluations.} 
        \label{fig:h4_de_b2_conv} 
    \end{subfigure}
    \begin{subfigure}[t]{0.49\textwidth} 
        \centering
        \includegraphics[width=\linewidth]{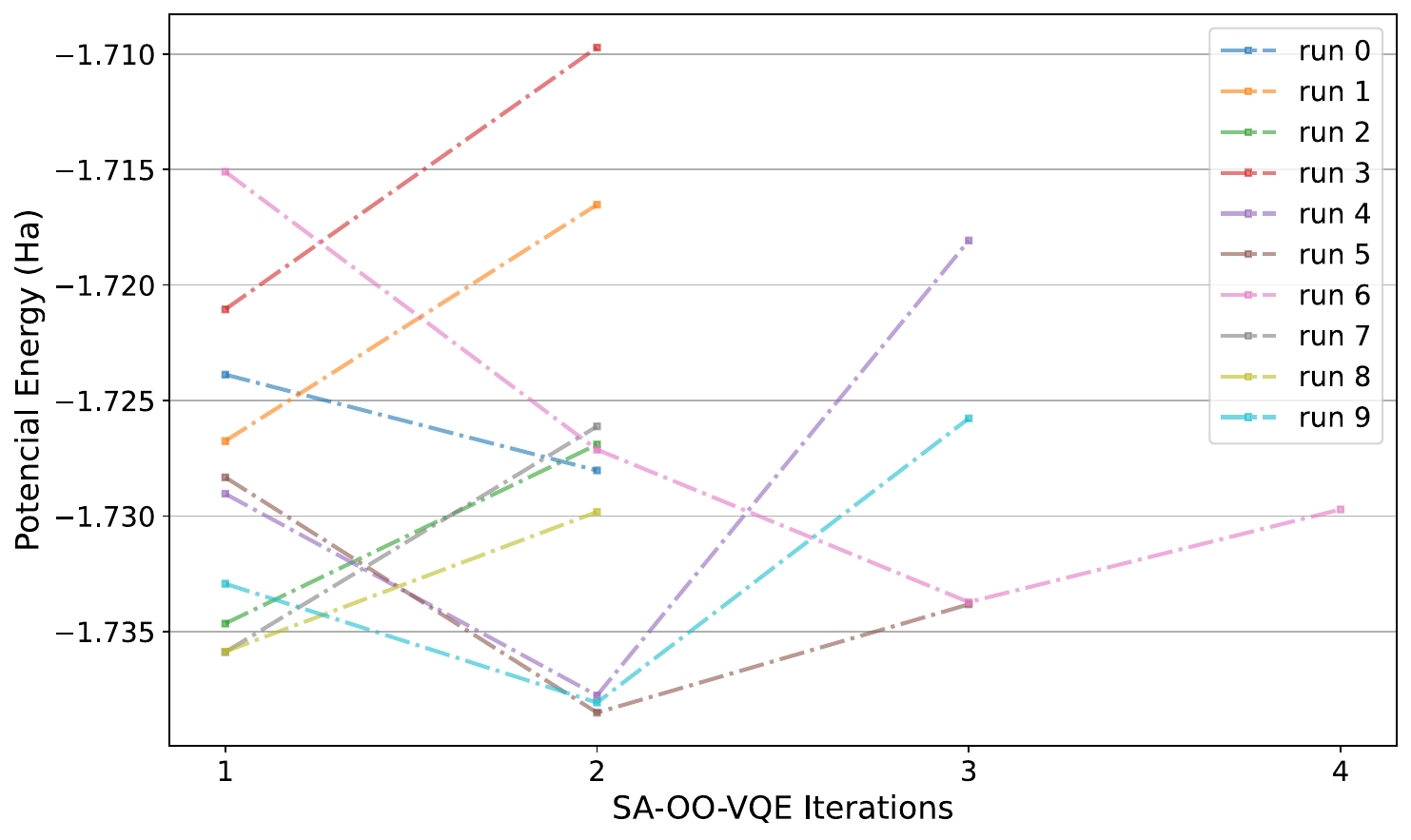}
        \caption{Energy vs. SA-OO-VQE iteration number.} 
        \label{fig:h4_de_b2_conv_iters} 
    \end{subfigure}
    \hfill 
    \begin{subfigure}[t]{0.49\textwidth} 
        \centering
        \includegraphics[width=\linewidth]{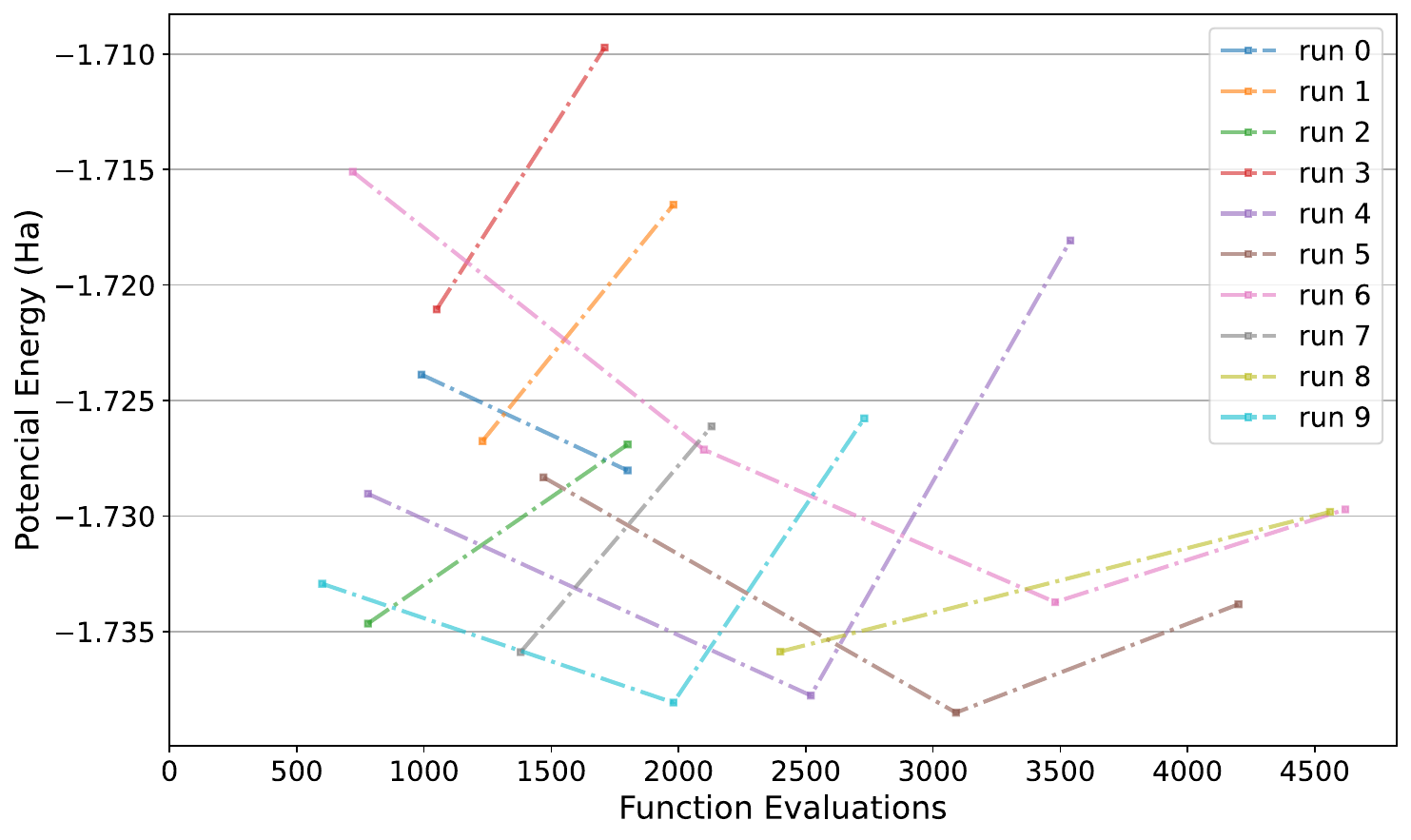}
        \caption{Energy after full SA-OO-VQE iterations vs. cumulative evaluations.} 
        \label{fig:h4_de_b2_conv_evals}
    \end{subfigure}

    \caption[Convergence plots of the DE/Best/2/bin optimizer within the SA-OO-VQE framework for the
H$_4$ molecule.]{Convergence analysis of the DE/Best/2/bin optimizer within the SA-OO-VQE framework for the H$_4$ molecule, based on 10 independent runs (shown in different colors/styles, see legend in plots). The plots display the state-average energy (Hartrees) progression viewed against different metrics: 
    (\subref{fig:h4_de_b2_conv}) Energy evaluated at the end of each internal Gradient Descent optimizer iteration, plotted against the cumulative number of function evaluations consumed up to that iteration point
    (\subref{fig:h4_de_b1_conv_iters}) Energy plotted at the end of each completed SA-OO-VQE iteration against the iteration number. 
    (\subref{fig:h4_de_b2_conv_evals}) Energy plotted at the end of each completed SA-OO-VQE iteration against the cumulative number of function evaluations consumed up to that iteration.}
\label{fig:h4_de_b2} 
\end{figure}

\begin{figure}[!htbp] 
    \centering
    \begin{subfigure}[t]{0.7\textwidth} 
        \centering
        \includegraphics[width=\linewidth]{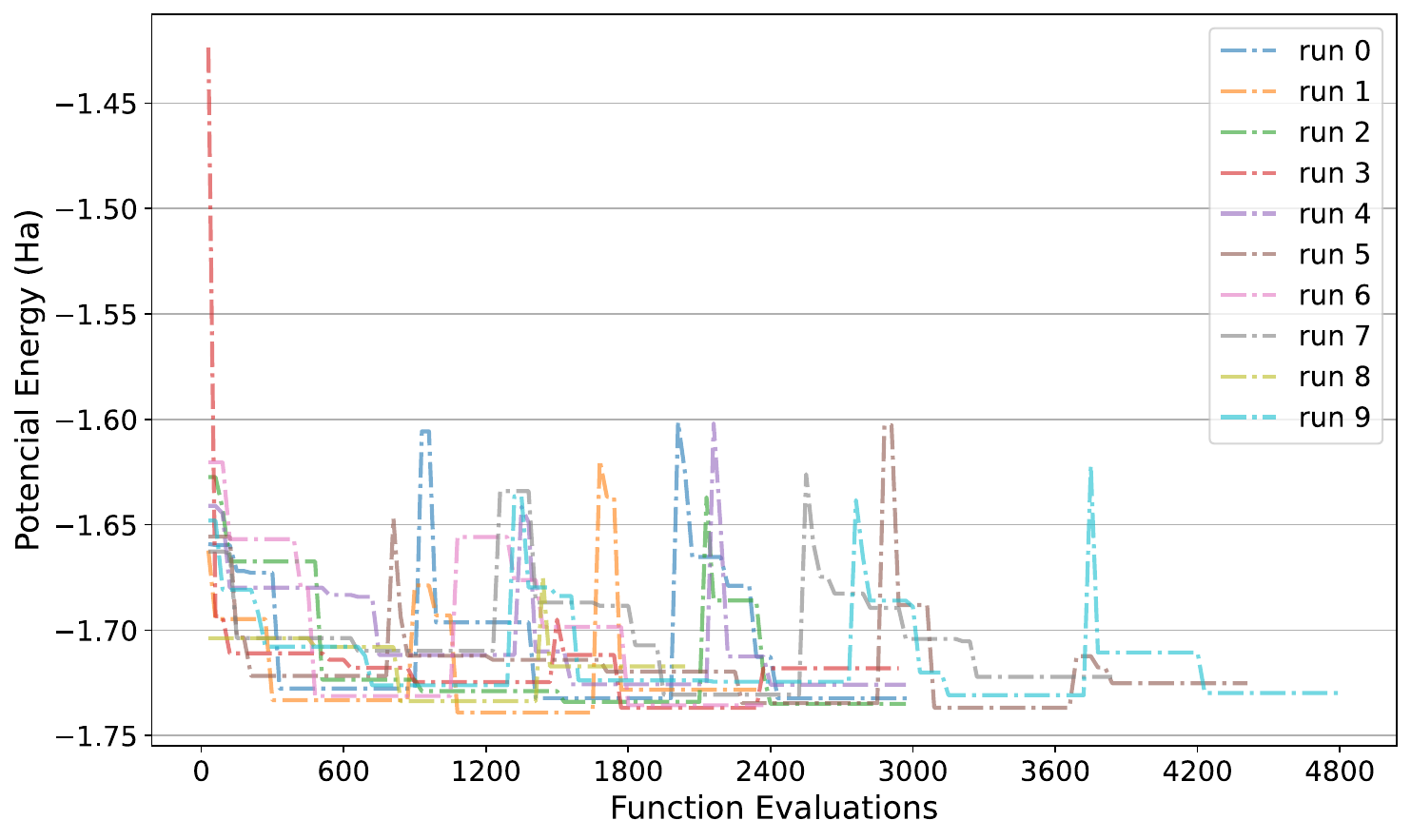}
        \caption{Energy after optimizer iterations vs. cumulative evaluations.} 
        \label{fig:h4_de_ctb_conv} 
    \end{subfigure}
    \begin{subfigure}[t]{0.49\textwidth} 
        \centering
        \includegraphics[width=\linewidth]{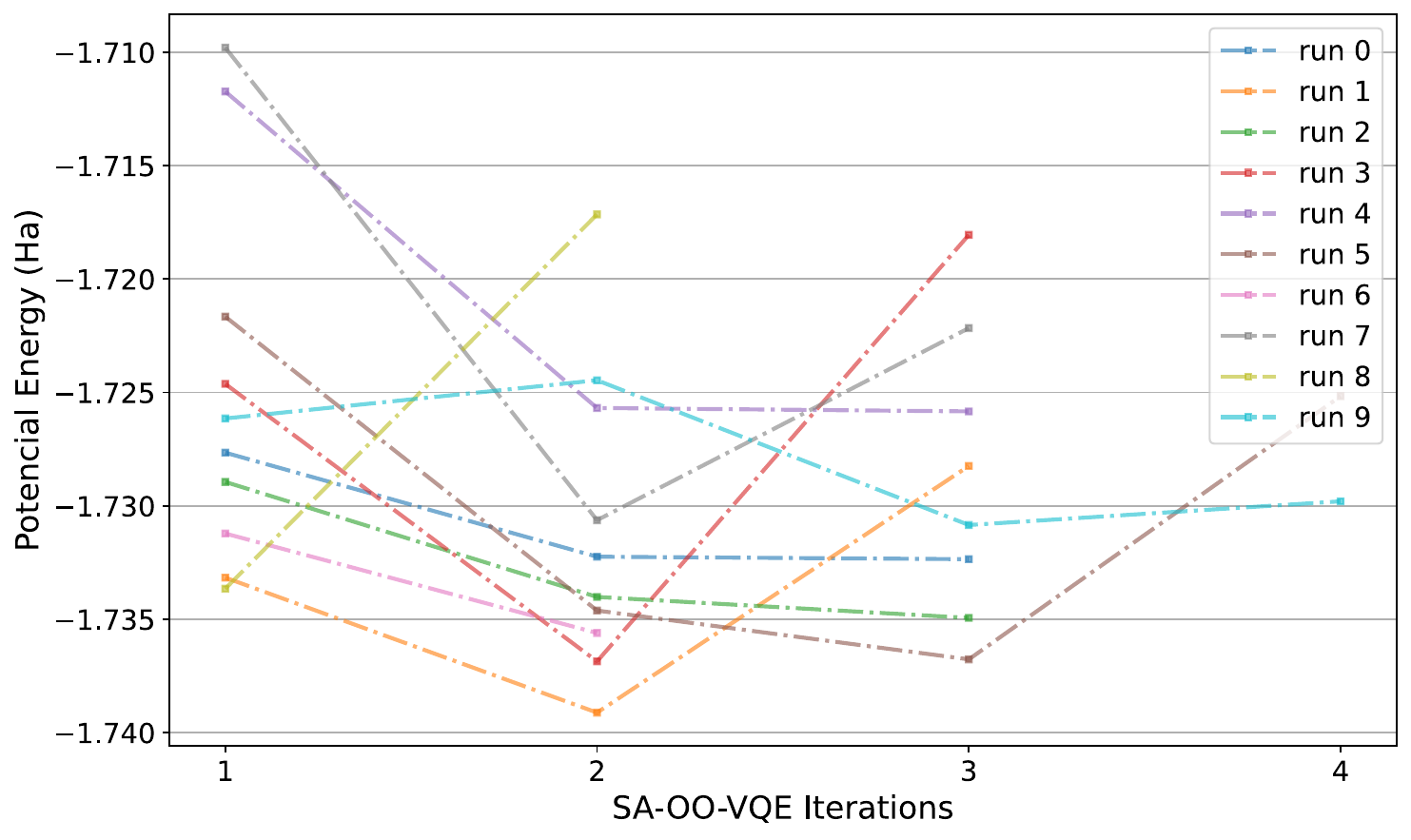}
        \caption{Energy vs. SA-OO-VQE iteration number.} 
        \label{fig:h4_de_ctb_conv_iters} 
    \end{subfigure}
    \hfill 
    \begin{subfigure}[t]{0.49\textwidth} 
        \centering
        \includegraphics[width=\linewidth]{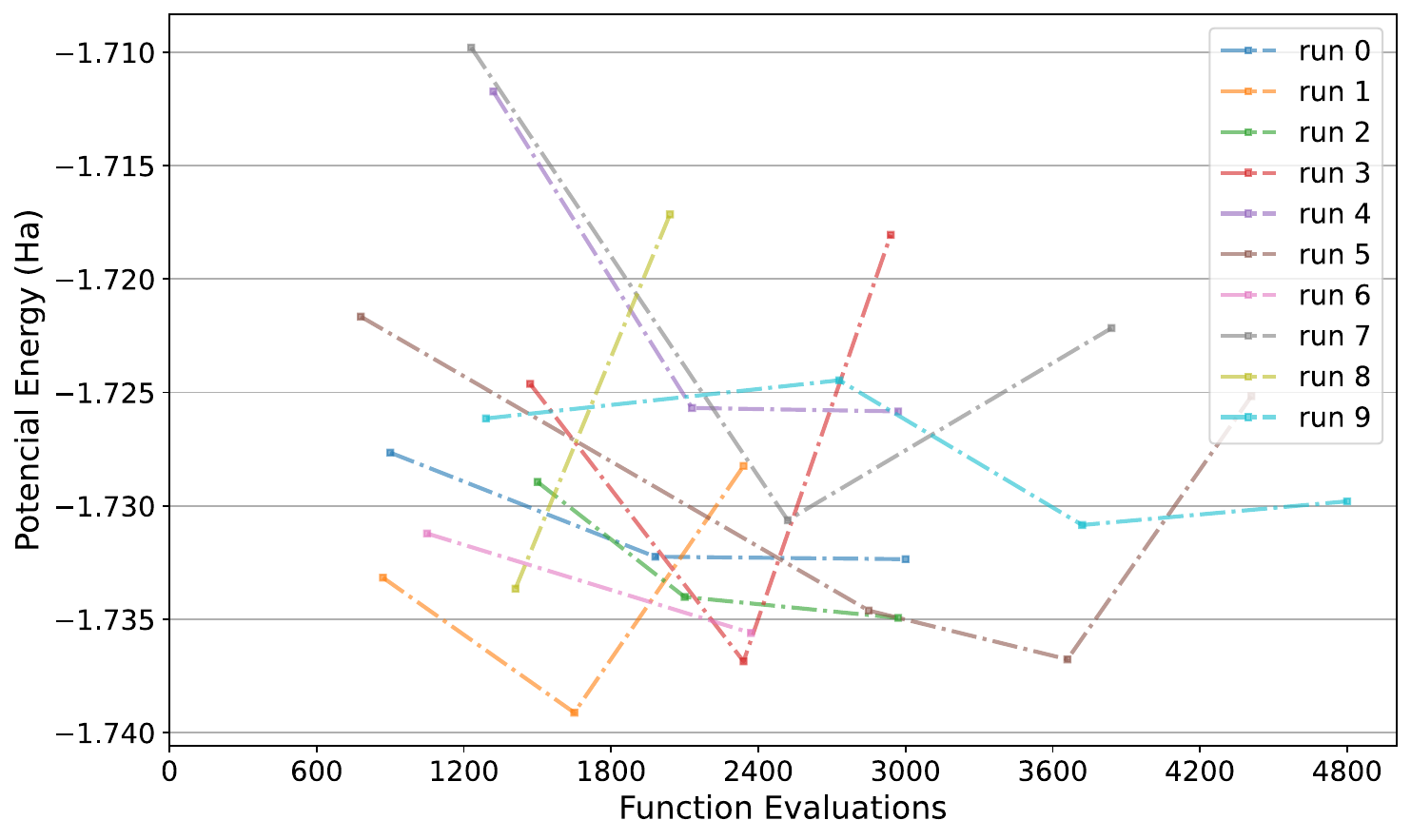}
        \caption{Energy after full SA-OO-VQE iterations vs. cumulative evaluations.} 
        \label{fig:h4_de_ctb_conv_evals}
    \end{subfigure}

    \caption[Convergence plots of the DE/Current-to-Best/1/bin optimizer within the SA-OO-VQE framework for the
H$_4$ molecule.]{Convergence analysis of the DE/Current-to-Best/1/bin optimizer within the SA-OO-VQE framework for the H$_4$ molecule, based on 10 independent runs (shown in different colors/styles, see legend in plots). The plots display the state-average energy (Hartrees) progression viewed against different metrics: 
    (\subref{fig:h4_de_ctb_conv}) Energy evaluated at the end of each internal Gradient Descent optimizer iteration, plotted against the cumulative number of function evaluations consumed up to that iteration point
    (\subref{fig:h4_de_ctb_conv_iters}) Energy plotted at the end of each completed SA-OO-VQE iteration against the iteration number. 
    (\subref{fig:h4_de_ctb_conv_evals}) Energy plotted at the end of each completed SA-OO-VQE iteration against the cumulative number of function evaluations consumed up to that iteration.}
\label{fig:h4_de_ctb} 
\end{figure}

\begin{figure}[!htbp] 
    \centering
    \begin{subfigure}[t]{0.7\textwidth} 
        \centering
        \includegraphics[width=\linewidth]{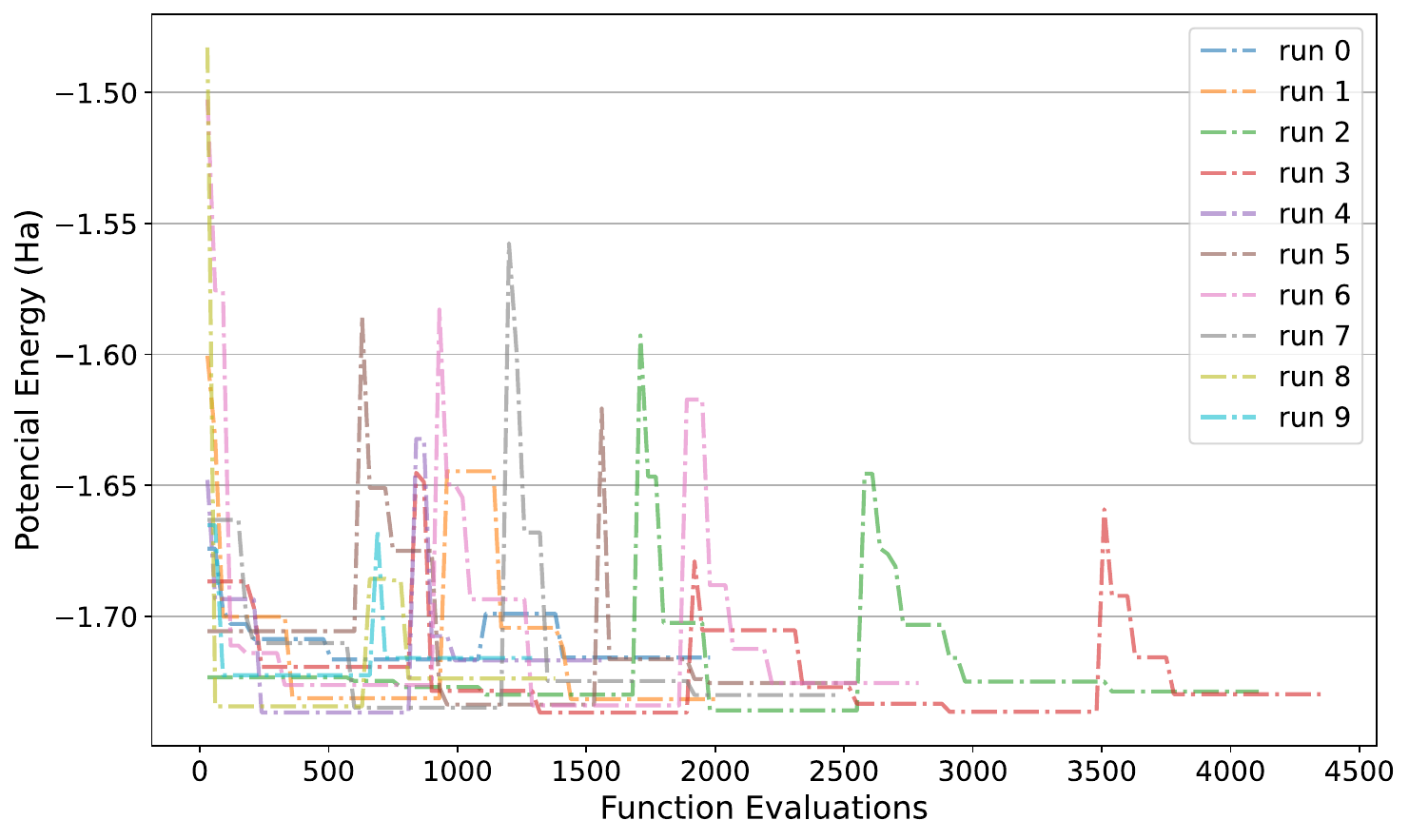}
        \caption{Energy after optimizer iterations vs. cumulative evaluations.} 
        \label{fig:h4_de_ctr_conv} 
    \end{subfigure}
    \begin{subfigure}[t]{0.49\textwidth} 
        \centering
        \includegraphics[width=\linewidth]{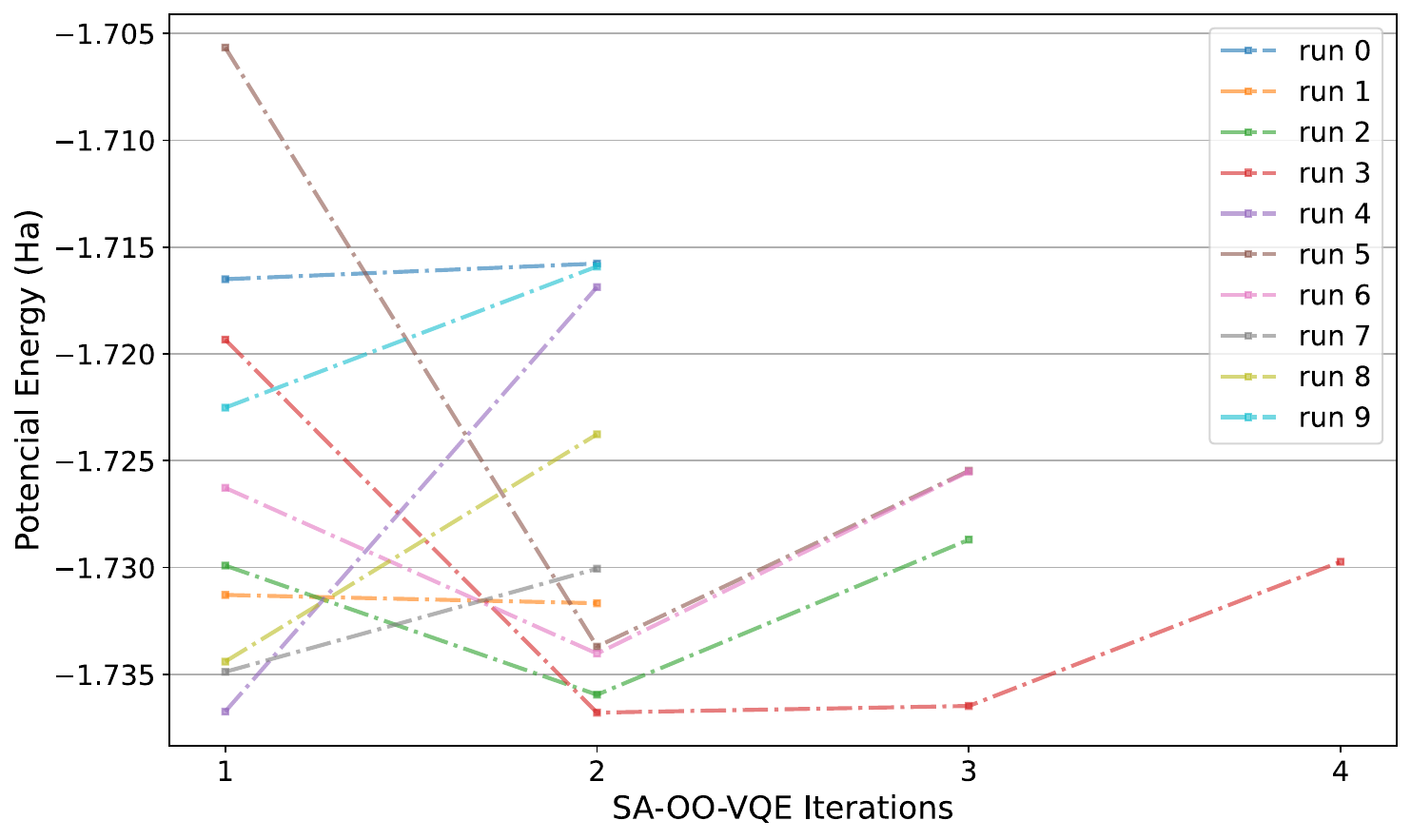}
        \caption{Energy vs. SA-OO-VQE iteration number.} 
        \label{fig:h4_de_ctr_conv_iters} 
    \end{subfigure}
    \hfill 
    \begin{subfigure}[t]{0.49\textwidth} 
        \centering
        \includegraphics[width=\linewidth]{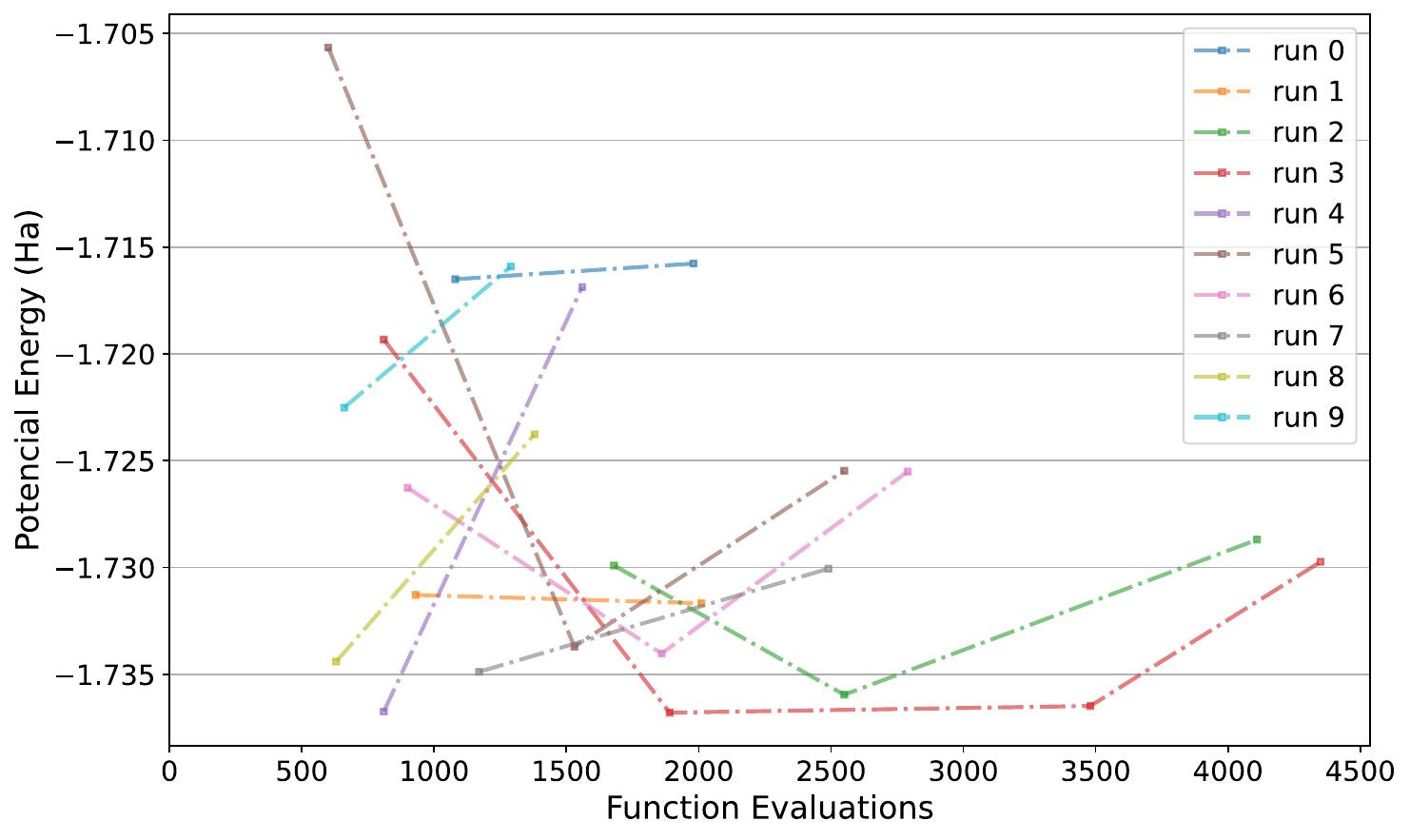}
        \caption{Energy after full SA-OO-VQE iterations vs. cumulative evaluations.} 
        \label{fig:h4_de_ctr_conv_evals}
    \end{subfigure}

    \caption[Convergence plots of the DE/Current-to-Random/1/bin optimizer within the SA-OO-VQE framework for the
H$_4$ molecule.]{Convergence analysis of the DE/Current-to-Random/1/bin optimizer within the SA-OO-VQE framework for the H$_4$ molecule, based on 10 independent runs (shown in different colors/styles, see legend in plots). The plots display the state-average energy (Hartrees) progression viewed against different metrics: 
    (\subref{fig:h4_de_ctr_conv}) Energy evaluated at the end of each internal Gradient Descent optimizer iteration, plotted against the cumulative number of function evaluations consumed up to that iteration point
    (\subref{fig:h4_de_ctr_conv_iters}) Energy plotted at the end of each completed SA-OO-VQE iteration against the iteration number. 
    (\subref{fig:h4_de_ctr_conv_evals}) Energy plotted at the end of each completed SA-OO-VQE iteration against the cumulative number of function evaluations consumed up to that iteration.}
\label{fig:h4_de_ctr} 
\end{figure}

\begin{figure}[!htbp] 
    \centering
    \begin{subfigure}[t]{0.7\textwidth} 
        \centering
        \includegraphics[width=\linewidth]{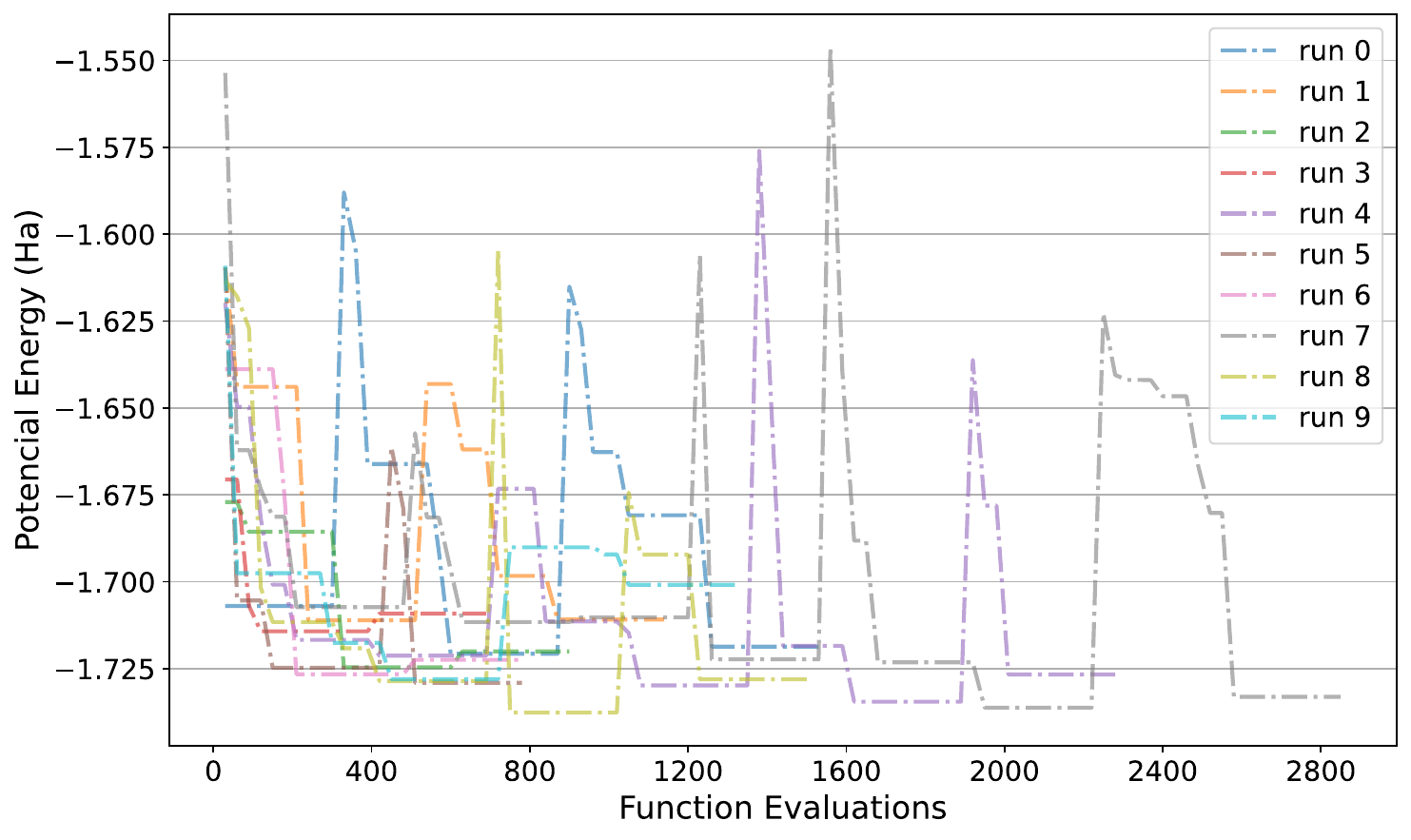}
        \caption{Energy after optimizer iterations vs. cumulative evaluations.} 
        \label{fig:h4_de_r1_conv} 
    \end{subfigure}
    \begin{subfigure}[t]{0.49\textwidth} 
        \centering
        \includegraphics[width=\linewidth]{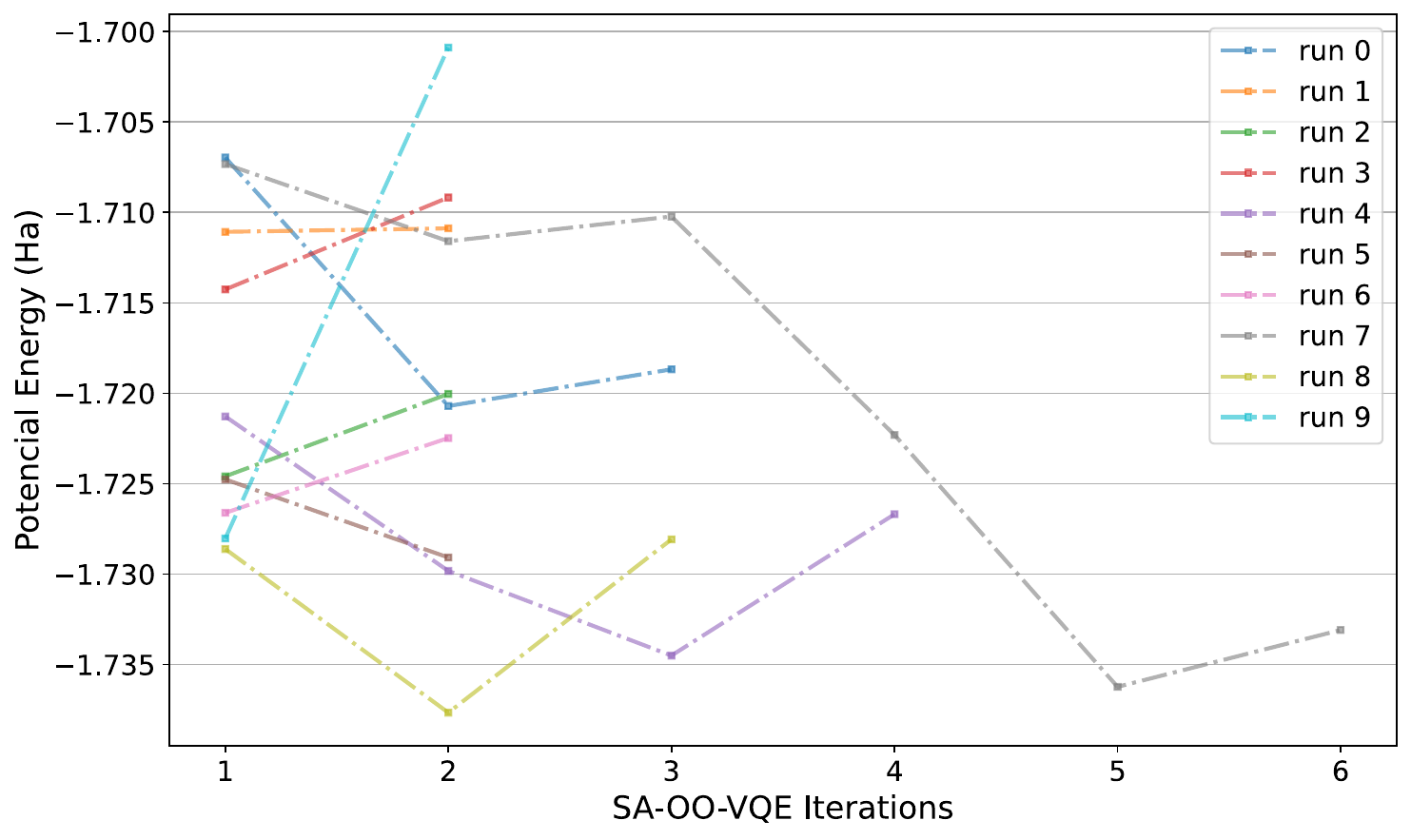}
        \caption{Energy vs. SA-OO-VQE iteration number.} 
        \label{fig:h4_de_r1_conv_iters} 
    \end{subfigure}
    \hfill 
    \begin{subfigure}[t]{0.49\textwidth} 
        \centering
        \includegraphics[width=\linewidth]{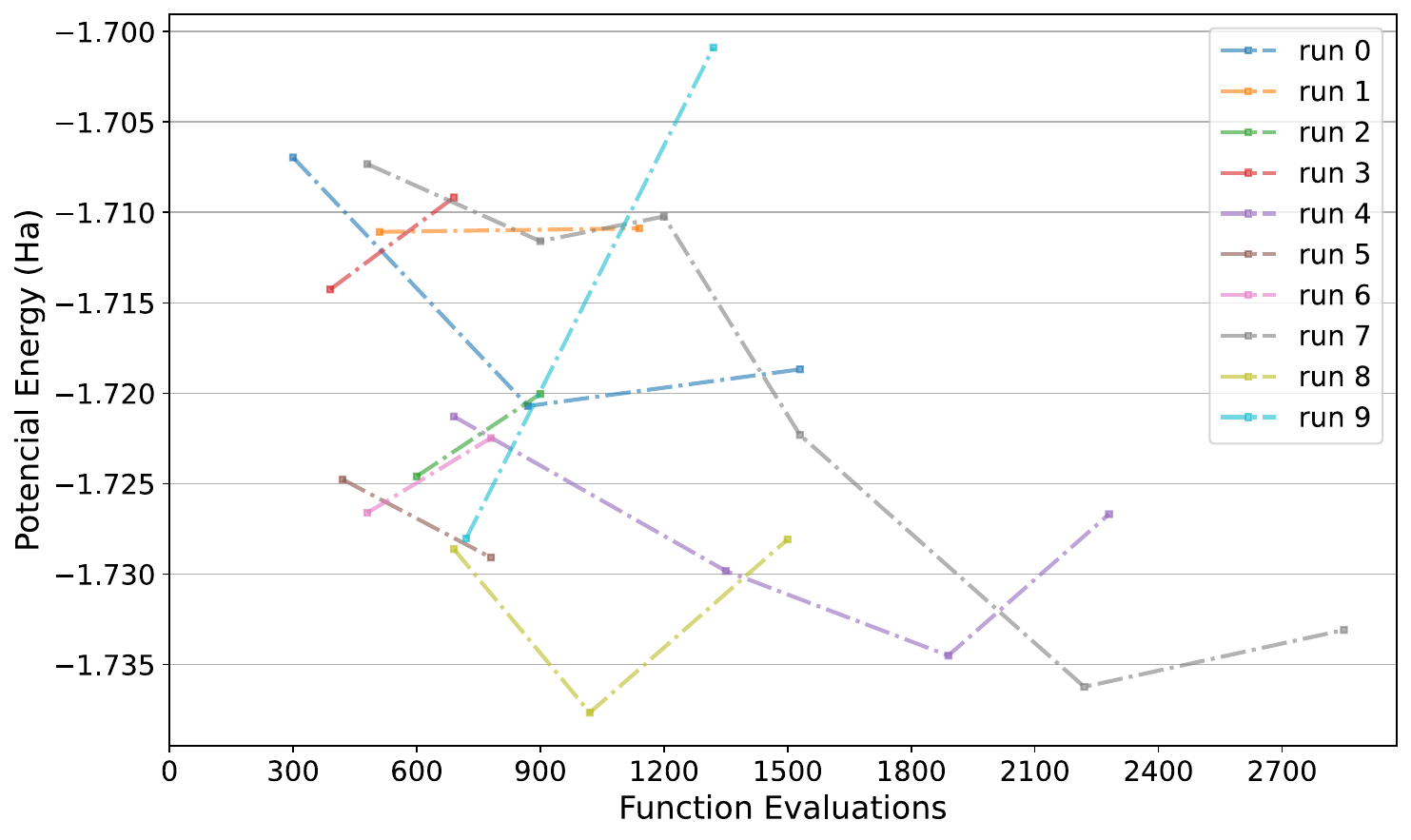}
        \caption{Energy after full SA-OO-VQE iterations vs. cumulative evaluations.} 
        \label{fig:h4_de_r1_conv_evals}
    \end{subfigure}

    \caption[Convergence plots of the DE/Rand/1/bin optimizer within the SA-OO-VQE framework for the
H$_4$ molecule.]{Convergence analysis of the DE/Rand/1/bin optimizer within the SA-OO-VQE framework for the H$_4$ molecule, based on 10 independent runs (shown in different colors/styles, see legend in plots). The plots display the state-average energy (Hartrees) progression viewed against different metrics: 
    (\subref{fig:h4_de_r1_conv}) Energy evaluated at the end of each internal Gradient Descent optimizer iteration, plotted against the cumulative number of function evaluations consumed up to that iteration point
    (\subref{fig:h4_de_r1_conv_iters}) Energy plotted at the end of each completed SA-OO-VQE iteration against the iteration number. 
    (\subref{fig:h4_de_r1_conv_evals}) Energy plotted at the end of each completed SA-OO-VQE iteration against the cumulative number of function evaluations consumed up to that iteration.}
\label{fig:h4_de_r1} 
\end{figure}

\begin{figure}[!htbp] 
    \centering
    \begin{subfigure}[t]{0.7\textwidth} 
        \centering
        \includegraphics[width=\linewidth]{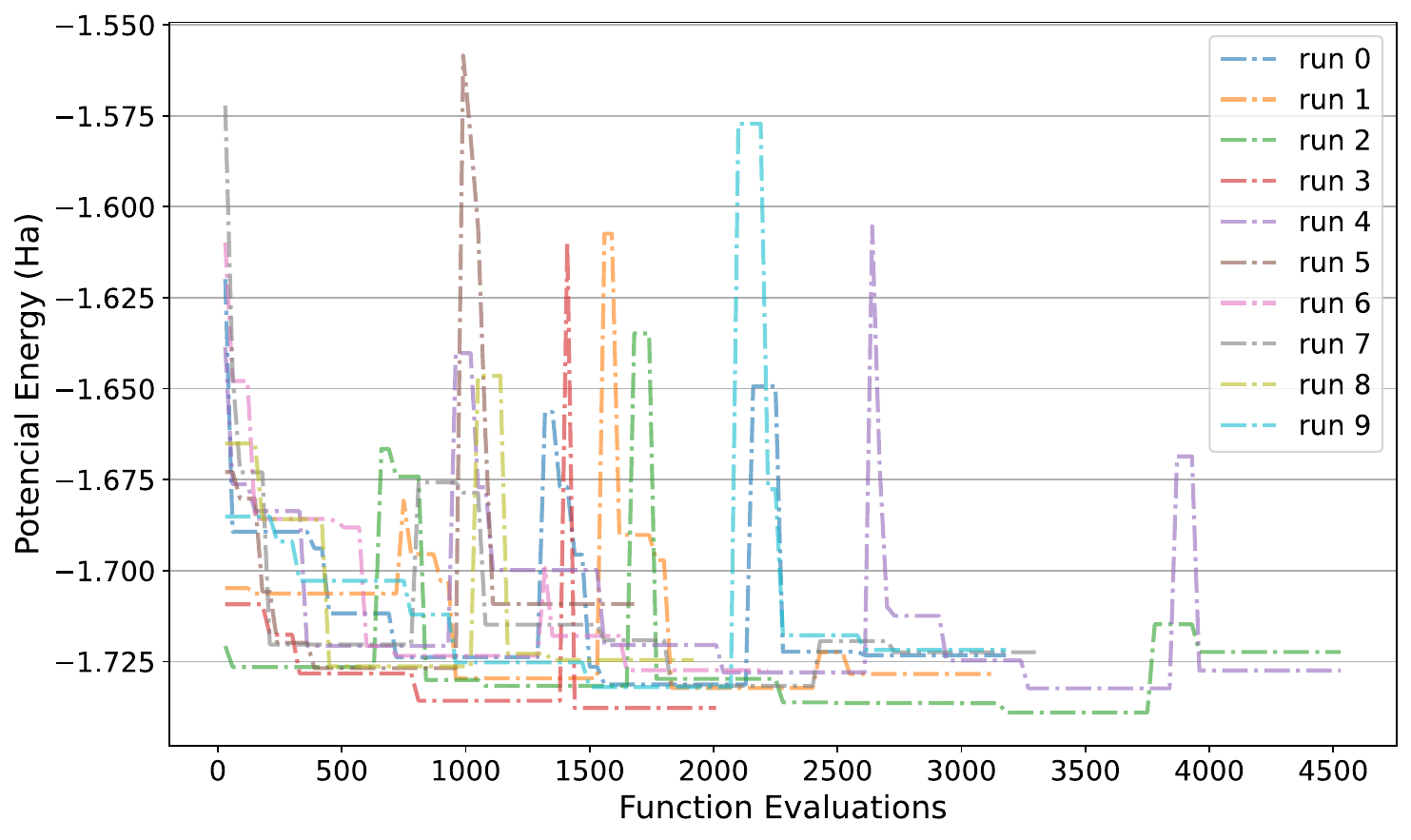}
        \caption{Energy after optimizer iterations vs. cumulative evaluations.} 
        \label{fig:h4_de_r2_conv} 
    \end{subfigure}
    \begin{subfigure}[t]{0.49\textwidth} 
        \centering
        \includegraphics[width=\linewidth]{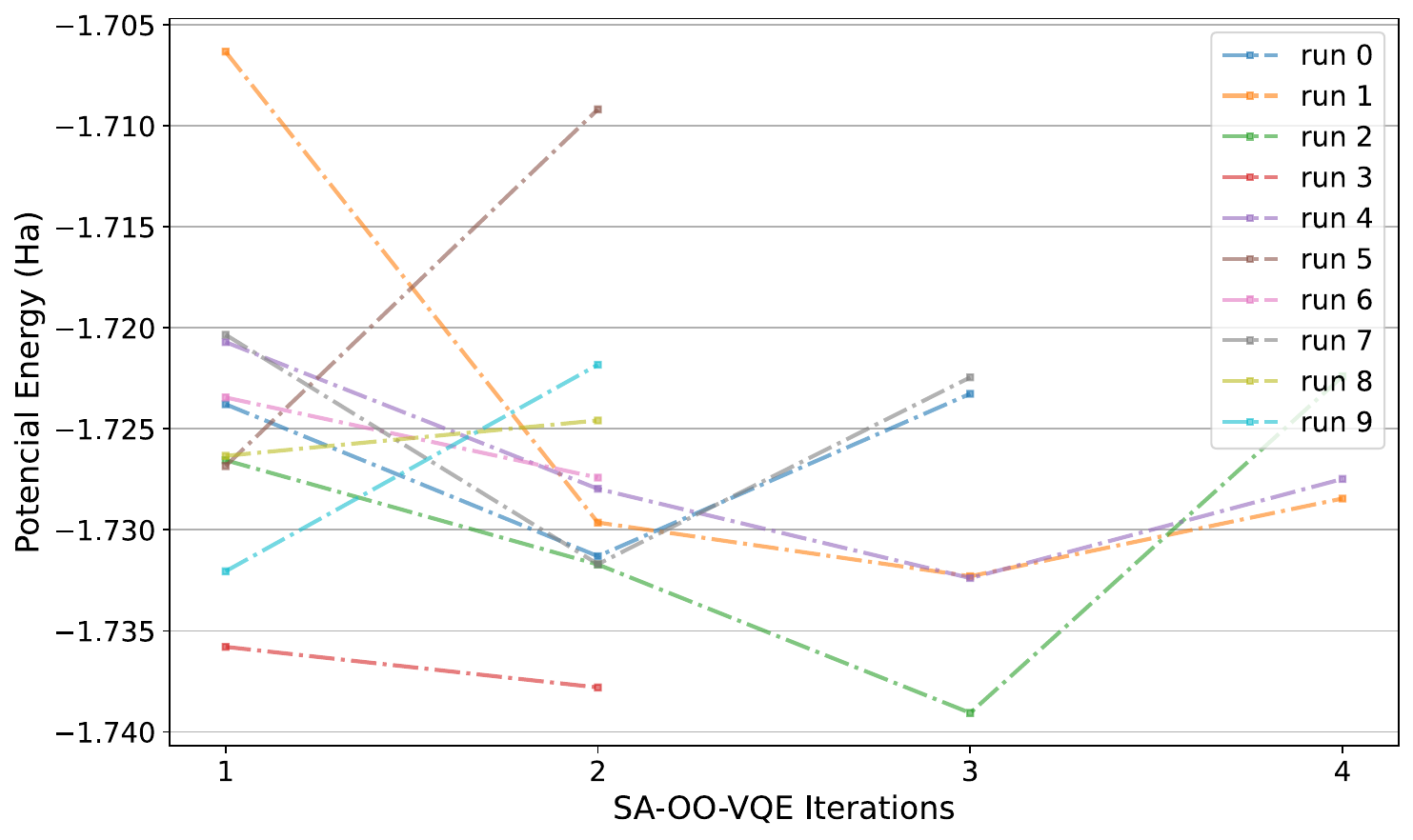}
        \caption{Energy vs. SA-OO-VQE iteration number.} 
        \label{fig:h4_de_r2_conv_iters} 
    \end{subfigure}
    \hfill 
    \begin{subfigure}[t]{0.49\textwidth} 
        \centering
        \includegraphics[width=\linewidth]{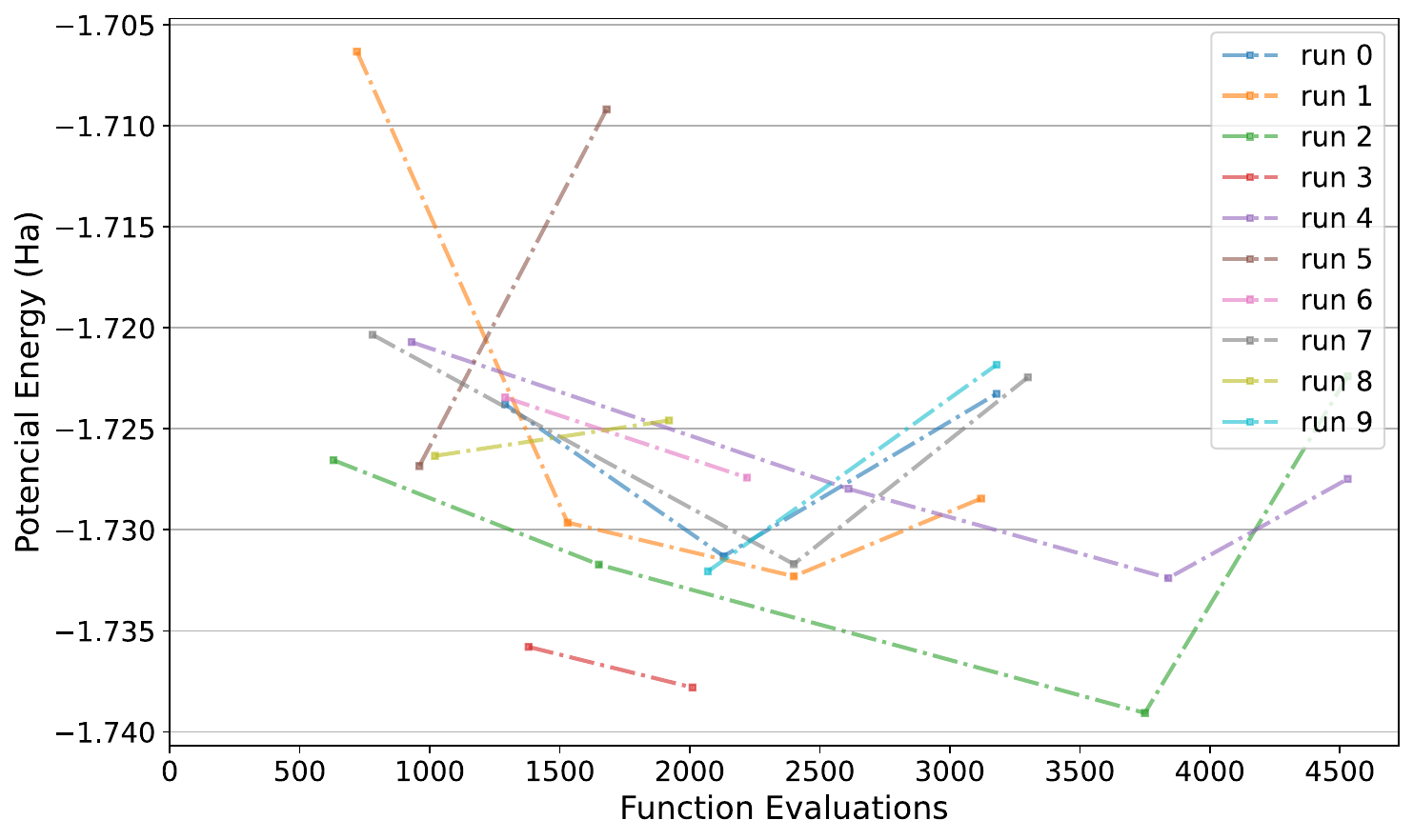}
        \caption{Energy after full SA-OO-VQE iterations vs. cumulative evaluations.} 
        \label{fig:h4_de_r2_conv_evals}
    \end{subfigure}

    \caption[Convergence plots of the DE/Rand/2/bin optimizer within the SA-OO-VQE framework for the
H$_4$ molecule.]{Convergence analysis of the DE/Rand/2/bin optimizer within the SA-OO-VQE framework for the H$_4$ molecule, based on 10 independent runs (shown in different colors/styles, see legend in plots). The plots display the state-average energy (Hartrees) progression viewed against different metrics: 
    (\subref{fig:h4_de_r2_conv}) Energy evaluated at the end of each internal Gradient Descent optimizer iteration, plotted against the cumulative number of function evaluations consumed up to that iteration point
    (\subref{fig:h4_de_r2_conv_iters}) Energy plotted at the end of each completed SA-OO-VQE iteration against the iteration number. 
    (\subref{fig:h4_de_r2_conv_evals}) Energy plotted at the end of each completed SA-OO-VQE iteration against the cumulative number of function evaluations consumed up to that iteration.}
\label{fig:h4_de_r2} 
\end{figure}

\begin{figure}[!htbp] 
    \centering
    \begin{subfigure}[t]{0.7\textwidth} 
        \centering
        \includegraphics[width=\linewidth]{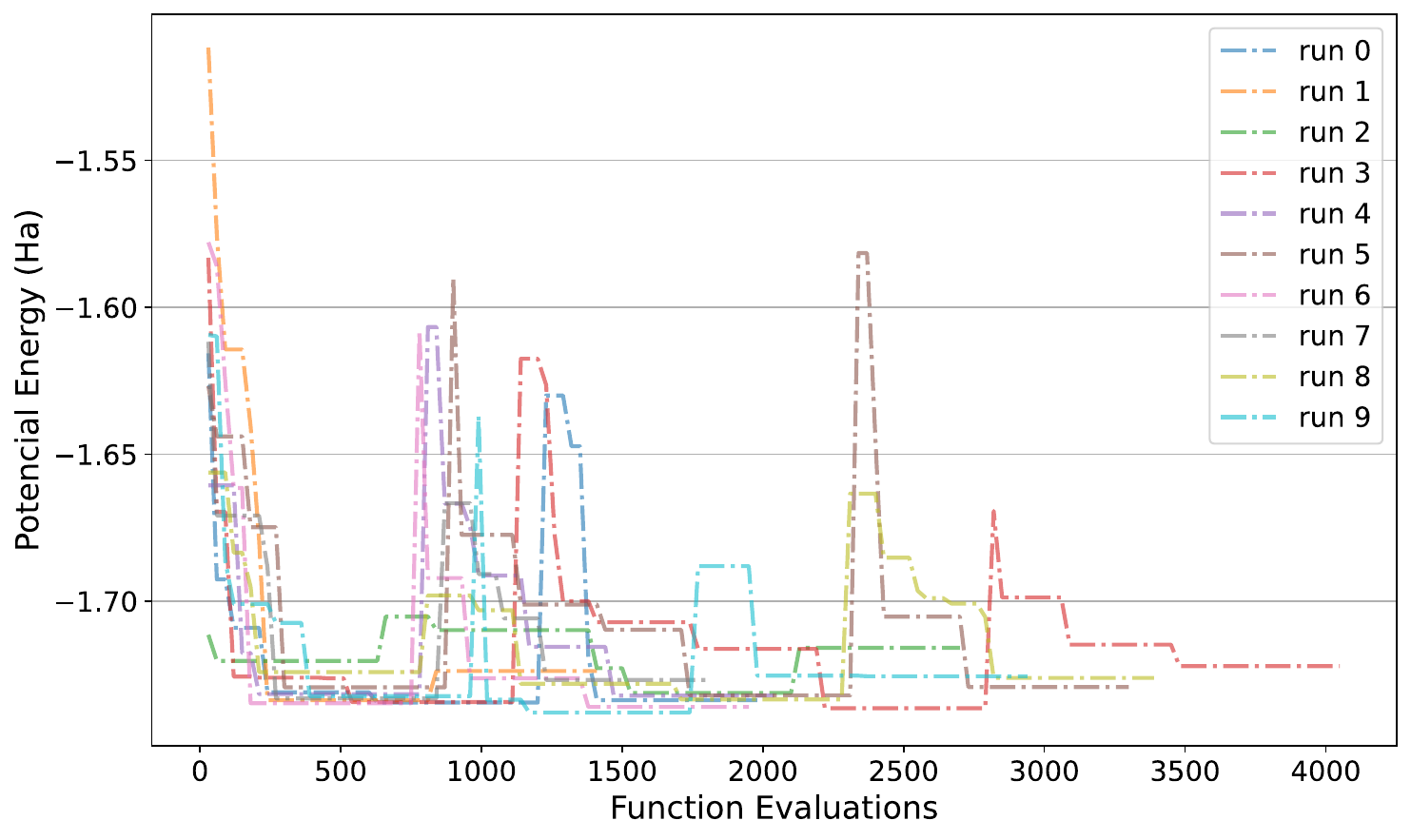}
        \caption{Energy after optimizer iterations vs. cumulative evaluations.} 
        \label{fig:h4_de_rtb_conv} 
    \end{subfigure}
    \begin{subfigure}[t]{0.49\textwidth} 
        \centering
        \includegraphics[width=\linewidth]{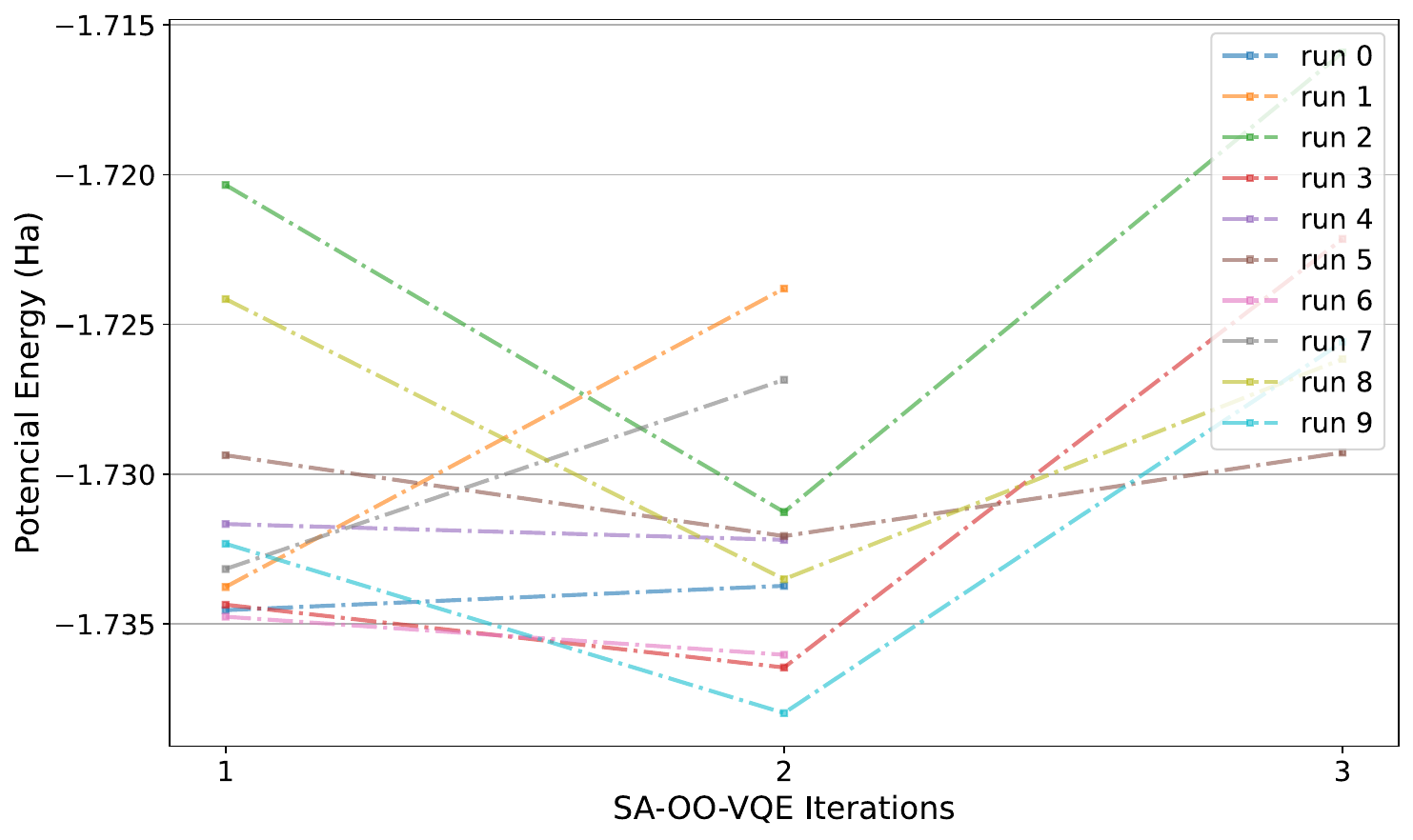}
        \caption{Energy vs. SA-OO-VQE iteration number.} 
        \label{fig:h4_de_rtb_conv_iters} 
    \end{subfigure}
    \hfill 
    \begin{subfigure}[t]{0.49\textwidth} 
        \centering
        \includegraphics[width=\linewidth]{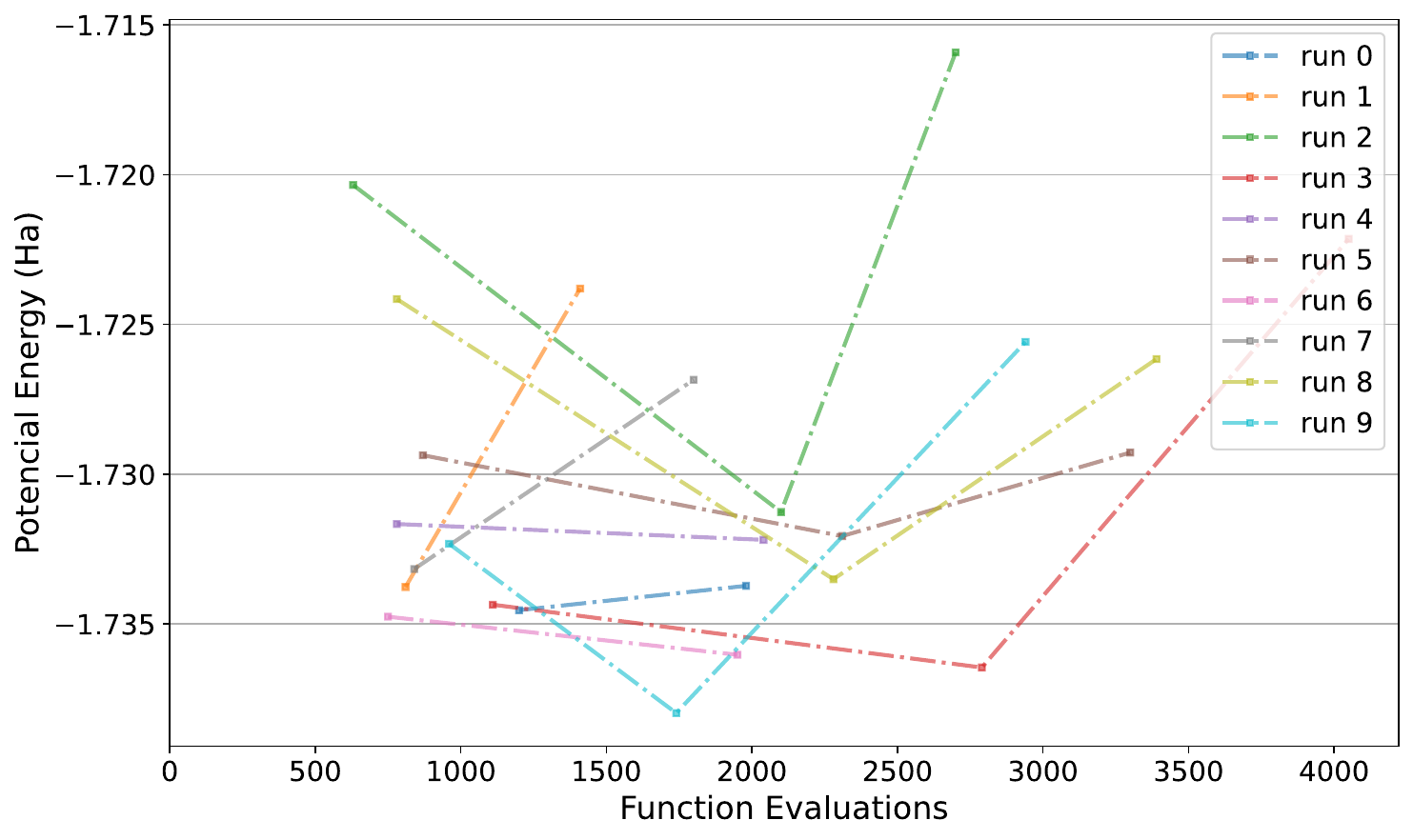}
        \caption{Energy after full SA-OO-VQE iterations vs. cumulative evaluations.} 
        \label{fig:h4_de_rtb_conv_evals}
    \end{subfigure}

    \caption[Convergence plots of the DE/Random-to-Best/1/bin optimizer within the SA-OO-VQE framework for the
H$_4$ molecule.]{Convergence analysis of the DE/Random-to-Best/1/bin optimizer within the SA-OO-VQE framework for the H$_4$ molecule, based on 10 independent runs (shown in different colors/styles, see legend in plots). The plots display the state-average energy (Hartrees) progression viewed against different metrics: 
    (\subref{fig:h4_de_rtb_conv}) Energy evaluated at the end of each internal Gradient Descent optimizer iteration, plotted against the cumulative number of function evaluations consumed up to that iteration point
    (\subref{fig:h4_de_rtb_conv_iters}) Energy plotted at the end of each completed SA-OO-VQE iteration against the iteration number. 
    (\subref{fig:h4_de_rtb_conv_evals}) Energy plotted at the end of each completed SA-OO-VQE iteration against the cumulative number of function evaluations consumed up to that iteration.}
\label{fig:h4_de_rtb} 
\end{figure}

\begin{figure}[!t] 
    \centering
    \begin{subfigure}[t]{0.49\textwidth} 
        \centering
        \includegraphics[width=\linewidth]{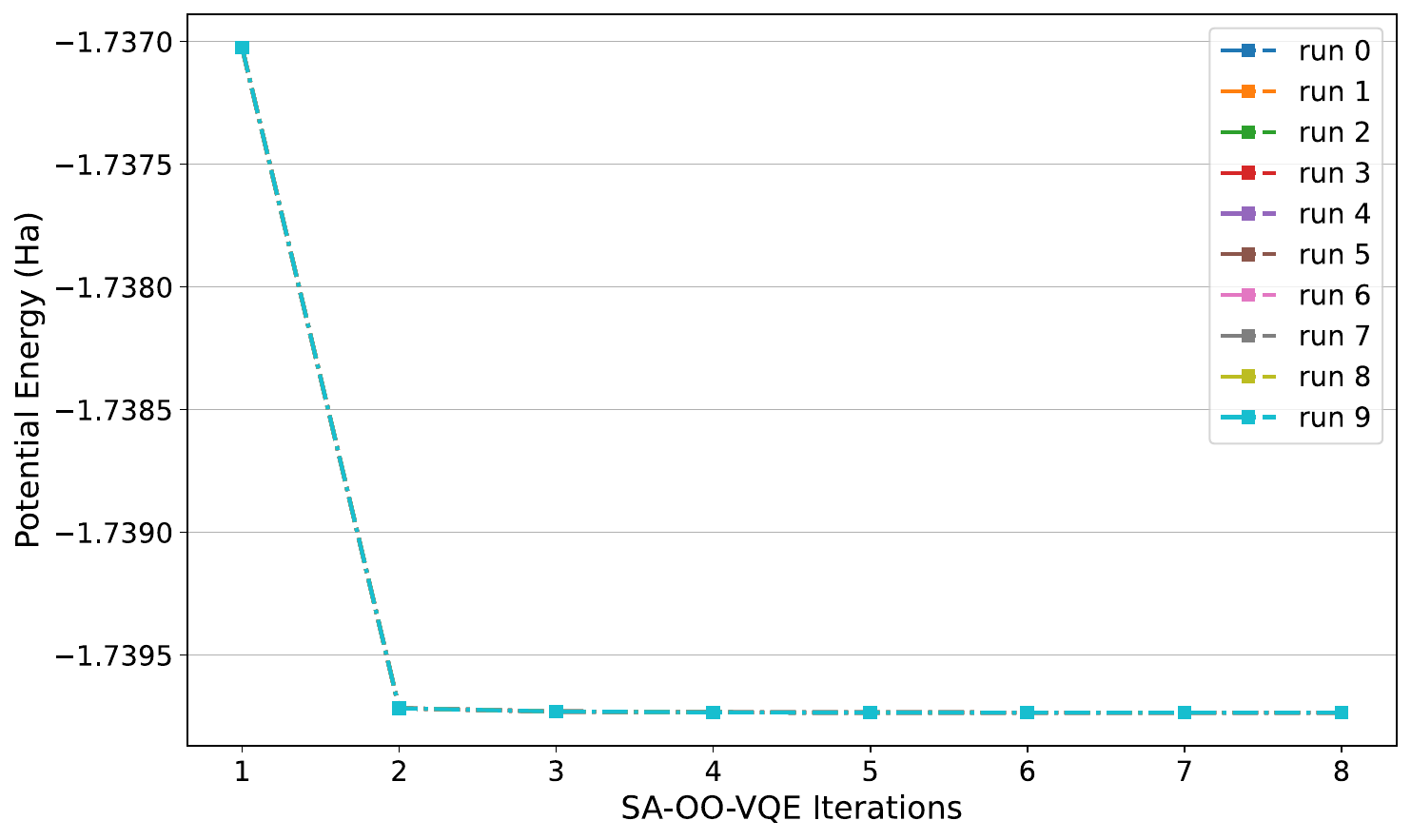}
        \caption{Energy vs. SA-OO-VQE iteration number.} 
        \label{fig:h4_slsqp_conv_iters} 
    \end{subfigure}
    \hfill 
    \begin{subfigure}[t]{0.49\textwidth} 
        \centering
        \includegraphics[width=\linewidth]{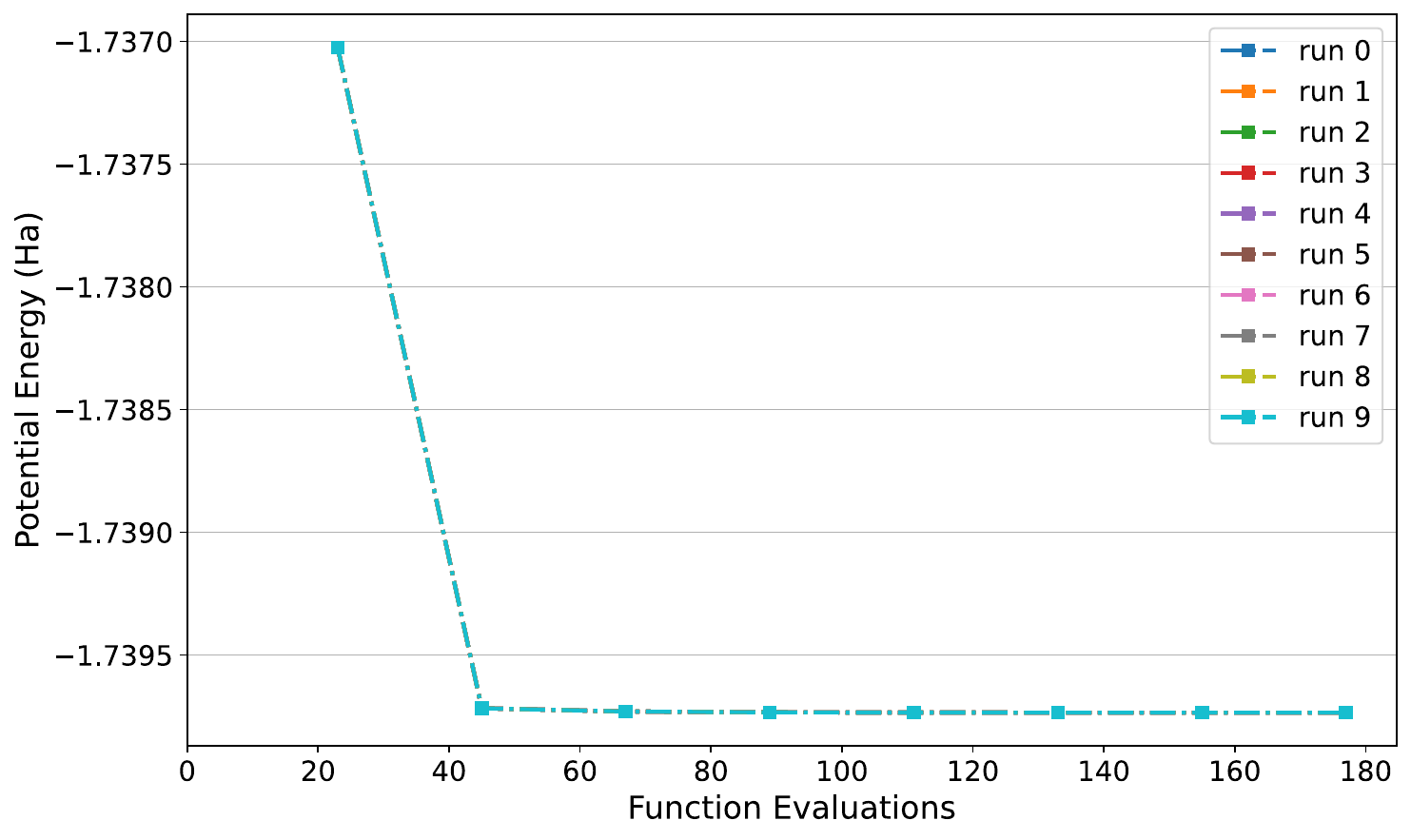}
        \caption{Energy after full SA-OO-VQE iterations vs. cumulative evaluations.} 
        \label{fig:h4_slsqp_conv_evals}
    \end{subfigure}

    \caption[Convergence plots of the SLSQP optimizer within the SA-OO-VQE framework for the
H$_4$ molecule.]{Convergence analysis of the SLSQP optimizer within the SA-OO-VQE framework for the H$_4$ molecule, based on 10 independent runs (shown in different colors/styles, see legend in plots). The plots display the state-average energy (Hartrees) progression viewed against different metrics: 
    (\subref{fig:h4_slsqp_conv_iters}) Energy plotted at the end of each completed SA-OO-VQE iteration against the iteration number. 
    (\subref{fig:h4_slsqp_conv_evals}) Energy plotted at the end of each completed SA-OO-VQE iteration against the cumulative number of function evaluations consumed up to that iteration.}
\label{fig:h4_slsqp} 
\end{figure}
\clearpage
 
\section{Convergence plots for LiH molecule}
\label{ap_sec:lih_conv}
\begin{figure}[!htbp] 
    \centering
    \begin{subfigure}[t]{0.49\textwidth} 
        \centering
        \includegraphics[width=\linewidth]{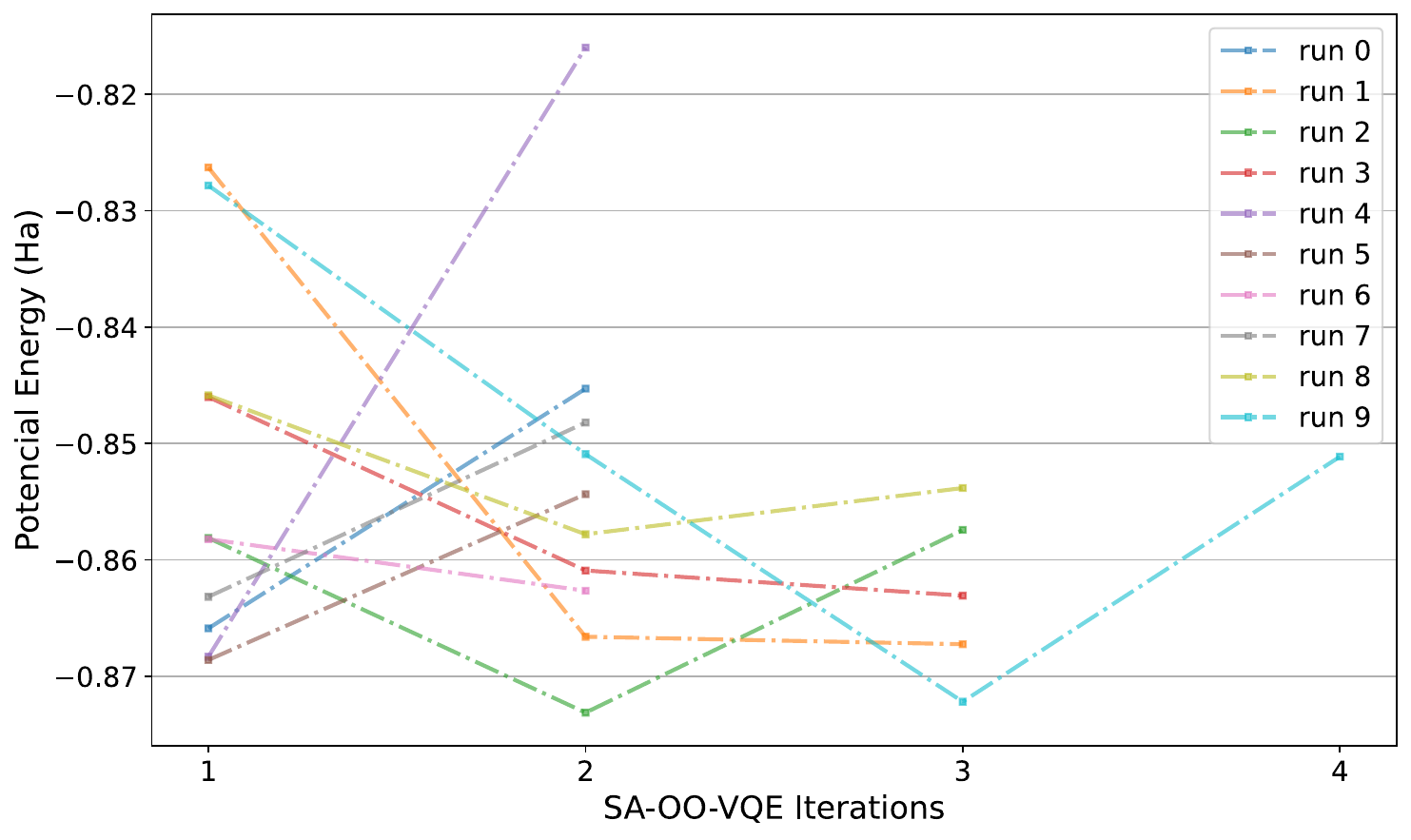}
        \caption{Energy vs. SA-OO-VQE iteration number.} 
        \label{fig:lih_bfgs_conv_iters} 
    \end{subfigure}
    \hfill 
    \begin{subfigure}[t]{0.49\textwidth} 
        \centering
        \includegraphics[width=\linewidth]{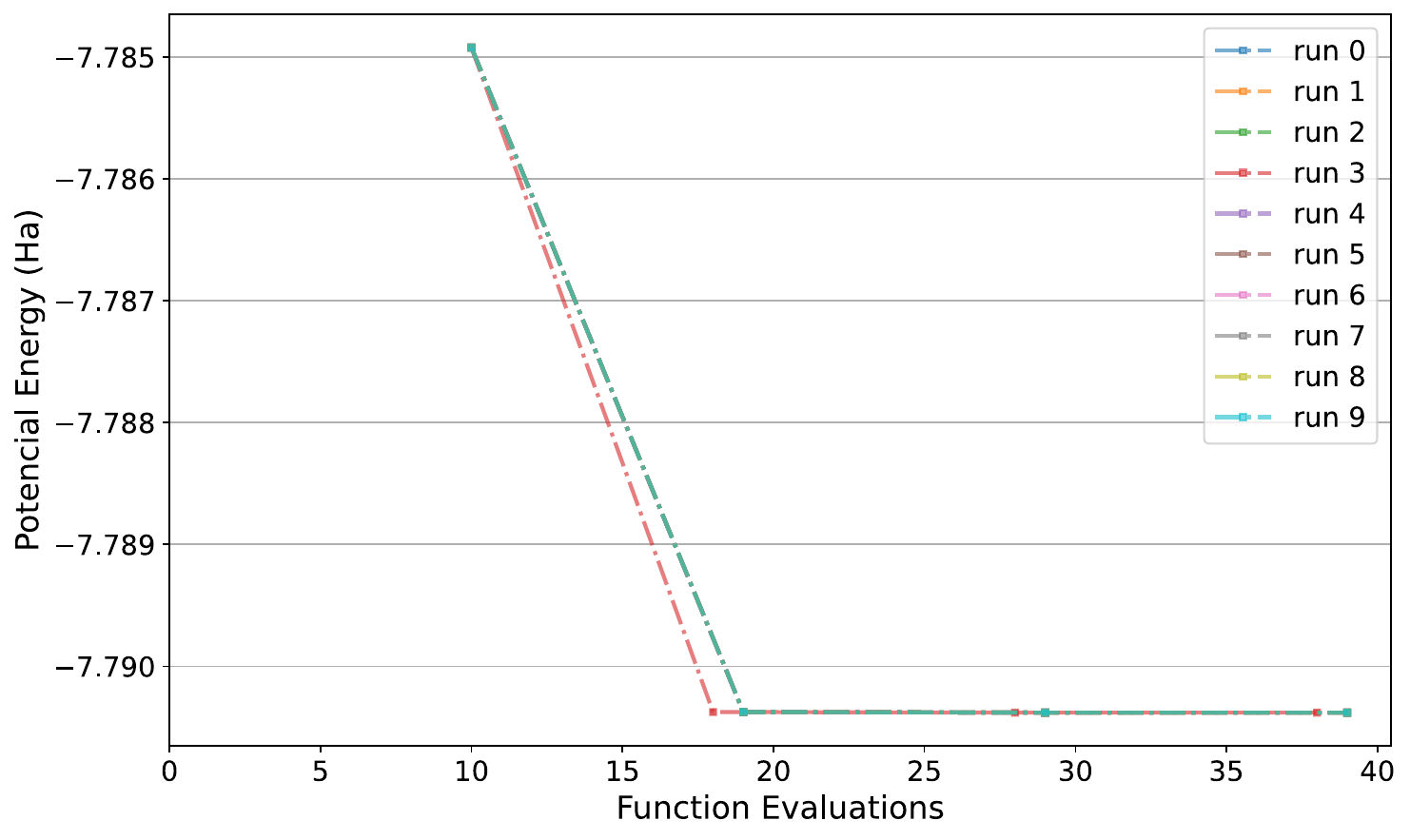}
        \caption{Energy after full SA-OO-VQE iterations vs. cumulative evaluations.} 
        \label{fig:lih_bfgs_conv_evals}
    \end{subfigure}

    \caption[Convergence plots of the BFGS optimizer within the SA-OO-VQE framework for the
LiH molecule.]{Convergence analysis of the BFGS optimizer within the SA-OO-VQE framework for the LiH molecule, based on 10 independent runs (shown in different colors/styles, see legend in plots). The plots display the state-average energy (Hartrees) progression viewed against different metrics: 
    (\subref{fig:lih_bfgs_conv_iters}) Energy plotted at the end of each completed SA-OO-VQE iteration against the iteration number. 
    (\subref{fig:lih_bfgs_conv_evals}) Energy plotted at the end of each completed SA-OO-VQE iteration against the cumulative number of function evaluations consumed up to that iteration.}
\label{fig:lih_bfgs} 
\end{figure}

\begin{figure}[!htbp] 
    \centering
    \begin{subfigure}[t]{0.49\textwidth} 
        \centering
        \includegraphics[width=\linewidth]{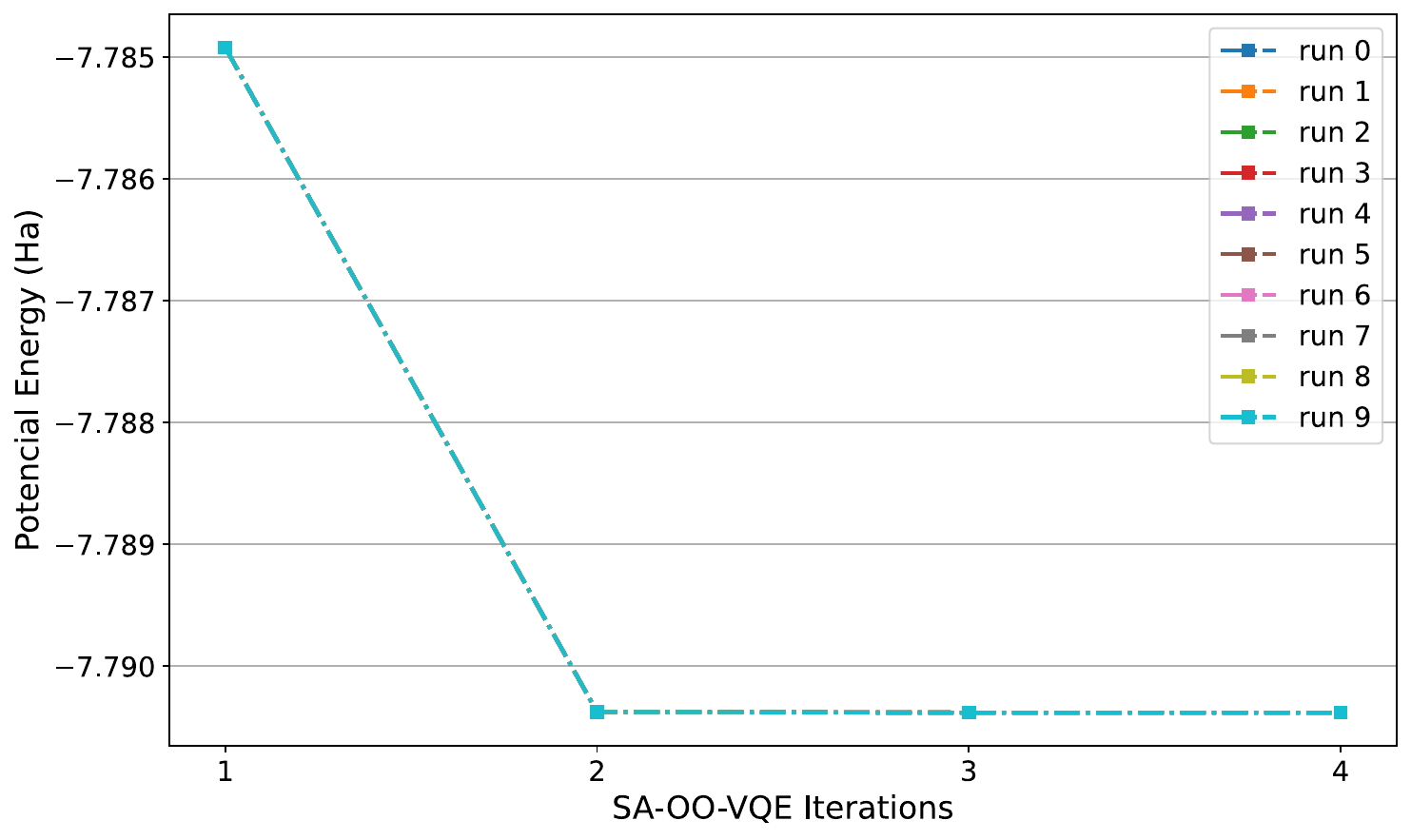}
        \caption{Energy vs. SA-OO-VQE iteration number.} 
        \label{fig:lih_cobyla_conv_iters} 
    \end{subfigure}
    \hfill 
    \begin{subfigure}[t]{0.49\textwidth} 
        \centering
        \includegraphics[width=\linewidth]{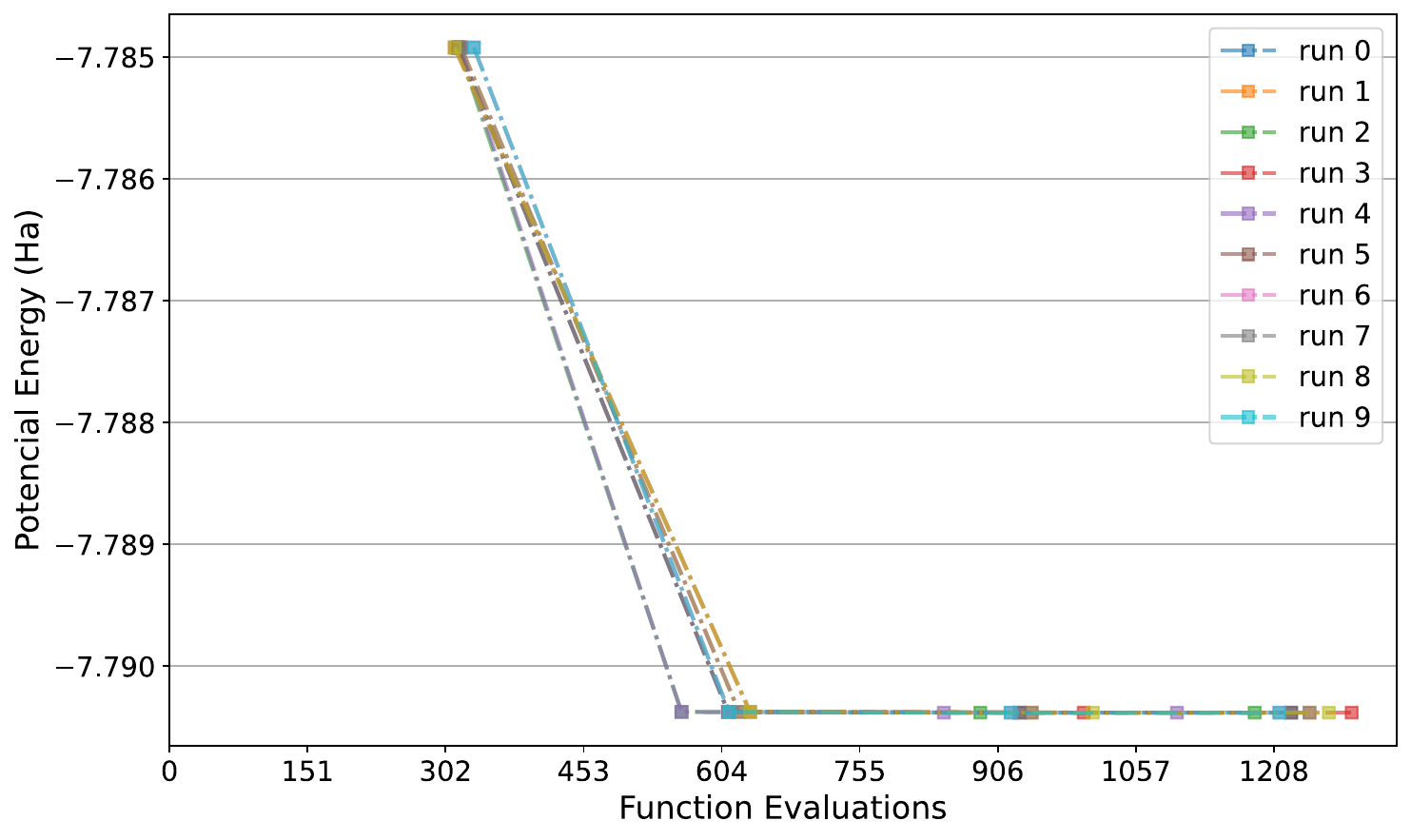}
        \caption{Energy after full SA-OO-VQE iterations vs. cumulative evaluations.} 
        \label{fig:lih_cobyla_conv_evals}
    \end{subfigure}

    \caption[Convergence plots of the COBYLA optimizer within the SA-OO-VQE framework for the
LiH molecule.]{Convergence analysis of the COBYLA optimizer within the SA-OO-VQE framework for the LiH molecule, based on 10 independent runs (shown in different colors/styles, see legend in plots). The plots display the state-average energy (Hartrees) progression viewed against different metrics: 
    (\subref{fig:lih_cobyla_conv_iters}) Energy plotted at the end of each completed SA-OO-VQE iteration against the iteration number. 
    (\subref{fig:lih_cobyla_conv_evals}) Energy plotted at the end of each completed SA-OO-VQE iteration against the cumulative number of function evaluations consumed up to that iteration.}
\label{fig:lih_cobyla} 
\end{figure}

\begin{figure}[!htbp] 
    \centering
    \begin{subfigure}[t]{0.7\textwidth} 
        \centering
        \includegraphics[width=\linewidth]{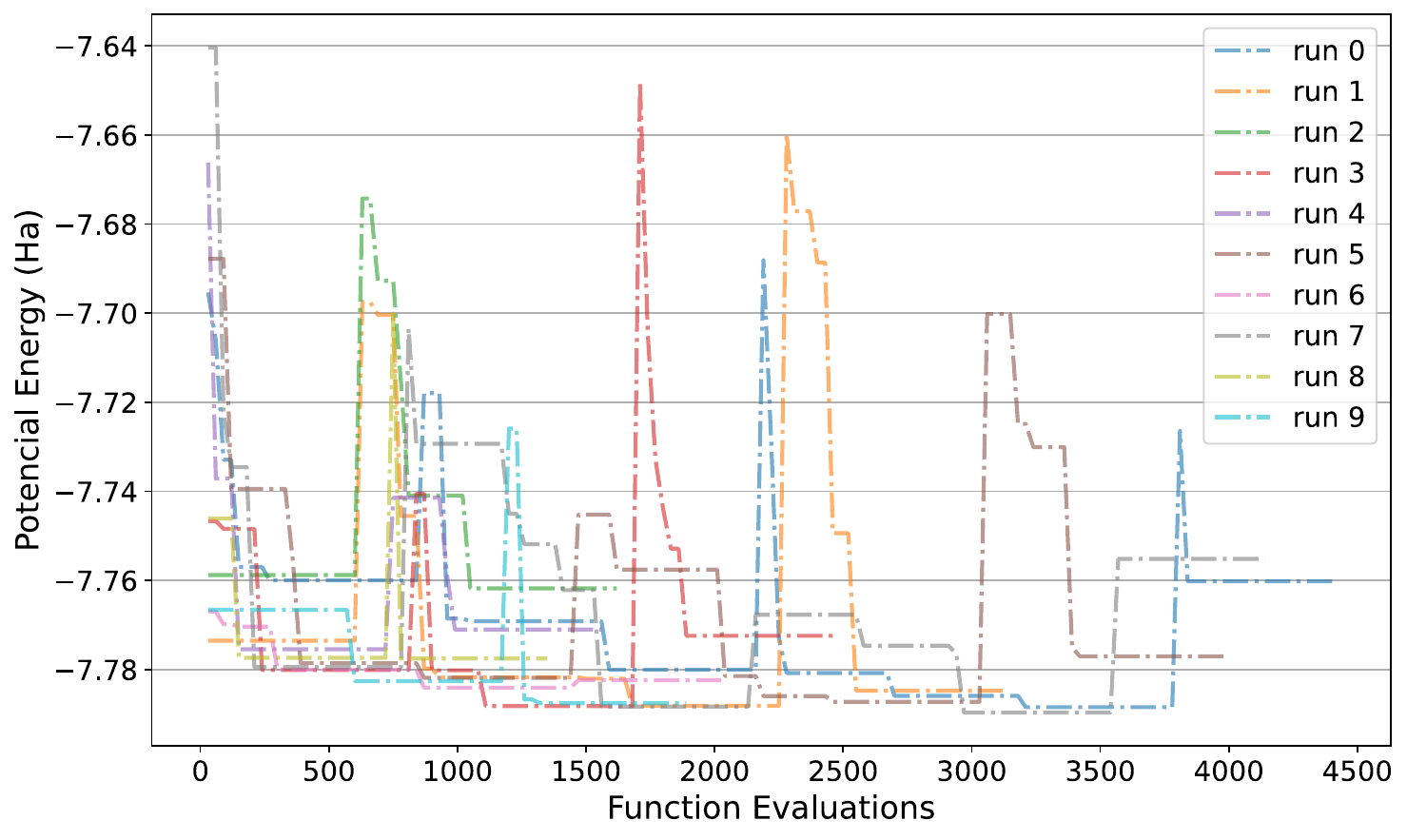}
        \caption{Energy after optimizer iterations vs. cumulative evaluations.} 
        \label{fig:lih_de_b1_conv} 
    \end{subfigure}
    \begin{subfigure}[t]{0.49\textwidth} 
        \centering
        \includegraphics[width=\linewidth]{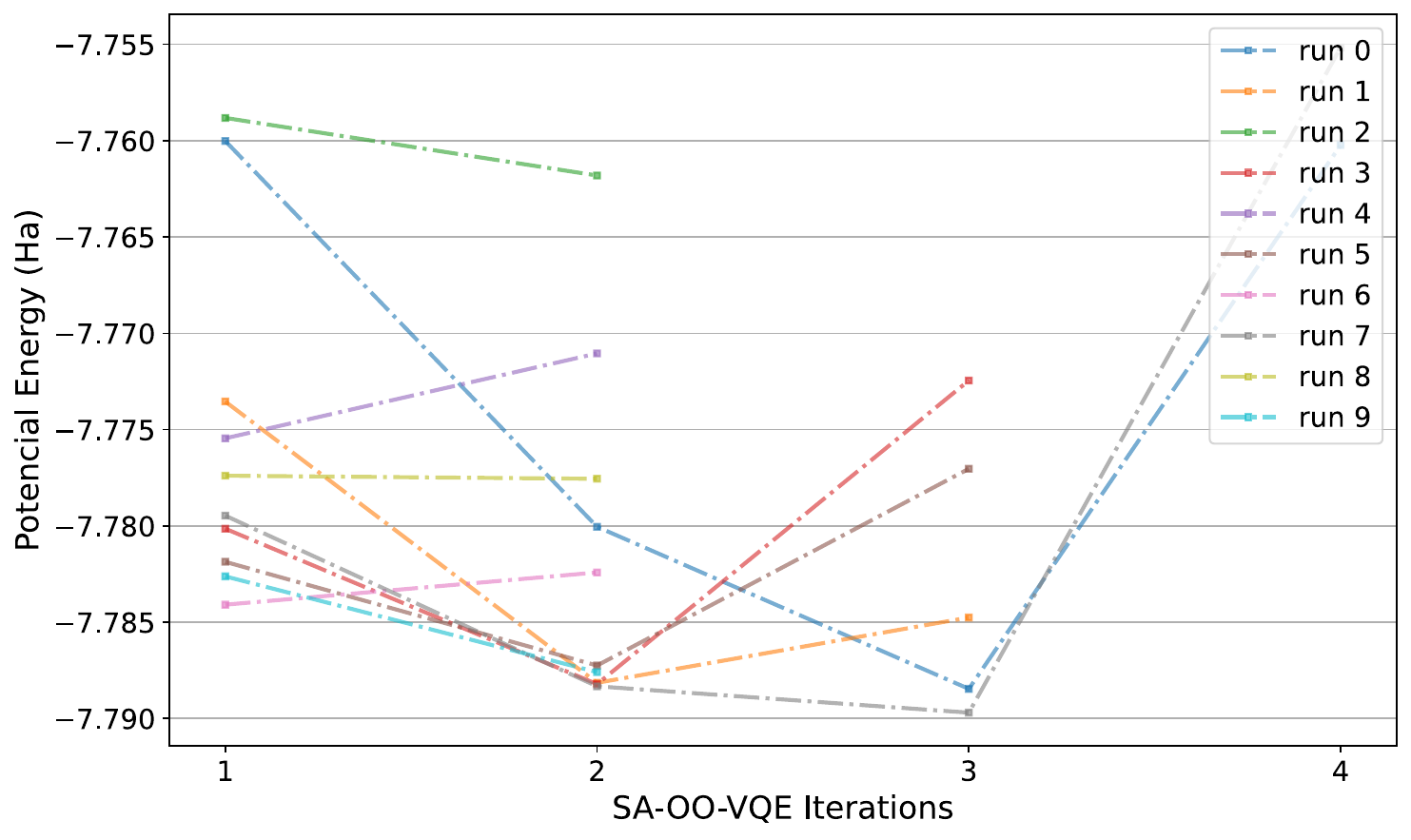}
        \caption{Energy vs. SA-OO-VQE iteration number.} 
        \label{fig:lih_de_b1_conv_iters} 
    \end{subfigure}
    \hfill 
    \begin{subfigure}[t]{0.49\textwidth} 
        \centering
        \includegraphics[width=\linewidth]{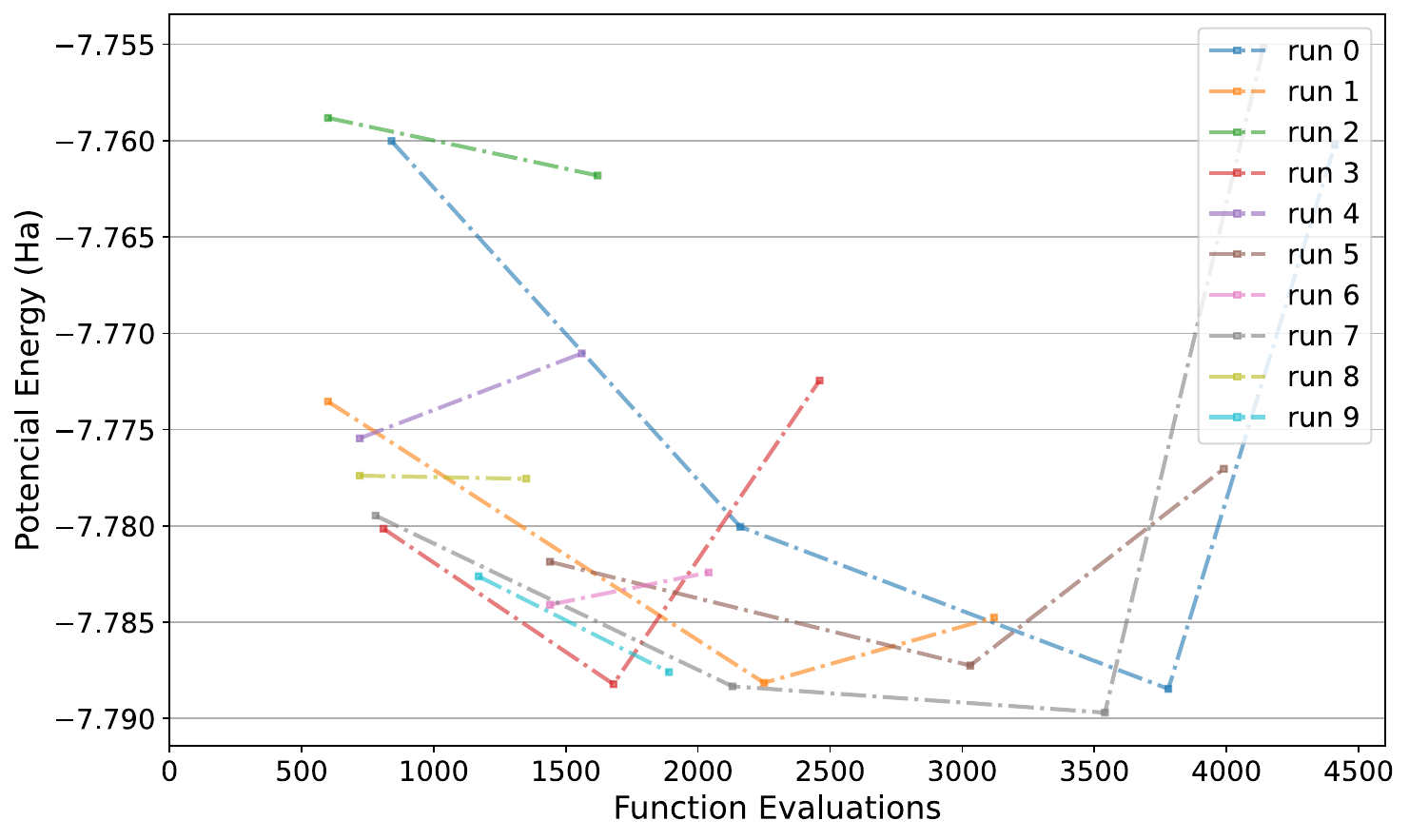}
        \caption{Energy after full SA-OO-VQE iterations vs. cumulative evaluations.} 
        \label{fig:lih_de_b1_conv_evals}
    \end{subfigure}

    \caption[Convergence plots of the DE/Best/1/bin optimizer within the SA-OO-VQE framework for the
LiH molecule.]{Convergence analysis of the DE/Best/1/bin optimizer within the SA-OO-VQE framework for the LiH molecule, based on 10 independent runs (shown in different colors/styles, see legend in plots). The plots display the state-average energy (Hartrees) progression viewed against different metrics: 
    (\subref{fig:lih_de_b1_conv}) Energy evaluated at the end of each internal Gradient Descent optimizer iteration, plotted against the cumulative number of function evaluations consumed up to that iteration point
    (\subref{fig:lih_de_b1_conv_iters}) Energy plotted at the end of each completed SA-OO-VQE iteration against the iteration number. 
    (\subref{fig:lih_de_b1_conv_evals}) Energy plotted at the end of each completed SA-OO-VQE iteration against the cumulative number of function evaluations consumed up to that iteration.}
\label{fig:lih_de_b1} 
\end{figure}

\begin{figure}[!htbp] 
    \centering
    \begin{subfigure}[t]{0.7\textwidth} 
        \centering
        \includegraphics[width=\linewidth]{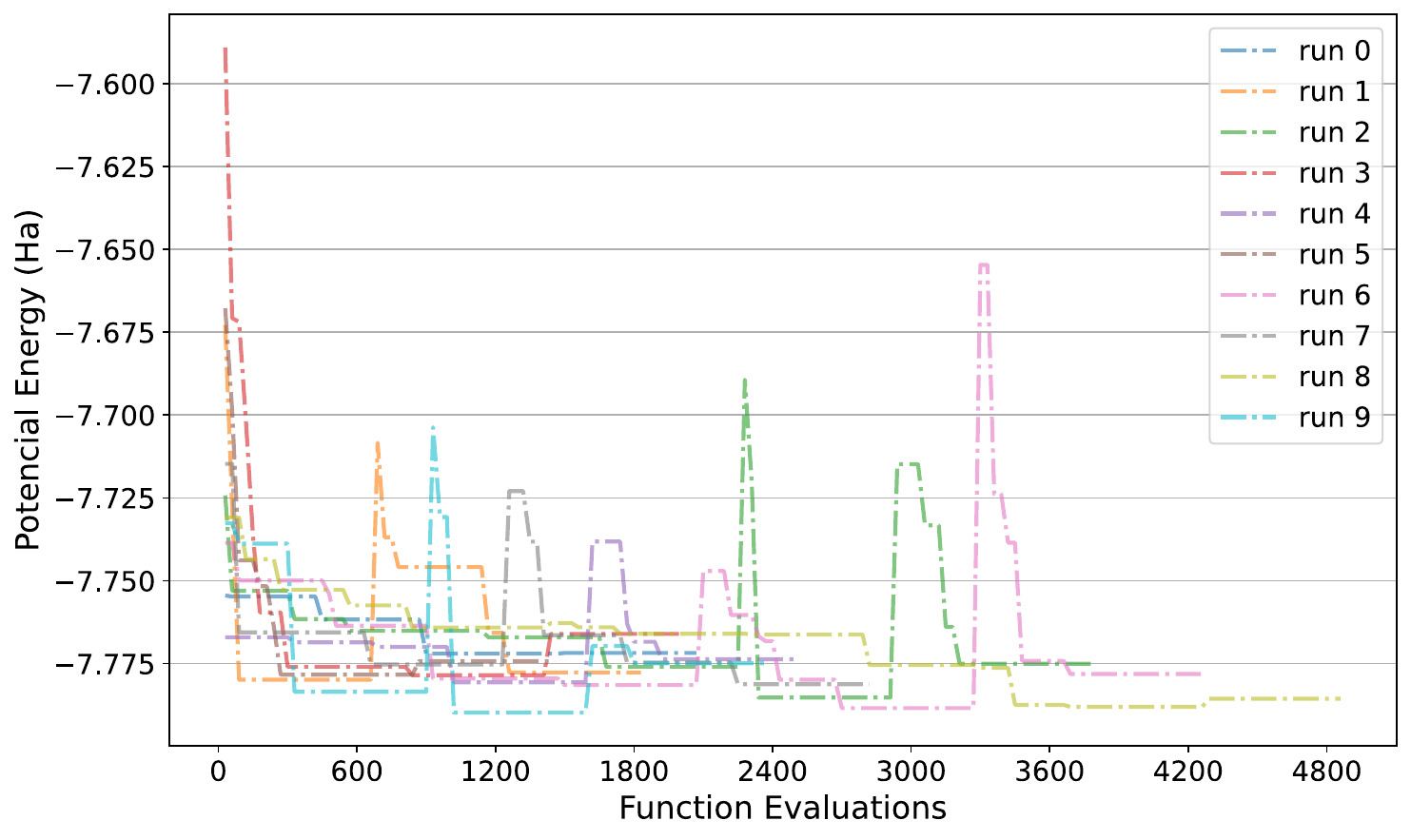}
        \caption{Energy after optimizer iterations vs. cumulative evaluations.} 
        \label{fig:lih_de_b2_conv} 
    \end{subfigure}
    \begin{subfigure}[t]{0.49\textwidth} 
        \centering
        \includegraphics[width=\linewidth]{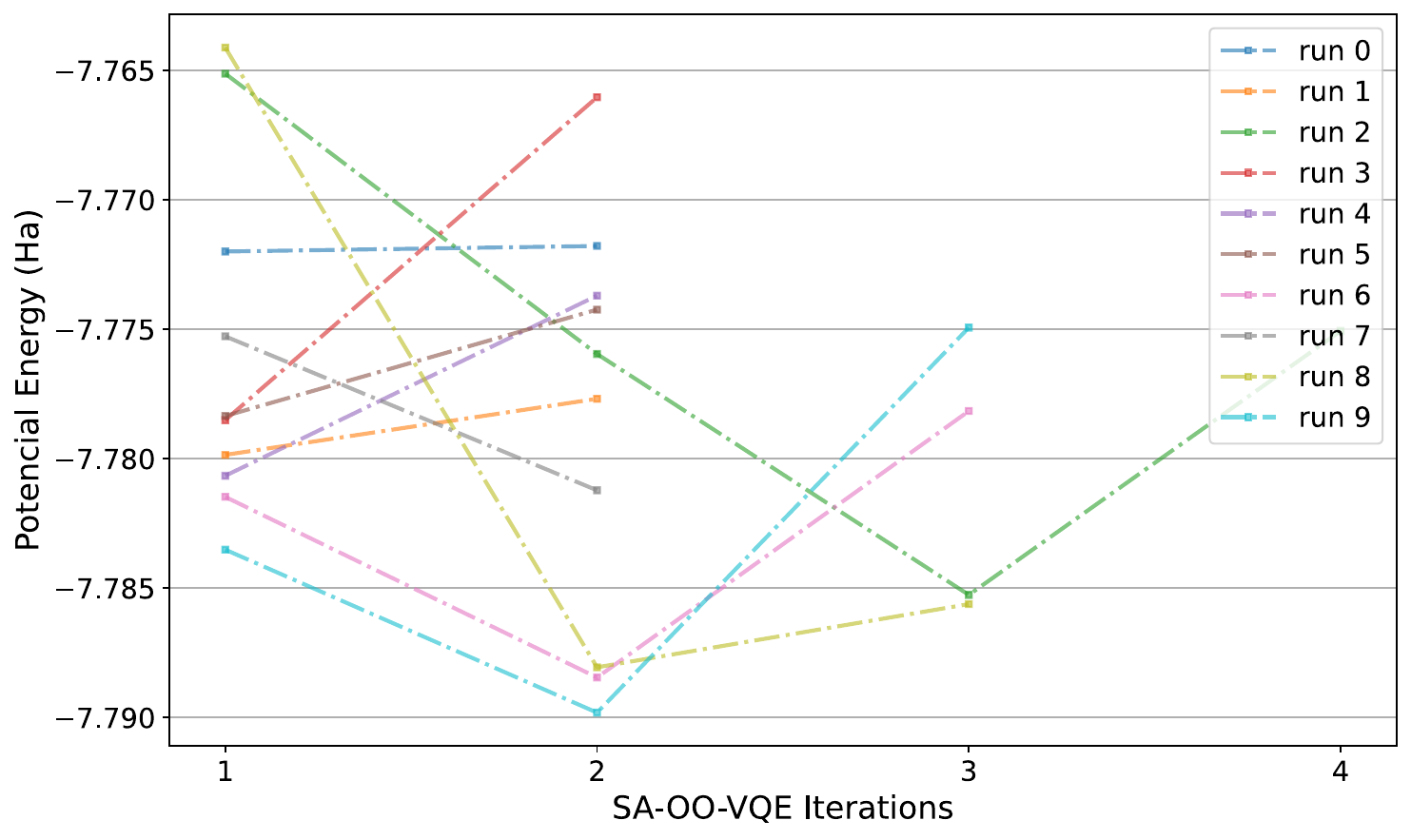}
        \caption{Energy vs. SA-OO-VQE iteration number.} 
        \label{fig:lih_de_b2_conv_iters} 
    \end{subfigure}
    \hfill 
    \begin{subfigure}[t]{0.49\textwidth} 
        \centering
        \includegraphics[width=\linewidth]{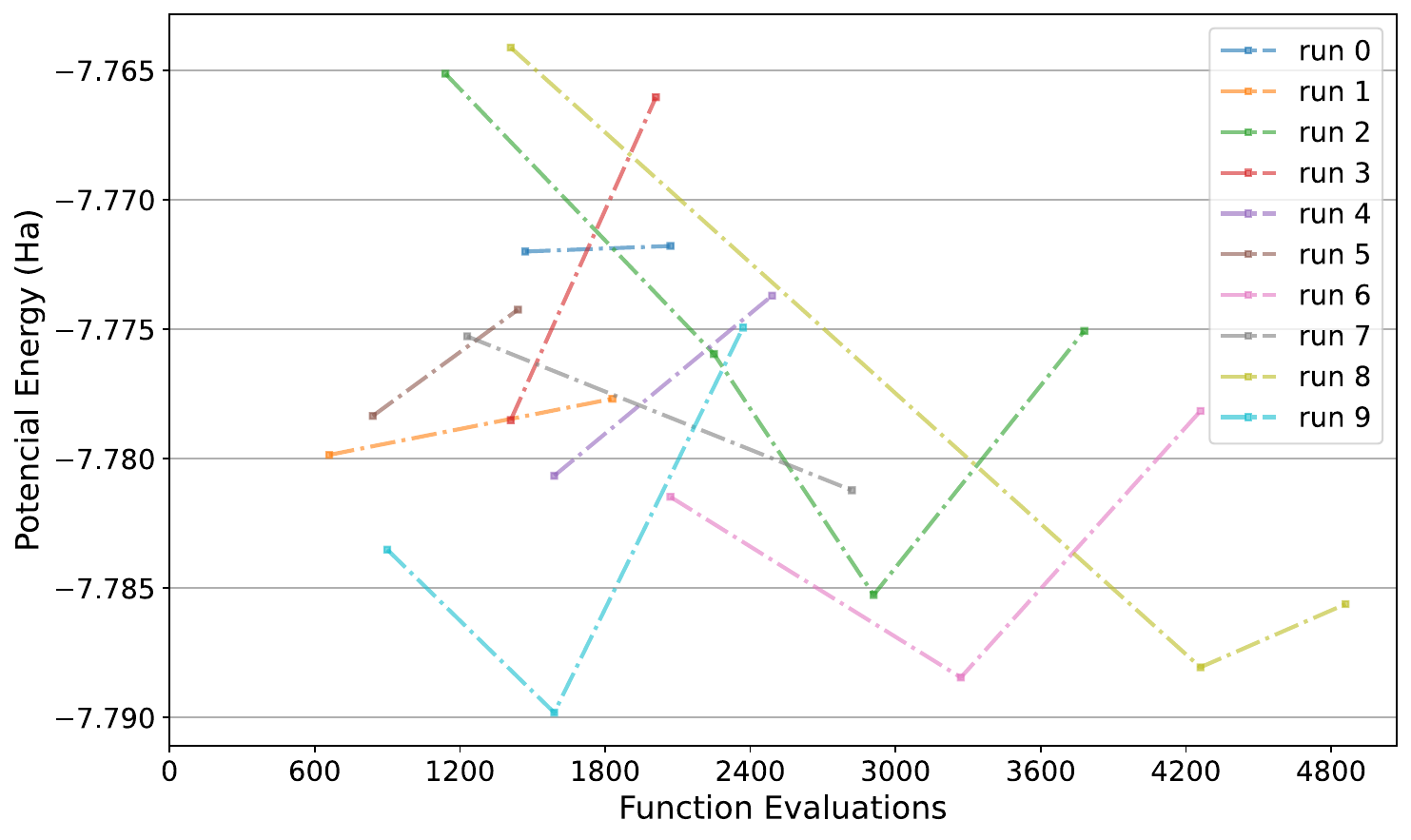}
        \caption{Energy after full SA-OO-VQE iterations vs. cumulative evaluations.} 
        \label{fig:lih_de_b2_conv_evals}
    \end{subfigure}

    \caption[Convergence plots of the DE/Best/2/bin optimizer within the SA-OO-VQE framework for the
LiH molecule.]{Convergence analysis of the DE/Best/2/bin optimizer within the SA-OO-VQE framework for the LiH molecule, based on 10 independent runs (shown in different colors/styles, see legend in plots). The plots display the state-average energy (Hartrees) progression viewed against different metrics: 
    (\subref{fig:lih_de_b2_conv}) Energy evaluated at the end of each internal Gradient Descent optimizer iteration, plotted against the cumulative number of function evaluations consumed up to that iteration point
    (\subref{fig:lih_de_b1_conv_iters}) Energy plotted at the end of each completed SA-OO-VQE iteration against the iteration number. 
    (\subref{fig:lih_de_b2_conv_evals}) Energy plotted at the end of each completed SA-OO-VQE iteration against the cumulative number of function evaluations consumed up to that iteration.}
\label{fig:lih_de_b2} 
\end{figure}

\begin{figure}[!htbp] 
    \centering
    \begin{subfigure}[t]{0.7\textwidth} 
        \centering
        \includegraphics[width=\linewidth]{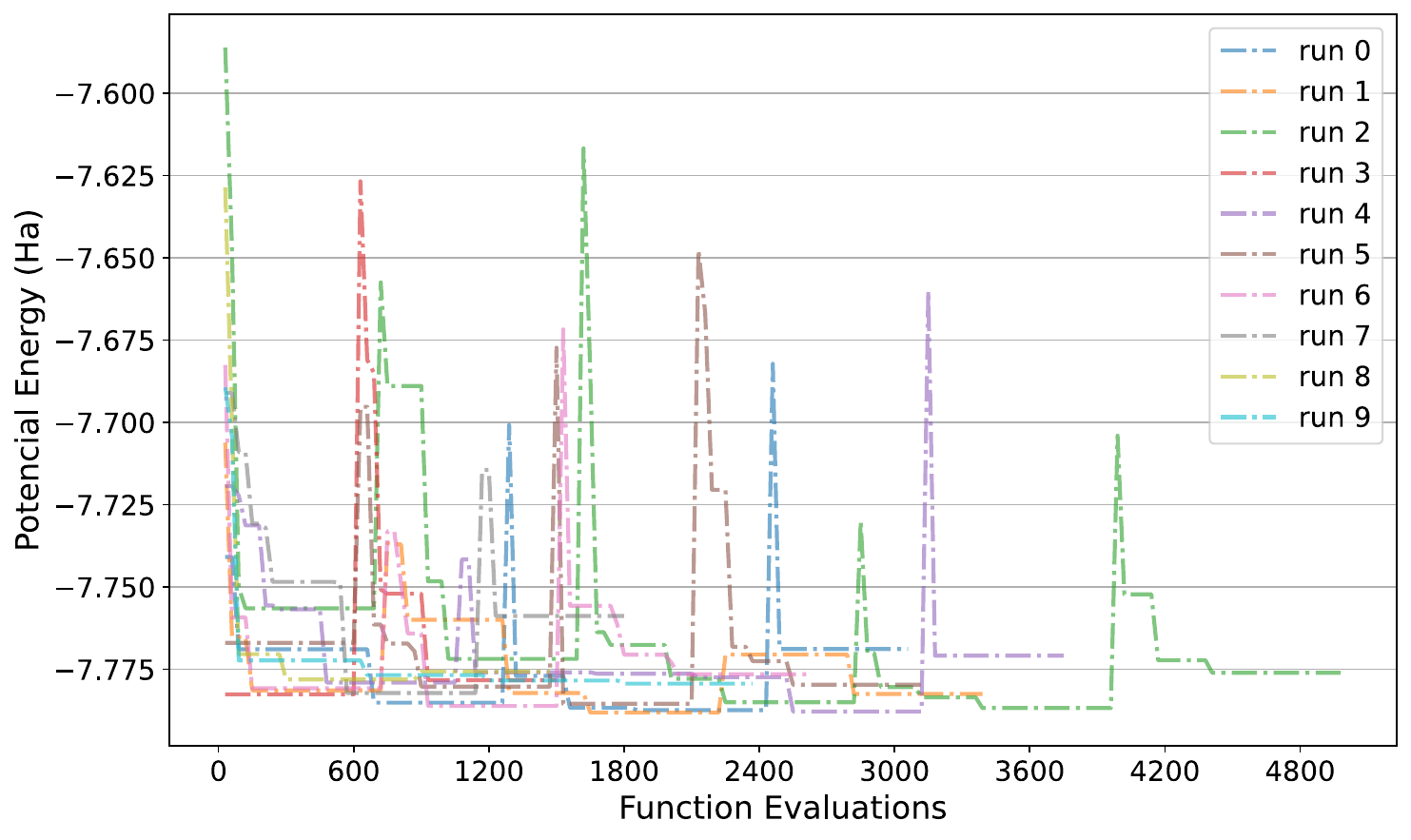}
        \caption{Energy after optimizer iterations vs. cumulative evaluations.} 
        \label{fig:lih_de_ctb_conv} 
    \end{subfigure}
    \begin{subfigure}[t]{0.49\textwidth} 
        \centering
        \includegraphics[width=\linewidth]{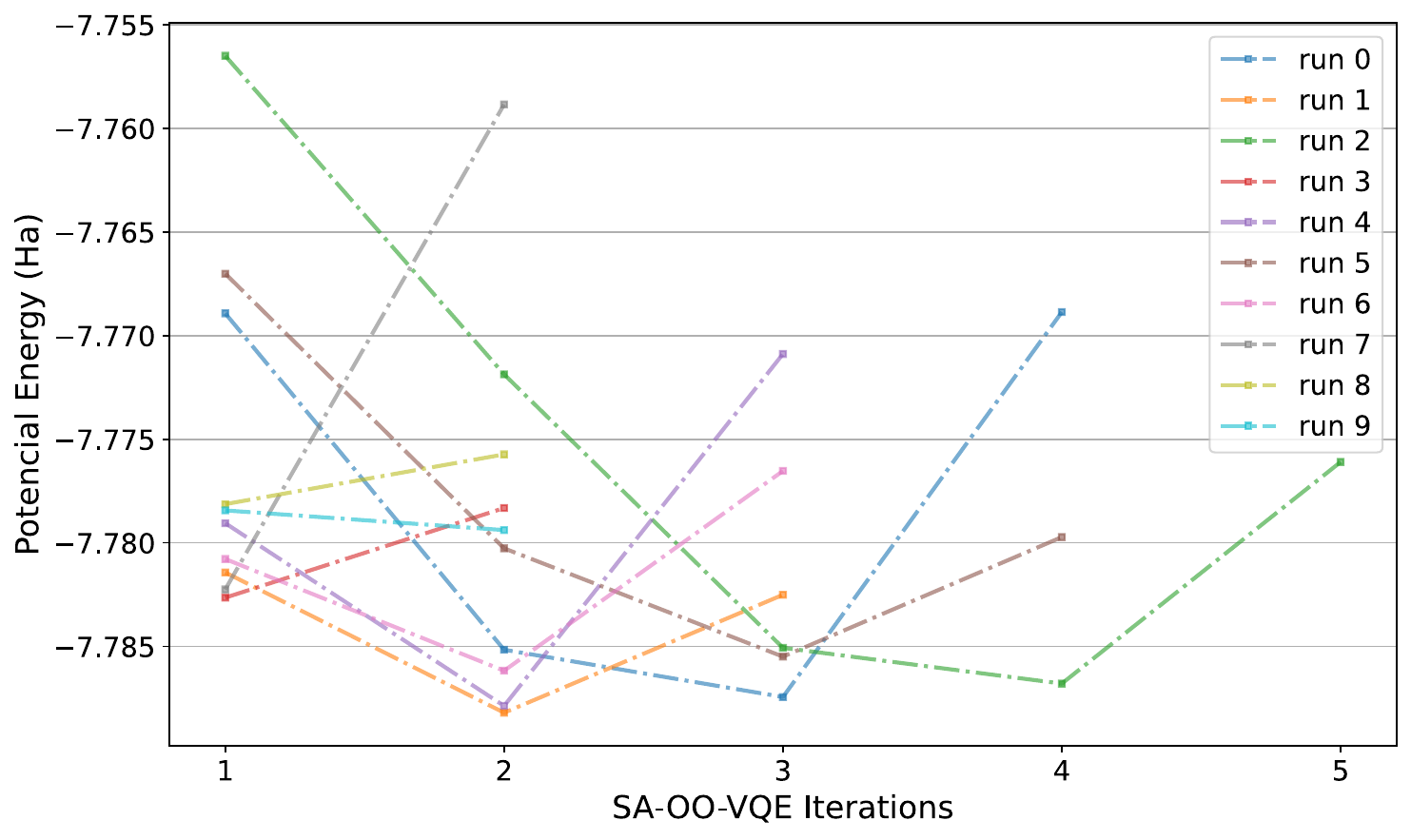}
        \caption{Energy vs. SA-OO-VQE iteration number.} 
        \label{fig:lih_de_ctb_conv_iters} 
    \end{subfigure}
    \hfill 
    \begin{subfigure}[t]{0.49\textwidth} 
        \centering
        \includegraphics[width=\linewidth]{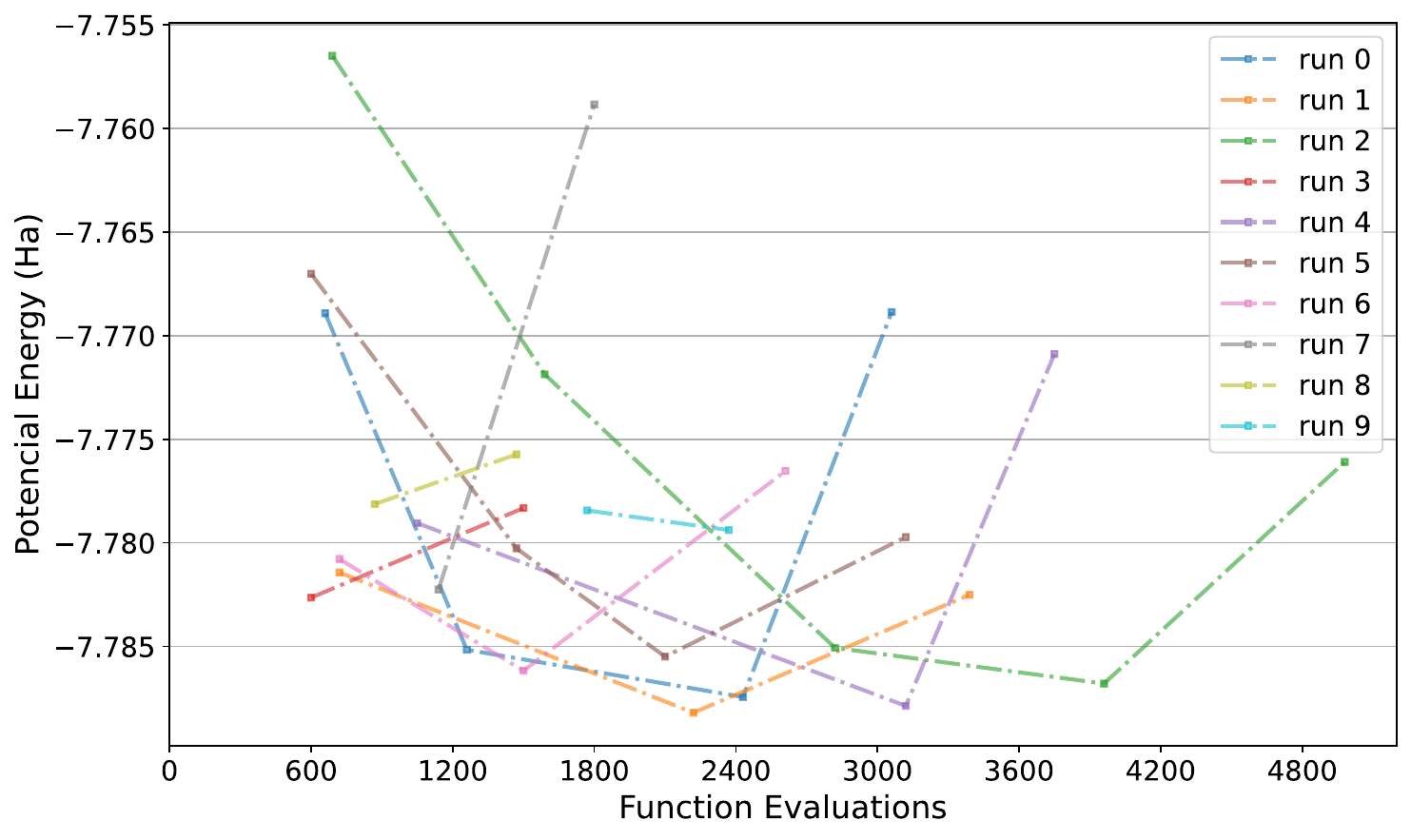}
        \caption{Energy after full SA-OO-VQE iterations vs. cumulative evaluations.} 
        \label{fig:lih_de_ctb_conv_evals}
    \end{subfigure}

    \caption[Convergence plots of the DE/Current-to-Best/1/bin optimizer within the SA-OO-VQE framework for the
LiH molecule.]{Convergence analysis of the DE/Current-to-Best/1/bin optimizer within the SA-OO-VQE framework for the LiH molecule, based on 10 independent runs (shown in different colors/styles, see legend in plots). The plots display the state-average energy (Hartrees) progression viewed against different metrics: 
    (\subref{fig:lih_de_ctb_conv}) Energy evaluated at the end of each internal Gradient Descent optimizer iteration, plotted against the cumulative number of function evaluations consumed up to that iteration point
    (\subref{fig:lih_de_ctb_conv_iters}) Energy plotted at the end of each completed SA-OO-VQE iteration against the iteration number. 
    (\subref{fig:lih_de_ctb_conv_evals}) Energy plotted at the end of each completed SA-OO-VQE iteration against the cumulative number of function evaluations consumed up to that iteration.}
\label{fig:lih_de_ctb} 
\end{figure}

\begin{figure}[!htbp] 
    \centering
    \begin{subfigure}[t]{0.7\textwidth} 
        \centering
        \includegraphics[width=\linewidth]{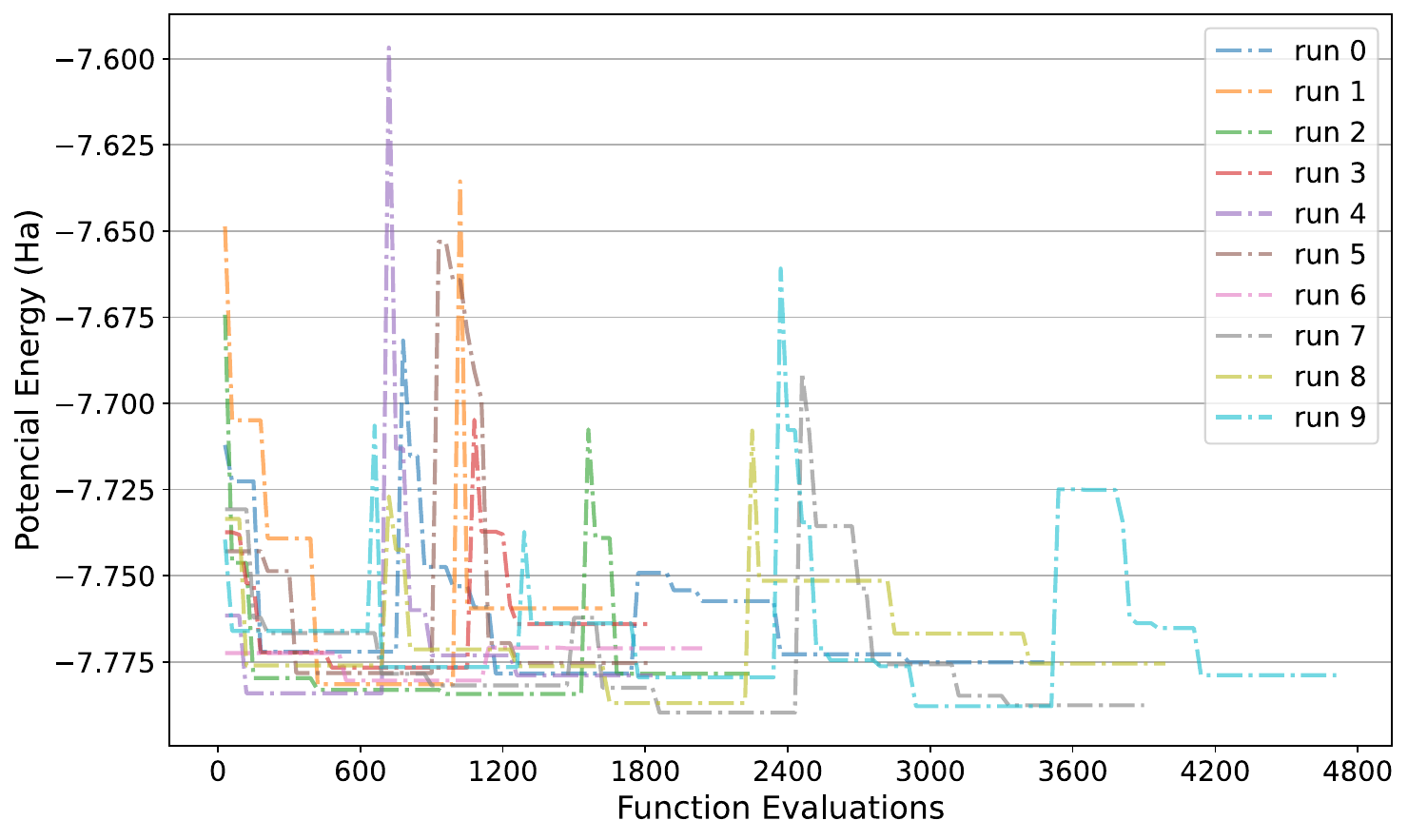}
        \caption{Energy after optimizer iterations vs. cumulative evaluations.} 
        \label{fig:lih_de_ctr_conv} 
    \end{subfigure}
    \begin{subfigure}[t]{0.49\textwidth} 
        \centering
        \includegraphics[width=\linewidth]{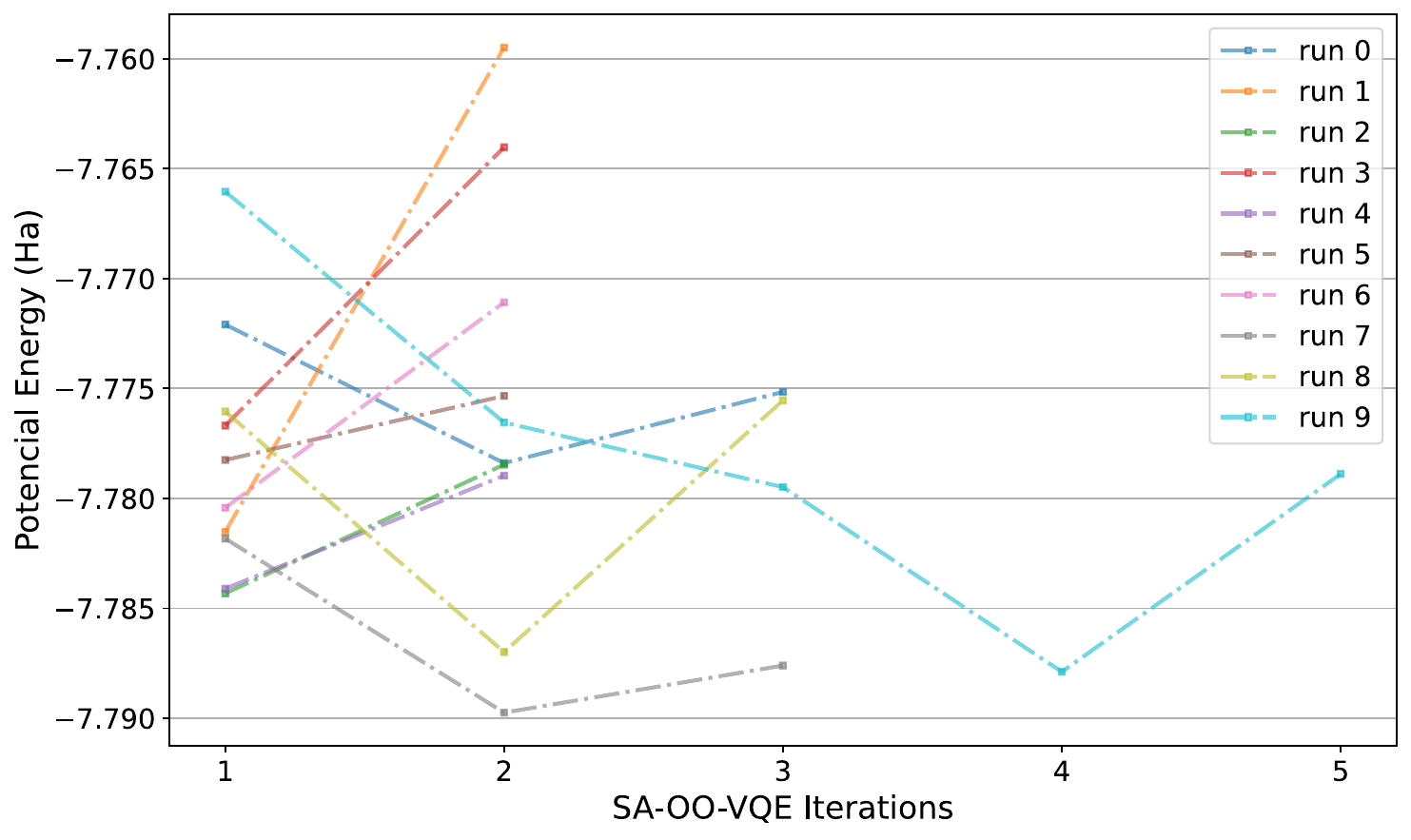}
        \caption{Energy vs. SA-OO-VQE iteration number.} 
        \label{fig:lih_de_ctr_conv_iters} 
    \end{subfigure}
    \hfill 
    \begin{subfigure}[t]{0.49\textwidth} 
        \centering
        \includegraphics[width=\linewidth]{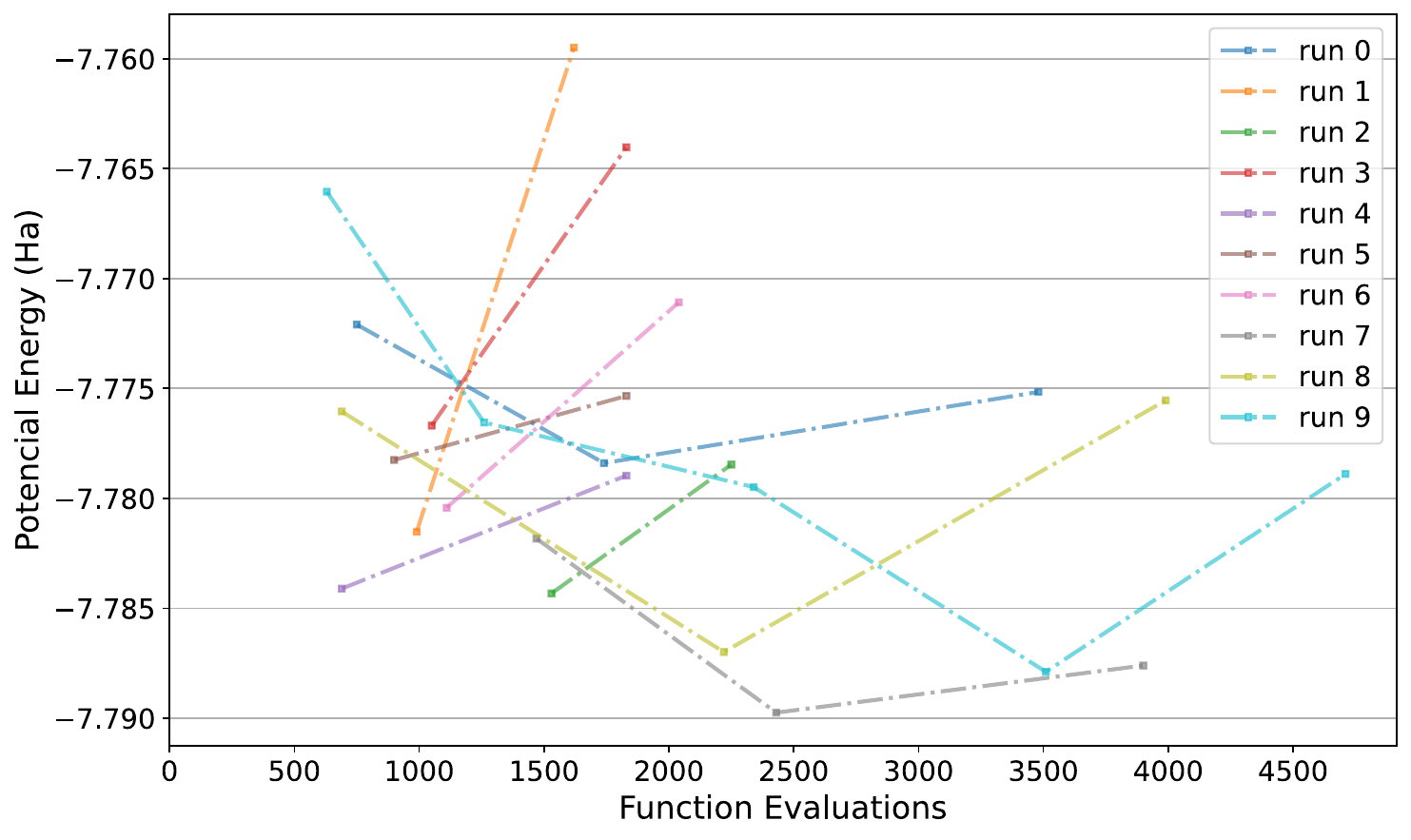}
        \caption{Energy after full SA-OO-VQE iterations vs. cumulative evaluations.} 
        \label{fig:lih_de_ctr_conv_evals}
    \end{subfigure}

    \caption[Convergence plots of the DE/Current-to-Random/1/bin optimizer within the SA-OO-VQE framework for the
LiH molecule.]{Convergence analysis of the DE/Current-to-Random/1/bin optimizer within the SA-OO-VQE framework for the LiH molecule, based on 10 independent runs (shown in different colors/styles, see legend in plots). The plots display the state-average energy (Hartrees) progression viewed against different metrics: 
    (\subref{fig:lih_de_ctr_conv}) Energy evaluated at the end of each internal Gradient Descent optimizer iteration, plotted against the cumulative number of function evaluations consumed up to that iteration point
    (\subref{fig:lih_de_ctr_conv_iters}) Energy plotted at the end of each completed SA-OO-VQE iteration against the iteration number. 
    (\subref{fig:lih_de_ctr_conv_evals}) Energy plotted at the end of each completed SA-OO-VQE iteration against the cumulative number of function evaluations consumed up to that iteration.}
\label{fig:lih_de_ctr} 
\end{figure}

\begin{figure}[!htbp] 
    \centering
    \begin{subfigure}[t]{0.7\textwidth} 
        \centering
        \includegraphics[width=\linewidth]{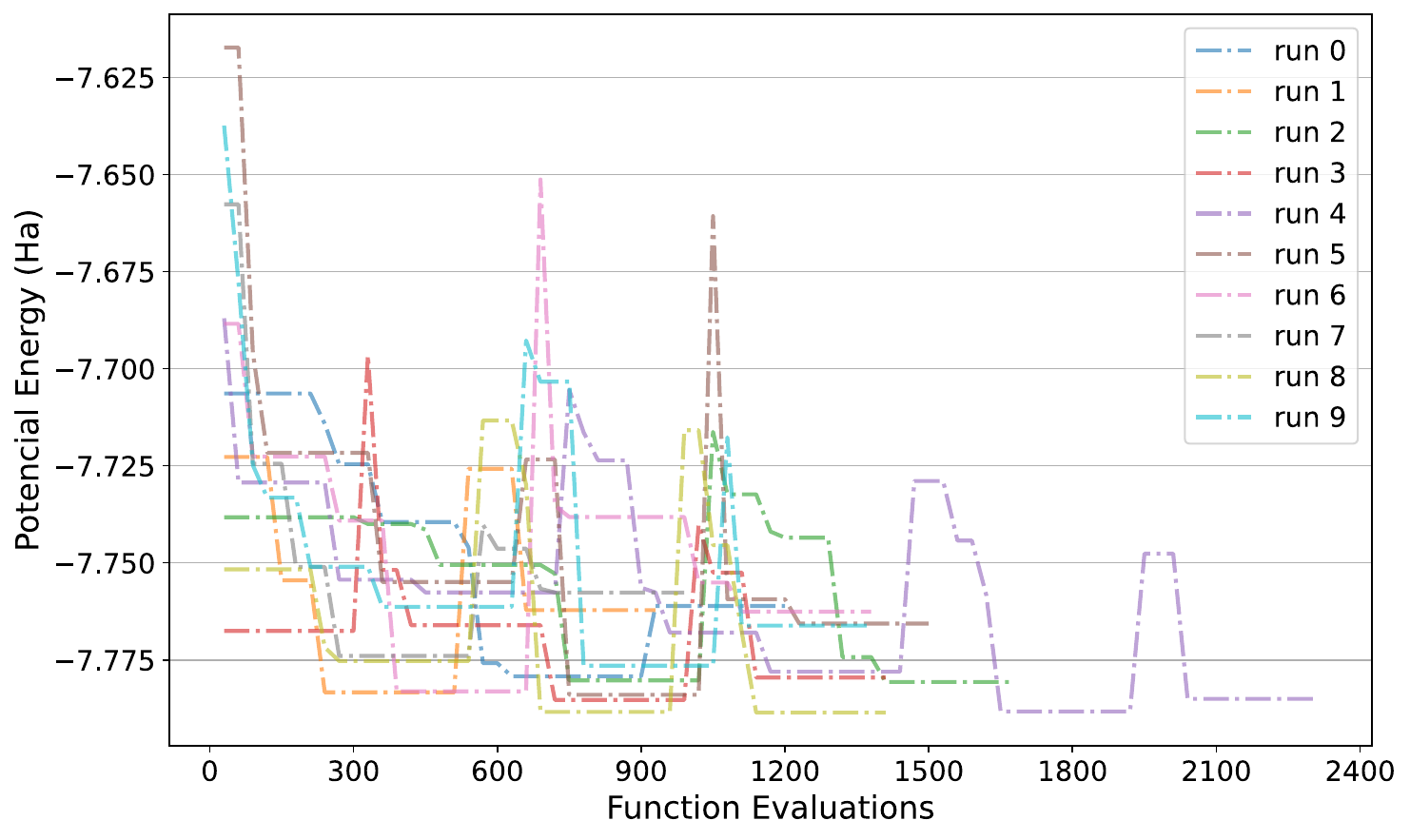}
        \caption{Energy after optimizer iterations vs. cumulative evaluations.} 
        \label{fig:lih_de_r1_conv} 
    \end{subfigure}
    \begin{subfigure}[t]{0.49\textwidth} 
        \centering
        \includegraphics[width=\linewidth]{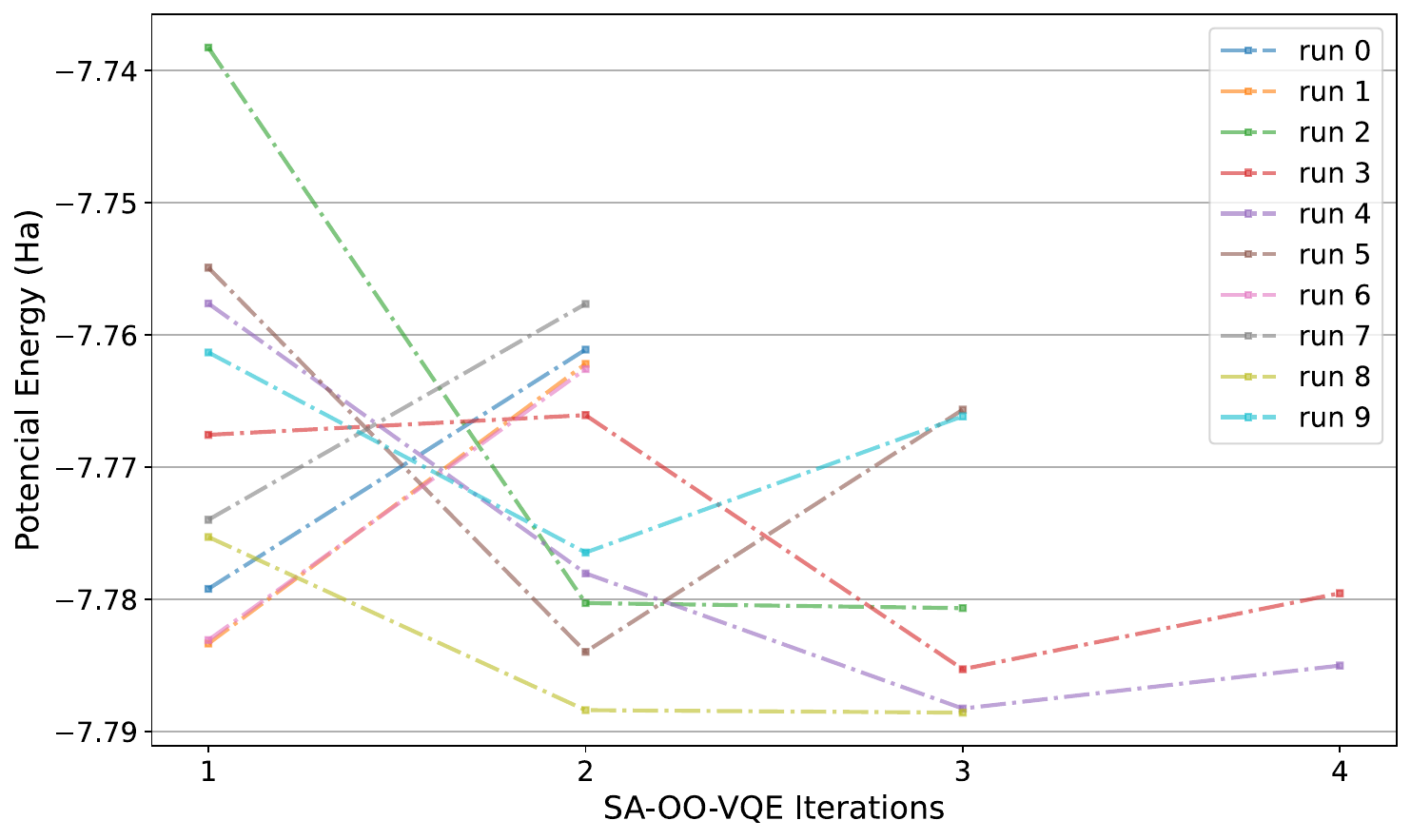}
        \caption{Energy vs. SA-OO-VQE iteration number.} 
        \label{fig:lih_de_r1_conv_iters} 
    \end{subfigure}
    \hfill 
    \begin{subfigure}[t]{0.49\textwidth} 
        \centering
        \includegraphics[width=\linewidth]{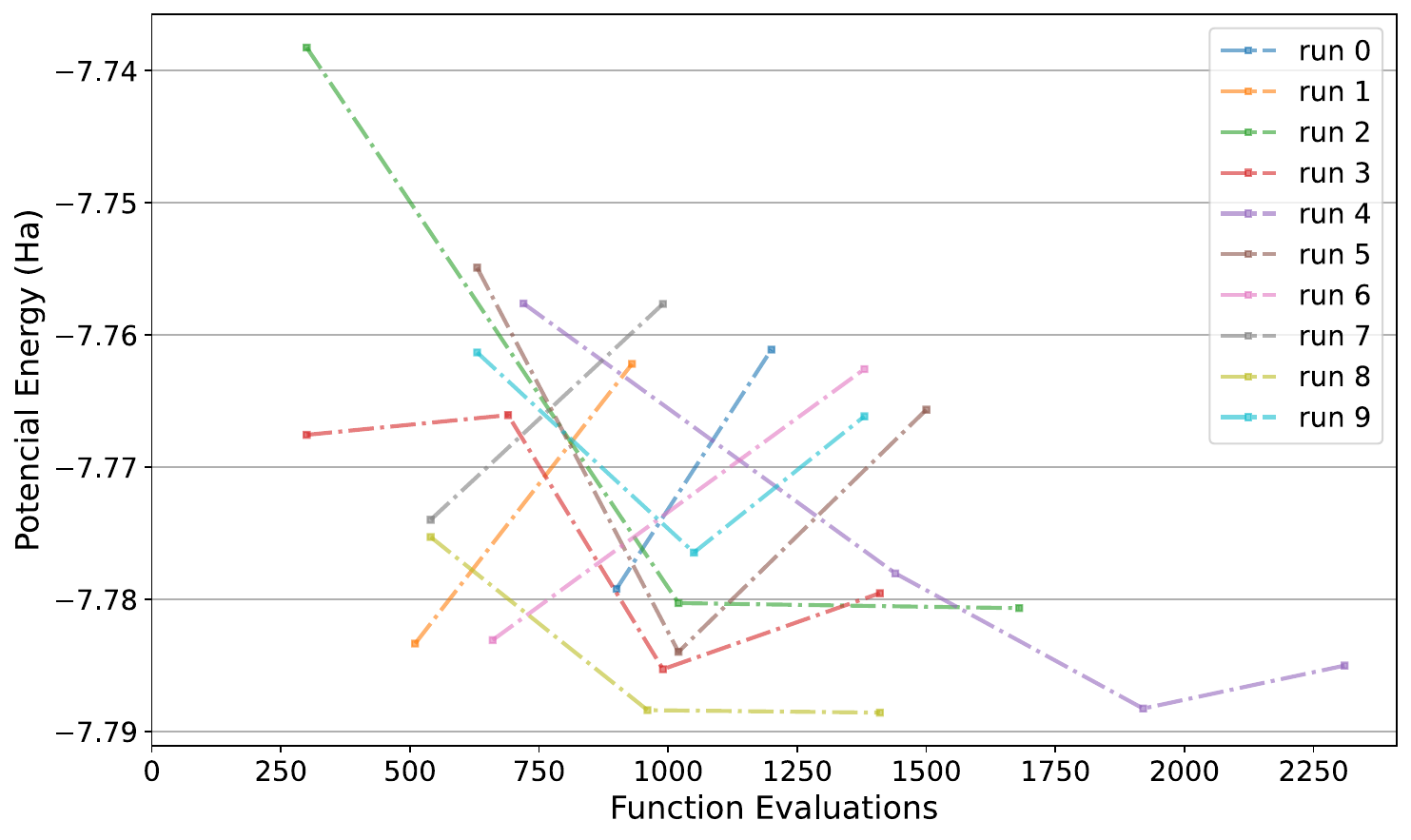}
        \caption{Energy after full SA-OO-VQE iterations vs. cumulative evaluations.} 
        \label{fig:lih_de_r1_conv_evals}
    \end{subfigure}

    \caption[Convergence plots of the DE/Rand/1/bin optimizer within the SA-OO-VQE framework for the
LiH molecule.]{Convergence analysis of the DE/Rand/1/bin optimizer within the SA-OO-VQE framework for the LiH molecule, based on 10 independent runs (shown in different colors/styles, see legend in plots). The plots display the state-average energy (Hartrees) progression viewed against different metrics: 
    (\subref{fig:lih_de_r1_conv}) Energy evaluated at the end of each internal Gradient Descent optimizer iteration, plotted against the cumulative number of function evaluations consumed up to that iteration point
    (\subref{fig:lih_de_r1_conv_iters}) Energy plotted at the end of each completed SA-OO-VQE iteration against the iteration number. 
    (\subref{fig:lih_de_r1_conv_evals}) Energy plotted at the end of each completed SA-OO-VQE iteration against the cumulative number of function evaluations consumed up to that iteration.}
\label{fig:lih_de_r1} 
\end{figure}

\begin{figure}[!htbp] 
    \centering
    \begin{subfigure}[t]{0.7\textwidth} 
        \centering
        \includegraphics[width=\linewidth]{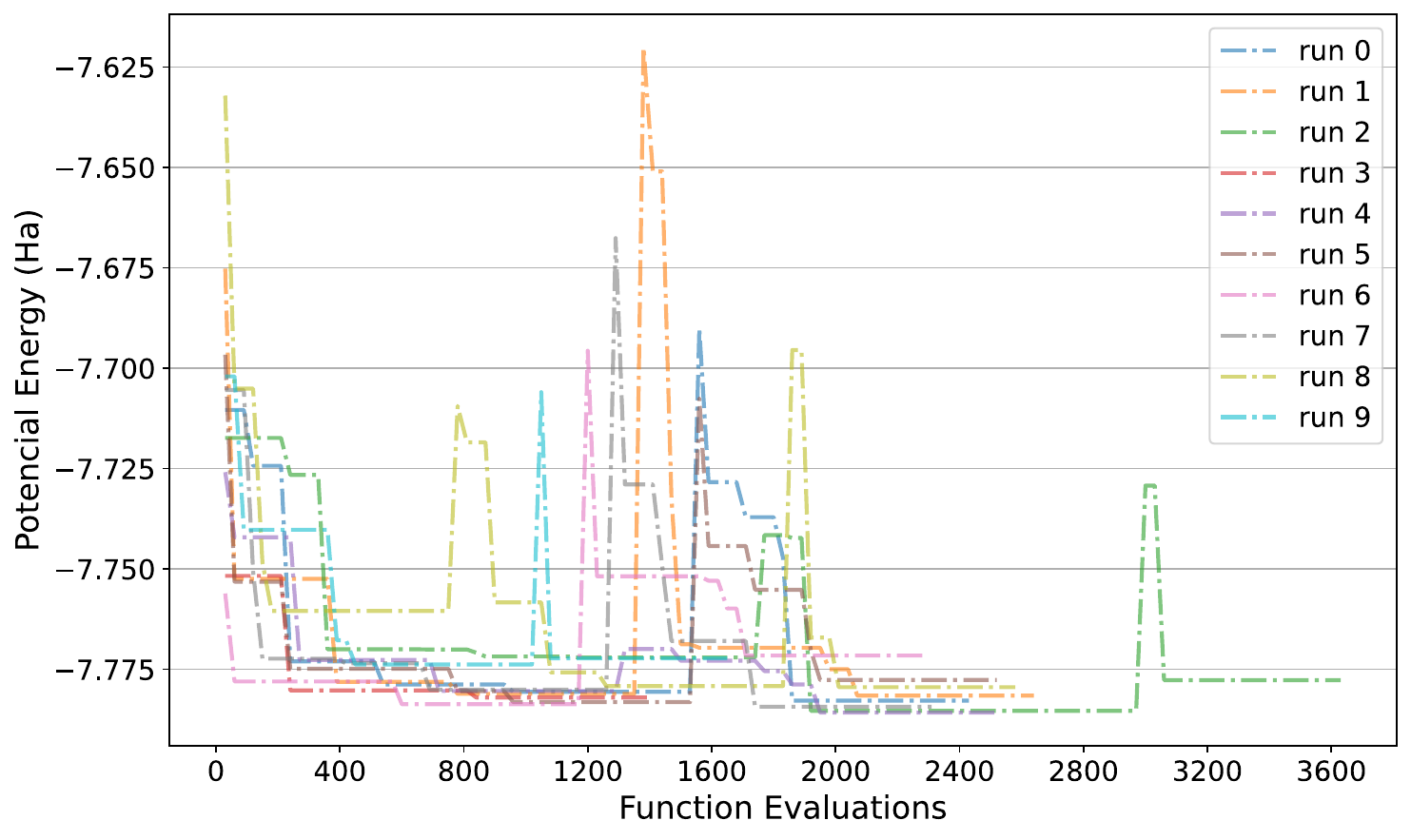}
        \caption{Energy after optimizer iterations vs. cumulative evaluations.} 
        \label{fig:lih_de_r2_conv} 
    \end{subfigure}
    \begin{subfigure}[t]{0.49\textwidth} 
        \centering
        \includegraphics[width=\linewidth]{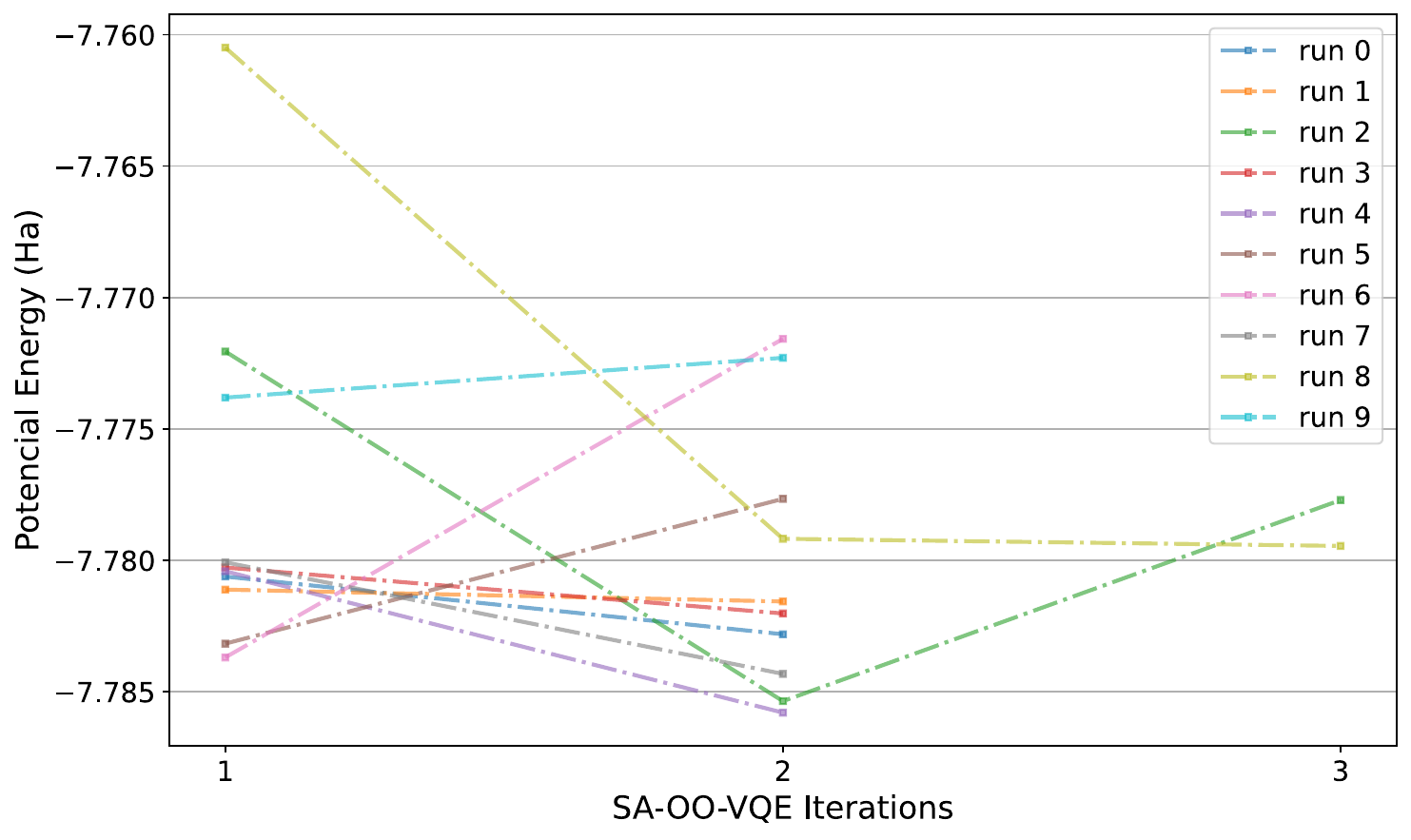}
        \caption{Energy vs. SA-OO-VQE iteration number.} 
        \label{fig:lih_de_r2_conv_iters} 
    \end{subfigure}
    \hfill 
    \begin{subfigure}[t]{0.49\textwidth} 
        \centering
        \includegraphics[width=\linewidth]{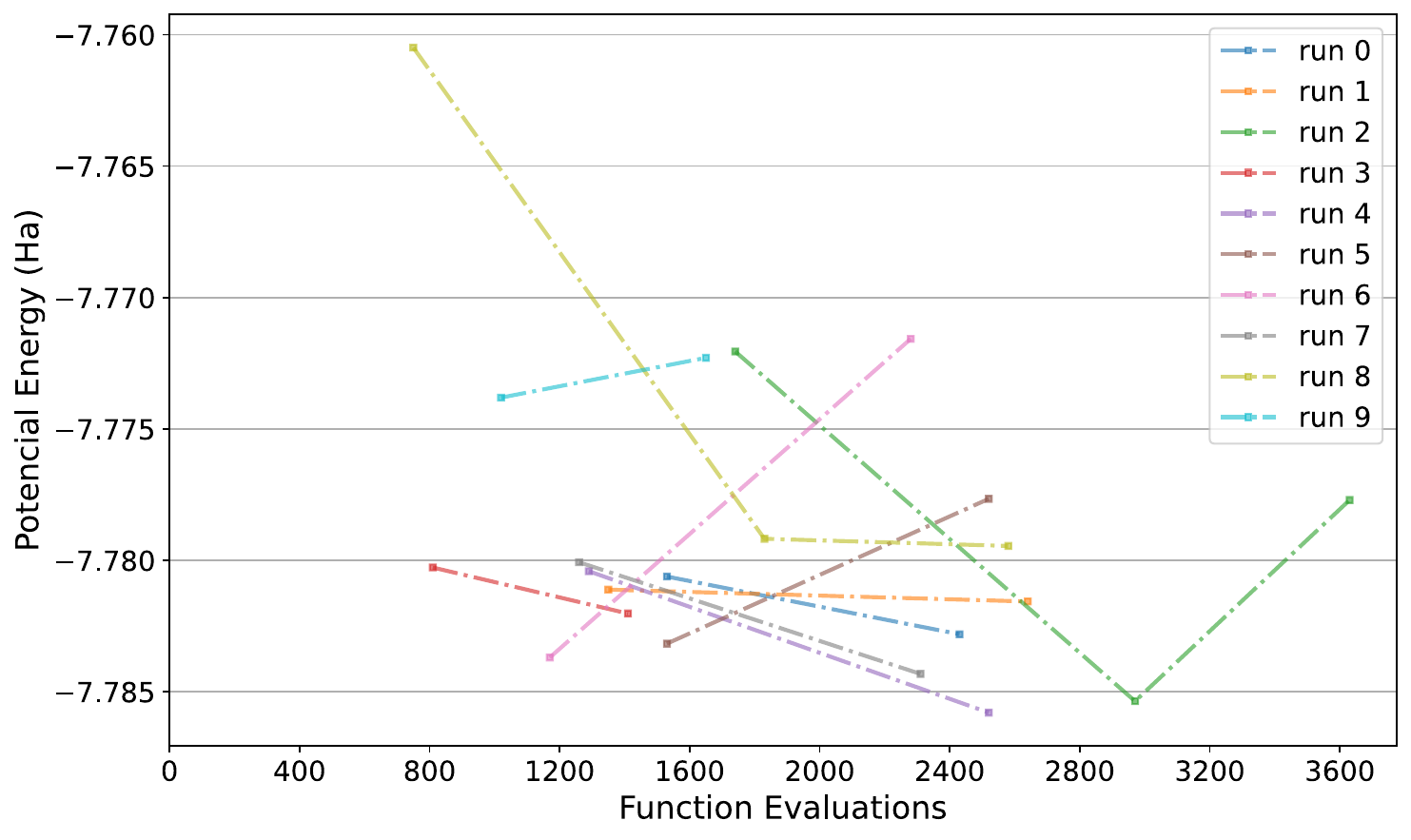}
        \caption{Energy after full SA-OO-VQE iterations vs. cumulative evaluations.} 
        \label{fig:lih_de_r2_conv_evals}
    \end{subfigure}

    \caption[Convergence plots of the DE/Rand/2/bin optimizer within the SA-OO-VQE framework for the
LiH molecule.]{Convergence analysis of the DE/Rand/2/bin optimizer within the SA-OO-VQE framework for the LiH molecule, based on 10 independent runs (shown in different colors/styles, see legend in plots). The plots display the state-average energy (Hartrees) progression viewed against different metrics: 
    (\subref{fig:lih_de_r2_conv}) Energy evaluated at the end of each internal Gradient Descent optimizer iteration, plotted against the cumulative number of function evaluations consumed up to that iteration point
    (\subref{fig:lih_de_r2_conv_iters}) Energy plotted at the end of each completed SA-OO-VQE iteration against the iteration number. 
    (\subref{fig:lih_de_r2_conv_evals}) Energy plotted at the end of each completed SA-OO-VQE iteration against the cumulative number of function evaluations consumed up to that iteration.}
\label{fig:lih_de_r2} 
\end{figure}

\begin{figure}[!htbp] 
    \centering
    \begin{subfigure}[t]{0.7\textwidth} 
        \centering
        \includegraphics[width=\linewidth]{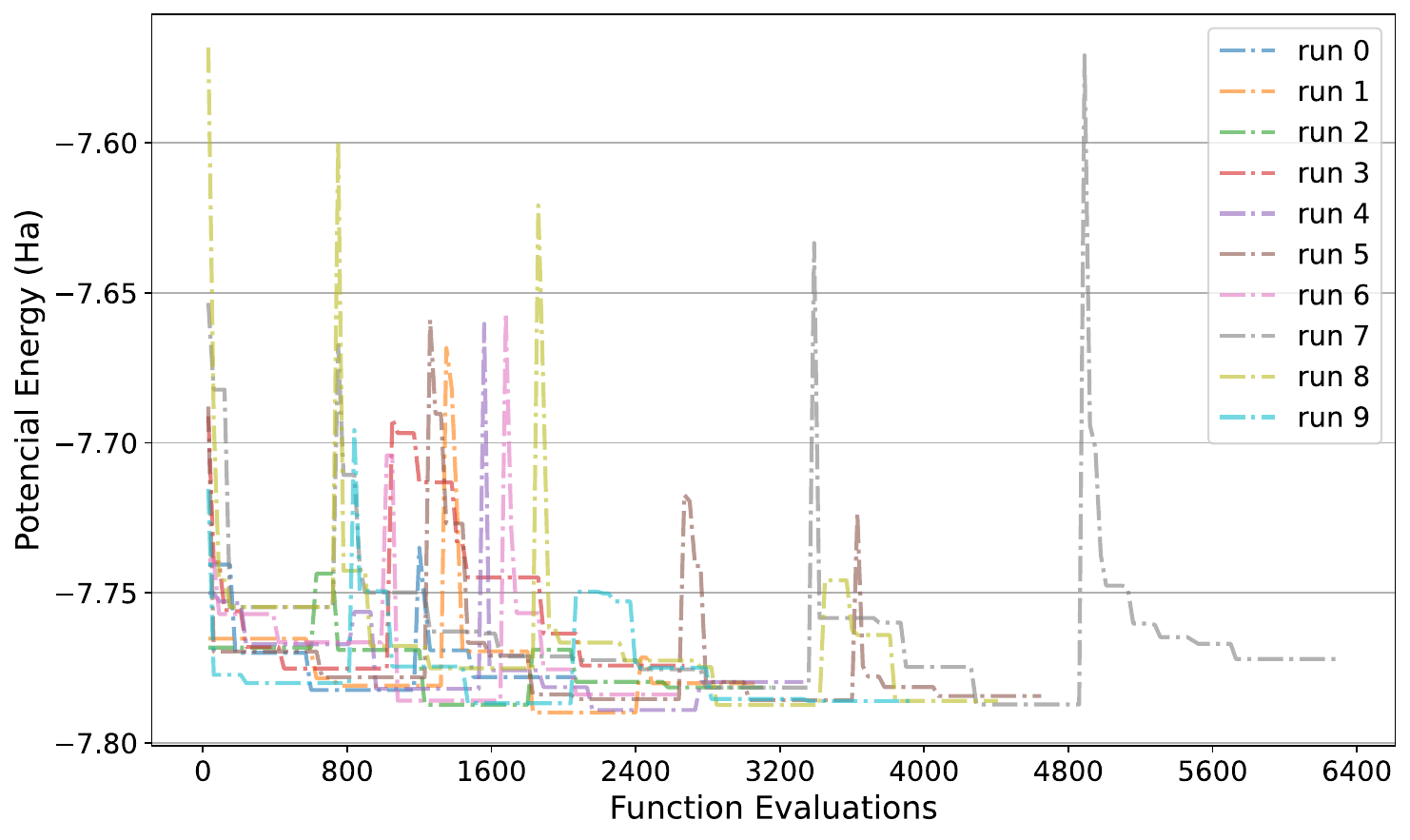}
        \caption{Energy after optimizer iterations vs. cumulative evaluations.} 
        \label{fig:lih_de_rtb_conv} 
    \end{subfigure}
    \begin{subfigure}[t]{0.49\textwidth} 
        \centering
        \includegraphics[width=\linewidth]{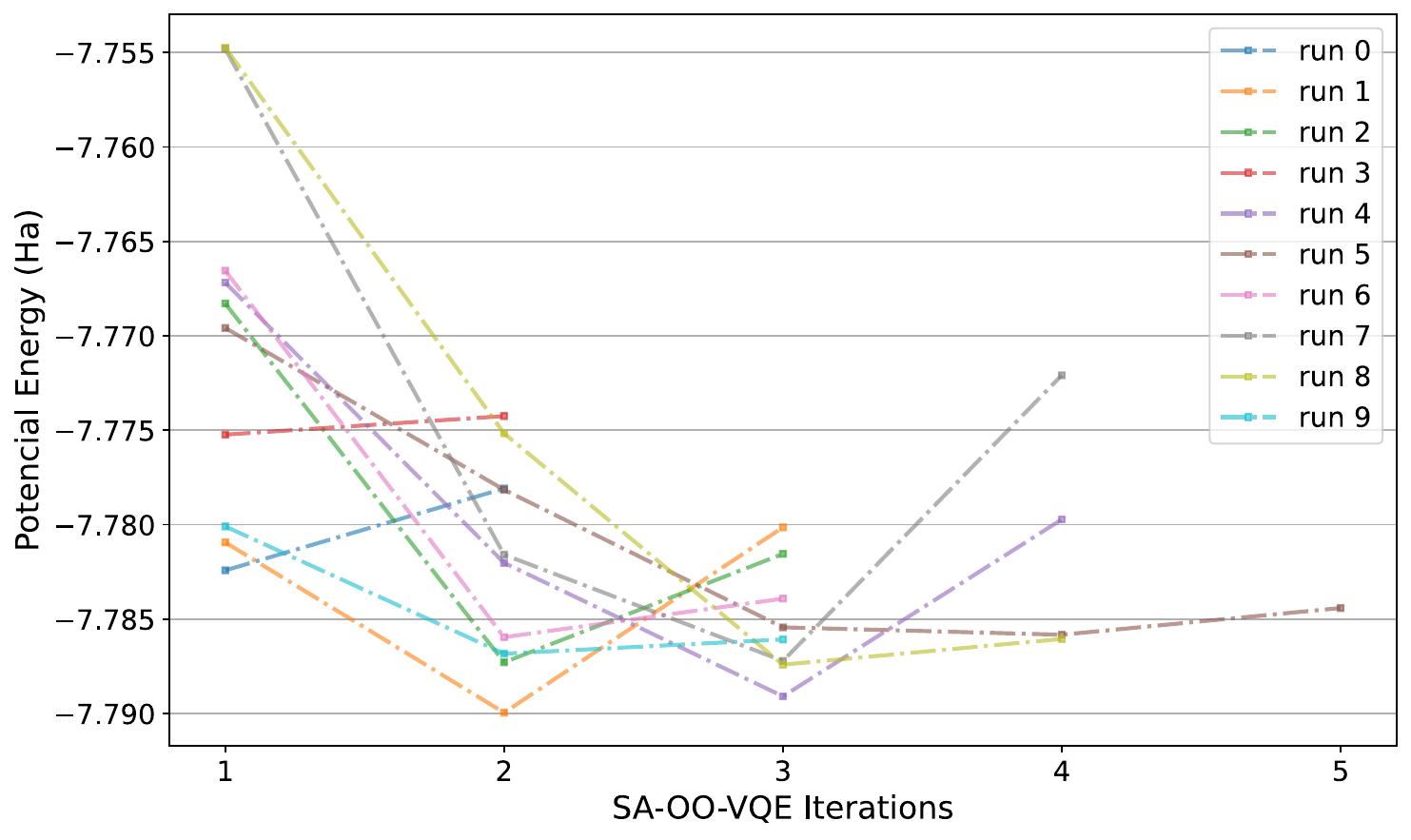}
        \caption{Energy vs. SA-OO-VQE iteration number.} 
        \label{fig:lih_de_rtb_conv_iters} 
    \end{subfigure}
    \hfill 
    \begin{subfigure}[t]{0.49\textwidth} 
        \centering
        \includegraphics[width=\linewidth]{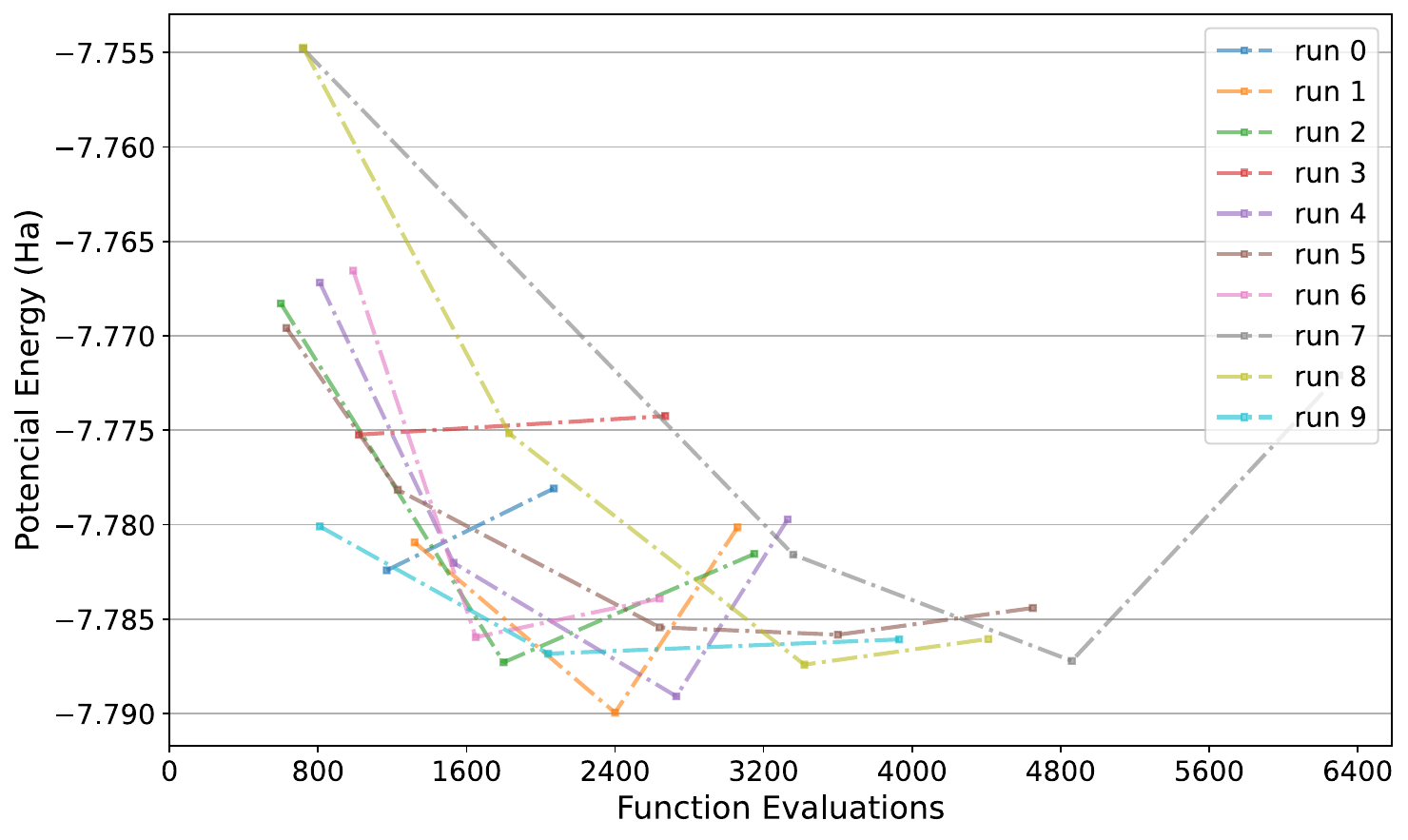}
        \caption{Energy after full SA-OO-VQE iterations vs. cumulative evaluations.} 
        \label{fig:lih_de_rtb_conv_evals}
    \end{subfigure}

    \caption[Convergence plots of the DE/Random-to-Best/1/bin optimizer within the SA-OO-VQE framework for the
LiH molecule.]{Convergence analysis of the DE/Random-to-Best/1/bin optimizer within the SA-OO-VQE framework for the LiH molecule, based on 10 independent runs (shown in different colors/styles, see legend in plots). The plots display the state-average energy (Hartrees) progression viewed against different metrics: 
    (\subref{fig:lih_de_rtb_conv}) Energy evaluated at the end of each internal Gradient Descent optimizer iteration, plotted against the cumulative number of function evaluations consumed up to that iteration point
    (\subref{fig:lih_de_rtb_conv_iters}) Energy plotted at the end of each completed SA-OO-VQE iteration against the iteration number. 
    (\subref{fig:lih_de_rtb_conv_evals}) Energy plotted at the end of each completed SA-OO-VQE iteration against the cumulative number of function evaluations consumed up to that iteration.}
\label{fig:lih_de_rtb} 
\end{figure}

\begin{figure}[t] 
    \centering
    \begin{subfigure}[t]{0.49\textwidth} 
        \centering
        \includegraphics[width=\linewidth]{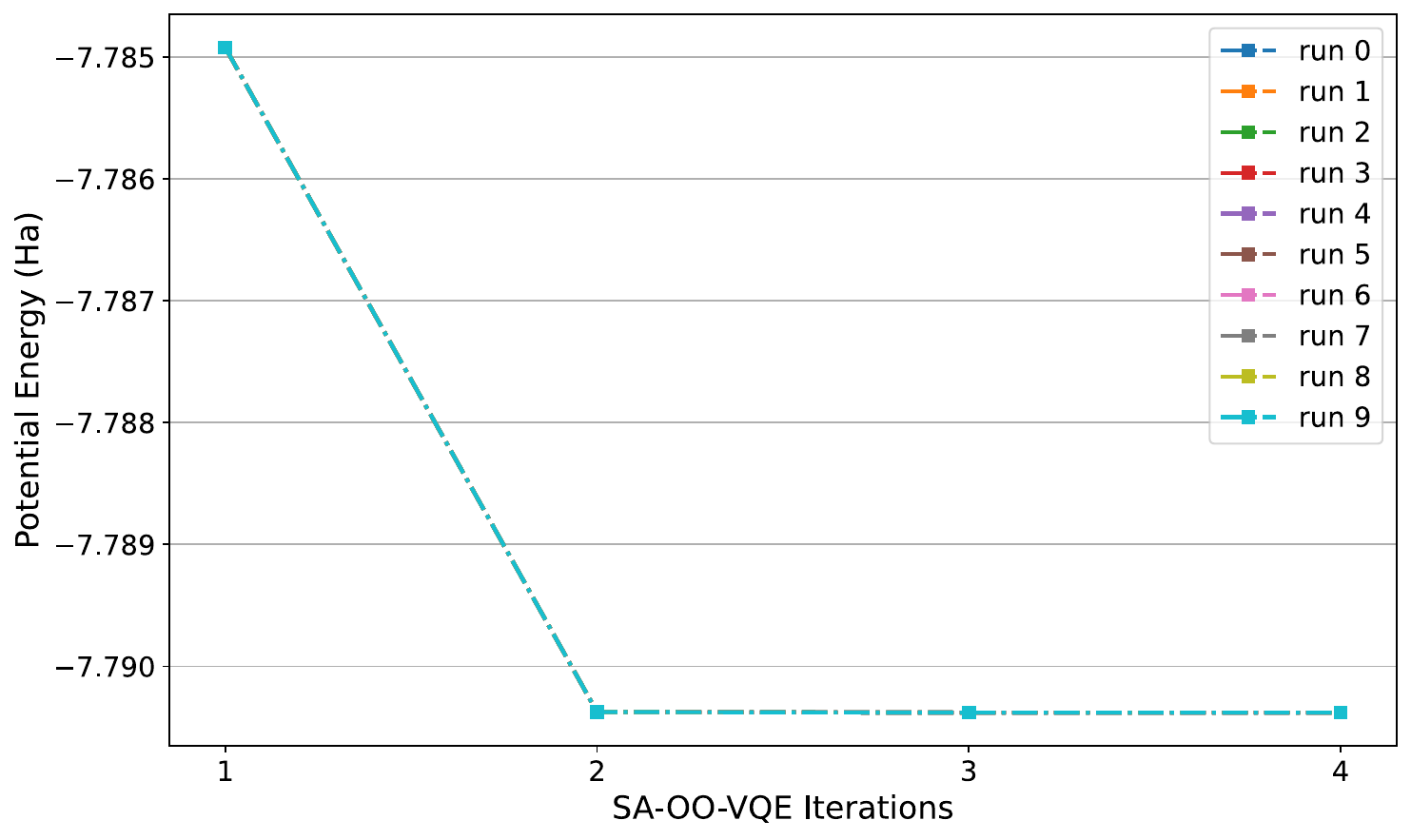}
        \caption{Energy vs. SA-OO-VQE iteration number.} 
        \label{fig:lih_slsqp_conv_iters} 
    \end{subfigure}
    \hfill 
    \begin{subfigure}[t]{0.49\textwidth} 
        \centering
        \includegraphics[width=\linewidth]{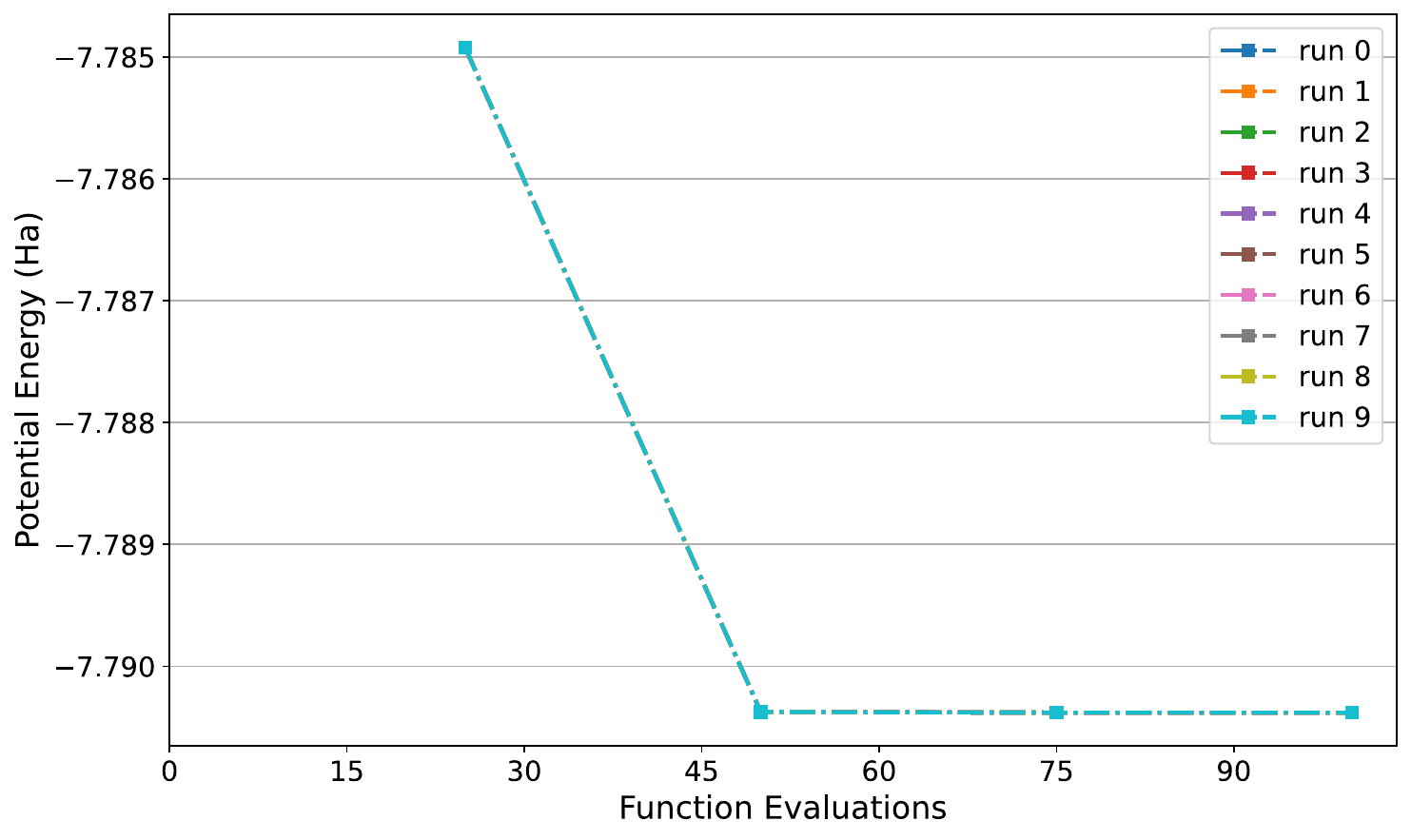}
        \caption{Energy after full SA-OO-VQE iterations vs. cumulative evaluations.} 
        \label{fig:lih_slsqp_conv_evals}
    \end{subfigure}

    \caption[Convergence plots of the SLSQP optimizer within the SA-OO-VQE framework for the
LiH molecule.]{Convergence analysis of the SLSQP optimizer within the SA-OO-VQE framework for the LiH molecule, based on 10 independent runs (shown in different colors/styles, see legend in plots). The plots display the state-average energy (Hartrees) progression viewed against different metrics: 
    (\subref{fig:lih_slsqp_conv_iters}) Energy plotted at the end of each completed SA-OO-VQE iteration against the iteration number. 
    (\subref{fig:lih_slsqp_conv_evals}) Energy plotted at the end of each completed SA-OO-VQE iteration against the cumulative number of function evaluations consumed up to that iteration.}
\label{fig:lih_slsqp} 
\end{figure}
\clearpage

\end{document}